\documentclass[oneside,12pt]{Classes/CUEDthesisPSnPDF}

\usepackage{setspace}

\usepackage{tocloft}
\ifpdf
    \pdfcatalog { /PageMode (/UseOutlines)
                  /OpenAction (fitbh)  }
\fi
\title{QCD STRUCTURE OF NUCLEAR INTERACTIONS}

\ifpdf
  \author{\href{mailto:cgran005@fiu.edu}{Carlos G. Granados}}
  \collegeordept{\href{http://www.fiu.edu/Physics}{Department of Physics}}
  \university{\href{http://www.fiu.edu}{Florida International University}}
\else
  \author{Carlos G. Granados}
  \collegeordept{PHYSICS}
  \university{FLORIDA INTERNATIONAL UNIVERSITY}
\fi

\luniversity{Florida International University}
\degree{DOCTOR OF PHILOSOPHY}
\major{PHYSICS}
\degreedate{2011}
\city{Miami, Florida}
\advisor{Professor Misak Sargsian, Major Professor} 

\usepackage{StyleFiles/watermark}
 \usepackage[ labelfont={}, margin=1cm]{caption}

\newcommand{\notocchapter}[1]{% 
    \refstepcounter{chapter}% 
    \chapter*{\thechapter \quad #1}}%   
 
\newcommand{\notocsection}[1]{% 
\normalsize\refstepcounter{section}% 
    \section*{\thesection \quad #1}}%  
    
\newcommand{\notocsubsection}[1]{% 
    \refstepcounter{subsection}% 
    \subsection*{\thesubsection \quad #1}}%     

\newcommand{\sh} {/ \hskip-7pt }
\renewcommand{\normalsize}{\msnormsizet} 
\DeclareMathSizes{13}{10}{6}{4}    
\usepackage{times}
\usepackage[toc,page]{appendix}

\def\thechapter{\Roman{chapter}}

\makeatletter
\renewcommand*\@makechapterhead[1]{%
  {\parindent \z@ \centering \reset@font
        \scshape \@chapapp{} \thechapter
    \par\nobreak
    \interlinepenalty\@M
    \bf    \vspace*{10\p@}%
      #1\par\nobreak
    \par
    \vspace*{20\p@}%
     }}
\makeatother

\makeatletter
\renewcommand\section{\@startsection{section}{1}{\z@}%
                                  {-3.5ex \@plus -1ex \@minus -.2ex}%
                                  {2.3ex \@plus.2ex}%
                                  {\normalfont\bfseries}}
\makeatother

\begin{document}

\doublespacing
{

\renewcommand\baselinestretch{1.2}

\maketitle

%set the number of sectioning levels that get number and appear in the contents

\setcounter{page}{2}
\setcounter{secnumdepth}{1}
\setcounter{tocdepth}{1}
\frontmatter

% Thesis Dedictation ---------------------------------------------------

\begin{dedication} %this creates the heading for the dedication page
DEDICATION\\
I would like to dedicate this thesis to my loving parents Aura and Campo and to my brothers Omar and Sebastian.

\end{dedication}

% ----------------------------------------------------------------------

%%% Local Variables: 
%%% mode: latex
%%% TeX-master: "../thesis"
%%% End: 

% Thesis Acknowledgements ------------------------------------------------

%\begin{acknowledgementslong} %uncommenting this line, gives a different acknowledgements heading
\begin{acknowledgements}      %this creates the heading for the acknowlegmen
\begin{center}
AKNOWLEDGMENTS
\end{center}
\vspace{-2ex}
I would like to acknowledge the contributions of the people and institutions that were crucial for the completion of this work. I thank the members of my committee for their support and advice at various stages during the writing of this dissertation. I'm particularly indebted to my adviser, Professor Misak Sargsian, for his guidance and for the resolute confidence that he has shown in my talent and in my work ethics throughout the course of my graduate program. I would also like to acknowledge the significant input that he has on the research work described in this document which resulted in the publication of three journal papers that I coauthored with him.
 
I am very grateful to the staff and administration of the Florida International University Graduate School for their assistance and for the support provided by their Doctoral Evidence Acquisition award program. I also thank  Jefferson Science Associates for their support provided through the JSA Graduate Fellowship award  which was fundamental in carrying out my research work.

Lastly, I thank the faculty, the very helpful staff and my colleagues of the FIU Physics Department for their encouragement and support that greatly enriched my years as a graduate student.

\end{acknowledgements}
%\end{acknowledgmentslong}

% ------------------------------------------------------------------------

%%% Local Variables: 
%%% mode: latex
%%% TeX-master: "../thesis"
%%% End: 

\begin{abstractseparate}
   
% Thesis Abstract -----------------------------------------------------

%\begin{abstractslong}    %uncommenting this line, gives a different abstract heading
%\begin{abstracts}        %this creates the heading for the abstract page

The research presented in this dissertation investigated selected processes involving baryons and nuclei in hard scattering reactions. These processes are characterized by the production of particles with large energies and transverse momenta. 
%This kinematic feature implies that the  particles (hadrons) participating in these reactions interact at distances much smaller than the typical hadrons' size, which put a quantitative description of these processes beyond the reach of theoretical frameworks in which the hadrons' internal structure is not taken into account. 
Through these processes, this work explored both, the constituent (quark) structure of baryons (specifically nucleons and $\Delta$-Isobars), and the mechanisms through which the interactions between these constituents ultimately control the selected reactions.  
 
 The first of such reactions is the hard nucleon-nucleon elastic scattering, which was studied here considering the quark exchange between the nucleons to be the dominant mechanism of interaction in the constituent picture. In particular, it was found that an angular asymmetry exhibited by proton-neutron elastic scattering data is explained within this framework if a quark-diquark picture dominates the nucleon's structure instead of a more traditional $SU(6)$ three quarks picture. The latter yields an asymmetry around 90$^o$ center of mass scattering with a sign opposite to what is experimentally observed.

The second process is the hard breakup by a photon of a nucleon-nucleon system in light nuclei. Proton-proton ($pp$) and proton-neutron ($pn$) breakup in $^3$He, and $\Delta\Delta$-isobars production in deuteron breakup were analyzed in the hard rescattering model (HRM), which in conjunction with the quark interchange mechanism provides a Quantum Chromodynamics (QCD) description of the reaction.  Through the HRM, cross sections for both channels in $^{3}$He photodisintegration were computed without the need of a fitting parameter. The results presented here for $pp$ breakup show excellent agreement with recent experimental data. 

In $\Delta\Delta$-isobars production in deuteron breakup, HRM angular distributions for the two $\Delta\Delta$ channels were compared to the $pn$ channel and to each other. An important prediction from this study is that the $\Delta^{++}\Delta^-$ channel consistently dominates $\Delta^+\Delta^0$, which is in contrast with models that unlike the HRM consider a $\Delta\Delta$ system in the initial state of the interaction. For such models both channels should have the same strength. These results are important in developing a QCD description of the atomic nucleus.

%\end{abstracts}
%\end{abstractlongs}

% ----------------------------------------------------------------------

%%% Local Variables: 
%%% mode: latex
%%% TeX-master: "../thesis"
%%% End: 

 \end{abstractseparate}
\clearpage 
\renewcommand{\cftbeforetoctitleskip}{0.0cm}
\renewcommand{\cfttoctitlefont}{\hspace*{2.25in}} 
\renewcommand{\cftaftertoctitle}{%
\\[\baselineskip]\mbox{}{\normalfont CHAPTER}\hfill{\normalfont PAGE}\vspace{-2ex}}
\renewcommand{\cftaftertoctitleskip}{\setlength{2ex}}
\renewcommand{\contentsname} {\normalsize TABLE OF CONTENTS}
\renewcommand{\cftchapaftersnum}{.}
\renewcommand{\cftchapnumwidth}{\setlength{0.5in}}
\renewcommand{\cftchapfont}{\normalfont}
\renewcommand{\cftbeforechapskip}{\setlength{2ex}}
\renewcommand{\cftchappagefont}{\normalfont}
\begin{singlespace}
\tableofcontents
\pagebreak
\renewcommand{\cftbeforeloftitleskip}{0.0cm}
\renewcommand{\cftloftitlefont}{\hspace*{2.4in}} 
\renewcommand{\cftafterloftitle}{%
\\[\baselineskip]{\normalfont FIGURE}\hfill{\normalfont PAGE}\vspace{-2ex}}

\renewcommand{\cftafterloftitleskip}{\setlength{-2ex}}
\renewcommand{\listfigurename}{\normalsize LIST OF FIGURES}
\renewcommand{\cftbeforefigskip}{\setlength{2ex}}
\listoffigures
\pagebreak

\renewcommand{\cftbeforelottitleskip}{0.0cm}
\renewcommand{\cftlottitlefont}{\hspace*{2.55in}\Large} 
\renewcommand{\cftafterlottitle}{%
\\[\baselineskip]\mbox{}{\normalfont TABLE}\hfill{\normalfont PAGE}\vspace{-2ex}}
\renewcommand{\cftafterlottitleskip}{\setlength{-2ex}}\renewcommand{\listtablename}{\normalsize LIST OF TABLES}
\renewcommand{\cftbeforetabskip}{\setlength{2ex}}
\listoftables
\end{singlespace}

\mainmatter

 %%% Thesis Introduction --------------------------------------------------
\chapter{INTRODUCTION}
\label{Intro}
\ifpdf
    \graphicspath{{Introduction/IntroductionFigs/PNG/}{Introduction/IntroductionFigs/PDF/}{Introduction/IntroductionFigs/}}
\else
    \graphicspath{{Introduction/IntroductionFigs/EPS/}{Introduction/IntroductionFigs/}}
\fi
 This introduction presents an overview of the progress made in nuclear physics towards achieving an unified description of the strong nuclear force; a description that can match the phenomenological successes of the standard model of electroweak interactions. In line with the specific goals of the studies presented in this dissertation, the  following sections place emphasis on the search for an elementary particle description of nucleon nucleon, and more generally hadron hadron interactions. Section \ref{TSI} addresses the discovery of the nucleus and of its constituents (nucleons) the proton and the neutron as well as the concerns that led to the discovery of the nuclear force. Section \ref{FHQCD} describes the road towards formulating a field theory of the strong nuclear force motivated by the quantum electrodynamics (QED) description of electromagnetic interactions. It also outlines the experimental and phenomenological work that lead to the discovery of quarks and gluons and to the formulation of quantum chromodynamics (QCD) as the fundamental field theory of the strong interaction. Section \ref{HEP} focuses on general features of hard exclusive processes, and on how the results from QCD phenomenology explored in  section \ref{FHQCD} can be used here to investigate the QCD mechanisms through which the particles involved in these processes interact. This section introduces the framework and some aspects of the methodology in which the processes of interest in this dissertation are studied in the chapters that follow. 

\section{The Strong Interaction}
\label{TSI} 
The discovery of subatomic particles in the late XIX and early XX century in conjunction with the formulation of special relativity and Quantum mechanics steered major efforts and developments in physics towards identifying the fundamental structure of matter. These developments resulted in the formulation of the Standard Model (SM) of particle physics, which has met outstanding phenomenological success in describing electromagnetic and weak interactions. All the fundamental elementary particles (fermions) are subject to either of these two interactions;  more generally, they are subject to the electroweak interaction which unifies the two interactions under a consistent mathematical framework. The validation of the SM has been achieved through the experimental programs that test phenomenological predictions emanating from this framework. Such tests are facilitated by major advances in accelerator technologies that allow studies at ever increasing energy regimes and much improved detection and particle identification capabilities. These conditions have made  possible the production and direct detection of leptons and of the otherwise hypothetical weak bosons which mediate the weak interaction and that are predicted by the standard model.

 On the other hand, the standard model picture of the strong interaction faces a more challenging phenomenological treatment. In the SM, the strong interaction is described by quantum chromodynamics (QCD), a quantum field theory in which quarks and gluons are respectively the elementary fermions and bosons. But these quarks and gluons are confined in hadrons (a family of particles of which protons and neutrons are part of) and have not been directly observed as free particles. However, as is discussed in the following sections, the existence of the strong force was postulated to explain the stability of the nucleus as a system of protons and neutrons (nucleons); thus in a first approach, a theory of the strong interaction would have nucleons as its fundamental elementary particles to later on include mesons as the elementary bosons. The origins of this description are summarized below  and in  section \ref{FHQCD}, in which it's also emphasized that its validity is limited to separation distances at which the interacting nucleons can still be considered elementary particles. At separation distances smaller than 1fm (10$^{-15}$m), the nucleons' internal structures (quark-gluon distributions) play a more active role in the interaction, and approaches assuming elementary (structureless) nucleons are no longer suited for describing phenomena at this scale. 
 
 After introducing QCD, the theory of interactions of quarks and gluons,  in section \ref{FHQCD}, section \ref{HEP} comments on general features that allow the use of QCD approaches for developing a quantitative description of the strong nuclear force at separation distances in this regime ($<$1fm). In the chapters that follow, the reactions of interest in this dissertation are investigated within the framework of QCD approaches.

\subsection*{The Atomic Nucleus}
\label{TANc}

The atomic nucleus concentrates most of the atomic mass in a radius about 10$^6$ times smaller than the atomic radius. It also carries the net positive charge that balances the electronic cloud surrounding it in a neutral atom. This accepted picture of the atom was partially completed by Rutherford in 1911 \cite{Rutherford:1911zz} when results from a scattering experiment   of alpha particles incident upon a thin film of gold  showed that some small, but larger than expected, portion of the beam was deflected at large angles while most of the beam passed through the film with almost no deflection \cite{geiger:1909}. Rutherford showed that under a coulomb interaction, the observed deflection of alpha particles at large angles could only be explained by a large charge and mass concentration in a very small volume at the center of each gold atom. The picture of the atom then switched from the one in which the positive charge was uniformly distributed throughout the atom (J. Thompson's model) and according to which the large angle deflection of the alpha beam would had been much more suppressed, to the one in which the mass and the net positive charge were concentrated in a  small volume (nucleus) surrounded by orbiting electrons that expand to the atom's size \cite{Rutherford:1911zz}.

This `planetary' model of the atom was still plagued with flaws. The most important being the fact that the atom's stability under this model would contradict classical electrodynamics. The orbiting electrons are accelerating and therefore should radiate electromagnetic waves losing kinetic energy to their electromagnetic field and therefore decaying by spiraling down towards the nucleus under the Coulomb force. This issue was then solved by the emerging quantum mechanics picture of the atom, in which the orbital structure of the electronic distribution was replaced by discrete quantum states of the electronic field surrounding the nucleus. An atom could only radiate or absorb energy when transitioning between these states, while the lowest energy state has still a definite finite energy.

\subsection*{The Strong Nuclear Force}
\label{TSNF}
Concerns about the composition of the nucleus were still to be addressed as well. One of the early models of the nucleus in which it consisted of protons and electrons has stability issues. The huge repulsing force that  positive charges would experience at the nuclear distances could not be balanced by the overall charge distribution of the atom. The nucleus would quickly disintegrate under the Coulomb repulsion. Therefore, a force of a different nature was thought to be responsible for keeping the nucleus' components in place. This nuclear force should be much stronger than the electromagnetic force at nuclear distances, but it should have a range limited to the nuclear distances as well since the evidence of this force is not found anywhere else outside the nucleus.

\subsection*{Nuclear Composition} 
\label{NuCo}

A lot of progress was made in understanding the nuclear structure of atoms by assuming the existence of this force, but without a detailed theory of its origin the successes were only confined to some general features. This picture was much improved by the discovery of the neutron by James Chadwick in 1932~\cite{Chadwick:1932ma}. The nucleus was now thought of as a system of nucleons (protons (positive charge carriers) and neutrons (of neutral charge)) confined by an effective nuclear potential. Both neutrons and protons are the sources of the force field generated by this potential which at the nuclear scale is attractive. Because neutrons are not meaningful sources of an electromagnetic field, their presence  in nuclei helped explain the stability trend of the atoms  with increasing atomic number. As the atomic number (the number of protons in the nucleus) increases, the ratio of the number of neutrons to protons  has to increase as well in order to balance the increasing electromagnetic repulsion felt by the protons with an increasing number of attractive nuclear force sources. 
The short range of the nuclear force will also constrain the size of the stable nuclei. 

The existence of neutrons in nuclei also solved the contradiction regarding the observed spin of nuclei inherent to the early (proton-electron) nuclear models. Under these models the $^{14}$N nucleus was made of 14 protons and 7 electrons, both particles with total spin 1/2. Their corresponding spins will add up to a system with a total half integer spin. However, it was observed that $^{14}$N obeys Bose statistics which implies that its total spin was of integer value. Because, like protons and electrons, neutrons are fermions (particles with half integer spin ), the $^{14}$N nucleus made up of 7 protons and 7 neutrons  would naturally account  for expected total integer spin of the $^{14}$N nucleus.

Just as for the atom, a quantum mechanics picture of the nucleus emerged in which discrete nuclear states correspond to the different configurations of protons and neutrons filling different nuclear energy levels. Transitions between nuclear states could involve energy absorption or energy emission in the form of radiation. A nucleus with  fixed atomic and mass numbers would have a unique photon emission spectrum coming from the decays of its allowed exited states just as an atom of a given element would. A decay to a different nuclear state in which the atomic and/or the mass numbers are altered is signaled by the emission of massive charged or neutral particles such as neutrons, alpha particles and beta particles. At this stage, it was understood that the two (and more) body nucleon-nucleon interactions had to be the source of the nuclear potentials which makes a nucleus a bound system.

\subsection*{Force Carriers}
\label{FoCa}

A picture of nucleon-nucleon (NN) interaction was however yet to be reached to the level at which for instance the electromagnetic interaction was then understood. Through quantum electrodynamics (QED), the relativistic quantum field theory of electromagnetism, charged particles interact by transferring energy and momentum to one another  through the exchange of field quanta (photons). In 1935, H. Yukawa proposed that just as in electromagnetism the Coulomb force was the result of the exchange of virtual photons, the strong nuclear force as well results from the exchange of corresponding virtual bosons (mesons)\cite{Yukawa:1935xg}. The short range nature of this strong force comes from these mesons having  nonzero masses. While in the non-relativistic limit the photon exchange interaction among charged particles can be effectively described by the classic form of the Coulomb potential ($\sim$1/r), the strong interaction mediated by massive mesons is in the nonrelativistic limit  dominated by the Yukawa potential ($e^{-\mu r}/r$), with $\mu$ being the mass of the exchanged meson.   Therefore, the mass of the exchange meson would define the range of the force. A candidate for the particle mediating the nucleon nucleon interaction was found in 1947. Known as the $\pi$ meson or pion, it has zero spin (scalar) and could have a charge of +e ($\pi^+$), -e($\pi^-$) or no charge ($\pi^0$) with a mass $\mu\sim$140MeV. Later on, other heavier mesons of spin 1 (vector mesons) $\rho$ and $\omega$ were discovered \cite{Erwin:1961ny}\cite{Alston:1961nx}. Contribution from the exchange of these vector mesons to the (NN) interaction are thought to be relevant in understanding the behavior of the (NN) force at distances far smaller than those were this force is acceptably described by either the pion exchange model or the Yukawa potential \footnote{See e.g. Ref.\cite{Machleidt:1989tm} for an extended overview of the meson theory of nuclear interactions.}. However, as the distance between the interacting nucleons keeps getting smaller, the particle exchange mechanisms required to explain the empirical behavior of the strong force grow in number and complexity. 

\section{From Hadrons to QCD}
\label{FHQCD}
The meson field theory for the strong interaction was  motivated by the phenomenological success of QED. This theory tries to include all possible interactions between nucleons in such a way that preserves the increasing number of global symmetries. Evidence of these symmetries\footnote{Each symmetry of the strong interaction was inferred from experimental evidence of certain conservation laws, some of which included the conservation of isospin and strangeness.}  
  arose from the discovery not only of new mesons, but also of new fermions that seem to interact via the strong interaction. The latter  was concluded from the observed high rates ( half lives of  $\sim10^{-24}$s) at which these fermions were decaying into nucleons that could only be explained by strong interaction couplings. These fermions along with the lighter and more stable nucleons came to form a new category of particles named baryons. Mesons and baryons form the group of particles that can interact through the strong interaction also known as hadrons. While baryons played a role analogous to that of fermions in QED, mesons as mentioned earlier assumed a role analogous to that of photons, as the carriers of the nuclear force.

\subsubsection{Isospin}
\label{IsoS} 
The nuclear force for instance contributes with the same strength in both proton proton ($pp$) and neutron neutron ($nn$) interaction. This experimental fact suggests that under the strong interaction protons and neutrons are different states of the same particle. In analogy with the spin formalism, this set of two states is labeled by a quantum number called isospin ($I$). Each state is in this set is labeled by another quantum number ($T$) analogous to the spin projection ($m$). Each state then can be transformed into the other by a rotation in this internal 'isospace'.  By convention, for a proton (neutron) $I=1/2$ and $T=+1/2$($-1/2$). Then a $pp$ system is turned into a $nn$ system  through a rotation in the isospace. As mentioned earlier the strength of the force remains the same after this rotation is performed. Therefore, the strong interaction is symmetric under rotations in the isospace. %This is consistent with Noether's theorem which associates a symmetry with a conservation law, because  in strong reaction processes total isospin and total isospin projections are conserved quantities. 
For now, the known mesons, $\pi,\rho$ and $\omega$ have integer isospins, and the isospin symmetry of the strong interaction  holds on reactions involving them as well. Isospin conservation puts specific restrictions on the structure of the $NN$ interactions through meson exchange. 

\subsubsection{Meson field theoretic approach}
\label{MFTA}
The isospin invariance of the strong interaction is utilized to obtain an appropriate representation of the elements of a meson exchange theory of the strong interaction. This invariance is also taken into account when writing the equations that govern the general dynamics under this interaction. In this framework, protons and neutrons are represented by Pauli spinors from $SU(2)_I$ fundamental representation, while mesons (the field quanta) belong to the triplet representation of $SU(2)$ (see e.g. Refs. \cite{Machleidt:1989tm} and \cite{bjorken:1964}):
\begin{eqnarray}
\Psi_N=\left(\begin{array}{c}
                  \psi_p\\
                  \psi_n
                  \end{array}\right),
\end{eqnarray}
and
\begin{eqnarray}
\Phi_\pi=\left(\begin{array}{c}
                  {\phi_{\pi^+}+\phi_{\pi^-}\over\sqrt2}\\
                  i{\phi_{\pi^+}-\phi_{\pi^-}\over\sqrt2}\\
                  \phi_{\pi^0}
                  \end{array}\right),
\end{eqnarray}
which transform under an infinitesimal $SU(2)_I$ rotation (parametrized by an isovector $\epsilon$) following,
\begin{eqnarray}
\Psi_N\to (1-i\epsilon\cdot\tau/2)\Psi_N\\
\Phi_\pi\to \Phi_\pi-\epsilon\times\Phi_\pi.
\label{SU2T}
\end{eqnarray} 
$\tau$ is a vector formed by Pauli matrices (the generators of transformations in $SU(2)$).

For a nucleon $\Psi_N$ in a pion field $\Phi_\pi$,

\begin{eqnarray}
(i\sh{\nabla}-m_N)\Psi_N(x)=g_0i\gamma_5(\tau\cdot\Phi_\pi(x))\Psi_N(x), 
\label{IsDeq}
\end{eqnarray}
which corresponds to the Dirac equation modified by a canonical transformation (``minimal substitution'') of the momentum operator, $i\sh{\nabla}\to i\sh{\nabla}-g_0i\gamma_5(\tau\cdot\Phi_\pi(x))$, in analogy to what is done for a charged particle in a electromagnetic field to account for interactions.  $\gamma_5$ ensures parity conservation, and $g_o$ is a coupling constant to be determined experimentally. Similarly, as in electromagnetism in the presence of a charged current $J^{em}$ the electromagnetic field $A(x)$ obeys $\nabla^2A(x)=J^{em}(x)$, the pion field in the strong interaction obeys,
\begin{eqnarray}
(\nabla^2-m^2_\pi)\Phi_\pi(x)=-g_0{\bar\Psi_N(x)}i\gamma_5\tau\Psi_N(x),
\label{IsKGeq}
\end{eqnarray} 
in which ${\bar\Psi_N(x)}i\gamma_5\tau\Psi_N(x)$ is identified as the isotopic nucleon current. Eqs. (\ref{IsDeq}) and (\ref{IsKGeq}) are covariant under $SU(2)_I$. The homogeneous versions of these equations correspond to the Dirac and Klein-Gordon equations for a free nucleon and pion field respectively. In a more formal treatment Eqs. (\ref{IsDeq}) and (\ref{IsKGeq}) can be obtained after minimizing the action under the Lagrangian density

\begin{eqnarray}
{\cal L}_s={1\over2}\left[(\partial_\mu\Phi_\pi)(\partial^\mu\Phi_\pi)-m^2\Phi_\pi^2\right] + \left[i{\bar\Psi}_N\sh{\nabla}\Psi_N-m{\bar\Psi}_N\Psi_N\right]-ig_0{\bar\Psi}_N\gamma^5\tau\cdot\Phi_\pi\Psi_N,\nonumber\\
\label{isoL}
\end{eqnarray}
i.e., from
$$\delta S=\delta\int d^4x {\cal L}_s(\Psi_N,\partial_\mu\Psi_N,{\bar\Psi_N},\Phi_\pi,\partial_\mu\Phi_\pi)=0,$$
which yields the Euler-Lagrange Equations
$$\partial_\mu{\partial{\cal L}_s\over\partial \partial_\mu{\bar\Psi_N}}-{\partial{\cal L}_s\over\partial{\bar\Psi_N}}=0,$$
resulting in Eq.\ref{IsDeq}, and
$$\partial_\mu{\partial{\cal L}_s\over\partial \partial_\mu{\Phi_\pi}}-{\partial{\cal L}_s\over\partial{\Phi_\pi}}=0,$$
resulting in Eq.\ref{IsKGeq}.

The Lagrangian density in the form of Eq.(\ref{isoL}) is invariant under the transformations in Eq.(\ref{SU2T}). This symmetry in conjunction with Euler-Lagrange equations leads to
$$\partial_\mu\left({\partial{\cal L}_s\over\partial\partial_\mu\Psi_N}\left(-i{\tau\over2}\right)\Psi_N-{\partial{\cal L}_s\over\partial\partial_\mu\Phi_\pi}\times\Phi_\pi\right)=0$$
in which the quantity in parentheses is identified as the conserved current $J^\mu$ corresponding to the $SU(2)_I$ symmetry. Explicitly, from Eq.(\ref{isoL}),
\begin{eqnarray}
J_\mu=1/2{\bar\Psi_N}\gamma_\mu\tau\Psi_N+\left(\Phi_\pi\times\partial_\mu\Phi_\pi\right).
\label{isoJ}
\end{eqnarray}
From $J^0$ we then obtain the  three components of isotopic charge of the system,
\begin{eqnarray}
I=\int d^3x J^0=\int d^3x\left[1/2\Psi_N^\dagger\tau\Psi_N+(\Phi_\pi\times\partial_0\Phi_\pi)\right],
\label{isoQ}
\end{eqnarray}
which is a constant of motion.  Because the Lagrangian density in Eq.(\ref{isoL}) should satisfy the conservation of the electric charge, it should be also symmetric under $U(1)$ transformations on electrically charged fields,
\begin{eqnarray}
\Psi_N\to\Psi-i|\epsilon|\left({1+\tau_3\over2}\right)\Psi\\
\Phi_\pi\to\Psi_\pi+\epsilon\times\Phi_\pi.
\end{eqnarray}
It can be shown that from this symmetry it follows that the electrical charge of the system $Q$ is a constant of motion as well. Likewise, the baryonic number $B$, $$B=2(Q-I_3)$$  is a constant of motion  in interactions under ${\cal L}_s$ in Eq.(\ref{isoL}).
\paragraph{Scattering matrix}
The fact that the conservation of $B$, $I$, and $Q$ charges in strong interactions is well established experimentally drew efforts into developing a field theory of pions and nucleons based on a Lagrangian such as Eq.(\ref{isoL}) and on the observed symmetries. The free parameters of such a theory  according to Eq.(\ref{isoL}) would in principle be the nucleon and the pion masses, and the pion nucleon coupling constant $g$ which are experimental observables. A quantitative analysis is then developed to evaluate the phenomenological accuracy of ${\cal L}_s$.  This is usually done by studying scattering processes for which the cross section can be calculated from entries of a scattering matrix $S_{if}$.

The scattering matrix is the probability amplitude of a system (generally a multiparticle system)in an initial state in which there's no interaction to evolve through interaction into a final state also away from the region of interaction, i.e.,
\begin{eqnarray}
S_{if}=Lim_{T\to\infty}\left\langle\phi_f,T|\phi_i,-T\right\rangle, 
\label{Smtrx}
\end{eqnarray}
in which $\phi_i$ and $\phi_f$ are solutions of equations such as the homogeneous versions of Eqs. (\ref{IsDeq}) and (\ref{IsKGeq}). In the Heisenberg picture (ses e.g. Ref~\cite{Peskin:1995ev}) Eq.(\ref{Smtrx}) can be written making explicit the role of a time evolution operator:
\begin{eqnarray}
S_{if}=Lim_{T\to\infty}\left\langle\phi_f|T\left[e^{-iH(2T)}\right]|\phi_i\right\rangle, 
\label{Smtrx1}
\end{eqnarray}
in which the Hamiltonian $H$ is given as a function of the field operators $\phi(x)$ and $\pi(x)=\partial_0\phi(x)$ and $T[...]$ stands for the time ordering of the product of operators in [...], e.g.,
$$T[\phi(x)\phi(y)]=\theta(x_0-y_0)\phi(x)\phi(y)+\theta(y_0-x_0)\phi(y)\phi(x)$$
These products come from the expansion  of $e^{-iH({\phi})t}$. 
Because the initial and final states in an scattering reaction are generally different, the nontrivial entries of the $S$ matrix  become the entries of a transition matrix ${\cal T }$ which relate to $S$ through
$$S=1+i{\cal T}.$$ An invariant matrix element ${\cal M}$ is then defined by removing the condition of 4-momentum conservation, i.e.,
$$
\langle \{p_f\} |i{\cal T}|\{p_i\}\rangle=2\pi^4\delta(\sum p_i-\sum p_f)i{\cal M}(\{p_i\}\to\{p_f\}).
$$ 
If the interaction is weak, the invariant matrix elements can be computed from the lowest order terms of an expansion of  Eq.(\ref{Smtrx1}). The terms of such expansion can be graphically represented by Feynman Diagrams such as the one shown in Fig.\ref{fig:piex}. The vertex factors are determined from the interaction terms of the Lagrangian or Hamiltonian, while the propagators or internal lines are obtained from the equations of motion derived from the free Lagrangian. e.g., from
$$
(q^2-m^2)G_\pi^{ij}(x,x')=\delta(x-x')\delta^{ij}
$$
we obtain the pion  propagator,
\begin{eqnarray}
G_\pi^{ij}(x,x')=\int {d^4q\over(2\pi)^4} e^{-iq(x-x')}G_\pi^{ij}(q),
\label{pprop1}
\end{eqnarray}
in which 
\begin{eqnarray}
G_\pi^{ij}(q)={i\delta^{ij}\over q^2-m_\pi^2}
\label{pprop2}
\end{eqnarray}
enters as a factor in calculating ${\cal M}$ in  momentum space, as illustrated in Fig.\ref{fig:piex}.
External lines emerge from taking the $T\to\infty$ limit in defining the scattering matrix as in Eq.(\ref{Smtrx1}). The corresponding factors for external lines are corresponding representations in momentum space of solutions for the free equations of motion.

\subparagraph{Feynman Diagrams}

\begin{figure}[ht]
	\centering
		\includegraphics[width=0.5\textwidth]{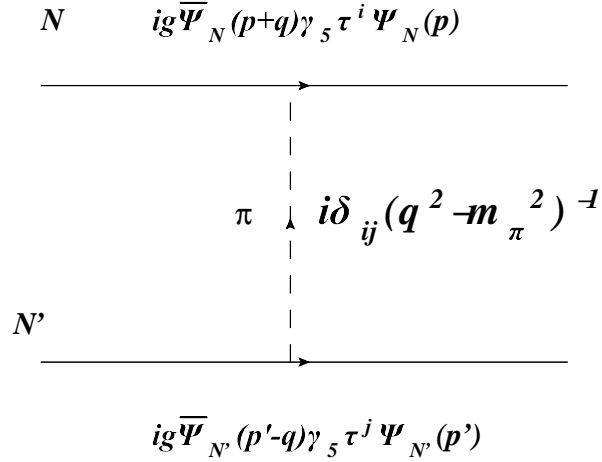}
	\caption{Nucleon Nucleon scattering through the one pion exchange mechanism.}
	\label{fig:piex}
\end{figure}

The matrix elements ${\cal M}$ for $NN$ scattering through  the pion exchange mechanism are then computed based on an expansion in which the lowest order term in $g$ (of second order) is illustrated in Fig.(\ref{fig:piex}). This term corresponds to the one pion  exchange (OPE) mechanism which is assumed in OPE models to sufficiently well describe the interaction at large distances (r$>$2fm). From Fig.(\ref{fig:piex}), ${\cal M}$ for $NN$ scattering within OPE can be expressed as,
\begin{eqnarray}
-i{\cal M_{OPE}}=ig_0{\bar\Psi_N(p+q)}\gamma_5\tau^i\Psi_N(p){i\delta^{ij}\over q^2-m_\pi^2}ig_0{\bar\Psi_N'(p'-q)}\gamma_5\tau^i\Psi_N'(p'),
\end{eqnarray}
in which ${i\delta^{ij}\over q^2-m_\pi^2}$ is the pion's propagator according to Eq. (\ref{pprop1}). 

It is possible to obtain the interaction potential from Feynman diagrams in the nonrelativistic limit. For instance, a radial potential can be derived from an angular integration of the Fourier transform of the s-wave component of this scattering amplitude. It yields (see e.g. \cite{bjorken:1964}),
\begin{eqnarray}
V_s(r)=-2f^2\left[{e^{m_\pi r}\over r} -{4\pi\over m^2_\pi}\delta^3(r)\right],
\label{opep}
\end{eqnarray}
in which $f^2={g_0^2\over4\pi}\left(m_\pi\over2M\right)^2$. The first term in Eq.(\ref{opep}) corresponds to the Yukawa potential which properly describes the long-range attraction. A short-range repulsion is accounted for by the Dirac delta term in Eq.(\ref{opep}). However, because of its nonrelativistic character, Eq.(\ref{opep}) is only accurate for distances at which the Yukawa potential is accurate. In such a region, the parameter $f$ and consequently $g_0$ can be set. It is found that experimentally ${g\over4\pi}^2\sim14$, i.e., the $\pi N$ coupling constant is much larger than 1. The use of a traditional expansion of $\cal M$ in orders of $g_0$ is not well justified in this case; the diagrams of higher order in $g_0$ higher are not necessarily suppressed. Such diagrams involve multiple pion exchanges and become more significant at smaller distances with a contribution to the $NN$ potential behaving as $\sim e^{-nm_\pi r}$ in which $n$ corresponds to the number of pions exchanged.

Calculations for diagrams with two or more pion exchanges are considerably more tedious and complicated and did not correlate well with experimental data.  Alternatively, it was argued that multiple uncorrelated pion exchanges contribute little to the the $NN$ force, and that instead additional heavier mesons in place of correlations between the exchanging pions would explain the $NN$ interaction at smaller distances. Particularly, the short range repulsion and a spin orbit force arise naturally by considering one vector-meson exchange diagrams. The $\rho$ and $\omega$  vector mesons were discovered in the early 60s as resonances of 2$\pi$ and 3$\pi$ states respectively. Both mesons have spin 1 and isospins 1 and 0 respectively. Accordingly, their couplings to nucleon currents are of the form $${\hat\Psi}_N\left(-g_v\gamma^{\mu}\phi^{(v)}_\mu - {f_v\over4M}\sigma^{\mu\nu}(\partial_\mu \phi^{(v)}_\nu-\partial_\nu\phi^{(v)}_\mu)\right)\Psi_N,$$ which are added to the Lagrangian and/or Hamiltonian as interaction terms along with the corresponding kinetic energy terms.

The strong attraction at the mid range of the $NN$ force however requires the inclusion of yet another kind of meson. It was named $\sigma$ meson and should have 0 spin and 0 isospin, and a mass in the range of 400-800 MeV. The existence of such a meson has not yet been established at least as a field particle in meson field theories. However, many other mesons were discovered increasing the density of degrees of freedom of the theory and consequently the number of interaction terms in the Lagrangian. Baryon resonances of nucleons and mesons such as $\Delta$-isobars needed to be included as well which contributed largely to the previously neglected uncorrelated multipion exchange diagrams. The number of free parameters such as coupling constants grew dramatically and soon the Isospin invariance together with the more fundamental symmetries (parity and Lorentz invariance) did not seem to sufficiently constrain the Lagrangian of the theory to a limited form desired for a field theory of fundamental degrees of freedom such as in QED (that has only one coupling constant).  

\paragraph{Renormalization and the Landau pole}
The large number of free parameters meant that the developing theory lacked predictive power.  Specially in the short range in which contributions other than the one meson exchange become relevant. This issues and the fact that hadrons were shown to have internal structure restricted the validity of the meson field theoretic approach to distances at which deviations from conceptual structureless hadrons are negligible. 

Furthermore, limitations to the perturbative approach also arise from divergences in meson meson interaction diagrams such as the one shown in Fig. (\ref{fig:phi4c}a). Such divergences are dealt through renormalization schemes by isolating a divergent term   from a convergent part of the diagram and then introducing counter terms into the interaction Lagrangian. These counter terms generate extra diagrams that cancel the divergent term in the original diagram.

The renormalized Lagrangian is then used to redefine parameters such as mass, coupling constant and wave function normalizations. Diagrams are then calculated using the redefined parameters in following Feynman rules. A scale dependence for instance is introduced in the renormalized or effective coupling constant ${\bar g}$. The value of this constant is now going to depend on the four momentum transfered at each vertex of a diagram. It also depends on its value $g$  measured at some other scale. In the case of QED $g=e=\sqrt{4\pi\alpha}$, with $\alpha\approx{1\over 137}$ measured at almost zero momentum transfer ($Q^2=-q^2=\mu^2$). The renormalized QED coupling constant is shown to deviate little from this value for a large range of energies. At the energy of the Z boson mass (90GeV), $\alpha\approx 1/127$. These small values justify the use of perturbative expansions in powers of $\alpha$ in the study of numerous phenomena at many energy scales in QED. Nevertheless, in the asymptotic limit ($-q^2\to\infty$), the value of the running coupling constant is given by:

\begin{equation}
\alpha(Q^2)={\alpha(\mu^2)\over 1-{\alpha(\mu^2)\over3\pi}log\left({Q^2\over\mu^2}\right)}
\label{Runcc}
\end{equation}
which is not very useful at energy scales in which $\alpha$ becomes greater than one, and of not use at all once $-q^2$ reaches the pole value (see e.g. Ref.~\cite{Landau:1955ip}). For QED however, $\alpha(Q^2)$ increases very slowly from $1/137$, and the pole on $\alpha(Q^2)$ is estimated to occur at energies$\sim10^{100's}GeV$, far beyond the Plank scale ($\sim 10^{19}$GeV) and outside the energy domain where physical phenomena can be currently studied.
Just as in QED, in a meson field theoretic approach to the strong interaction, treating divergent diagrams such as Fig. (\ref{fig:phi4c}a) requires adding counter terms to the Lagrangian that generate diagrams such as Fig. (\ref{fig:phi4c}b), and consequently a redefinition of the coupling constant (see Eq.(\ref{runCC})) that in the asymptotic limit  just as Eq.(\ref{Runcc}), grows with $Q^2$ and has a pole at some scale. However, the measured coupling constant at low energies($\sim100s$MeV) is already much larger than 1 ($g^2/4\pi\sim14$), which leads to a near pole also at $Q\sim100's$ MeV, making an issue of the convergence of an expansion of the $S$ matrix in series of Feynman diagrams. 
\begin{figure}[ht]
	\centering
		\includegraphics[width=0.5\textwidth]{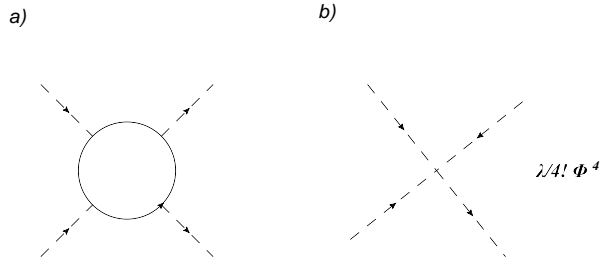}
	\caption{(a) Divergent fermion loop. (b) Counter term }
	\label{fig:phi4c}
\end{figure} 
\begin{eqnarray}
g={g(\mu)\over1-{3g(\mu)\over16\pi^2}log{Q\over\mu}}
\label{runCC}
\end{eqnarray}
\subparagraph{Effective field Theories} 
Due to the singularities discussed abuve, a meson field theoretic approach to the $NN$ force is not well suited for phenomena at energy scales of few GeV. At these energies the internal structures of nucleons and mesons play more explicit roles in their interaction that cannot be accounted for  by using hadronic degrees of freedom. More fundamental variables are sought for a theory of the strong interaction  that can be extended to all scales where the strong force is dominant. These variables were later associated with the internal constituents of hadrons (quarks and gluons).

 A new Lagrangian, the QCD Lagrangian, is constructed from terms involving quark and gluonic fields, and from such a Lagrangian a consistent quantitative description of strong interaction phenomena is expected to emerge for all energy scales of interest. It turns out that  in working which such a Lagrangian, there is not an analytic way of effectively computing observables at low energies. Instead, an effective field theory is developed constructing a new Lagrangian with the same global properties of the QCD Lagrangian, such as its invariance under chiral transformations in the massless quark limit. In this limit, these transformations flip the helicity of the quark fields. The spontaneous breaking of such symmetry gives rise to a set of  massless pseudoscalar bosons which are identified with the lightest pseudoscalar mesons. It is also shown that the masses of these mesons come from the explicit breaking of the chiral symmetry when quark mass terms are added to the fermionic part of the QCD Lagrangian. Hence, hadronic degrees of freedom emerge through spontaneous and explicit symmetry breaking of the free quark Lagrangian.
 
An effective Lagrangian is accordingly written to group these effective hadronic degrees of freedom and observables are computed through the resulting Feynman diagrams, which now are ordered in powers of (${p^2\over\Lambda_\chi}$) where p is some external momentum, and $\Lambda_\chi\sim$1GeV is known as the chiral scale. A perturbative approach is then valid as long as this parameter remains small, i.e., for energies of the order of hundreds MeV. Thus, again, the use of the hadronic degrees of freedom in the analytical computation of observables based on a field theory is limited to a certain energy regime.  
\medskip
\medskip

While the quark mass term in the effective Lagrangian break chiral symmetry, mass differences between quark species breaks $SU(2)$ isospin invariance. A larger, $SU(3)$ flavor symmetry, is also broken by these mass differences. This symmetry is suggested by the discovery of hadrons with `strange' decaying properties as it is described in the next section. `Strange' pseudoscalar mesons named kaons compliment pions to form the group of eight light pseudoscalar  mesons associated with the eight Goldstone bosons emerging from the spontaneous breaking of chiral $SU(3)_R\times SU(3)_L$ symmetry.

\subsubsection{Strangeness}
\label{Stran}
In 1947 a new particle was discovered with  the property of decaying into two hadrons $p$ and $\pi^-$ at a rate much slower (half life $\sim 10^{-10}$s) than typical decays through the strong interaction ($10_{20}$s). It was named $\Lambda^0$ and since the net number of baryons or baryon number ($B$) was thought to be a conserved quantity  regardless of the nature of an interaction, the $\Lambda^0$ particle was concluded to be a baryon with a `strange' decaying property accordingly named ``strangeness''. This property was also observed around the same time in the discovery of a meson named $K$ which would decay into two pions, and in the discovery of a set of three baryons called $\Sigma^+$, $\Sigma^-$ and $\Sigma^0$ that `slowly' decay into a nucleon and a pion. 

The half lives  of $\Lambda$, $\Sigma$ and $K$ are typical in weak interaction processes such as beta and pion decays. Evidence of them taking part in strong interaction processes  was found through the discovery of a resonance ($\Sigma(1385)$)  in the reaction
$K^- p\to\Lambda^0\pi^+\pi^-$. The Resonance $\Sigma(1385)$ decays into  a final $\Lambda\pi$ system with a half live of $\sim10^{-22}$s which is characteristic of the strong interaction. This reaction and many others involving strange particles that followed were found to occur at these high rates. The common property of these reactions was that if one strange particle was present in the initial state, one strange particle will be present in the final state of the reaction as well. This is not the case for the weak decays explored earlier.

However, in some cases in which two or more strange particles were present in the initial state of a strong reaction, the final state did not necessarily have the same number of strange particles. This is the case for instance in the reaction
\begin{equation}
K^0\Lambda^0\rightarrow\pi^-p.
\label{klpip}
\end{equation} 
To infer a conservation law related to strangeness,  a new additive quantum number $S$ was introduced. As the baryonic number is positive for baryons and negative for antibaryons, the strangeness quantum number can also take positive or negative values. While for nucleons  $S=0$, for K mesons (kaons) $S=+1$, and for $\Lambda$ and $\Sigma$ baryons $S=-1$ with corresponding antihadrons having opposite strangeness. With this convention the empirical evidence shows that reactions under the strong interactions conserve strangeness.

Naturally, to preserve isospin as a good quantum number, values for both $S$ and $T$ should be assigned to the discovered strange baryons supported by the empirical evidence. Table \ref{tab:HadronFamilies} shows these assignments along with corresponding masses($M$) and spin quantum numbers $J$. It also includes the heavier baryon named ``cascade'' ($\Xi$) in reference to the two step weak strange decay ($\Xi\to\Lambda \pi\to N\pi\pi$) observed in its discovery  that generated a cascade of particles. This behavior is understood if for $\Xi$, $S=-2$ which is also consistent with $S$ conservation in strong reactions involving $\Xi$.
\begin{table*}[ht]
	\centering
		\begin{tabular}{|c|c|c|c||c|||c||c|}\hline
				Hadron&$M$(GeV$^2$)     &$J$         &$B$  &$S$  &         $I$&           $T$\\\hline
				p     &0.938&${1\over2}$&1  &0  &${1\over2}$&+${1\over2}$\\
				n     &0.940&${1\over2}$&1  &0  &${1\over2}$&$-{1\over2}$\\\hline		
		$\Lambda$ &1.116&${1\over2}$&1  &-1  &$     0   $&$    0     $\\\hline		
		$\Sigma^+$&1.189&${1\over2}$&1  &-1  &$     1   $&$    +1    $\\		
		$\Sigma^0$&1.192&${1\over2}$&1  &-1  &$     1   $&$    0     $\\		
		$\Sigma^-$&1.197&${1\over2}$&1  &-1  &$     1   $&$    -1    $\\\hline	
		$\Xi^0$  &1.315 &${1\over2}$&1  &-2  &$1\over2$  &$-{1\over2}$\\        	
		$\Xi^-$  &1.322 &${1\over2}$&1  &-2  &$1\over2$  &$+{1\over2}$\\\hline\hline        	
		$\pi^+$  &0.140 &$     0    $&0  &0   &$     1   $&$    +1    $\\		
		$\pi^0$  &0.135 &$     0    $&0  &0   &$     1   $&$     0    $\\		
		$\pi^-$  &0.140 &$     0    $&0  &0   &$     1   $&$    -1   $\\\hline
		$K^+$&0.494 &$     0    $&0  &1   &${1\over2}$&$+{1\over2}$\\     	
		$K^0$&0.498 &$     0    $&0  &1   &${1\over2}$&$-{1\over2}$\\     	
		$\bar{K^0}$&0.494 &$     0    $&0  &-1   &${1\over2}$&$+{1\over2}$\\     	
		$K^-$&0.494 &$     0    $&0  &-1   &${1\over2}$&$-{1\over2}$\\\hline     	
\end{tabular}
	\caption{Hadron Families}
	\label{tab:HadronFamilies}
\end{table*}

\subsection{$SU(3)$ Flavor and the Quark Model}
\label{SU3QM}

 Table \ref{tab:HadronFamilies} suggests the classification of the shown hadrons in families of isospin ($I$) multiplets of a given strangeness ($S$). These multiplets coincide with dimensional representations of the $SU(2)$ group of unitary transformations. The dimension of the representation would correspond to the number of particles in the multiplet. For instance, the family of pions is a three dimensional representation of $SU(2)$, while the dimension of the $\Xi$ multiplet is {\bf2}. 
 
The conservation of strangeness hinted to a larger symmetry group for the strong interaction. The proposed group was the $SU(3)$-flavor  symmetry group, of which $SU(2)$ was a subgroup. While the fundamental representation of $SU(2)$ is two dimensional,  which is realized in the nucleon and cascade duplets in Table \ref{tab:HadronFamilies}, the dimension of the fundamental representation for $SU(3)$ is $\bf3$. And while the two states in the $SU(2)_I$-fundamental representation are labeled $u$ (or $up$ for $T=+1/2$) and $d$ (or $down$ for $T=-1/2$), the states in the fundamental representation of $SU(3)_f$ are labeled $u$, $d$ and $s$.

The baryons in Table \ref{tab:HadronFamilies} can be arranged in a eight-dimensional representation of $SU(3)$ as represented in the in $I_3,Y$-plane (with $Y=S+B$ and $I_3=T$) Fig.\ref{fig:octet}.  Because of the mass differences between baryons, this symmetry is actually broken. This octet is part of the representations from arising from combining three $SU(3)$ fundamental representations,
{\bf3}$\otimes${\bf3}$\otimes${\bf3}={\bf3$\otimes$(6$\oplus$3*)}={\bf10$_S\oplus$8$_{MS}\oplus$8$_{MA}\oplus$1} (see e.g. \cite{close:1979}). Sub-indexes  $S$, $A$ and $M$ stand for symmetric, antisymmetric and mixed under the interchange of flavors.
\begin{figure}[!htbp]
	\centering
		\includegraphics[width=0.4\textwidth]{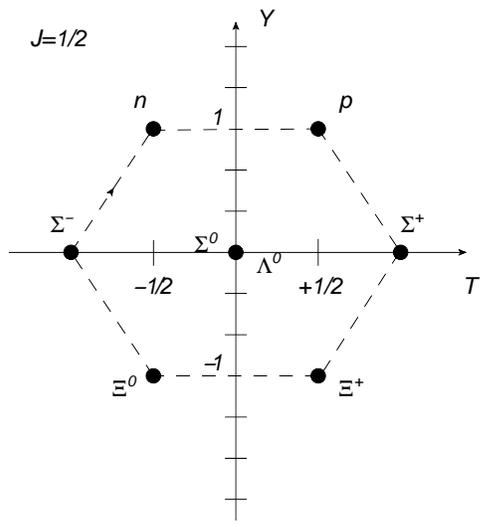}
	\caption{Baryon octet}
	\label{fig:octet}
\end{figure}

The $J=3/2$ baryons were later presumed to form the 10-dimensional representation of $SU(3)$ allowed in {\bf3$\otimes$3$\otimes$3}. This picture was completed by the discovery of a baryon with $S=-3$ ($\Omega^-$) in 1964.

\begin{figure}[!htbp]
	\centering
		\includegraphics[width=0.64\textwidth]{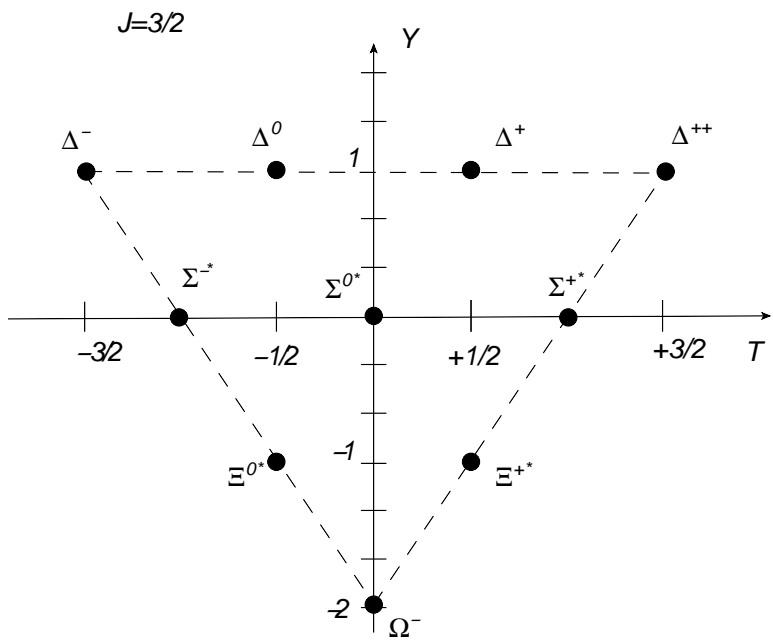}
	\caption{Baryon decuplet}
	\label{fig:decuplet}
\end{figure}

The pion triplet in Table \ref{tab:HadronFamilies} is built from a combination of the isospin doublet {\bf2}, and  its conjugate representation which explains the presence of antimesons to complete the triplet,{\bf2}$\otimes${\bf2*}={\bf3$\oplus$1}. In $SU(3)$ with the same idea, the mesons  should be arranged in  multiplets from {\bf3}$\otimes${\bf3*}={\bf8$\oplus$1}. This structure corresponds to a 8-dimensional multiplet and a singlet. The mesons $\eta$ and $\eta'$ were discovered in 1961, and join the mesons in Table \ref{tab:HadronFamilies} to complete  the {\bf8} and {\bf1} representations as shown in  Fig.\ref{fig:nonet}. Here, $SU(3)$ is broken by the mass difference between the mesons.
\begin{figure}[!htbp]
	\centering
		\includegraphics[width=0.5\textwidth]{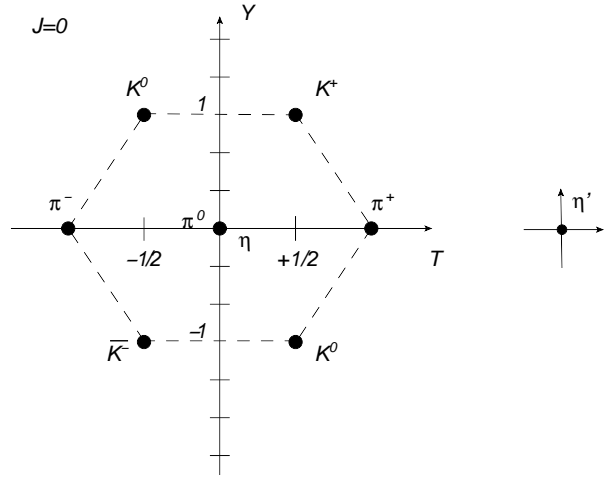}
	\caption{Meson nonet}
	\label{fig:nonet}
\end{figure}

Despite the elegance of associating the hadron families with representations of a symmetry group,  there was not a concrete explanation of why only some of the $SU(3)$ representations were realized in nature. In particular, why there was not a hadron family associated with the fundamental representation of $SU(3)$ (i.e. {\bf3}) as it was the case for $SU(2)_I$.

Gell-Mann \cite{GellMann:1964} and Zweig \cite{Zweig:1964} suggested in 1964 that this family of particles exists. They were named quarks, and it was further proposed that hadrons were actually systems of confined quarks from which they attain their intrinsic properties. The particles in this fundamental $SU(3)$ triplet are named after their flavors $u,$ $d$ and $s$, and as it's illustrated in Fig.\ref{fig:su3fund}, their baryon numbers are $B=1/3$,  and $B=-1/3$ for their corresponding antiparticles in the conjugate representation. Then in accordance with the group theory formalism of representations, baryons are now states of a system made up of three quarks {\bf3$\otimes$3$\otimes$3}, and mesons are states of a system of a quark and and antiquark (from the conjugate of the fundamental representation)  {\bf3$\otimes$3$^*$}. This way the hadrons' baryonic numbers result from the sum of their respective constituent quarks' baryonic numbers.
\begin{figure}[ht]
	\centering
		\includegraphics[width=0.7\textwidth]{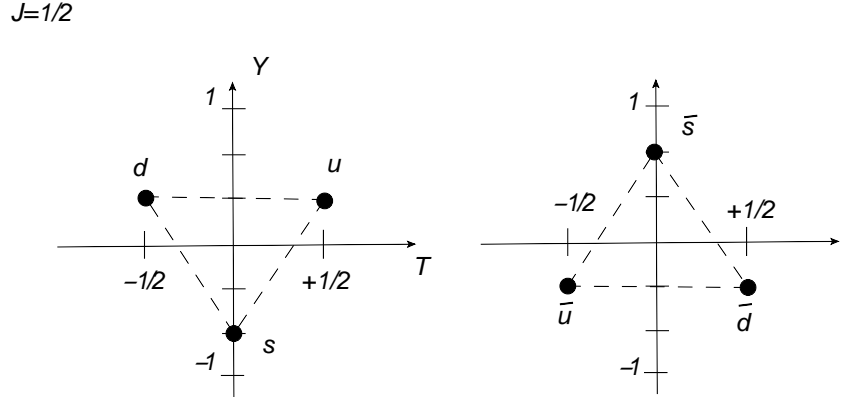}
	\caption{Quarks in $SU(3)_f$ fundamental representation}
	\label{fig:su3fund}
\end{figure}

The hadrons' spin states should now come from the combination of the spin states of their two or three constituent quarks. In the language of representations, they are multiplets of dimensions allowed by {\bf2$\otimes$2} for mesons, and {\bf2$\otimes$2$\otimes$2} for baryons, where {\bf2} is the fundamental representation of $SU(2)$-spin. The allowed multiplets are {\bf1$\oplus$3} for mesons and {\bf2$_{M_S}\oplus$2$_{M_A}\oplus$4} for baryons ($M_S$ and $M_A$ means symmetric and antisymmetric with respect to the exchange of the spins of the first two quarks) . This means that mesons can be grouped  in multiplets of spin $J=0$ and $J=1$, while baryons are grouped in two multiplets of spin $J=1/2$ and one with $J=3/2$. 

These $SU(2)$-spin representations are then combined with the $SU(3)_f$ representations to fully characterize a hadronic ($|h\rangle$) state in terms of its constituent quarks' states ($|qq...\rangle$). For instance in the Bras-Kets notation a meson $|m\rangle$ or a baryon $|b\rangle$ with  defined spin  and flavor numbers is expanded in $|q{\bar q}\rangle$ or $|qqq\rangle$ respectively  according to:
\begin{eqnarray}
|h(J,J_z,I,T,Y)\rangle=\sum_{J_{z1},J_{z2},T_1,T_2,Y_1,Y_2}C^{J,J_z}_{J_{z1},J_{z2}}C^{I,T,Y}_{T_1,T_2,Y_1,Y_2}\times
|q_1{\bar q}_2\rangle,\nonumber\\
%|q(J_{z1},T_1,Y_1){\bar q}(J_{z2},T_2,Y_2)\rangle\nonumber\\
\label{mqexp}
\end{eqnarray}
and
\begin{eqnarray}
|b(J,J_z,I,T,Y)\rangle=\sum_{J_{z1},J_{z2},J_{z3},T_1,T_2,T_3,Y_1,Y_2,Y_3}C^{J,J_z}_{J_{z1},J_{z2},J_{z3}}C^{I,T,Y}_{T_1,T_2,T_3,Y_1,Y_2,Y_3}
\times |q_1q_2q_3\rangle,\nonumber\\
%\times |q(J_{z1},T_1,Y_1)\bar q(J_{z2},T_2,Y_2)q(J_{z3},T_3,Y_3)\rangle\nonumber\\
\label{bqexp}
\end{eqnarray}
where $$q_i=q(J_{zi},T_i,Y_i),$$ and the coefficients $C$ are Clebsch-Gordan coefficients of the expansion. The labeling of $C^{I,T,Y}$ is simplified understanding that there is an additional dependency from the labeling of the hadron's multiplet, i.e., a pair of numbers (p,q) analogous to $J$ in $SU(2)$ that define a representation in $SU(3)_f$. The values of the $C$ coefficients for both $SU(2)$-spin and $SU(3)$-flavor  also depend on the symmetry properties of the spin and flavor representations of $|h\rangle$, i.e., the sub indexes $S$, $A$, and $M$. Then, the combined representation ({\bf SU(2),SU(3)}) of which $|h\rangle$ is part of inherits a symmetry subindex as well from the product of the symmetries of the spin and flavor representations:
\begin{table}[ht]
	\centering
		\begin{tabular}{cccccc}
			Symmetric:&$(S,S)$&$(M,M)$&$(A,A)$&&\\
			Antisymmetric:&$(S,A)$&$(M,M)$&$(A,S)$&&\\
			Mixed:&$(S,M)$&$(M,S)$&$(M,M)$&$(M,A)$&$(A,M)$
		\end{tabular}
	\caption{Symmetry of ({\bf SU(2), SU(3)}) representations.}
	\label{tab:SymmetrySU2SU3}
\end{table}
For instance, the baryon decuplet is symmetric only in the spin-flavor combination ({\bf 4,10}). In fact, the only symmetric combined representations of ({\bf(2$\otimes$2$\otimes$2),(3$\otimes$3$\otimes$3)}) are in ({\bf 2,8}) and ({\bf 4,10}), which correspond to the observed baryon $J=1/2$ octet and $J=3/2$ decuplet. However, if as assumed quarks are fermions, according to Pauli's principle they should only form antisymmetric states, therefore in principle a combination such as ({\bf 4,10}) should not exists for baryons. The contradiction is more explicit in baryons such as the $\Delta^{++}$ resonance. In its ground state $|J=3/2,J_z=+3/2\rangle$, all its constituent quarks are in the same state ($|u(J=1/2,J_z=+1/2)\rangle$), which is forbidden for identical fermions regardless of the exactness of $SU(3)_f$. The anomaly was later resolved by introducing the concept of color charge which eventually led to the formulation of quantum chromodynamics (QCD), the fundamental theory of the strong interaction.

\subsection{The Parton Model}
\label{TPMo}
Aside from the conflict between the idea of fermionic quarks and the exclusion principle, at the time there was not yet direct empirical evidence of their existence as real particles. The phenomenological successes of the $SU(3)_f$ model in describing the hadron spectrum did not require quarks to be real particles. However, if in fact they exist and are the building blocks of hadrons, the strong interaction should be more fundamentally described through a theory of interactions between quarks. 
In this light, nucleons, and hadrons in general are composite systems of more elemental particles. The discrete nature of their structure was later revealed through experiments in deep inelastic scattering. 

From inelastic proton-proton scattering experiments it was observed that most of the hadrons emerging from the collision were vastly produced collinear with the collision axis. The strong suppression at large angles hints that the collision evolves in a very weakly interacting or dilute medium.

Deep inelastic electron proton scattering experiments however  showed  that high energy electrons have a  large probability of scattering off protons with a significant energy and momentum transfer. This meant that  electrons were being deflected through the interaction with very localized concentrations of charge within the proton.
\begin{figure}[ht]
	\centering
		\includegraphics[width=0.7\textwidth]{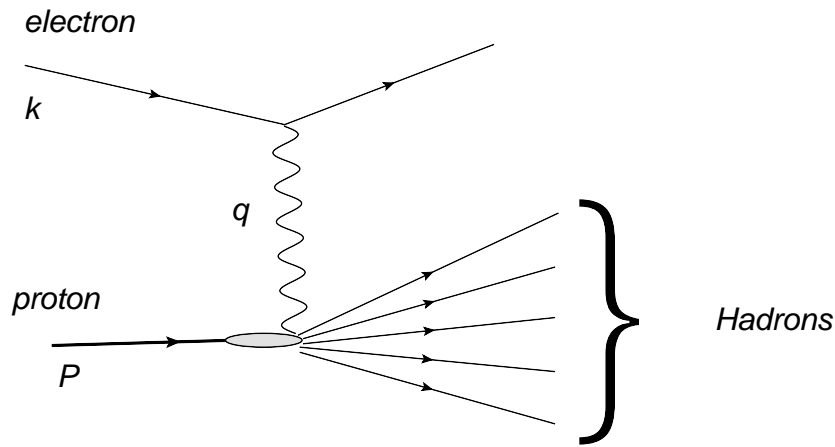}
	\caption{Deep inelastic scattering}
	\label{fig:dis}
\end{figure} 						
In the parton model proposed by Feynman in 1969 (see Refs.\cite{Feynman:1969wa} and \cite{Feynman:1973xc}), the electrons interact incoherently with nearly structureless particle-like entities named partons inside the hadrons. The incoherence of this interaction is better understood in a reference frame where the hadron's longitudinal momentum $P\to\infty$ accordingly named infinite momentum frame. Having that in the rest frame, partons inside the hadron interact by exchanging finite energy and momentum, their interaction times are finite as well. Then, when boosted to this ``infinite momentum'' frame  the interaction time between partons is extremely dilated such that they appear basically free to the electromagnetic probe.
\begin{figure}[ht]
	\centering
		\includegraphics[width=0.7\textwidth]{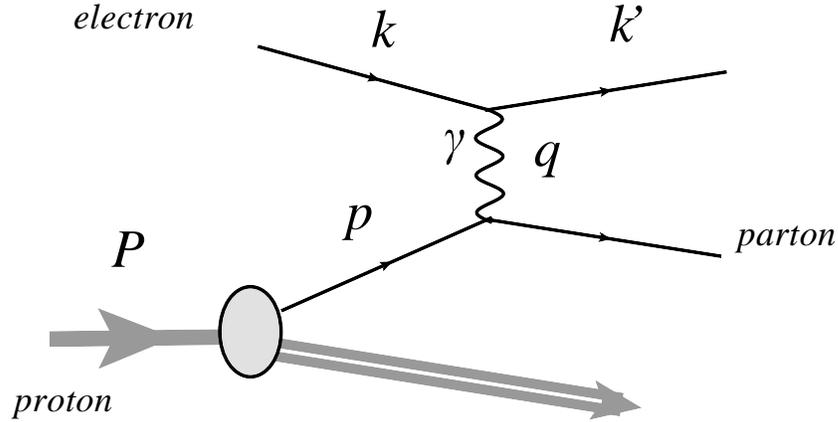}
	\caption{Parton picture of electron hadron  interaction in DIS}
	\label{fig:parton}
\end{figure} 
\subsubsection{Scaling in DIS}
\label{ScDIS}	
In the parton model developed by Bjorken and Paschos \cite{Bjorken:1969ja}, the cross section for $e+p\to e'+X$ is constructed from the cross section of an elementary electron parton elastic scattering (see Fig.(\ref{fig:parton})). Assuming that such parton of some specie (flavor, spin,...) $i$ is a fermion ($q_i$), the invariant amplitude for this elementary scattering is obtained using Feynman rules and  for the spin averaged squared amplitude we have,
\begin{eqnarray}
{\bar{\cal |M^2}_{eq_i\to eq_i}|}={8e^4Q_i^2\over t^2}\left({\hat{s}^2+\hat{u}^2\over4} \right), 
\end{eqnarray}
in which $eQ_i$ is the electric charge of the struck parton $q_i$, and the invariants $\hat{s},\hat{t},$ and $\hat{u}$ are the  Mandelstam variables for this subprocess:
\begin{eqnarray}
\hat{s}=(p+k)^2&&\hat{t}=(k-k')^2=q^2\nonumber\\
&\hat{u}=(p-k')^2&
\label{hatmand}
\end{eqnarray}
such that in the massless limit they satisfy $\hat{s}+\hat{t}+\hat{u}=0$.
Then, for the differential cross section we have,
\begin{eqnarray}
{d\sigma_{eq_i\to eq_i}\over d\hat{t}}&=&{1\over 16\pi s}{\bar{\cal |M_{i}|}^2}\nonumber\\
&=&{2\pi\alpha^2Q^2_i\over \hat{s}^2}\left({\hat{s}^2+(\hat{s}+\hat{t})^2\over\hat{t}^2}\right).
\label{prtdcrs}
\end{eqnarray}

The invariants $\hat{s}$, $\hat{t}$, and $\hat{u}$ defined in Eq. (\ref{mandhat})  can be related to the Mandelstam variables $s=(P+k)^2$, $t=(k-k')^2$ and $u=(P-k')^2$ defined for the $e+p\to e+ X$ reaction by assuming that the parton's momentum is collinear with the proton's momentum. Then, defining $\xi$ such that $p=\xi P$ in the infinite momentum frame, we have that,
\begin{eqnarray}
\hat{s}=\xi s&& \hat{t}=t,
\end{eqnarray}
in which $\xi$ can be related to experimental variables. In the massless limit, for the parton we have that, $$0\approx(p+q)^2=2p\cdot q +q^2=2\xi P\cdot q+q^2.$$ 
Then with $Q^2=-q^2$,
\begin{eqnarray} 
\xi=x\equiv{Q^2\over2P\cdot q}, 
\end{eqnarray}  
that in the lab reference frame corresponds to $x={Q^2\over 2M_pq_0}$. 
Equation (\ref{prtdcrs}) then corresponds to the differential cross section for an electron scattering off a quasi free parton of specie $i$ concentrating $x$ times the momentum of its parent proton (measured in the infinite momentum frame).

Eq.(\ref{prtdcrs}) can now be written in terms of measurable $s$, $Q^2$, and $x$ as,

\begin{eqnarray}
{d\sigma_{eq_i\to eq_i}\over dQ^2}={2\pi\alpha^2Q^2_i\over Q^4}\left(1+\left(1-{Q^2\over xs}\right)^2\right)
\label{prtdcrs2}
\end{eqnarray}

Thus,  Eq.(\ref{prtdcrs2}) is the contribution to $e+p\to e+X$ from the scattering of the electron from any single parton of specie $i$ and momentum fraction $x$. The total contribution of partons of specie $i$ and momentum fraction $x$ is going to be weighed by the probability $f_i(x)$ of finding one of such parton in the proton.

To obtain the differential cross section for the $e+p\to e+X$ reaction, Eq.(\ref{prtdcrs2}) is weighed by $f_i(x)$, summed  over all parton species and integrated over all momentum fractions $x$ to yield,

\begin{eqnarray}
{d\sigma\over dQ^2}&=&\int dx\sum_i f_i(x){d\sigma_{eq_i\to eq_i}\over dQ^2}\nonumber\\
&=&\int dx\sum_i f_i(x)Q^2_i{2\pi\alpha^2\over Q^4}\left(1+\left(1-{Q^2\over xs}\right)^2\right),
\label{prtdis2}
\end{eqnarray}
On the other hand, from general principles the differential cross section of the inclusive reaction  $e+p\to e+X$ can be expressed as follows,
\begin{eqnarray}
{d\sigma\over dQ^2}=\int dx{4\pi\alpha^2\over Q^4}\left(y^2F_1(x,Q^2)+\left(1-y-{M_p^2x^2y^2\over Q^2}\right){F_2(x,Q^2)\over x}\right),
\label{prtdis3}
\end{eqnarray}
in which $F_1$ and $F_2$ are inelastic structure functions, and $y={P\cdot q\over P\cdot k}$. In the large $s$ limit ($s>>M_p$, $y\approx {Q^2\over s x}$), comparing Eq.(\ref{prtdis2}) and Eq.(\ref{prtdis3}) one obtains
\begin{eqnarray}
2xF_1(x,Q^2)=F_2(x,Q^2)=\sum_iQ_i^2xf_i(x).
\label{f1f2prt}
\end{eqnarray}
The first equality is well supported by experiments. It is known as the Callan-Gross relation, and is a consequence of the partons having spin ${1\over2}$. 
Then, as it follows from Eq.(\ref{f1f2prt}), the inelastic structure functions, that can be extracted experimentally, are expected to become independent of $Q^2$ within the parton model. This independence is referred as Bjorken scaling. Such prediction was confirmed in 1969 by the SLAC-MIT experiment in which the scaling behavior expected from Eq.(\ref{prtdis2}) was confirmed for $1$GeV$^2<Q^2<8$GeV$^2$.
\begin{figure}[!htbs]
	\centering
		\includegraphics[width=0.8\textwidth]{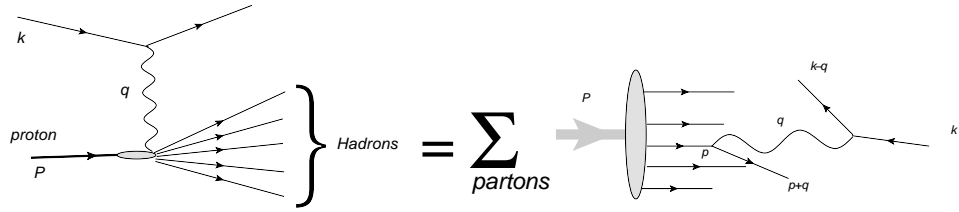}
	\caption{Bjorken Scaling emerges from the the electron's incoherent scattering of the proton's partons in DIS}
	\label{fig:scaling}
\end{figure} 			 

The experimental confirmation of the Bjorken scaling in $ep\to eX$ scattering seemed to validate the picture of the nucleon to be a collection of almost free constituents (partons) as seen by a hard electromagnetic probe. Partons are then identified with the quarks introduced in the previous section, and the parton's specie $i$ is associated with the quark's flavor. However, the observed scaling contradicts a quantum field theory that would describe the interaction between these quarks. Since they do interact to be bound in nucleons, even at the smallest distances they are expected to exchange field quanta and  deviations from a scaling pattern were to be expected instead.

The above mentioned  conflict was resolved in the 1970's by the discovery of field theories in which the interaction is characterized by asymptotic freedom (the strength of the interaction decreases as $Q^2$ increases). Such behavior was discovered  as a result of the regularization of diverging diagrams that, similar to QED and $\phi^4$ field theories, introduces a $Q^2$ dependence in the redefinition of the coupling constant. However, unlike  QED and $\phi^4$, the  value of the running coupling constant in these new theories can decrease to zero as $Q^2\to\infty$. This feature favors the main assumption of the parton model, namely that at very short distances the interaction between the partons in the nucleon is negligible. Nevertheless, deviations from scaling in DIS are still expected from a finite probability of field quanta emission by the probed quark. They become significant for   larger $Q^2$ or for   smaller $x$ than those reached by the SLAC-MIT experiment. Such violations of Bjorken scaling can be systematically studied through perturbation theory thanks to the smallness of the coupling constant.

\medskip
Quarks now become the fundamental fermions in a quantum field theory that describes the strong force, while hadrons  become multiquark systems bounded by their interaction under such force. As described in the following section such interaction not only binds quarks inside hadrons, but it also solves the apparent contradiction between the fermionic nature of quarks and the exclusion principle.

\subsection{Color Charge and QCD}
\label{CCQCD}
In order to keep quarks as fermions and for them to obey Pauli's statistics, Greenberg proposed in 1964 \cite{Greenberg:1964pe} that a group of seemingly identical quarks should differ at least in a hidden quantum number for them to form a ground state hadron. This hidden quantum number was named `color' or color charge, and its hidden property means that it cannot be observed, or that it cannot be probed in any observed particle. This proposition is in agreement with the fact that the $SU(3)_f$ fundamental representation is not realized in nature, if it's assumed that quarks have a nonzero color charge. Since quarks are `colored' objects, they have to be confined within colorless systems such as hadrons. 

In the language of representations, quarks are color non-singlets that combine to form the color singlet hadrons. A meson for instance is formed by a quark of color $c$ and an antiquark of color $-c$ such that the total color of the system is 0 . For a baryon where the color charges of its quarks are $c_1$, $c_2$, and $c_3$  we have that $c_1+c_2+c_3=0.$ Then, a quark within the baryon could carry any color charge $c$ provided that the color charges of the other two balance it through the above relation. The smaller number of independent options for $c$ is 3, thus a quark may be on one of three color states or on a linear combination of three states  traditionally labeled $R$, $G$ and $B$ after $red$, $green$ and $blue$ in analogy with theory of colors where the combination of red, green and blue yields white. 

The special unitary group $SU(3)$ is then the simpler choice to represent a quark color state. These states then belong to the fundamental representation of $SU(3)$-color or {\bf3}. Now, a system of a colored quark and an anticolored antiquark can be represented by {\bf3$\otimes$3$^*$}={\bf1$\oplus$8} as discussed earlier regarding $SU(3)$-flavor. Likewise, three colored quark states are represented by {\bf3$\otimes$3$\otimes$3}={\bf1$\oplus$8$\oplus$8$\oplus$10}. Unlike  $SU(3)$-flavor, only the singlets ({\bf1}) are observed in nature.

\subsubsection{The $e+e^+\to $ Hadrons reaction}
\label{ee2H}

\begin{figure}[!htbs]
	\centering
		\includegraphics[width=0.8\textwidth]{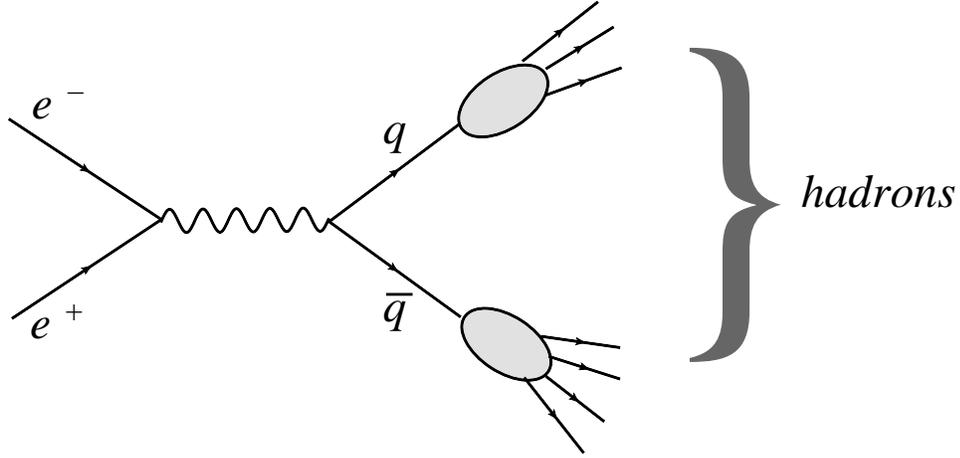}
	\caption{Elementary particle picture of $e^-e^+\to Hadrons$}
	\label{fig:eehad}
\end{figure} 			 

 As opposed to $SU(3)$-flavor, $SU(3)$-color is an exact symmetry. While a quark in a flavor state has specific mass and charge, it's color state is independent of any other intrinsic property. The absence of color nonsinglets  constrains direct evidence for this color degree of freedom hence direct evidence of this symmetry. Indirect evidence however is found in experiments in electron positron collisions where the energy from the annihilation creates multiple hadrons emerging from such collisions. The mechanism that creates these hadrons should start with the creation of a quark ($q$) and an antiquark (${\bar q}$) from a photon to which the $e,e^+$ pair fused to. The $q$ and ${\bar q}$ pair is created with enough center of mass  (c.m.) energy, that it can create more $q{\bar q}$ pairs from the vacuum ultimately hadronizing before reaching the detectors.
As it is assumed in the parton model of DIS, for $e^-e^+\to hadrons$ the dynamics of the reaction is mainly dictated by the elementary subprocess $e^-e^+\to q_i\bar{q_i}$, whose total cross section in the massless limit differs from that of $e^-e^+\to \mu^-\mu^+$ only by a electric charge factor $Q_i$ and by a color charge factor. Summing over all the possible color states of $q$ in the colorless $q\bar{q}$ system, this color charge factor is 3, thus,
\begin{eqnarray}
\sigma(e^-e^+\to q_i\bar{q_i})=3Q_i^2\sigma(e^-e^+\to \mu^-\mu^+),
\end{eqnarray}
and summing over all possible quark flavors $i$ in order to obtain the total cross section for $e^-e^+\to hadrons$,
\begin{eqnarray}
\sigma(e^-e^+\to hadrons)&=&\sum_i \sigma(e^-e^+\to q_i\bar{q_i})\nonumber\\
&=&3\sum_iQ_i^2\sigma(e^-e^+\to \mu^-\mu^+).
\end{eqnarray}
Then, the existence of a $SU(3)$ color degree of freedom implies that
\begin{eqnarray}
R\equiv{\sigma(e^-e^+\to hadrons)\over\sigma(e^-e^+\to \mu^-\mu^+)}=3\sum_iQ_i^2
\label{Rhad}
\end{eqnarray}
The ratio $R$ then depends on the number of quark flavors that can be produced in the elementary pair creation. The number depends on the center of mass energy of the $e^-e^+$ system. If the center of mass (c.m.) energy equals the mass of the lightest $q\bar{q}$ bound state for a given quark flavor, that flavor enters in the sum in Eq.(\ref{Rhad}). Then, for different flavor families we have,
\begin{eqnarray}
R=\left\{\begin{array}{cccc}3\left[\left({2\over3}\right)^2+\left({1\over3}\right)^2+\left({1\over3}\right)^2\right]&=2&for&u,d,s\\
2+3\left({2\over3}\right)^2&={10\over3}&for& u,d,s,c\\
{10\over3}+3\left({1\over3}\right)^2&={11\over3}&for&u,d,s,c,b
\end{array}\right.
\label{fl_R}
\end{eqnarray}
A comparison of Eq.(\ref{fl_R}) with experimental data is illustrated in Fig.(\ref{fig:sigma_R}). The agreement improves as the center of mass energy $\sqrt{s}$ increases far from the flavor threshold resonances regions. A description for such regions would involve more complex mechanisms than the one discussed here.
\begin{figure}[!htbs]
	\centering
		\includegraphics[width=0.5\textwidth]{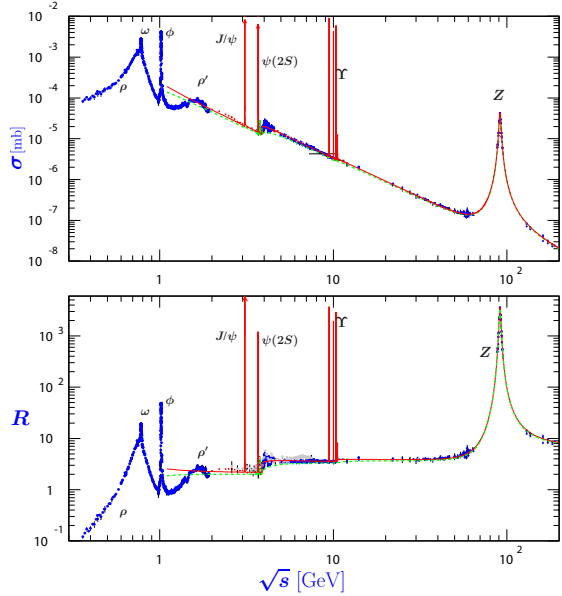}
	\caption{ World data on $e^+e^-\to hadrons$ cross sections (top) and $R$ (botton) as a function of the center of mass energy  ($\sqrt(s)$) of the system \cite{pdg:2008}.}
	\label{fig:sigma_R}
\end{figure} 
\medskip
The results displayed in Fig.(\ref{fig:sigma_R}) strongly suggest the existence of a 3-degenerated color degree of freedom which consequently reinforces the parton-quark picture of hadronic structure.	

However, the main motivation for the study of hadronic structure, understanding the fundamental origin of the strong nuclear force, requires a description of how quarks, the elementary constituents of hadrons, interact. As described so far, evidence of the existence of these elementary particles is based on the notion of noninteracting multiquark configurations, but as mentioned earlier such configurations are possible as an asymptotic limit in a field theory describing interactions between quarks. Just as QED describes interactions involving electrical charged particles and photons, this field theory describes interactions of color-charged quarks and the corresponding field quanta. Like photons in QED, these field quanta are bosons and they couple to corresponding color-charged currents, thus the name Quantum Chromodynamics (QCD) for this field theory. 

\subsubsection{Gauge Field Theory of Quarks and Gluons}
\label{FTQG}
The QCD field quanta are known as gluons, from the expectation that the force emerging from quarks emitting/absorbing gluons binds (glues) them to form hadrons. Just as pions, the field quanta of the previously studied $SU(2)_{isospin}$ meson field theory, carried isospin charge and formed a 3-dimensional representation of $SU(2)_{isospin}$, gluons carry color-charge and correspond to a 8-dimensional representation of $SU(3)_{color}$. Unlike pions which are pseudo-scalar (spin 0) fields,  and like photons, gluons are vector (spin 1) fields. 
%The idea of quarks interacting through the exchange of vector gluons explains mass splittings in hadrons of equal quark content but different intrinsic angular momentum. In analogy with QED, this is associated with a spin spin interaction term proportional to a `color' magnetic moment of the interacting quarks. Such term is not present if the interaction is mediated by a scalar field. 
Taking these considerations  into account when writing representations of the elements of the quark gluon field theory, we have that for a quark field of a given flavor, taking into account color degrees of freedom, 
\begin{eqnarray}
q(x)=\phi_q(x)\left(\begin{array}{c}
           q_1 \\
           q_2\\
           q_3\end{array}\right),
\label{colq}           
\end{eqnarray}   
while a gluon field is represented by,
\begin{eqnarray}
G^\mu(x)=\{G_i^\mu(x)\} 
\end{eqnarray}
in which (i=1,...8).
Under a global  $SU(3)_{color}$ transformation,
\begin{eqnarray}
q(x)\to q'(x)=e^{-i\theta_i\lambda_i\over2}q\nonumber\\
\bar{q(x)}\to \bar{q'(x)}=\bar{q(x)}e^{i\theta_i\lambda_i\over2},
\label{qlsu_3}
\end{eqnarray}
in which $\{\theta_i\}$ is a set of 8 real constant parameters. $\{\lambda_i\}$ are the Gell-Mann matrices, the generators of the $SU(3)$ group of transformations that as Pauli matrices in $SU(2)$ satisfy the commutation relations
\begin{eqnarray}
\left[{\lambda_i\over2},{\lambda_j\over2}\right]=if^{ijk}{\lambda_k\over2}
\label{lamcom}
\end{eqnarray}
and the normalization condition,
\begin{eqnarray}
tr\left({\lambda_i}{\lambda_j}\right)=2\delta^{ij}
\end{eqnarray}.

The Lagrangian density for a noninteracting quark field,
\begin{eqnarray}
{\cal L_q}=\bar{q}(x)\left(i\sh{\nabla}-m\right)q(x)
\label{fqlag}
\end{eqnarray}
is invariant under the transformations in Eq.(\ref{qlsu_3}). 
Quark-gluon dynamics arises however if we require the Lagrangian of the theory to be invariant under local $SU(3)$ transformations, $U(x)$, i.e., if a space-time dependency is introduced in the set of parameters $\theta$ in Eq.(\ref{qlsu_3}). Then,
\begin{eqnarray}
q(x)\to q'(x)=U(x)q(x)=e^{-i\theta_i(x)\lambda_i\over2}q\nonumber\\
\bar{q(x)}\to \bar{q'(x)}=\bar{q(x)}U^{-1}(x)=\bar{q}e^{i\theta_i(x)\lambda_i\over2}.
\label{lqlsu_3}
\end{eqnarray}
The Lagrangian density $\cal L_q$ in Eq.(\ref{fqlag}) is not invariant under these transformations. This is because of the differential operator $\nabla$ acting on the parameters $\theta(x)$ which generates extra terms in the transformed Lagrangian. However, if a new Lagrangian is built from Eq.(\ref{fqlag}) by replacing $\nabla$ with 
\begin{eqnarray}
D_\mu\equiv\partial_\mu-ig{\lambda^i\over2} G^i_\mu(x),
\label{covder}
\end{eqnarray}
thus obtaining,
\begin{eqnarray}
{\cal L_{qI}}=\bar{q}(x)\left(i\sh{D}-m\right)q(x).
\label{Iqlag}
\end{eqnarray}
Having that under the local $SU(3)$ transformation of Eq.(\ref{lqlsu_3}),
\begin{eqnarray}
\lambda^iG^i_\mu(x)\to U(x)\lambda^iG^i_\mu(x)U^{-1}(x)-{i\over g}[\partial_\mu U(x)]U^{-1}(x),
\label{gtrf}
\end{eqnarray}
the new Lagrangian density $\cal L_{qI}$ in Eq.(\ref{Iqlag}) is then invariant under local $SU(3)$ transformations, since Eq.(\ref{gtrf}) guarantees that $Dq(x)$, known as the covariant derivative of $q(x)$,  transforms by the same rule as $q(x)$.
Equation (\ref{gtrf}) is known as a $SU(3)$ gauge transformation on $G^\mu(x)$, and $G_i^\mu(x)$ are known as gauge fields which then are chosen to represent gluons. Thus, $Dq$ introduces a  quark-gluon interaction term ($\bar{q}\gamma\lambda G q$) in a local $SU(3)$ invariant Lagrangian. A special case, $U(1)$ gauge transformations  correspond to the local gauge symmetry of the QED Lagrangian that gives rise to the photon fermion interaction\footnote{See e.g. Refs, \cite{Peskin:1995ev} and \cite{Cheng:1985bj} for a  complete introduction of gauge field theories and their quantization}. 

As it stands, Eq.(\ref{Iqlag}) accounts for the dynamics of the quark field. However just as in QED $G^{mu}$ fields by themselves should contribute to the total Lagrangian. They do so through the gauge invariant term,
\begin{eqnarray}
{\cal L}_g=-{1\over4}F^i_{\mu\nu}F_i^{\mu\nu}
\label{glag}
\end{eqnarray} 
in which, 
\begin{eqnarray}
{\lambda ^i\over2}F^i_{\mu\nu}=\partial_\mu {\lambda ^i\over2}G^i_\mu-\partial_\nu{\lambda ^i\over2}G^i_{\mu}-ig\left[{\lambda ^i\over2},{\lambda ^j\over2}\right]G^i_\mu G^j_\nu.
\label{rotg}
\end{eqnarray}
Because of the Abelian nature of $U(1)$, in QED the third term in Eq.(\ref{rotg}) is zero. In QED, $F_{\mu\nu}$ is recognized as the electromagnetic tensor, and classically the equations of motion obtained from Eq.(\ref{glag}) yield the homogeneous or free Maxwell equations for electric and magnetic fields.

For QCD on the other hand, from Eq.(\ref{lamcom}), the commutator term in Eq.(\ref{rotg}) is not zero and $F_{\mu\nu}$ is not a linear function of $G_\mu$. Consequently, Eq.(\ref{glag}) introduces $gGG\partial_\mu G$ and $g^2GGGG$ terms in the total Lagrangian which now takes the following form,
\begin{eqnarray}
{\cal L}=\bar{q}(x)\left(i\gamma^\mu\partial_\mu-m\right)q(x)+g\bar{q}(x)\gamma^\mu{\lambda^i\over2} G^i_\mu(x)q(x)-{1\over4}F^i_{\mu\nu}F_i^{\mu\nu}.
\label{QCDLag}
\end{eqnarray}
These three-gluon and four-gluon fields terms give then rise to elementary interactions between gluons at the lowest an the next to lowest order expansions in the coupling constant ($\alpha_s={g^2\over4\pi}$) in the perturbative approach to QCD's elementary processes.  Fig.(\ref{qcdver}) illustrates the building blocks for such interactions according to Eqs.(\ref{rotg}) and (\ref{QCDLag}). In contrast, from a QED Lagrangian, only an analogous to the first vertex in Fig.(\ref{qcdver}) is present.
\begin{figure}[!htbs]
	\centering
		\includegraphics[width=0.5\textwidth]{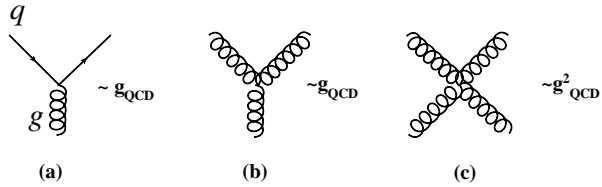}
	\caption{Elementary interactions in ${\cal L}_{QCD}$. (b) and (c) are introduced by ${\cal L_g}$ (Eq.(\ref{glag}) from the nonabelian nature of $SU(3)_{color}$}
	\label{qcdver}
\end{figure} 

\subsubsection{Asymptotic Freedom and pQCD}
\label{AFpQCD} 
In perturbation theory, the expansion of QCD observables in Feynman diagrams involves loop contributions such as those to the gluon propagator displayed in Fig.(\ref{qcdren}). Factors corresponding to the tree-gluon vertex in Fig.(\ref{qcdver}) generate the gluon loop diagram correction to the gluons Green's function. As it is the case for QED, for QCD, these contributions diverge individually, with such divergences emerging from the integration over all momentum space of the loop momentum variable not fixed by 4-momentum conservation. 
Thus, as discussed earlier, through renormalization schemes these divergences can be absorbed in the redefinition of the parameters of the theory such that the methodology of perturbative expansions in Feynman diagrams  can be safely applied. 
\begin{figure}[!htbs]
	\centering
		\includegraphics[width=0.5\textwidth]{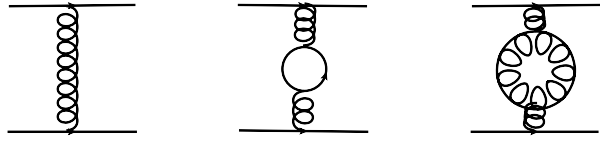}
	\caption{Contributions to the gluon propagator. The loop diagrams are divergent, and their regularization leads to the $Q^2$ dependence in the redefinition of the coupling constant (Eq.(\ref{runqcd})). }
	\label{qcdren}
\end{figure} 

A scale dependence is then introduced in the redefinition of the QCD's coupling constant. In the large $Q^2$ limit it's found that at the one loop correction,
\begin{eqnarray}
\alpha_s(Q^2)={\alpha(\mu^2)\over1+{\alpha(\mu^2)\over12\pi}(33-2n_f)log\left(Q^2\over\mu^2\right)},
\label{runqcd}
\end{eqnarray}
in which $\alpha(\mu^2)$ is the value of the coupling constant measured at some scale $\mu^2$, and $n_f$ is the number of quark flavors in the theory.
With the substitution,
\begin{eqnarray}
\Lambda^2\equiv\mu^2exp\left[{-12\pi\over(33-2n_f)\alpha_s(\mu^2)}\right],
\label{lamqcd}
\end{eqnarray}
Eq.(\ref{runqcd}) can be rewritten,
\begin{eqnarray}
\alpha_s(Q^2)={12\pi\over(33-2n_f)log\left({Q^2\over\Lambda^2}\right)}.
\label{runqcd2}
\end{eqnarray}
Unlike $\alpha$ for QED in Eq.(\ref{Runcc}), provided that $n_f\leq16$, $\alpha_s$ for QCD in Eq.(\ref{runqcd2}) decreases towards zero as $Q^2\to\infty$. Thus, the interaction mediated by  gluon exchanges grows weaker as the distance between the interacting particles becomes smaller. This property is known as Asymptotic Freedom and its discovery by David Gross, Frank Wilczek \cite{Gross:1974cs} and David Politzer \cite{Politzer:1973fx} in the 1970's validated within the framework of a field theory the main assumption on which the earlier parton model successfully described phenomena such as scaling in $ep\to eX$. Namely, that to a electromagnetic probe with high resolution (large $Q^2$) quarks deep inside the nucleon behave basically as free particles.

 Quantum chromodynamics then provides an important feature in understanding properties of hadronic structure. It also provides a methodology for examining corrections to results predicted within the parton model approximation. Equation (\ref{runqcd}) is valid for momentum energy scales at which $\alpha_s<1$. For such momentum-energy regime, the use of perturbation theory at the elementary particle level is well justified. Therefore, corrections to results from the parton model arise from calculable diagrams of gluon emissions by the struck parton. For instance, the leading order correction to the structure function $F_2$ is of order $\alpha\alpha_s$ and it comes from the superposition of the gluon emission diagrams in Fig.(\ref{scavio}). 
\begin{figure}[!htbs]
	\centering
		\includegraphics[width=0.5\textwidth]{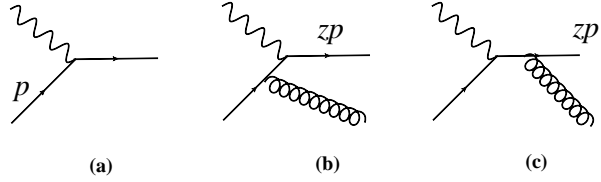}
	\caption{(a) quark-photon interaction controlling the $Q^2$ dependence in $ep$ DIS. (b) and (c) gluon emission diagrams introducing a $Q^2$ dependence in $F_2$ and braking the scaling in Eq.(\ref{f1f2prt}). }
	\label{scavio}
\end{figure} 
The quark emerging
from the second or third diagrams carries a fraction $z$ of the momentum of the initial parton. Unlike the situation in the first diagram, this quark can also carry transverse momentum relative to the virtual photon's in the infinite momentum frame. This transverse momentum is balanced by the transverse momentum of the emitted gluon to recover the zero transversity of the initial quark. Therefore, The elementary cross sections of these $\gamma^*q\to gq$ subprocesses are dependent on this $z$ momentum fraction, and on the transverse momentum that the quark acquires after emitting the gluon. After integrating over all possible values allowed for this transverse momentum, we have that;
\begin{eqnarray}
\hat{\sigma}(\gamma^*q\to gq)&=&\alpha Q_i^2\hat{\sigma_0}\int^{(p_\perp)_{max}}_{\mu^2}{dp_\perp^2\over p^2_\perp}{\alpha_s\over2\pi}P_{qq}(z)\nonumber\\
&=&\alpha Q_i^2\hat{\sigma_0}{\alpha_s\over2\pi}P_{qq}(z)log\left({Q^2\over\mu^2}\right)
\label{gecorr}
\end{eqnarray}
where $\hat{\sigma_0}=4\pi^2\alpha/\hat{s}$, and $P_{qq}(z)={4\over3}{1+z^2\over1-z}$ is called splitting function and represents the probability of a quark emitting a gluon that carries a $1-z$ fraction of the parent quark's momentum.

For the second line in Eq.(\ref{gecorr}) it is used that $$(p_\perp^2)_{max}={\hat{s}\over4}=Q^2{1-z\over 4z},$$ and a $\mu$ cut off is introduced to regularize the divergence when $p_\perp\to 0$.
Thus, as opposed to the parton model approximation, QCD predicted gluon emissions introduced a $Q^2$ dependence in the elementary parton electron interaction. As a result, the structure functions are corrected as follows,
\begin{eqnarray}
{F_2(x,Q^2)\over x}=\sum_q e_q^2\left(q(x)+\Delta q(x,Q^2)\right),
\end{eqnarray}
with
\begin{eqnarray}
\Delta q(x,Q^2)={\alpha_s\over2\pi}log\left({Q^2\over\mu^2}\right)\int^{1}_{x}{dy\over y}q(y)P_{qq}\left({x\over y}\right).
\label{difq}
\end{eqnarray}
The parton model limit is met when the above fluctuation is negligible.Then, $q(x)=f_q(x)$ and Eq.(\ref{f1f2prt}) is recovered from the general form,
\begin{eqnarray}
{F_2(x,Q^2)\over x}=\sum_q e_q^2 q(x,Q^2).
\end{eqnarray}
\begin{figure}[!htbs]
	\centering
		\includegraphics[width=0.8\textwidth]{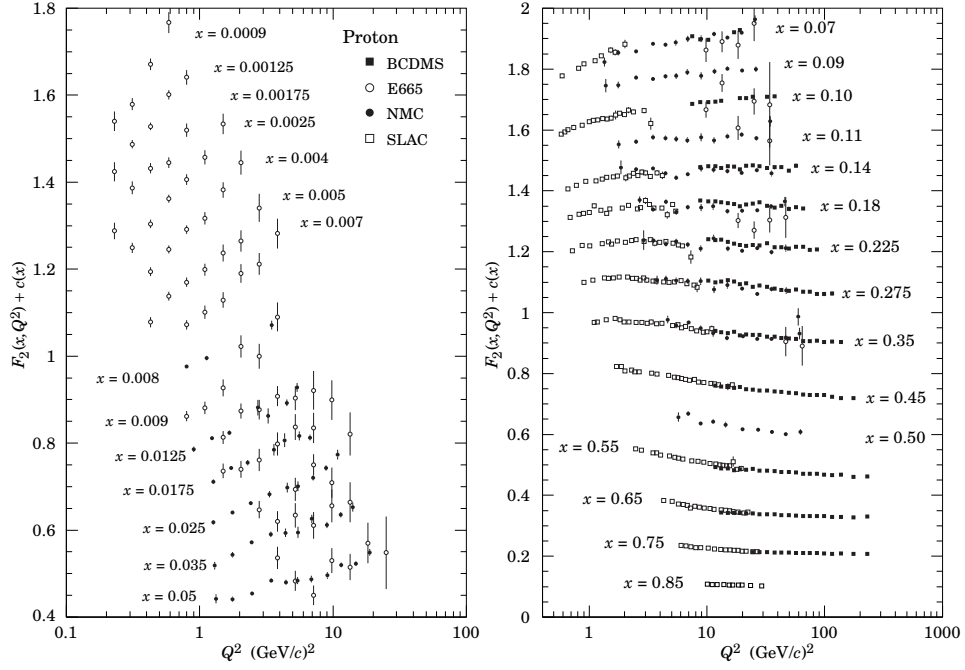}%{f2_smallx}
	\caption{$Q^2$ behavior of $F_2$. Figure taken from Ref. \cite{pdg:2000}}
	\label{f2allx}
\end{figure}
Experimental data on $F_2$ is shown in Fig.(\ref{f2allx}). Deviations from scaling behavior of $F_2$ are observed across values of $x$ being more dramatic at small $x$ at which as $Q^2$ increases the structure function increases as well. The structure function $F_2$ scales approximately in the region $0.2<x<0.3$, while for larger $x$ it slowly decreases with increasing $Q^2$.  
 Equation (\ref{difq}) can be rewritten to formally exhibit the $Q^2$ evolution of the distribution functions $q(x,Q^2)$ (see e.g. Ref. \cite{Halzen:1984mc}), 
\begin{eqnarray}
{d\over dlogQ^2}q(x,Q^2)={\alpha_s\over2\pi}\int^{1}_{x}{dy\over y}q(y,Q^2)P_{qq}\left({x\over y}\right).
\label{1DGLAP}
\end{eqnarray}
Thus, from the preexisting knowledge of an experimentally extracted distribution function for some fixed $Q^2=Q_0^2$, through Eq.(\ref{1DGLAP}), $q(x,Q^2)$ can be computed for any large value of $Q^2$. The evolution from this fixed $Q_0^2$ is dominated by the $Q^2$ dependence of $\alpha_s$, i.e., $q(x,Q^2)$ evolves logarithmically with $Q^2$, and thus, for large $Q^2$ the deviation from scaling is rather a subtle effect as can be seen in Fig.(\ref{f2allx}).

Equation (\ref{1DGLAP}) is further corrected by considering that the struck quark may have come from a quark-antiquark splitting of a prexisting gluon in the proton. The  contribution from pair creation to $q(x,Q^2)$ depends on a preexisting distribution of gluons ($g(x,Q^2)$). Then, Eq.(\ref{1DGLAP}) becomes
\begin{eqnarray}
{d\over dlogQ^2}q(x,Q^2)={\alpha_s\over2\pi}\int^{1}_{x}{dy\over y}\left(q(y,Q^2)P_{qq}\left({x\over y}\right)+g(y,Q^2)P_{qg}\left({x\over y}\right)\right),
\label{qDGLAP}
\end{eqnarray}
in which  the splitting function $P_{qg}={z^2+(1-z)^2\over2}$ represents the probability for the struck quark to carry a $z$ fraction of the momentum of the initial gluon. The function $P_{qg}(z)$ can be obtained from the cross section of the subprocess $\gamma^*g\to q\bar{q}$ in the same way that $P_{qq}(z)$ is obtained from the cross section of the subprocess $\gamma^*q\to gq$.

Solving Eq.(\ref{qDGLAP}) requires knowing $g(x,Q^2)$. Because gluon sources in the nucleon are both quark and preexistent gluons an evolution equation analogous to Eq.(\ref{qDGLAP}) can be derived through the same methodology for $g(x,Q^2)$ yielding
\begin{eqnarray}
{d\over dlogQ^2}g(x,Q^2)=\sum_{q}{\alpha_s\over2\pi}\int^{1}_{x}{dy\over y}\left(q(y,Q^2)P_{gq}\left({x\over y}\right)+g(y,Q^2)P_{gg}\left({x\over y}\right)\right),
\label{gDGLAP}
\end{eqnarray}
in which the splitting functions $P_{gq}(z)$ and $P_{gg}(z)$ can be obtained from  elementary diagrams of a gluon emitted with a $z$ momentum fraction of parent quark or a parent gluon respectively. In all we have for $z<1$ the following splitting functions
\begin{eqnarray}
P_{qq}(z)&=&P_{gq}(1-z)={4\over3}{1+z^2\over1-z}\nonumber\\
P_{qg}(z)&=&{z^2+(1-z)^2\over2}\nonumber\\
P_{gg}(z)&=&6\left({1-z\over z}+{z\over1-z}+z(1-z)\right).
\end{eqnarray} 

The equations (\ref{qDGLAP} and \ref{gDGLAP}) are known as the DGLAP (Dokshitzer \cite{YLD:1977},Gribov and Lipatov \cite{Gribov:1972ri}, and Altarelli-Parisi \cite{Altarelli:1977zs}) QCD evolution equations, and they describe the internal longitudinal momentum structure of the nucleon as it evolves in the large $Q^2$ region.

Going back to Eq.(\ref{prtdis2}), to account for the evolution of parton distributions with $Q^2$, the cross section for $ep\to eX$ is corrected to
\begin{eqnarray}
{d\sigma\over dQ^2}&=&\int dx\sum_q q(x,Q^2){d\sigma_{eq\to eq}(x,Q^2)\over dQ^2}.
\label{moddis}
\end{eqnarray}
The evolution equations show that the factorization implied in the parton model is validated by the slow evolution of $q(x,Q^2)$ at large $Q^2$. In contrast, as seen in Eq.(\ref{prtdcrs2})  the elementary subprocess's cross section has a large dependence on $Q^2$. Thus, for these kinematics the short distance physics of the reaction is mostly contained in this elementary subprocess' dynamics while the  long distance effects are isolated in factors such as the parton distributions $q(x)$ in Eq.(\ref{moddis}).

 These factors then describe the  dynamics of partons (quarks and gluons) as they interact to form hadrons. The full description of these factors however is beyond the reach of pQCD, because the behavior of the coupling constant is not longer described by Eq.(\ref{runqcd2}). How this effective  coupling behaves at long distances is not well understood; Eq.(\ref{runqcd2}) is no longer valid  for $Q^2<\Lambda\approx200MeV$, at which values $\alpha_s>1$, making perturbative  expansions in Feynman diagrams no longer convergent. 

The strength of the interaction between quarks is believed to grow as the distance between them increases which leads ultimately to the confinement of quarks into hadrons. Although the dynamics of this confinement cannot be described by pQCD, general features of hadrons' structure are expected to be described through the use of the QCD Lagrangian in nonperturbative approaches of which  Lattice QCD is the most widely accepted.

In addition to explaining the deviation from the scaling pattern of structure functions in DIS,  from the behavior of $\alpha_s$ dictated by Eq.(\ref{runqcd2}), QCD also introduces corrections to the results in Eq.(\ref{fl_R}) for $e, e^+$ annihilation into hadrons. Accounting for gluon emission by one of the quarks of the created quark antiquark pair introduces an additional energy dependency in $R$,
\begin{eqnarray}
R=3\sum_qe_q^2\left(1+{\alpha_s(Q^2)\over\pi}\right).
\end{eqnarray}
Again QCD predicts a scaling violation of order $logQ^2$ for the otherwise $Q^2$ independent  behavior of $R$. Current experimental data however do not reach a region where this difference is observable as seen in Fig.(\ref{fig:sigma_R}), thus making the effects of gluon emission negligible. 

If now within this approximation the reaction $ee^+\to h X$ (with $h$ a hadron of a given specie) is considered, its cross section is also predicted to scale according to
\begin{eqnarray}
{1\over\sigma}{d\sigma\over dz}(ee^+\to h X)={\sum_{q}e^2_q\left[D_q^h(z)+D^h_{\bar{q}}(z)\right]\over\sum_{q}e^2_q}
\label{lalhx}
\end{eqnarray}
in which $D_q^h(z)$ is known as a fragmentation function, and  similarly to the parton distribution functions in $ep\to eX$, it represents the probability that the hadron $h$ carries a fraction $z={2E_h\over Q}$ of the energy of the parent quark ($q$) in the quark antiquark pair from which it is produced. Likewise, the scaling predicted by Eq.(\ref{lalhx}) is broken by a $log Q^2$ evolution of the fragmentation functions arising from gluon emission. Another source of scale breaking arises from the  crossing of the charm threshold at which the heavy $c\bar{c}$ system is created almost at rest and subsequently decaying weakly into many low $z$ hadrons.

\paragraph{Drell-Yan processes}.

Hadron-hadron interactions  can be naturally approached through the same methodology applied to the electroproduction processes. One standard example is the reaction $pp\to l^-l^+ X$ in which a lepton-antilepton pair is created at large transverse momentum in a proton-proton collision. It is known as the Drell-Yan process and proceeds at the parton level as shown in fig.(\ref{DY}).
\begin{figure}[!htbs]
	\centering		\includegraphics[width=0.4\textwidth]{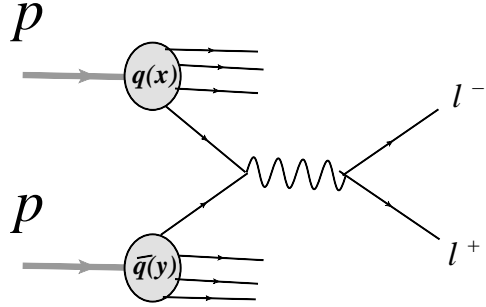}%{f2_smallx}
	\caption{Drell-Yan process. $pp\to l^-l^+$ in the QPM}
	\label{DY}
\end{figure}
The Drell-Yan cross section  within the quark-parton model (QPM) is,
\begin{eqnarray}
{d\sigma\over dQ^2}(pp\to l^-l^+X)=\sum_{q}\int dx\int dy q(x)\bar{q}(y){d\sigma\over dQ^2}(q\bar{q}\to l^+l^-),
\label{DYcrs}
\end{eqnarray}
in which 
\begin{eqnarray}
{d\sigma\over dQ^2}(q\bar{q}\to l^+l^-)=e_q^2{4\pi\alpha^2\over9Q^2}\delta(Q^2-(xp_1+yp_2)^2), 
\end{eqnarray}
with $x$ and $y$ being the fraction of the respective parent nucleons momenta carried by the quark and antiquark participating in the annihilation subprocess, and $Q^2$ being the invariant mass of the $l^-l^+$ system. 

For large $s$ and $Q^2$ we have that $Q^2=xys$. Making this condition explicit in Eq.(\ref{DYcrs}) yields
\begin{eqnarray}
{d\sigma\over dQ^2}(pp\to l^-l^+X)={4\pi\alpha^2\over9Q^4}\sum_{q}e_q^2\int dx\int dy q(x)\bar{q}(y)\delta\left(1-xy{s\over Q^2}\right),
\label{DYcrs2}
\end{eqnarray}
which suggest the introduction of a scaling variable $w=xy={Q^2\over s}$ resulting in the scaling law,
\begin{eqnarray}
Q^4{d\sigma\over dQ^2}(pp\to l^-l^+X)={\cal F}(w)
\label{DYcrs3}
\end{eqnarray}
This scaling is well satisfied by experimental data from which ${\cal F}(w)$ can be extracted. Also, just as it happens in DIS and $e^-e^+\to hX$ the scaling predicted in Eq.(\ref{DYcrs3}) is broken by $logQ^2$ coming from corrections to fig.(\ref{DY}) including quark gluon interactions.

Cross sections and other observables of numerous processes involving hadrons can be studied by making use of parton distribution functions and fragmentation functions. The universality of these functions further allows the study of QCD subprocesses controlling $\alpha^n$ order corrections in reactions such as those shown in fig.(\ref{incl})
\begin{figure}[!htbs]
	\centering		\includegraphics[width=0.8\textwidth]{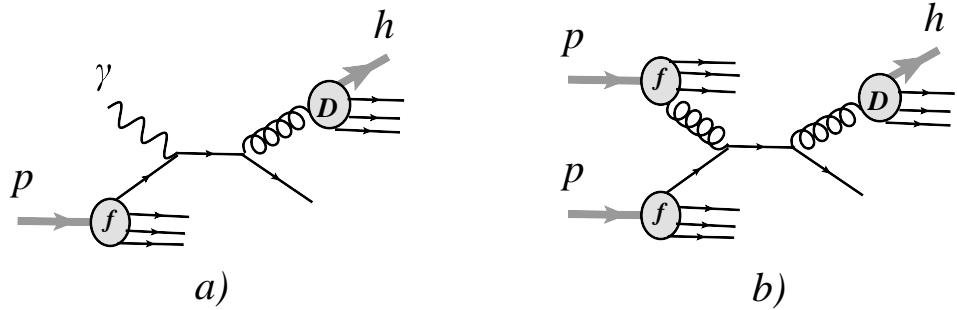}%{f2_smallx}
	\caption{Typical inclusive DIS processes involving gluons in the corresponding parton subprocesses. }
	\label{incl}
\end{figure}

These studies are made possible by the large kinematic variables involved in the reactions ($s,Q,p_T$) that allow the use of pQCD in describing the dynamics of the embedded subprocess. The kinematic ranges in which this description is applied is known as the hard kinematic regime, and the controlling subprocesses are known as hard subprocesses. Consequently, the short distance factors such as the cross sections of the hard subprocesses and the long distance factors such as parton distribution functions ($f$) and fragmentation functions ($D^h$) are known as hard factors and soft factors respectively.    

Asymptotic freedom has allowed the experimental study of QCD at the elementary particle level. Because of confinement, in experiments, the reactions studied do not produce free quarks or gluons which are the fields associated with the degrees of freedom in the QCD interaction Lagrangian. But because at large energy and momentum scales the QCD's coupling constant weakens and changes slowly, the elementary particle processes  at the core of these reactions dominate the kinematic dependence of the reaction and thus the short distance dynamics, making the study of the reaction an almost direct study of the hard subprocess. Corrections to this picture are also a test of QCD, since in the hard kinematics, such corrections come from the calculable pQCD evolution of the soft factors.

The kind of reactions considered  so far are of the form $ab\to (cd...) X$, i.e., in experiments only particles $(cd...)$ in the final state are detected, while $X$ sums up all those that are not. The final states of these reactions  then are not specified. The cross sections given are obtained including all possible final states, thus these kind of reaction are known as inclusive  or semi-inclusive reactions. The extensive study of inclusive reactions has reinforced QCD as the theory of the strong interaction that arises from the interactions between quarks and gluons. However, precisely because all the hadronic states are not completely resolved a  description of the strong force can't be extracted through QCD from inclusive processes alone. 

In what follows we focus on a more constrained kind of processes known as exclusive reactions. Such processes are of the form $AB\to (CDE...)$, thus all the particles in the final state are detected. The role of QCD degrees of freedom in hadronic structure and in hadronic interactions is more carefully investigated here because of the constrains of constructing the final hadron states from the scattered elementary  particles. This difficulty is of course avoided through the parton model in the previously studied inclusive reactions.

\section {Hard Exclusive Processes} 
\label{HEP}
In this section we introduce a methodology and a formalism that is closely followed in the study of the processes of interest in this dissertation.

With the purpose of developing a QCD description of interactions between hadrons (as bound states of quark systems) and at the same time gaining further insight into their quark structure, the studies of exclusive reactions at large momentum transfer have attracted particular interest. These hard processes facilitate the study of the short distance regime of the strong nuclear force in reactions such as hard proton proton $(pp)$ and proton neutron $(pn)$ elastic scattering. The characteristic kinematic region for hard exclusive processes is defined by large kinematic variables, i.e. $s,-t,-u>>m_N^2,$ and ${t\over s},$ and ${u\over s}$ fixed.

To illustrate the extent through which the analysis of  hard exclusive scattering processes potentially probes the quark-gluon dynamics that underlies hadronic processes we first consider the reaction $\gamma^*N\to N$.  
Fig.(\ref{nform}) illustrates how the reaction may proceeds at the constituent quark level.

\begin{figure}[!htbs]
	\centering		\includegraphics[width=0.5\textwidth]{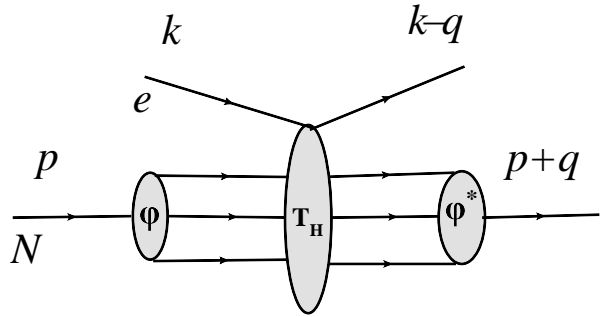}%{f2_smallx}
	\caption{$ep$ elastic scattering factorization according to Eq.(\ref{GMmat}).}
	\label{nform}
\end{figure}

The large momentum transfer by the virtual photon to the nucleon is distributed among the constituents through the hard subprocess $T_H$. Unlike the situation in inclusive DIS, here, coherence is required to form the outgoing nucleon, thus all the constituents are involved in the hard subprocess. Fig.(\ref{nform}) also suggests the factorization of the hard scattering amplitude ($T_H$) from the soft factors corresponding to the constituent distributions of the incoming ($\varphi$) and outgoing ($\varphi^*$) nucleons. These factors are convoluted in  the matrix element \cite{Lepage:1980fj} of the diagram in Fig.(\ref{nform}),

\begin{eqnarray}
G_M&=&{1\over2p^+}\left\langle N(\lambda'={1\over2})|J^+|N(\lambda={1\over2})\right\rangle\nonumber\\
&=&{1\over\alpha^2}\int[dx]\int[dy]\varphi^*(y_i)T_H(x_i,y_i,Q)\varphi(x_i).
\label{GMmat}
\end{eqnarray}
This matrix element in the infinite momentum frame corresponds to the nucleons's magnetic form factor which at large $-t=Q^2$ and $s>>-t$ can be extracted from experiments using the relation
\begin{eqnarray}
{d\sigma\over dt}(ep\to ep)=2\pi\alpha^2 {G_M^2(t)\over t^2}.
\label{epcrs}
\end{eqnarray}
In Eq.(\ref{GMmat}), $$[dx]=\delta(1-\sum^{n_N}_{j}x_j)\prod^{n_N}_{i}dx_i,$$ in which $x_i$ ($y_i$) is the fraction of the initial (final) nucleon's light cone momentum  carried by its constituent $i$. The  $\varphi(x)$ ($\varphi^*(y)$) is the amplitude for finding in the incoming (outgoing) nucleon the configuration of near on-shell partons  entering (leaving) the hard $T_H$ blob.

 %If such a state contains a small number of particles with large transverse momentum, the hard subprocesses can involve more than one parton from each hadron, thus through coherent scattering of partons with large momentum transfer they can deflect collinearly to form the emergent hadrons.
\subsubsection{Quark wave functions of hadrons}
\label{QWFH}
The nucleon shown in Fig.(\ref{nform}) is represented by a system of three partons, three quarks that in this case is the minimum number of quarks that are needed to reconstruct the helicity and isospin of the nucleon. The nucleon in Fig.(\ref{nform}) illustrates what is known as the minimal component of the Fock expansion of the nucleon wave function. In Fock space the nucleon is expanded in states with a definite number of constituents, i.e, $$\psi=qqq+qqqg+qqq\bar{q}q+...$$
Labeling each component by $N$ (number of constituents) the nucleon wave function in terms of its quark and gluon constituents is given by \cite{Carlson:1992ug} 
\begin{eqnarray}
|p,h\rangle=\sum_{N}\int[dx][d^2k_T]{1\over\sqrt{x_1...x_N}}\psi(x_i,k_{iT},h_i)|x_i,k_{iT},h_i\rangle,
\end{eqnarray}
in which $$[d^2k_T]=\delta^2\left(\sum_{j=i}^{N}k_{jT}\right)\prod_{i=1}^Nd^2k_{iT},$$  where $|x_i,k_{iT},h_i\rangle$ represents an individual constituent state. In the light cone gauge \cite{Lepage:1980fj}, the amplitudes $\varphi(x)$ in Eq.(\ref{GMmat}) is related to $\psi(x_i,k_{iT},h_i)$ by
\begin{eqnarray}
\varphi(x_1,...,x_N)=\int [d^2k_T]\psi(x_i,k_{iT},h_i,f_i),
\end{eqnarray}
in which it is understood that the additional helicity and flavor expansion coefficients such as those in Eq.(\ref{bqexp}) are contained in  $\varphi(x)$.

%\subsection{QCD Factorization}
\subsection{Scaling Laws} 
\label{ScLa}
The hard subprocess in Fig.(\ref{nform})involves the far off shell ($k_i^2\sim Q^2>>m_i^2$) propagation of the incoming nucleon's constituents into the outgoing constituents that form the final nucleon. A typical diagram contributing to the $eqqq\to eqqq$ subprocess is shown in Fig.(\ref{TH}). 
\begin{figure}[!htbs]
	\centering		\includegraphics[width=0.5\textwidth]{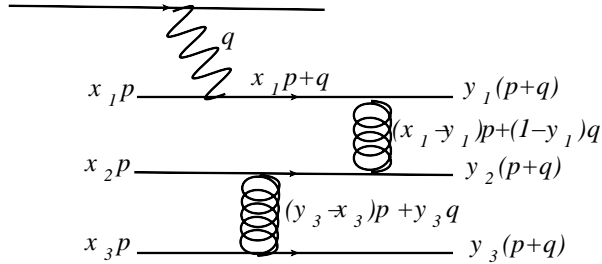}%{f2_smallx}
	\caption{Typical minimally connected diagram contributing to $T_H$ in $eN$ elastic scattering.}
	\label{TH}
\end{figure}
It is known as a minimally connected diagram and guarantees that the momentum $q$ transfered by the electron probe steers all the quarks  collinearly in the direction of the outgoing nucleon.  At $Q^2\to \infty$ this kind of diagrams dominate in $T_H$ since they represent the lowest order interaction in $\alpha_s$ among the constituent quarks. The $Q^2$ asymptotic dependence of $T_H$ can then be obtained from pQCD. Having that the quarks in $T_H$ interact through vector exchange, each pair of external lines contributes a factor  $\sim Q$. This is seen for instance in the vertex element, 
\begin{eqnarray}
\bar{u}(h'=h,k+q)\gamma^\mu u(h,k)|_{Q^2\to\infty}\propto Q
\label{qvert}
\end{eqnarray}
in which $u$ are quark helicity spinors. If $h'\neq h$ this factor behaves as $1\over Q$.  Each fermion propagator contributes a factor of $\sim{1\over Q}$ and each vector boson propagator contributes a factor ${\sim{1\over Q^2}}$, thus resulting in $$T_H\sim \alpha({\alpha_s\over Q^2})^2T_H(x,y)$$ which from Eq.(\ref{GMmat})and Eq.(\ref{epcrs}) leads to
\begin{eqnarray}
{d\sigma\over dt}(ep\to ep)\sim{(\alpha\alpha_s^2)^2\over t^6},
\label{epsc}
\end{eqnarray}
or for fixed c.m. angle of $ep\to ep$ scattering ($t/s$ fixed, $t,s\to\infty$, and $s>>-t$),
\begin{eqnarray}
{d\sigma\over dt}(ep\to ep)\to{(\alpha\alpha_s^2)^2\over s^6}f(t/s).
\label{epsc2}
\end{eqnarray}
The contributions to $T_H$ from diagrams with additional constituent lines taking part in the hard subprocess fall off as powers of ${\alpha_s\over Q}$ faster as compared to the contribution from Fig.(\ref{nform}) that contains the minimal number of constituents. Hence, the minimal Fock component of the nucleon contributes to the dominant term of the scattering amplitude. 

Through the same rules that lead to Eq.(\ref{epsc2}), for the $e\pi\to e\pi$ scattering reaction we have that, ${d\sigma\over dt}\sim {1\over s^4}$, while it is known that for  $e\mu\to e\mu$, ${d\sigma\over dt}\sim{1\over s^2}$. Thus in general we have that for  $eH\to eH$,
\begin{eqnarray}
s^2s^{2+N_H}{d\sigma\over dt}^{eH\to eH}\sim f_H({t\over s}),
\label{dimsc1}
\end{eqnarray}
in which  $N_H$ is the number of minimal constituents in particle $H$.
\begin{figure}[!htbs]
	\centering		\includegraphics[width=0.6\textwidth]{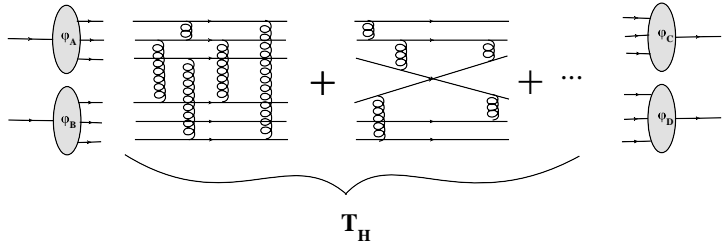}%{f2_smallx}
	\caption{$NN$ elastic scattering controlled by the $T_H$ expansion of connected constituent diagrams.}
	\label{NNsca}
\end{figure}

A hadron hadron scattering amplitude  in the hard kinematic regime can be factorized similarly to Eq.(\ref{GMmat}) in the following form:
\begin{eqnarray}
{\cal M}_{ab\to cd}=\int\prod_{i=a,b,c,d}[dx_i]\varphi^*_{C}(x_c)\varphi^*_D(x_c)T_H(x_i,s,t)\varphi_{A}(x_c)\varphi_B(x_c).
\label{nnfscm} 
\end{eqnarray}
Then in the above equation $T_H$  is dominated by  connected diagrams involving constituents from the minimal Fock components of the interacting hadrons such as those shown in Fig.(\ref{nnfscm}) for nucleon nucleon ($NN$) elastic scattering. Although there are at least thousands of such diagrams contributing to $T_H$, they contain the same number of fermion and vector boson  external and internal lines. Then as before, at fixed angle in the asymptotic limit there is a $\sqrt{s}$ factor per each pair of quark external lines,  a ${1\over\sqrt{s}}$ per each quark internal line, and a ${1\over s}$ factor per each vector gluon internal line. Then for $NN$ elastic scattering, $T_H\sim{1\over s^4}T(x,\theta_{c.m.})$, and for the $NN$ elastic scattering amplitude one obtains,
\begin{eqnarray}
{\cal M}_{NN\to NN}\sim {1\over s^4}M(t/s),
\label{MNNscal}
\end{eqnarray}
and
\begin{eqnarray}
{d\sigma\over dt}^{NN\to NN}&\sim& {1\over s^2}|{\cal M}|^2_{NN\to NN}\nonumber\\
&\sim&{1\over s^{10}}f_{NN}(t/s). 
\label{NNscal}
\end{eqnarray}
In general, the asymptotic energy dependence of the invariant amplitudes and of the cross sections for exclusive processes is given respectively by the forms (see Refs. \cite{BF75}, and \cite{MMT}),
\begin{eqnarray}
{\cal M}\sim s^{-{1\over2}(n_A+n_B+n_C+n_D+...-4)}{\cal M}(t/s),
\label{qcruls}
\end{eqnarray}
and as in Eq.(\ref{NNscal}),
\begin{eqnarray}
{d\sigma\over dt}\sim s^{-(n_A+n_B+n_C+n_D+...-2)}f(t/s),
\label{dimscal}
\end{eqnarray}
in which $n_H$ is the minimal number of constituents of particle $H$ ($H=ABCD...$) taking part in the hard subprocess, while $f(t/s)$ is an angular function that becomes invariant under a change of energy scale. The dimensions  of the scale invariant $f(t/s)$ are given by the number of constituents taking part of the elementary  hard subprocesses (for instance, $f(t/s)$ is dimensionless for $e\mu\to e\mu$) consequently determining the power fall-off  of the energy ($s$) dependence of the  differential cross section. 

Evidence of a power law fall-off in energy distributions at fixed angle has been observed in experiments for several hard processes involving hadrons, with many of them closely fitting the scaling behavior predicted by Eq. (\ref{dimscal}) (see e.g. Refs.\cite{Owen:1969tz,Landshoff:1970ff,Akerlof,Allaby,pnexp1,pnexp2}). Empirical agreement with Eq. (\ref{dimscal}) however does not necessarily implies the perturbative approach used here to arrive at this scaling law. Nonetheless, the correlation between the number of constituents and the energy dependence of the exclusive reactions evidenced by this result hints to the dominance of minimal Fock component of the partonic wave functions of the hadrons participating in the reactions.  %interactions between constituents to be controlling the reaction.

\subsection {Quark Interchange}
\label{QuIn}

In the constituent picture, the processes through which hadrons in an exclusive reaction interact include the mechanisms shown in Fig.(\ref{hhsc}). Unlike the situation involving reactions at low energy and low momentum transfer, in which the reactions proceed through the exchange of mesons, as illustrated in Fig.(\ref{hhsc}) hard exclusive processes involving hadrons proceed through the exchange of quarks or gluons. QCD excludes the mechanism of one gluon exchange in hadron-hadron exclusive scattering because hadrons are color singlets (of zero color charge), but the exchange of one gluon  would result in a net transferring of  color charge from one hadron to the other  producing two non-singlet particles in the final state. The latter scenario is not observed in nature, therefore gluon exchange mechanisms in hadron-hadron scattering involve the exchange of two gluons, as shown in Fig.(\ref{hhsc}a), or more such that it can ensure a net transfer of zero color charge between the interacting hadrons. The other mechanisms of interaction shown in Fig.(\ref{hhsc}) (b) and (c), correspond to quark-antiquark annihilation and quark interchange. The quark-antiquark annihilation mechanisms are considered in scattering reactions in which one of the hadrons participating contains antiquarks of corresponding quarks from the second hadron. The reaction then proceeds through the annihilation of a quark from one hadron with an antiquark from the other hadron. Quark interchange mechanisms are considered for instance in  elastic scattering reactions in which both hadrons contain quarks of common flavor.  
\begin{figure}[!htbs]
	\centering		\includegraphics[width=0.6\textwidth]{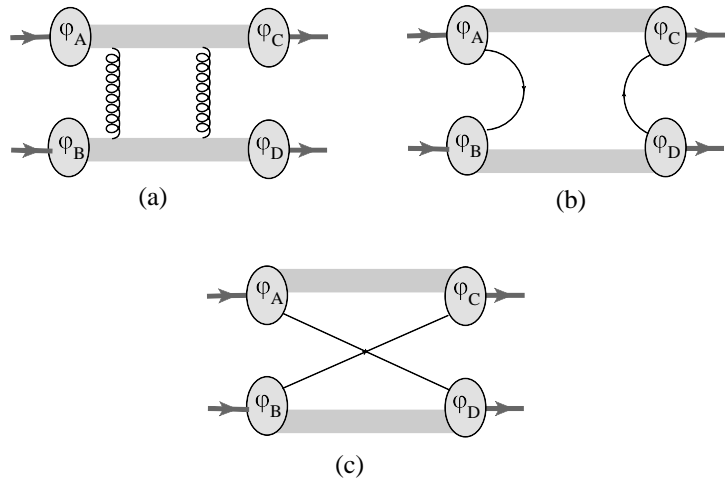}%{f2_smallx}
	\caption{QCD description of exclusive hadron-hadron scattering through (a) gluon exchange, (b) quark-antiquark annihilation and (c) quark interchange. }
	\label{hhsc}
\end{figure}
It has been argued that the quark interchange mechanism dominates the exclusive scattering  of hadrons that posses quarks of common flavor among their constituents \cite{Gunion:1973ex,Sivers:1975dg}. 

The dominance of the quark interchange mechanism has been experimentally concluded when comparing for instance data on large angle scattering for proton proton ($pp$) and proton antiproton ($p\bar{p}$) elastic reactions. It was found that at large angle, the $pp$ cross sections largely dominates $p\bar{p}$  \cite{h20}. At 90$^o$ c.m angle of scattering and 10GeV of beam momentum, the cross section of $pp$ elastic scattering is more than 24 times larger than that of $p\bar{p}$ \cite{h20}. If gluon exchange mechanisms ( Fig. (\ref{hhsc}a)) were the dominant mechanisms of interaction, the cross sections for both reaction would be expected to be similar because these mechanisms do not depend on the quark-antiquark composition of the interacting baryons. Because there are no quarks of common flavor between protons and antiprotons, quark interchange mechanisms do not contribute in $p\bar{p}$ elastic scattering while they do in $pp$ elastic scattering. Having that the cross section  of $pp$ elastic scattering is much larger than that of $p\bar{p}$, it is concluded that quark interchange dominates $pp$ elastic scattering \cite{h20}.
 
\begin{figure}[!htbs]
	\centering		\includegraphics[width=0.6\textwidth]{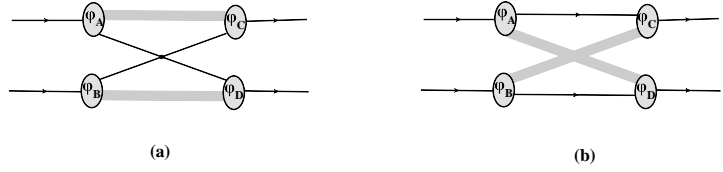}%{f2_smallx}
	\caption{Scattering channels in the quark exchange mechanism of baryon-baryon ($BB$) exclusive scattering. }
	\label{qimch}
\end{figure}

Quark exchange in baryon baryon exclusive scattering $B_AB_B\to B_CB_D$ proceeds through either of the two channels illustrated in Fig.(\ref{qimch})(a) and (b). The scenario (a) corresponds to the exchanging quarks scattering in the $u$ channel while the residual system scatters in the $t$ channel, while in (b) the residual system scatters in the $u$ channel while the `exchanging' quarks scatter in the $t$ channel. As it follows from the previous section,  at fixed angle, the energy dependence of these two contributions is the same. The angular dependence on the other hand goes as $C_tF(t/s)$ for Fig.(\ref{qimch})(a), and $C_uF(u/s)$ for Fig.(\ref{qimch})(b). While $F$ comes from the interaction among constituents, the coefficients $C_t$, and $C_u$ come from the quark wave functions of the baryons participating in the interaction. Appendix \ref{haqim} illustrates how these coefficients are obtained from a given flavor-helicity expansion of the quark wave functions of the interacting baryons. In such approach, the residual system consists of the two remaining quarks in the minimal Fock component of the wave function of the interacting baryon. 

\section{Hard Processes Involving Nuclei}
\label{HPIN}
The descriptions of nuclear processes based on QCD is met with the empirical difficulty of quarks and gluons not being experimentally observable particles. Up to certain scale they are instead `hidden' within the hadrons that mediate and take part of the interaction binding nucleons in the nucleus. The experimental constrains and the computational challenges brought on by confinement made the pursuing of a truly elementary particle description of the nucleus very impractical when compared to the alternative effective field theoretical frameworks.%{\footnote{This situation can be changed in the near future by increasing computing capabilities and improved techniques developed in lattice QCD}}. 

Nonetheless, reactions involving nuclei in the hard kinematic regime (at energy scales in which pQCD may be applicable) can probe the quark-gluon dynamics of small size nucleon-nucleon ($NN$) configurations ( small enough such that their overlapping dynamics takes the nucleons' internal structure into account). Such configurations are expected to be naturally present in dense nuclear matter.% in which scenarios of QCD degrees of freedom controlling the intra nuclear dynamics  are evidenced by phenomena in DIS experiments on heavy nuclear targets such as the EMC effect. It consists on  parton distributions  of nucleons in nuclei deviating from such distributions measured in free nucleons which indicates a phase in which the quarks and gluons inside the nucleon interact more directly (independently, incoherently) with the nuclear medium. 

Small $NN$ configurations can also be  probed in light nuclei for instance through exclusive reactions in which  nuclei absorb enough energy to break into nucleons emerging with large relative transverse momenta. This kind of reactions is the focus of the research work detailed in chapters~\ref{HBNNHe} and \ref{HBdDD} and Refs.\cite{Sargsian:2008zm} and \cite{Granados:2010cj} in which the explicit role of subnucleonic degrees of freedom is investigated in selected hard processes involving light nuclei. The following section lists the main features of characteristic approaches developed to incorporate QCD in the description of these processes.

\subsection{Hard breakup of $NN$ system in nuclei}
\label{HBNN}
In the research work described in this dissertation, we focused on the reaction $\gamma+A\to (NN)+(A-2)$ in which $(NN)$ is produced at large center of mass angle. 
Two-body breakup reactions involving nuclei at high momentum and energy transfer 
play an important role in studies of nuclear QCD. The uniqueness of these processes 
is in the effectiveness by which large values of  invariant 
energy are  produced at rather moderate values of beam energy. For a photodisintegration process of an $NN$ system we have that,
\begin{equation}
s_{\gamma NN} \approx   4m_N^2 + 4 E_\gamma m_N,
\label{s}
\end{equation}  
in which the produced invariant energy grows with the energy of the probe  twice as fast as compared, 
for example, to  hard processes involving two protons, in which case $s_{NN} = 2m_N^2 + 2E\cdot m_N$.
As it follows from Eq.(\ref{s}), already at photon energies of $2$~GeV the produced invariant 
mass on one nucleon, $M\sim {\sqrt{S_{\gamma NN}\over 2}}$, exceeds the threshold at which 
deep-inelastic processes become important, $M\gtrsim 2$~GeV.

Combining the above property with a requirement that the momentum 
transfer in the reaction exceeds the masses of the particles involved in the scattering 
($-t,-u\gg m_N^2$) in order for the reaction to reach the hard scattering kinematic regime. In this regime it is expected that  only the minimal Fock components dominate in the wave function of the particles 
involved in the scattering.  Assuming that all the constituents of minimal Fock component participate 
in a hard scattering, it is expected that the energy dependence of the reaction follows the constituent counting rule of Eqs.(\ref{qcruls}) and (\ref{dimscal}) \cite{BF75,MMT} . These predictions have been confirmed for  a wide 
variety of hard processes involving leptons and hadrons (see e.g. Refs.\cite{Owen:1969tz,Landshoff:1970ff,Akerlof,Allaby,pnexp1,pnexp2}).   

One of the most interesting aspects of the constituent-counting rule is that its application allows 
us to check the onset of quark degrees of freedom in hard reactions involving nuclei~\cite{BCh,Holt}.
This is essential for probing the quark-gluon structure of nuclei.
For example, if quarks are involved in hard photodisintegration of the deuteron then according 
to Eqs.(\ref{qcruls}) and (\ref{dimscal}) one expects that 
$\frac {d\sigma}{dt}\sim s^{-11}$~\cite{BCh}.

During the last decade there were several experiments in which 90$^0$ c.m.  photodisintegration of 
the deuteron had been studied at high photon energies 
\cite{NE8,NE17,E89012,Schulte1,gdpnpolexp1,Schulte2,Mirazita,gdpnpolexp2}.
These experiments clearly demonstrated the onset of $s^{-11}$ scaling for the differential 
cross section at 90$^0$ c.m., starting at $E_{\gamma}\ge 1$~GeV. Also, the polarization 
measurements\cite{gdpnpolexp1,gdpnpolexp2} were generally in agreement with the prediction of 
the helicity conservation -- a precursor of the dominance of the mechanism of hard gluon exchange 
involving quarks.

Even though two-body scattering experiments demonstrate clearly an onset of quark degrees of 
freedom in the reaction, they do not affirm the onset of the perturbative QCD (pQCD) regime.  
Indeed it has been argued   that  the validity of constituent-counting rule does not 
necessarily lead to the validity of pQCD(see e.g. Refs.\cite{Isgur_Smith,Rady}). In several measurements
in which the constituent quark rule works pQCD still underestimates  the observed 
cross sections sometimes by several orders of magnitude (see e.g. Refs.\cite{Farrar,BDixon}).  
The latter may indicate a substantial contribution because of nonperturbative effects although
one still may expect sizable contributions from pQCD due to generally unaccounted 
hidden color components in the hadronic and nuclear wave functions\cite{HColor}.

A similar situation also exists for the case of hard photodisintegration involving nuclei. 
Even though experiments clearly indicate the onset of $s^{-11}$ scaling for the cross section of for example $\gamma d\to pn$ reactions at 90$^0$c.m., one still expects sizable nonperturbative effects. 
Theoretical methods of calculation of these effects are very restricted.  They  
use different approaches to incorporate nonperturbative contributions in the process of 
hard photodisintegration of the deuteron.  The reduced nuclear amplitude~(RNA) formalism  
includes some of the nonperturbative effects through the nucleon form factors\cite{RNA1,RNA2}, 
while
in the quark-gluon string model~(QGS)\cite{QGS} nonperturbative effects are accounted for through the 
reggeization of scattering amplitudes. Also recently, large c.m. angle photodisintegration of the 
deuteron for photon energies up to 2~GeV was calculated within point-form relativistic quantum 
mechanics approximation\cite{MP} in which the strength of the reaction was determined by  short 
range properties of the $NN$ interaction potential.

In the QCD hard rescattering model~(HRM)\cite{gdpn,gdpnpc} it is assumed that the energetic 
photon knocks-out a quark from one nucleon in the deuteron which subsequently experiences hard 
rescattering with a quark of the second nucleon. The latter leads to the production of 
two nucleons with large relative momentum. The summation of all the relevant rescattering diagrams results 
in a scattering amplitude in that the hard rescattering is determined by the large-momentum transfer 
$pn$ scattering amplitude, which includes noncalculable nonperturbative contributions.  
Experimental data are used to estimate the  hard $pn$ scattering amplitude.  The HRM  
allows us to calculate the absolute cross section of   $90^0$ c.m.
hard photodisintegration of the deuteron  without using additional adjustable parameters.

Also,  within the QGS\cite{QGSpol} approximation and the HRM\cite{gdpnpol,hrmictp} 
rather reasonable agreement  has been  obtained for polarization 
observables\cite{gdpnpolexp2}.

\subsubsection {Hard photodisintegration of $^3$He}
\label{IHPHe}
Although all the above-mentioned models describe the major characteristics of hard 
photodisintegration  of the deuteron
they are based on very different approaches in the calculation of the nonperturbative 
parts of the photodisintegration reaction.  To investigate further the validity of these 
approaches it was suggested in  Ref.\cite{gheppn} to extend the studies of high energy two-body 
photodisintegration to the case of large angle c.m. breakup of two protons from the $^3$He target. 
In this case not only do the predictions of the above-described models (RNA, QGS, HRM) for 
absolute cross section diverge significantly, but also the  
two-proton breakup reaction from $^3$He  provides additional observables such as 
spectator-neutron momentum distributions that can be used to check further the 
validity of the models. 

Detailed  analysis of reactions involving  hard breakup of 
both $pp$ and $pn$ pairs from the $^3$He target is presented in chapter \ref{HBNNHe}.  New insight into the nature of large c.m. angle scattering is gained through the comparative study of 
$pp$ and $pn$ 
breakup processes.
One important observation is that the relative 
strength of $pp$ to $pn$ breakup is 
larger than the one observed in low energy reactions. This characteristic is related to 
the onset of quark degrees of freedom in hard breakup reactions in which effectively more 
charges are exchanged between two protons than between proton and neutron. 

Another signature of the HRM is that the shapes of the  energy dependencies of $s^{11}$-scaled differential cross sections 
of $pp$ and $pn$ breakup reactions  mirror the shapes of the energy dependencies of $s^{10}$-scaled differential 
cross sections of hard elastic $pp$ and $pn$  scatterings. 

Within the HRM  one observes also that $pp$ and $pn$ hard breakup processes are sensitive to different 
components of the $^3$He ground state wave function, resulting in different spectator-nucleon 
momentum dependencies for $pp$ and $pn$ hard breakup cross sections.

Because of the different ground state wave function components involved in $pp$ and $pn$ breakup reactions,
the HRM also predicts significantly different magnitudes for  transferred longitudinal polarizations for these 
two processes.

\medskip
\medskip

%\subsection{Deuteron Breakup}

% With the strong interaction being symmetric under the $SU(2)$-isospin group of transformations, a Lagrangian of a field theory of this interaction can be constrained to include only terms that as a whole make it invariant under $SU(2)$-isospin ($SU(2)_I$) as  well, terms such as,
 %\begin{eqnarray}
  %g{\hat\psi}\varphi\psi,
  %\label{lterm}
  %\end{eqnarray}
  %where ${\bar\psi}$ and $\psi$  represent two baryons, while $\varphi$ represents a meson. 

% An important observation from  this classification in Table \ref{tab:HadronFamilies} is that hadrons with increasing $S^2$ have consistently larger masses. This should have been an early indicator of an internal structure in hadrons that can be potentially resolved by the strong interaction.   

%%% ----------------------------------------------------------------------

%%% Local Variables: 
%%% mode: latex
%%% TeX-master: "../thesis"
%%% End: 

\chapter{PROTON NEUTRON ELASTIC SCATTERING}
\label{PNES}
\ifpdf
    \graphicspath{{Chapter1/Chapter1Figs/PNG/}{Chapter1/Chapter1Figs/PDF/}{Chapter1/Chapter1Figs/}}
\else
    \graphicspath{{Chapter1/Chapter1Figs/EPS/}{Chapter1/Chapter1Figs/}}
\fi

%\begin{abstract}
This chapter looks into an asymmetry in the angular distribution 
of hard elastic proton-neutron scattering with respect to 90$^0$ center of 
mass scattering angle. 
It will be  demonstrated that the magnitude of the angular asymmetry is related 
to  the helicity-isospin symmetry of the quark wave function of the nucleon.
An estimate of the asymmetry within the quark-interchange model
of hard scattering demonstrates that the quark wave function of a nucleon 
assuming the exact SU(6) symmetry predicts an angular asymmetry opposite 
to that of  experimental observations \cite{Granados:2009jh}. On the other hand the quark wave 
function derived from the diquark picture of  the  nucleon produces an 
asymmetry  consistent with the data. Comparison with the data allows 
extracting  the relative sign and the magnitude of the vector and scalar 
diquark components of the  quark wave function of the nucleon. These 
two quantities are essential in constraining QCD models of a nucleon.
Overall, it is concluded that the angular asymmetry of a hard elastic  
scattering of baryons provides a new venue in probing 
quark-gluon structure of baryons and should be considered as an 
important  observable in constraining the theoretical models.
%\end{abstract}

For several decades elastic nucleon-nucleon scattering at high momentum transfer 
($-t,-u\ge  M_N^2$~GeV$^2$) has been one  of the important testing grounds for 
QCD dynamics of the strong interaction between hadrons.   Two major observables 
considered were the energy dependence of the elastic cross section 
and the polarization properties of the reaction.

Predictions for energy dependence are determined by the underlying dynamics of 
the hard scattering of quark components of the  nucleons. 
One such prediction is derived from the quark-counting rule
\cite{BF75,MMT} according to which the differential 
cross section of two-body elastic scattering ($ab\rightarrow cd$) 
at high momentum transfer behaves like 
${d\sigma\over dt} \sim s^{-(n_a + n_b + n_c + n_d)}$, where 
$n_i$ represents the number of constituents in particle $i$ (i=a,b,c,d).

For elastic $NN$ scattering,  the quark-counting rule  predicts 
$s_{NN}^{-10}$ scaling which agrees reasonably well 
with experimental measurements 
(see e.g. Refs.\cite{Akerlof,Allaby,pnexp1,pnexp2}). 
In addition to energy dependence, the comparison~\cite{h20} of the cross 
sections of hard exclusive scattering of hadrons containing quarks  
with the same  flavor  with the scattering of hadrons that share no 
common flavor of quarks demonstrated that the quark-interchange represents the 
dominant mechanism of hard elastic scattering for up to ISR energies 
(see discussion in \cite{BCL}).

For polarization observables, the  major prediction of the QCD dynamics 
of hard elastic scattering 
is the conservation of helicities of  interacting hadrons. The latter 
prediction stems from the fact that the gluon exchange  
in massless quark limit conserves the helicity of interacting quarks.

The quark counting rule and helicity conservation however do not describe 
completely the features of hard scattering data. The energy dependence of 
$pp$ elastic cross section scaled by $s_{NN}^{10}$ exhibits an oscillatory 
behavior which indicates the existence of other possibly nonperturbative mechanisms 
for the scattering\cite{pire,BT}.   These expectations are reinforced also 
by the observed large asymmetry, $A_{nn}$ at some 
hard scattering kinematics\cite{Krabb} which 
indicates  an anomalously large contribution from double helicity flip 
processes.
These  observed discrepancies however do not represent the dominant 
features of the data and overall one can conclude that  the bulk of the 
hard elastic $NN$ scattering amplitude  is defined by the exchange mechanism of  
valence quarks which interact through the hard gluon exchange 
(see e.g. Refs.\cite{FGST,BCL}). Quark-interchange mechanism also reasonably well 
describes the $90$ c.m. hard  break-up  of two nucleons from the deuteron\cite{gdpn,gdpnpol}.

However,  the energy dependence of a hard scattering cross section, 
except for the verification of the   
dominance of  the minimal-Fock component of the quark wave function of 
nucleon,  provides rather limited information about the symmetry properties of 
the valence quark  component of the nucleon wave function.

In this chapter it is demonstrated that an  observable such as the asymmetry of 
a hard elastic proton-neutron scattering with respect to $90^0$ c.m. scattering may 
provide a new insight into the  helicity-flavor symmetry of 
the quark wave function  of the nucleon. Namely we consider
\begin{equation}
A_{90^0}(\theta) = 
{\sigma(\theta) - \sigma(\pi -\theta)\over \sigma(\theta) + \sigma(\pi-\theta)},
\label{Asym}
\end{equation}
where $\sigma(\theta)$ - is the differential cross section of the elastic 
$pn$ scattering. 
We will discuss this asymmetry in the hard kinematic regime 
in which the energy dependence of the cross section is $\sim s^{-10}$.  
Our working assumption is the dominance of the quark-interchange  
mechanism~(QIM) in the $NN$ elastic scattering at these kinematics. 

\begin{figure}[ht]
\centering\includegraphics[scale=0.6]{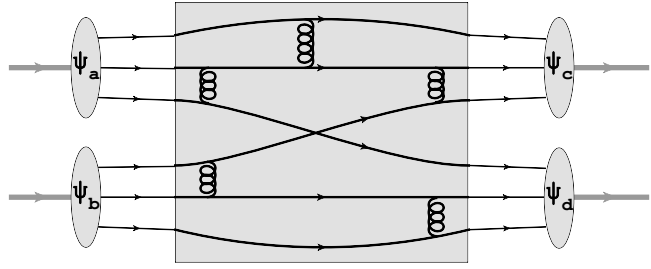}
\vspace{-0.2cm}
\caption{Typical diagram for quark-interchange mechanism of $NN\rightarrow NN$ scattering.}
\label{QIM}
\end{figure}

Within the QIM, the characteristic diagram for $pn$ elastic scattering can be  represented as it's shown in Fig.\ref{QIM}.
Here, one assumes a factorization of the soft part of the reaction 
in the form of the initial and final state wave functions of nucleons, and of the hard 
part which is characterized by the QIM scattering that proceeds with five hard gluon exchanges. 
This hard factor generates the energy dependence in accordance to the quark counting rule.  
In order  to calculate the absolute cross section of the reaction, 
one needs to sum hundreds of diagrams similar to one of Fig.\ref{QIM}. 
However, for the purpose of estimation of the asymmetry in Eq.(\ref{Asym}) the 
important observation is that the hard scattering kernel is flavor-blind and 
conserves helicity. As a result, one expects that the angular asymmetry will 
be generated mainly through the underlying spin-flavor symmetry of the quark wave functions 
of the  interacting nucleons.

The amplitude of the hard elastic $a+b\rightarrow c+d$ scattering of Fig.\ref{QIM}, 
within quark-interchange approximation, can  be presented as follows:
%\begin{widetext}
\begin{eqnarray}
\langle cd\mid T\mid ab\rangle & = &  
\sum\limits_{\alpha,\beta,\gamma} 
\langle  \psi^\dagger_c\mid\alpha_2^\prime,\beta_1^\prime,\gamma_1^\prime\rangle
\langle  \psi^\dagger_d\mid\alpha_1^\prime,\beta_2^\prime,\gamma_2^\prime\rangle 
\nonumber \\
& & \times
\langle \alpha_2^\prime,\beta_2^\prime,\gamma_2^\prime,\alpha_1^\prime\beta_1^\prime
\gamma_1^\prime\mid H\mid
\alpha_1,\beta_1,\gamma_1,\alpha_2\beta_2\gamma_2\rangle\cdot 
\langle\alpha_1,\beta_1,\gamma_1\mid\psi_a\rangle
\langle\alpha_2,\beta_2,\gamma_2\mid\psi_b\rangle,\nonumber\\
\label{ampl}
\end{eqnarray}
%\end{widetext}
in which ($\alpha_i, \alpha_i^\prime$), ($\beta_i,\beta_i^\prime$) and ($\gamma_i,\gamma_i^\prime$) 
describe the spin-flavor quark states before and after the hard 
scattering, $H$,  and 
\begin{equation}
C^{j}_{\alpha,\beta,\gamma} \equiv \langle\alpha,\beta,\gamma\mid\psi_j\rangle,
\label{Cs}
\end{equation}
describes the probability amplitude of finding the $\alpha,\beta,\gamma$ helicity-flavor 
combination of three valence quarks in the nucleon $j$\cite{FGST}.

To be able to calculate the $C^{j}_{\alpha,\beta,\gamma}$ factors, one  
represents the nucleon wave function through the helicity-flavor basis of the valence  
quarks.  We use a rather general form  separating  the wave function into two parts characterized 
by two (e.g. second and third)  quarks being  in spin zero - isosinglet and spin one - isotriplet states 
as follows:
%\begin{widetext}
\begin{eqnarray}
\psi^{i^3_{N},h_N} & = & {N\over \sqrt{2}}\left\{
\sigma (\chi_{0,0}^{(23)}\chi_{{1\over2},h_N}^{(1)})\cdot
(\tau_{0,0}^{(23)}\tau_{{1\over 2},i_N^{3}}^{(1)}) 
\right.  +   \nonumber \\
& & 
\rho \sum\limits_{i_{23}^3=-1}^{1} \ \ \sum\limits_{h_{23}^3=-1}^{1}
\langle 1,h_{23}; {1\over 2},h_{N}-h_{23}\mid {1\over 2},h_N\rangle
\langle 1,i^3_{23}; {1\over 2},i^3_{N}-i^3_{23}\mid {1\over 2},i^3_N\rangle \nonumber \\
& &\left. \times (\chi_{1,h_{23}}^{(23)}\chi_{{1\over2},h_N-h_{23}}^{(1)})\cdot
(\tau_{1,i^3_{23}}^{(23)}\tau_{{1\over 2},i_N^{3}-i^3_{23}}^{(1)})\right\},
\label{wf}
\end{eqnarray}
%\end{widetext}
in which $j_N^3$ and $h_N$ are the isospin component  and the helicity of the nucleon.
Here, the $k_i$'s are the light cone momenta of quarks. These momenta is represented by  
($x_i,k_{i\perp}$), in which $x_i$ is a light cone momentum fraction of the nucleon 
carried by the $i$-quark. 
We define $\chi_{j,h}$ and $\tau_{I,i^3}$ as helicity 
and isospin  wave functions, where $j$ is the spin, $h$ is the helicity, 
$I$ is the isospin and $i^3$ its third component.
The Clebsch-Gordan coefficients are 
defined as $\langle j_1,m_1;j_2,m_2\mid j,m\rangle$. 
Here, $\Phi_{I,J}$ represents the momentum dependent part of the wave function 
for ($I=0,J=0$) and ($I=1,J=1$) two-quark spectator states respectively.
Since the asymmetry in Eq.(\ref{Asym}) does not 
depend on the absolute normalization of the cross section,  
a more relevant quantity for us is
the relative strength of these two
momentum dependent wave functions.  For our discussion we introduce a 
parameter, $\rho$: 
\begin{equation}
\rho = {\langle \Phi_{1,1}\rangle \over \langle \Phi_{0,0}\rangle }
\label{rho}
\end{equation}
which characterizes an average relative  magnitude of 
the wave function components corresponding to 
($I=0,J=0$) and ($I=1,J=1$) quantum  numbers  of two-quark ``spectator'' states.
Note that the two extreme values of $\rho$ define two well know approximations:
$\rho=1$ corresponds to the exact SU(6) symmetric picture of the nucleon 
wave function  and 
$\rho=0$ will correspond to the contribution of only the scalar diquark configuration 
in the nucleon wave function (see e.g. Ref.\cite{Ansel,RJ,SW,BCR} in which this component is referred as 
a scalar or good diquark configuration~($[qq]$) as opposed to a vector or bad diquark 
configuration denoted by  $(qq)$).  
In further discussions $\rho$ is kept as a free parameter.

To calculate the scattering amplitude of Eq.(\ref{ampl}), we assume the 
conservation of the helicities of the quarks participating in the hard scattering. 
This allows us to approximate the hard scattering part of the amplitude, $H$,  
in the following form:
\begin{equation}
H \approx \delta_{\alpha_1\alpha_1^\prime}\delta_{\alpha_2\alpha_2^\prime}
\delta_{\beta_1,\beta_1\prime}
\delta_{\gamma_1,\gamma_1^\prime}
\delta_{\beta_2,\beta_2\prime}
\delta_{\gamma_2,\gamma_2^\prime} {f(\theta)\over s^4}.
\label{H}
\end{equation}
Inserting this expression into Eq.(\ref{ampl}) for the QIM 
amplitude, one obtains\cite{FGST}:
\begin{equation}
\langle cd\mid T\mid ab\rangle = Tr(M^{ac}M^{bd})
\label{Tampl2}
\end{equation}
with:
\begin{equation}
M^{i,j}_{\alpha,\alpha^\prime} = 
C^{i}_{\alpha,\beta\gamma}C^{j}_{\alpha^\prime,\beta\gamma} + 
C^{i}_{\beta\alpha,\beta}C^{j}_{\beta\alpha^\prime,\beta} + 
C^{i}_{\beta\gamma\alpha}C^{j}_{\beta\gamma\alpha^\prime},
\label{QIMMs}
\end{equation} 
where we sum over the all possible values of $\beta$ and $\gamma$.
Furthermore, we separate the energy dependence from the scattering amplitude as
follows:
\begin{equation}
\langle cd\mid T\mid ab\rangle = {\langle h_c,h_d\mid T(\theta)\mid h_a,h_b\rangle \over s^4},
\label{phidef}
\end{equation}
and define five independent angular parts of the helicity amplitudes as:  
\begin{eqnarray}
& & \phi_1 = \langle ++\mid T(\theta)\mid ++\rangle; \ \ 
 \phi_2 =  \langle --\mid T(\theta)\mid ++\rangle; \nonumber \\ 
& &  \phi_3 =  \langle +-\mid T(\theta)\mid +-\rangle; \ \
\phi_4 =  - \langle -+\mid T(\theta)\mid +-\rangle; \nonumber \\
& & \phi_5 =  \langle -+\mid T(\theta)\mid ++\rangle.
\label{phis}
\end{eqnarray}
Here, the ``-'' sign in the definition of $\phi_4$ follows from the 
Jacob-Wick helicity convention\cite{JW} according to which a (-1) phase 
is introduced if two quarks that scatter to $\pi-\theta_{cm}$ angle have
opposite helicity (see also Ref.\cite{FGST}).

Using Eqs.(\ref{Tampl2},\ref{QIMMs})  for the non-vanishing helicity 
amplitudes of Eq.(\ref{phis}) one obtains:\\
for $pp\rightarrow pp$:
\begin{eqnarray}
\phi_1 & = &    (3 + y)F(\theta)    +  (3 + y)  F(\pi-\theta) \nonumber \\
\phi_3 & = &    (2 - y)F(\theta)    +  (1 + 2y) F(\pi-\theta) \nonumber  \\
\phi_4 & = &   -(1 + 2y)F(\theta)  -  (2 - y)  F(\pi-\theta)
\label{pppp}
\end{eqnarray}
and for  $pn\rightarrow pn$:
\begin{eqnarray}
\phi_1 & = &  (2 - y)F(\theta)   +  (1 + 2y)  F(\pi-\theta) \nonumber \\
\phi_3 & = &  (2 + y)F(\theta)   +  (1 + 4y) F(\pi-\theta) \nonumber \\
\phi_4 & = &   2y F(\theta)      +     2y  F(\pi-\theta)
\label{pnpn}
\end{eqnarray}
with $\phi_2=\phi_5=0$ due to helicity conservation. Here:
\begin{equation}
y = x(x+1) \ \ \mbox{with } x = {2\rho \over 3(1+\rho^2)} 
\label{xy}
\end{equation}
and $F(\theta)$ is the angular function. Note that the 
$\rho=1$ case reproduces the SU(6) result of Refs.\cite{FGST} and \cite{BCL}.
The results of Eqs.(\ref{pppp}) and (\ref{pnpn}) could be obtained 
also through the formalism of the H-spin introduced in Ref.\cite{BCL}. 
In this case, the helicity amplitudes are expressed through the average number 
of quarks to be found in a given helicity-spin state. These numbers will 
be directly defined through the wave function  of Eq.(\ref{wf}).

Introducing the symmetric and antisymmetric parts of the angular function $F$ as follows:
\begin{equation}
s(\theta) = {F(\theta) + F(\pi-\theta)\over 2}; \ \ 
a(\theta) = {F(\theta) - F(\pi-\theta)\over 2},
\label{as}
\end{equation}
and using  Eq.(\ref{pnpn}) for the asymmetry as it is defined in Eq.(\ref{Asym}) one obtains:
\begin{equation}
A_{90^0}(\theta) = {6 a(\theta)s(\theta)(1-2y - 3y^2)\over 
a(\theta)^2 (1-3y)^2 + 3s(\theta)^2(3 + 6y + 7y^2)}.
%hayr mer
\label{asym}
\end{equation}

\begin{figure}[t]
\centering\includegraphics[scale=0.4]{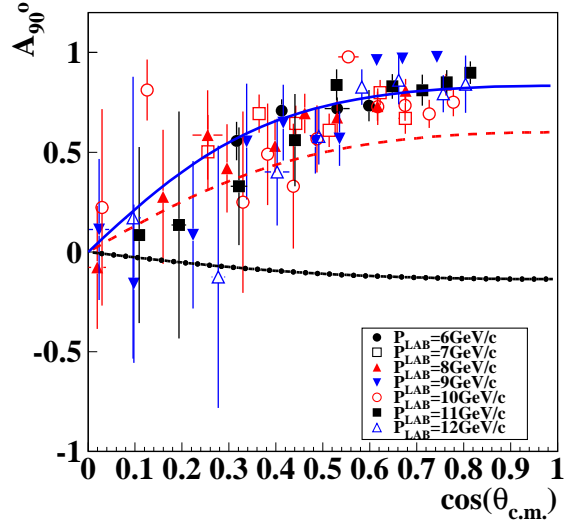}
\vspace{-0.1cm}
\caption{Asymmetry of $pn$ elastic cross section. Solid dotted line - SU(6), with $\rho=1$,
dashed line diquark-model with $\rho=0$, solid line - fit with $\rho=-0.3$.}
\label{Fig.2}
\end{figure}
One can make a rather general observation from  Eq.(\ref{asym}), that for the 
SU(6) model, ($\rho=1$, $y = {4\over 9}$) and for any positive function, $a(\theta)$ at 
$\theta \le {\pi\over 2}$, the  angular asymmetry  has a negative sign opposite 
to the  experimental asymmetry~(Fig.\ref{Fig.2}).  Note that one expects 
a positive $a(\theta)$ at $\theta \le {\pi\over 2}$ from general grounds based on 
the expectation that  in the hard scattering regime the number of $t$-channel 
quark scatterings  dominates the number of $u$-channel quark scatterings  in the forward 
direction.  

As it follows from Eq.(\ref{asym}), a positive asymmetry can be achieved 
only for $1-2y-3y^2 >0$, which according to Eq.(\ref{xy}) imposes the following restrictions 
on $\rho$: $\rho<0.49$ or $\rho>2.036$. 
The first condition indicates the preference for the scalar diquark-like configurations in the 
nucleon wave function, while the second one will indicate the strong dominance of the vector-diquark 
component which contradicts empirical observations\cite{Ansel,RJ,SW}.

In Fig.\ref{Fig.2} the asymmetry 
of $pn$ scattering calculated with SU(6)~($\rho=1$) and pure scalar-diquark~($\rho=0$) models are 
compared with the data.  In these estimates we use 
$F(\theta) = C\cdot sin^{-2}(\theta)(1-cos(\theta))^{-2}$ 
as the dependence of the angular function\cite{RS}, which is consistent with the 
picture of hard collinear QIM scattering of valence quarks with five gluon exchanges, and  
reproduces reasonably well the main characteristics  of the  angular dependencies  of 
both $pp$ and $pn$ elastic scatterings. Note that using a form of the angular function 
based on nucleon form-factor arguments\cite{BCL,FGST}, $F\approx (1-cos(\theta))^{-2}$
will result in the same angular asymmetry.

The comparisons show that the nucleon wave function~(\ref{wf}) with a scalar 
diquark component~($\rho=0$) produces the right sign for the angular asymmetry.
On the other hand,  even large errors of the data do not 
preclude to conclude that the exact SU(6) symmetry~($\rho=1$) of the quark wave 
function of nucleon is in qualitative disagreement with the experimental asymmetry.

Using the above defined angular function $F(\theta)$,  $A_{90^0}$ is fitted in  
Eq.(\ref{asym}) to the data at  $-t,-u\ge 2$~GeV$^2$ varying $\rho$ as a free parameter.  
We used the Maximal Likelihood method of fitting excluding those data points from the data set  
whose errors are too large for meaningful identification of the asymmetry. 
The best fit is found for 
\begin{equation}
\rho \approx  -0.3\pm 0.2.
\label{rhofit}
\end{equation}
The nonzero magnitude of $\rho$ indicates  the small but finite relative strength of 
a bad/vector diquark  configuration in the nuclear wave function as compared 
to the scalar diquark component.   
It is intriguing that the obtained magnitude of $\rho$ is consistent with  the $10\%$ 
probability of ``bad'' diquark configuration discussed in Ref.\cite{SW}.

Another interesting property of Eq.(\ref{rhofit}) is the negative sign of the  
parameter $\rho$.

Within  a qualitative  quantum-mechanical picture, the 
negative sign of $\rho$  may indicate for example the 
existence of a repulsion in the quark-(vector- diquark) channel as opposed to 
the attraction in the quark - (scalar-diquark) channel.  It is rather surprising that 
both the magnitude and sign agree with the result of the phenomenological interaction 
derived in the one-gluon exchange quark model discussed in Ref.\cite{RJ}.

In conclusion, we demonstrated that the angular asymmetry of hard elastic $pn$ 
scattering can be used to probe the symmetry structure of the valence quark wave 
function of the nucleon. We demonstrated that the exact SU(6) symmetry does not reproduce 
the experimental angular asymmetry of hard elastic $pn$ scattering.
The use of nucleon wave functions consistent with the diquark structure results in an asymmetry in better agreement with the empirically observed. The fit to the data  indicates  $10\%$ probability 
for the existence of bad/vector diquarks in the wave function of nucleons.
It also  shows that the vector  and scalar $qq$ components of the  wave function  may be 
in  opposite phase.  This will indicate a different dynamics for 
$q-[qq]$ and $q-(qq)$ interactions.

The relative magnitude and the sign of the 
vector $(qq)$ and scalar $[qq]$
components can be used to constrain different QCD  predictions which require the existence 
of diquark components in the nucleon wave function. These quantities in  principle  can be 
checked in Lattice calculations.  The angular asymmetry studies can be extended also to include 
the scattering of other baryons such as  $\Delta$-isobars (which may have a larger fraction of 
vector diquark component), as well  as strange baryons which will allow us to study the relative 
strength of $(qq)$ and $[qq]$ configurations involving strange quarks.
% ------------------------------------------------------------------------

%%% Local Variables: 
%%% mode: latex
%%% TeX-master: "../thesis"
%%% End: 

\chapter{HARD BREAKUP OF A NUCLEON NUCLEON SYSTEM IN THE $^3$He NUCLEUS}
\label{HBNNHe}
\ifpdf
    \graphicspath{{Chapter2/Chapter2Figs/PNG/}{Chapter2/Chapter2Figs/PDF/}{Chapter2/Chapter2Figs/}}
\else
    \graphicspath{{Chapter2/Chapter2Figs/EPS/}{Chapter2/Chapter2Figs/}}
\fi

%\begin{abstract}
This chapter  investigates the large angle photodisintegration of 
two nucleons from the $^3$He nucleus within the framework of 
the hard rescattering model (HRM). In the HRM a  quark of one nucleon 
knocked out by an incoming photon rescatters with a quark of the 
other nucleon leading to the production of two nucleons with large
relative momentum.  Assuming the dominance of the quark-interchange mechanism 
in a hard nucleon-nucleon scattering, the HRM allows the expression of the amplitude of a  two-nucleon break-up reaction through the convolution of photon-quark 
scattering, $NN$ hard scattering amplitude and  nuclear spectral function 
which can be calculated using a nonrelativistic  $^3$He wave function.  
The photon-quark scattering amplitude can be explicitly calculated in the high energy regime, 
whereas for $NN$ scattering one uses the fit of the available experimental data.  
The HRM predicts several specific features for the hard breakup reaction. First, the 
cross section will approximately scale as $s^{-11}$. Secondly, the $s^{11}$ weighted 
cross section will have the shape of energy dependence similar to that of $s^{10}$ weighted $NN$ 
elastic scattering cross section. 
Also one predicts an enhancement of the $pp$ breakup relative to  
the $pn$ breakup cross section as compared to the results from low energy kinematics.
Another result is the prediction of different spectator momentum dependencies of $pp$ 
and $pn$ breakup cross sections. This is because of the fact that same-helicity $pp$-component 
is strongly suppressed in the ground state wave function of $^3$He. Because of this suppression 
the HRM predicts significantly different asymmetries for the cross section of polarization 
transfer $NN$ breakup  reactions for circularly polarized photons.  For the $pp$ breakup this 
asymmetry is predicted to be zero while for the $pn$ it is close to ${2\over 3}$.

%\end{abstract}

\medskip
\medskip

This chapter is organized as follows:  Section~\ref{II}, within the HRM,  presents a detailed 
derivation of the differential cross section of   the reaction of
hard breakup of two-nucleons from a $^3$He target.  In Section~\ref{III}  the formulas derived in the previous section are used to calculate  the differential cross section of a 
proton-neutron breakup reaction, while in Section~\ref{IV} calculations are done for a 
two-proton breakup reaction. Section~\ref{V} considers 
the relative contribution of two- and three-body processes for hard breakup reactions involving 
$A\ge 3$ nuclei. In Section~\ref{VI}  numerical estimates 
are presented for  differential cross sections of $pn$ and $pp$ breakup reactions. 
In Section~\ref{VII}  the polarization transfer mechanism of the HRM is discussed, and estimates of the 
asymmetry of the cross section with respect to the helicity of the 
outgoing proton are presented. Results are summarized in Section~\ref{VIII}.  

The details of the derivation of the 
hard rescattering amplitude are given in Appendix~\ref{app1}. The quark-interchange 
contribution to the hard $NN$ elastic scattering amplitude is discussed in Appendix~\ref{app2}.  Appendix~\ref{haqim} describes a method for calculating 
quark-charge factors within quark-interchange mechanism of $NN$ hard elastic scattering.
A complete list of  HRM helicity amplitudes for high energy two-nucleon breakup is presented in Appendix~\ref{HAP} for both, deuteron and $^3$He photodisintegration.

\section{Hard Photodisintegration of Two Nucleons from $^3$He}
\label{II}

\subsection{ Reference frame and kinematics}
\label{RFrame}

We are considering a hard photodisintegration of two nucleons  from the $^3$He  
target through the reaction:
\begin{equation}
\gamma + ^3\textnormal{He} \rightarrow (NN) + N_s,
\label{gheNN_N}
\label{reaction}
\end{equation} 
in which two nucleons $(NN)$ are produced at large angles in the ``$\gamma$-$NN$'' center of 
mass reference frame with momenta comparable to the momentum of the initial photon, 
$q$~($>$$1$~GeV/$c$). 
The third nucleon, $N_s$, is produced  with very small momentum $p_s\ll m_N$. 
(Definitions of four-momenta involved in the reaction are given in Fig.\ref{Fig.1}.)

We consider  ``$\gamma$-$NN$'' in a ``$q_+=0$'' reference frame, where 
the light-cone momenta\footnote{The light-cone four-momenta are 
defined as $(p_+,p_-,p_\perp)$, where $p_{\pm} = E\pm p_z$. Here the $z$ axis is defined 
in the direction opposite to the incoming photon momentum.}  of the photon and the $NN$ pair
are defined as follows:
\begin{eqnarray}
 && q^\mu \equiv (q_+,q_-,q_\perp) =   (0, \ \sqrt{s_{NN}'},\ 0), \nonumber \\
 && p_{NN}^\mu \equiv (p_{NN+},p_{NN-},p_{NN\perp}) = (\sqrt{s_{NN}^\prime}, \ 
{M_{NN}^2\over \sqrt{s_{NN}^\prime}}, \ 0),
\label{q+=0}
\end{eqnarray}
where $p_{NN}^\mu=p_{^3\textnormal{He}}^\mu-p_s^\mu$, $M_{NN}^2 = p_{NN}^\mu p_{NN,\mu}$, 
and   $s_{NN}^\prime = s_{NN} - M_{NN}^2$. Here the invariants, $s_{NN}$ and 
$t_{NN}$ are defined as follows:
\begin{eqnarray}
s_{NN} = (q+p_{NN})^2 = (p_{f1}+p_{f2})^2 \nonumber \\ 
t_{NN} = (q - p_{f1})^2 = (p_{f2}-p_{NN})^2.
\label{mans_st}
\end{eqnarray}

As it follows from Eq.(\ref{q+=0}) in the limit of ${M_{NN}^2\over s_{NN}^\prime}\rightarrow 0$ the 
``$q_{+}=0$'' reference frame coincides with  the center of mass frame of the $\gamma$-$NN$ system.
As such it is maximally close to the 
reference frame used  for the $\gamma d\rightarrow p+n$ reaction in Refs.\cite{gdpn} and \cite{gdpnpol}.

\begin{figure}[t]
\centering\includegraphics[height=8.6cm,width=8.6cm]{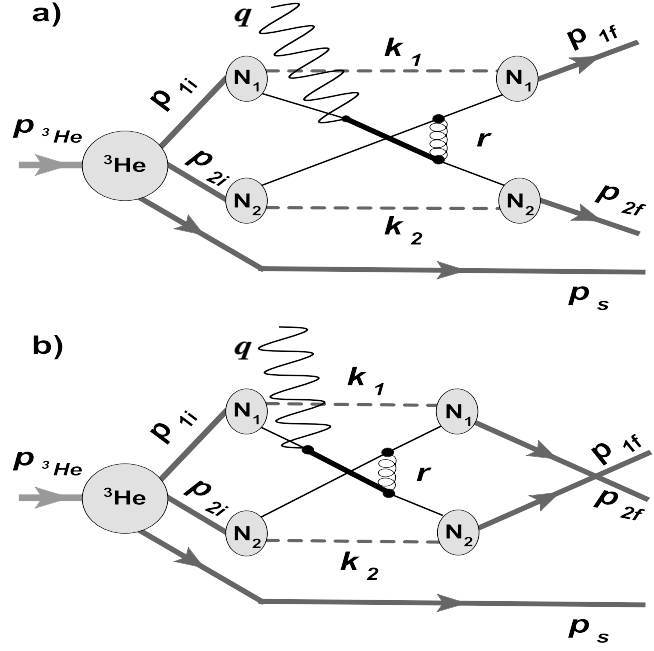}
\caption{ Typical hard rescattering diagram for the  $NN$ photodisintegration from the $^3$He target.}
\label{Fig.1}
\end{figure}

\subsection{Hard rescattering model}
\label{Sec.2B}

The hard rescattering model is based on the assumption that in the hard two-nucleon photodisintegration 
reaction, two nucleons with large relative momenta are produced  because the hard rescattering of a fast quark from one nucleon with a quark from the other nucleon.
In this scenario the fast quark is knocked out from a low-momentum nucleon in the nucleus
by an incoming  photon.  
This approach is an alternative to the models in which it is assumed that the incoming 
photon breaks the preexisting  two-nucleon state which has very large relative momentum in 
the nucleus.

The validity of the HRM is derived from the observation that the ground state wave functions of light nuclei 
peak strongly at small momenta of bound nucleons, $p\sim 0$. Thus,  diagrams in which an energetic 
photon interacts
with  bound nucleons of small momenta will strongly dominate the diagrams in which  the photon 
interacts with bound  nucleons that have  relative momenta $p\ge  m_N$.

The resulting scenario that the HRM sketches out is as follows (see e.g. Fig.\ref{Fig.1}): first, the  
incoming photon will  knock out 
a quark from one of the nucleons in the nucleus and then the struck quark that now carries almost 
the whole
momentum of the photon  will share its momentum 
with a quark from the other nucleon  through the exchanged gluon.  The resulting two energetic 
quarks will recombine with the residual quark-gluon systems to produce two nucleons with large relative 
momentum ~($\sim$$q$). This recombination will contain  gluon exchanges and also 
incalculable nonperturbative interactions.

Note that for the quark-gluon picture discussed above to be relevant the intermediate masses 
$m_{\textnormal{\scriptsize int}}$ produced after 
the photon absorption should exceed the  mass scale  characteristic for deep inelastic scattering, 
$W\sim 2.2$~GeV.
Using the relation $m_{\textnormal{\scriptsize int}}\approx \sqrt{m_N^2 + 2E_{\gamma}m_N}$, from the requirement that 
$m_{\textnormal{\scriptsize int}}\ge W$ 
one obtains the condition $E_\gamma\ge 2$~GeV. Additionally, to ensure the validity of quark degrees 
of freedom in the final state rescattering, one requires  $k_{\textnormal{\scriptsize rel}}\ge 1$~GeV/$c$
for the relative momentum $k_{\textnormal{\scriptsize rel}}$, of two outgoing 
nucleons. All of these imposes a restriction on the incoming photon energy,
$E_\gamma \ge 2$~GeV, and for transferred momenta $-t,-u\ge 2$~GeV$^2$. 
Note that provided a smooth transition from hadronic to quark-gluon degrees of freedom 
in nuclei one expects the validity of the HRM to extend to even lower values of $E_\gamma$~
($\gtrsim$$1$~GeV). This expectation was confirmed in recent measurements of angular dependencies 
of the $\gamma d\rightarrow pn$ cross section for a wide range of incoming photon energies\cite{Mirazita}.

\medskip
\medskip

To calculate the differential cross section of the hard photodisintegration reaction of 
Eq.(\ref{gheNN_N}) within  the HRM one needs to evaluate the sum of   hard rescattering diagrams 
similar to the one presented in Fig.\ref{Fig.1}.
We start with analyzing the scattering amplitude corresponding to the 
diagrams of Fig.\ref{Fig.1}. Using Feynman rules and applying the light-cone 
wave function reduction described in Appendix~\ref{app1},  we obtain
\begin{eqnarray}
& & \langle\lambda_{1f},\lambda_{2f},\lambda_s\mid A \mid \lambda_\gamma, \lambda_A\rangle = 
\sum\limits_{(\eta_{1f},\eta_{2f}),(\eta_{1i},\eta_{2i}),(\lambda_{1i},\lambda_{2i})} \int 
\left\{ {\psi^{\dagger\lambda_{2f},\eta_{2f}}_N(p_{2f},x'_2,k_{2\perp})\over 1-x'_2}\bar 
u_{\eta_{2f}}(p_{2f}-k_2) \right. 
\nonumber \\ & & 
[-igT^F_c\gamma^\nu]
{i[\sh p_{1i}-\sh k_1 + \sh q + m_q]\over (p_{1i}-k_1+q)^2-m_q^2 + i\epsilon}
[-iQ_ie{\bf \epsilon^{\lambda_\gamma}_\perp\gamma^\perp}]u_{\eta_{1i}}(p_{1i}-k_1)
\left. {\psi_N^{\lambda_{1i},\eta_{1i}}(p_{1i},x_1,k_{1\perp})\over (1-x_1)}\right\}_1\times
\nonumber \\ & & 
\left\{ 
{\psi^{\dagger\lambda_{1f},\eta_{1f}}_N(p_{1f},x'_1,k_{1\perp})\over 1-x'_1}
\right.  
 \bar u_{\eta_{1f}}(p_{1f}-k_1)[-igT^F_c\gamma^\mu]u_{\eta_{2i}}(p_{2i}-k_2) 
\left . {\psi_N^{\lambda_{2i},\eta_{2i}}(p_{2i},x_2,k_{2\perp})\over (1-x_2)} \right\}_2\times
\nonumber \\ & &
G^{\mu,\nu}(r) {dx_1\over x_1}{d^2k_{1\perp}\over 2 (2\pi)^3}
{dx_2\over x_2}{d^2k_{2\perp}\over 2 (2\pi)^3} 
{\Psi_{^3\textnormal{\scriptsize {He}}}^{\lambda_A,\lambda_{1i},\lambda_{2i},\lambda_s}(\alpha,{p_\perp},p_s)
\over (1-\alpha)}{d\alpha\over \alpha}
{d^2p_{\perp}\over 2(2\pi)^3}  -  \left(p_{1f}\longleftrightarrow p_{2f}\right),
\label{ampl0}
\end{eqnarray}
where the $\left(p_{1f}\longleftrightarrow p_{2f}\right)$ part accounts for the diagram in 
Fig.\ref{Fig.1}(b).
Here the four-momenta, $p_{1i}$, $p_{2i}$, $p_s$, $k_1$, $k_2$, $r$, $p_{1f}$ and $p_{2f}$ 
are defined in Fig.\ref{Fig.1}. Note that $k_1$ and $k_2$ define
the four-momenta of residual quark-gluon system of the nucleons without specifying 
their actual composition. We also define  $x_1$, $x'_1$, 
$x_2$ and $x'_2$ as the light-cone  momentum fractions of initial and final nucleons 
carried by their respective residual quark-gluon systems: $x_{1(2)} = {k_{1(2)+}\over p_{1(2)i+}}$ 
and $x'_{1(2)} = {k_{1(2)+}\over p_{1(2)f+}}$. For the $^3$He wave function,
$\alpha={p_{2+}\over p_{NN+}}$ is 
the light-cone momentum fraction of the $NN$ pair carried by one of the nucleons in the pair, and 
$p_{\perp}$ is their relative transverse momentum.
The scattering process in Eq.(\ref{ampl0}) can be described through the combination 
of the following blocks: 
(a)~$\Psi_{^3\textnormal{\scriptsize {He}}}^{\lambda_A,\lambda_{1i},\lambda_{2i},\lambda_s}(\alpha,{p_{\perp}},p_s)$, 
is the light-cone $^3$He-wave function that describes a transition of the $^3$He nucleus with 
helicity $\lambda_A$ into three nucleons with $\lambda_{1i}$ , $\lambda_{2i}$, and $\lambda_s$ helicities, 
respectively. 
(b)~ The term in $\{...\}_{1}$ describes the ``knocking out'' of 
an $\eta_{1i}$-helicity quark  from a $\lambda_{1i}$-helicity nucleon by an incoming 
photon with helicity $\lambda_\gamma$.
Subsequently, the knocked-out  quark exchanges a gluon,  
($[-igT^F_c\gamma^\nu]$), with a quark from the second nucleon producing a final 
$\eta_{2f}$-helicity quark that combines into the nucleon ``${2f}$'' with helicity $\lambda_{2f}$. 
(c)~The term in $\{...\}_{2}$ describes the emerging $\eta_{2i}$-helicity
quark from the $\lambda_{2i}$ -helicity nucleon
then  then exchanges a gluon, ($[-igT^F_c\gamma^\mu]$), with the knocked-out 
quark and produces a final $\eta_{1f}$-helicity quark that combines into the  
nucleon ``${1f}$'' with helicity $\lambda_{1f}$.  
d)~The propagator of the exchanged gluon is 
$G^{\mu\nu}(r) = {d^{\mu\nu}\over r^2+i\varepsilon}$ with polarization matrix, 
$d^{\mu\nu}$ (fixed by the light-cone gauge), and $r=(p_2-k_2+l)-(p_1-k_1+q)$, with 
$l = (p_{2f}-p_{2i})$. In Eq.(\ref{ampl0}) the $\psi^{\lambda,\eta}_N$ represents everywhere an 
$\eta$-helicity  single quark  wave function of a $\lambda$-helicity nucleon 
as defined in Eq.(\ref{nwf}) and $u_\tau$ is the quark spinor defined in the helicity basis.

The denominator of the struck quark's propagator  can be  represented as follows:
\begin{equation}
(p_{1i}-k_1 +q)^2 - m_q^2+i\varepsilon = 
(1-x_1)s_{NN}^\prime(\alpha_c - \alpha+i\epsilon),
\label{denom}
\end{equation}
where
\begin{equation}
\alpha_c = 1 + {1\over s_{NN}'}\left[\tilde m_N^2 - 
{ m_s^2(1-x_1)+m_q^2x_1+(k_1-x_1p_1)^2\over x_1(1-x_1)}\right]
\label{alphac}
\end{equation}
Here $m_s^2$ and $\tilde m_N^2\approx m_N^2$ are defined in Eqs.(\ref{k1onshell}) and (\ref{tilmass}), 
and  $m_q$ represents the current quark mass of the knocked out quark. 
In what follows, we use the fact 
that the  $^3$He wave function  strongly peaks at $\alpha={1\over 2}$, which corresponds to the
kinematic situation in which two constituent nucleons have equal share of the $NN$ pair's 
light-cone momentum.  Thus one expects that the integral in Eq.(\ref{ampl0}) is dominated by 
the value of the integrand at $\alpha=\alpha_c = {1\over 2}$.   This allows us to perform 
$\alpha$-integration in Eq.(\ref{ampl0}) through the pole of the denominator (\ref{denom}) 
at $\alpha=\alpha_c$, i.e. keeping only the $-i\pi \delta(\alpha-\alpha_c)$ part of 
the relation 
$${1\over \alpha_c-\alpha+i\epsilon} = -i\pi\delta(\alpha-\alpha_c) + P{1\over \alpha_c-\alpha},$$ 
and later replacing $\alpha_c$ by ${1\over 2}$.
Using this relation 
to estimate the propagator of the struck quark at its on-mass shell value ($\alpha=\alpha_c$)  allows 
to write,
$$(\sh p_{1i} - \sh k_1 + \sh q)^{\textnormal{\scriptsize {on \ shell}}} + m_q = \sum_\zeta u_\zeta 
(p_1-k_1+q)\bar u_\zeta(p_1-k_1+q)$$. Then  for the 
scattering amplitude of Eq.(\ref{ampl0}) one obtains
\begin{eqnarray}
& & \langle\lambda_{1f},\lambda_{2f},\lambda_s\mid A_i \mid \lambda_\gamma, \lambda_A\rangle = 
\nonumber \\ & & 
\sum\limits_{(\eta_{1f},\eta_{2f}),(\eta_{1i},\eta_{2i}),(\lambda_{1i},\lambda_{2i}),\zeta} \int 
\left\{ {\psi^{\dagger\lambda_{2f},\eta_{2f}}_N(p_{2f},x'_2,k_{2\perp})\over 1-x'_2}\bar 
u_{\eta_{2f}}(p_{2f}-k_2) [-igT^F_c\gamma^\nu]\right. \nonumber \\ & & 
{i\cdot u_\zeta(p_1-k_1+q)\bar u_\zeta(p_1-k_1+q)\over (1-x_1)s'}
[-iQ_ie{\bf \epsilon^{\lambda_\gamma}_\perp\gamma^\perp}]u_{\eta_{1i}}(p_{1i}-k_1)
\left. {\psi_N^{\lambda_{1i},\eta_{1i}}(p_{1i},x_1,k_{1\perp})\over (1-x_1)}\right\}_1
\nonumber \\ & & 
\left\{ {\psi^{\dagger\lambda_{1f},\eta_{1f}}_N(p_{1f},x'_1,k_{1\perp})\over 1-x'_1}\right.  
 \bar u_{\eta_{1f}}(p_{1f}-k_1)[-igT^F_c\gamma^\mu]u_{\eta_{2i}}(p_{2i}-k_2) 
\left . {\psi_N^{\lambda_{2i},\eta_{2i}}(p_{2i},x_2,k_{2\perp})\over (1-x_2)} \right\}_2 
\nonumber \\ & & 
G^{\mu,\nu}(r) {dx_1\over x_1}{d^2k_{1\perp}\over 2 (2\pi)^3}
{dx_2\over x_2}{d^2k_{2\perp}\over 2 (2\pi)^3} 
{\Psi_{^3\textnormal{\scriptsize {He}}}^{\lambda_A,\lambda_{1i},\lambda_{2i},\lambda_s}(\alpha_c,p_{\perp},p_s)\over (1-\alpha_c) 
\alpha_c}
{d^2p_{\perp}\over 4(2\pi)^2}-     \left(p_{1f}\longleftrightarrow p_{2f}\right).
\label{ampl1}
\end{eqnarray}

Next, we evaluate the matrix element of the photon-quark interaction using
on-mass shell spinors for the struck quark.
Taking into account the fact that $(p_{1i}-k)_{+}\gg |k_{\perp}|, m_q$, for this matrix element 
we obtain
\begin{eqnarray}
\bar u_\zeta(p_{1i} - k_{1} + q)[-iQ_ie \epsilon^{\lambda_\gamma}_\perp \gamma^\perp]u_{\eta_{1i}}(p_{1i}-k_1) = 
ieQ_i2\sqrt{2 E_2E_1}(\lambda_\gamma)\delta^{\lambda_\gamma\zeta}\delta^{\lambda_\gamma\eta_{1i}}
\label{melement}
\end{eqnarray}
where $E_{1}= (1-\alpha)(1-x_1){\sqrt{s_{NN}^\prime}\over 2}$ and $E_2 = (1-(1-\alpha)(1-x_1)){\sqrt{s_{NN}^\prime}\over 2}$.

Further explicit calculations of Eq.(\ref{ampl1}) require the knowledge of  quark wave functions 
of the nucleon. Also, one needs to sum over the multitude of the amplitudes representing 
different topologies of  quark knock-out  rescattering and recombinations into 
two final nucleon states.

This difficulty can be circumvented  by again using  that 
the $^3$He wave function strongly peaks at $\alpha = {1\over 2}$  .
We  evaluate Eq.(\ref{ampl1}) setting everywhere  $\alpha_c = {1\over 2}$. 
Such approximation significantly simplifies further derivations. 
As it follows from Eq.(\ref{alphac}) the $\alpha_c={1\over 2}$ condition restricts the
values of $x_1$ of the recoil quark-gluon system to  $x_1 \sim {k_{1\perp}^2\over s_{NN}^\prime}$, thereby 
ensuring that the quark-interchange happens for the valence quarks with $x_q= 1-x_1\sim 1$.
The latter allows us to simplify Eq.(\ref{melement}) setting $E_{1}=E_{2} = {\sqrt{s^\prime_{NN}}\over 4}$.
Using these approximations and substituting Eq.(\ref{melement}) into Eq.(\ref{ampl1}) one obtains
\begin{eqnarray}
& & \langle\lambda_{1f},\lambda_{2f}\lambda_s\mid {\cal M}_i \mid \lambda_\gamma, \lambda_A\rangle = 
i[\lambda_\gamma]e\sum\limits_{(\eta_{1f},\eta_{2f}),(\eta_{2i}),(\lambda_{1i},\lambda_{2i})} \int 
{Q_i \over \sqrt{2s'}}
\times  \nonumber \\& &
\left[ \left\{ {\psi^{\dagger\lambda_{2f},\eta_{2f}}_N(p_{2f},x'_2,k_{2\perp})\over 1-x'_2}\bar 
u_{\eta_{2f}}(p_{2f}-k_2) [-igT^F_c\gamma^\nu]\right. \right. 
u_{\lambda_\gamma}(p_1-k_1+q)
\left. 
{\psi_N^{\lambda_{1i},\lambda_\gamma}(p_{1i},x_1,k_{1\perp})\over (1-x_1)} 
\right\}  
\nonumber \\
& & \left\{ 
{\psi^{\dagger\lambda_{1f},\eta_{1f}}_N(p_{1f},x'_1,k_{1\perp})\over 1-x'_1}
\bar u_{\eta_{1f}}(p_{1f}-k_1)[-igT^F_c\gamma^\mu] \right.  
\left. u_{\eta_{2i}}(p_{2i}-k_2) 
 {\psi_N^{\lambda_{2i},\eta_{2i}}(p_{2i},x_2,k_2)\over (1-x_2)} \right\} \nonumber \\
& & \left. G^{\mu,\nu}(r) {dx_1\over x_1}{d^2k_{1\perp}\over 2 (2\pi)^3}
{dx_2\over x_2}{d^2k_{2\perp}\over 2 (2\pi)^3} \right]_{\textnormal{\scriptsize {QIM}}}
\Psi_{^3\textnormal{\scriptsize {He}}}^{\lambda_A,\lambda_{1i},\lambda_{2i}}(\alpha={1\over 2},p_{2\perp)}
{d^2p_{2\perp}\over (2\pi)^2} \ \ \ \ \   - \left(p_{1f}\leftrightarrow p_{2f}\right).
\label{ampl2}
\end{eqnarray}
Note that due to the $\delta$ factors in Eq.(\ref{melement}) the helicity of the knocked out 
quark in Eq.(\ref{ampl2}) is equal to the helicity of incoming photon, that is 
$\eta_{1i}=\lambda_{\gamma}$.

To proceed, we observe that the kernel, $[\dots ]_{\textnormal{\scriptsize QIM}}$  representing the quark-interchange 
mechanism (QIM) of the rescattering in  Eq.(\ref{ampl2}) can 
be identified with the quark-interchange contribution in the $NN$ scattering amplitude (see Appendix~\ref{app2}).
Such identification can be done by observing that in the chosen reference frame, $q_+=0$, and 
the quark wave function of the nucleon depends on the quark's light-cone momentum fraction and  transverse 
momentum only, which are the same in both Eqs.(\ref{ampl2}) and (\ref{NN_qim}).
For our derivation we also use the above-discussed observation that the  
$\alpha=\alpha_c={1\over 2}$ condition ensures that 
the quark-interchange happens for the valence quarks with $x_q= 1-x_1\sim 1$. This justifies
our next assumption, that valence quarks carry  the helicity of their parent nucleon 
(i.e. $\eta_{1i} = \lambda_i$).   The last assumption allows us to perform the summation of 
Eq.(\ref{ampl2}) over the helicities of the exchanged quarks ($\eta_{2i},\eta_{1f},\eta_{2f}$)  
and to use Eq.(\ref{NN_qim}) to express  
the QIM part   in Eq.(\ref{ampl2}) through the corresponding QIM amplitude of $NN$ scattering. 
Summing for all possible topologies of quark-interchange  diagrams we arrive at
\begin{eqnarray}
& & \langle\lambda_{1f},\lambda_{2f},\lambda_s\mid {\cal M} \mid \lambda_\gamma, \lambda_A\rangle 
= ie[\lambda_\gamma]\times  \nonumber \\
& & \left\{ \sum\limits_{i \in N_1}\sum\limits_{\lambda_{2i}} \int 
{Q_i^{N_1}\over \sqrt{2s'}}
\langle \lambda_{2f};\lambda_{1f}\mid T_{NN,i}^{\textnormal{\scriptsize {QIM}}}(s,l^2)\mid \lambda_\gamma;\lambda_{2i}\rangle 
\Psi_{^3\textnormal{\scriptsize {He}}}^{\lambda_A}(p_1,\lambda_\gamma;p_2,\lambda_{2i},p_s,\lambda_s)
{d^2p_\perp \over (2\pi)^2} \right. \nonumber \\ 
& + &\left.
\sum\limits_{i \in N_2}\sum\limits_{\lambda_{1i}} \int 
{Q_i^{N_2}\over \sqrt{2s'}}
\langle \lambda_{2f};\lambda_{1f}\mid T_{NN,i}^{\textnormal{\scriptsize {QIM}}}(s,l^2)\mid \lambda_{1i};\lambda_{\gamma}\rangle 
\Psi_{^3\textnormal{\scriptsize {He}}}^{\lambda_A}(p_1,\lambda_{1i};p_2,\lambda_{\gamma},p_s,\lambda_s)
{d^2p_\perp \over (2\pi)^2} \right\}
\label{ampl4}
\end{eqnarray}
where nucleon momenta $p_1$ and $p_2$ have half of their c.m.  momentum fractions  and $p_\perp$ is their 
relative transverse momentum with respect to the direction of the photon momentum [see Eq.(\ref{momrel})].
Here, for example,  $Q_i^{N}\cdot \langle \lambda_{2f};\lambda_{1f}\mid T_{NN,i}^{\textnormal{\scriptsize {QIM}}}(s,l^2)\mid 
\lambda_1;\lambda_{2}\rangle$ 
represents the quark-interchange amplitude of $NN$ interaction weighted with the charge of those  
interchanging quarks $Q^{N}_i$ that are struck from a nucleon $N$ by the incoming photon.
The sum ($\sum\limits_{i \in N}$) can be performed within the quark-interchange model of  $NN$ interaction 
which allows us to represent the $NN$ scattering amplitude as follows~\cite{BCL}:
\begin{equation} 
\langle a'b' |T_{NN}^{\textnormal{\scriptsize {QIM}}}|ab\rangle = 
{1\over 2}\langle a'b'| \sum\limits_{i\in a\ , \ j\in b} 
[I_iI_j + \vec \tau_i\vec\cdot\tau_j] F_{i,j}(s,t)|ab\rangle
\label{NNQIM}
\end{equation}
where $I_i$ and  $\tau_i$ are the identity and  Pauli matrices 
defined in the SU(2) flavor (isospin) space of the interchanged 
quarks. The kernel $F_{i,j}(s,t)$ describes an interchange of $i$ and $j$ 
quarks.~\footnote{The additional assumption of helicity conservation allows us to 
express the kernel in the form\cite{BCL} 
$F_{i,j}(s,t) = {1\over 2}[I_iI_j + \vec \sigma_i\vec\cdot\sigma_j]\tilde F_{i,j}(s,t)$, 
where $I_i$ and  $\sigma_i$ operate in the SU(2) helicity ($H$-spin) space of 
exchanged ($i,j$) quarks\cite{BCL}. However for our discussion the assumption of 
helicity conservation is not required.}.

Using Eq.(\ref{NNQIM}) one can calculate the quark-charge weighted QIM 
amplitude, $Q_i\cdot \langle a'b' |T_{NN,i}^{\textnormal{\scriptsize {QIM}}}|ab\rangle$, as follows:
\begin{eqnarray}
 \sum\limits_{i \in N} Q^{N}_i\langle a'b' |T_{NN,i}^{\textnormal{\scriptsize {QIM}}}|ab\rangle   &  = & 
{1\over 2}\langle a'b'| \sum\limits_{i\in a\ , \ j\in b} 
[I_iI_j + \vec \tau_i\vec\cdot\tau_j] (Q_i)F_{i,j}(s,t)|ab\rangle \nonumber\\
&  = & Q^N_F\cdot \langle a'b' |T_{NN}^{\textnormal{\scriptsize {QIM}}}|ab\rangle,
\label{QNNQIM}
\end{eqnarray}
where $Q^N_F$ are the charge factor that are  explicitly calculated using the method described in 
Appendix C. These factors can be expressed through the combinations of  valence quark charges $Q_i$ of nucleon $N$ and 
the number of quark interchanges for each flavor of quark, $N_{Q_i}$, necessary to 
produce a given helicity $NN$ amplitude,  as follows,
\begin{equation}
Q^{N}_F= {N_{uu}(Q_u) + N_{dd}(Q_d) + N_{ud}(Q_u+Q_d)\over 
N_{uu} + N_{dd} + N_{ud}}.
\label{QF}
\end{equation}

\medskip
\medskip

Next we discuss the light-cone wave function of $^3$He that enters in Eq.(\ref{ampl4}). 
The important result 
that allows us to evaluate the wave function is the observation that two nucleons that interact with the photon 
share equally the $NN$ pair's c.m. momentum ($p_{NN}$), i.e., $\alpha = {1\over 2}$.  
If we constrain the third nucleon's light-cone momentum fraction 
$\alpha_s  = {3\cdot p_{s+}\over p_{^3\textnormal{\scriptsize He+}}} = {3(E_s + p_s^z)\over E_{^3\textnormal{\scriptsize {He}}}+ p_s^z + p_{NN}^z} \approx 1$ 
and  transverse momentum $p_{s\perp}\ll m_N$, then the momenta of all the nucleons in the nucleus 
are nonrelativistic. In this case one can use the calculation of triangle diagrams,
which provides the normalization of nuclear wave functions based on baryonic number 
conservation to relate LC and  nonrelativistic nuclear wave functions  as follows\cite{FS81,FS88}
\begin{equation}
\Psi_{^3\textnormal{\scriptsize {He}}}(\alpha,p_\perp, \alpha_s, p_{s,\perp}) =
\sqrt{2}(2\pi)^3m_N \Psi_{^3\textnormal{\scriptsize {He,NR}}}(\alpha,p_{\perp},\alpha_s,p_{s,\perp})
\label{wfnr}
\end{equation}
where for  $\Psi_{^3\textnormal{\scriptsize{He,NR}}}$ we can use known nonrelativistic $^3$He wave functions 
(see e.g.,\cite{Nogga}).

Substituting Eqs.(\ref{QNNQIM}) and (\ref{wfnr}) into Eq.(\ref{ampl4}) for the two-nucleon 
photodisintegration amplitude we obtain
\begin{eqnarray}
& & \langle\lambda_{1f},\lambda_{2f},\lambda_s\mid {\cal M} \mid \lambda_\gamma, \lambda_A\rangle =  
{i[\lambda_\gamma]e\sqrt{2}(2\pi)^3\over \sqrt{2S^\prime_{NN}}}\times \nonumber \\
& & \ \ \left\{ Q_F^{N_1} \sum\limits_{\lambda_{2i}} \int  
\langle \lambda_{2f};\lambda_{1f}\mid T_{NN}^{\textnormal{\scriptsize {QIM}}}(s_{NN},t_{N})\mid \lambda_\gamma;\lambda_{2i}\rangle 
\Psi_{^3\textnormal{\scriptsize {He,NR}}}^{\lambda_A}
(\vec p_1,\lambda_\gamma;\vec p_2,\lambda_{2i};\vec p_s,\lambda_s)m_N{d^2p_{\perp} \over (2\pi)^2} + \right. \nonumber \\
& & \ \ \left. Q_F^{N_2}  \sum\limits_{\lambda_{1i}} \int  
\langle \lambda_{2f};\lambda_{1f}\mid T_{NN}^{\textnormal{\scriptsize {QIM}}}(s_{NN},t_{N})\mid \lambda_{1i};\lambda_{\lambda}\rangle 
\Psi_{^3\textnormal{\scriptsize {He,NR}}}^{\lambda_A}
(\vec p_1,\lambda_{1i};\vec p_2,\lambda_{\gamma};\vec p_s,\lambda_s)m_N{d^2p_{\perp} \over (2\pi)^2}\right\}
\nonumber \\
\label{ampl5}
\end{eqnarray}
where in the Lab frame of the $^3$He nucleus, defining the $z$ direction along the 
direction of $q_{\textnormal{\scriptsize Lab}}$ one obtains
\begin{eqnarray}
&  \alpha = {E_2 - p_{2z}\over M_A-E_s - p_{sz}}; \ \ \ \  &   
p_{\perp} = {p_{1\perp}-p_{2\perp}\over 2}, \nonumber \\
&  \alpha_s = {E_s + p_{sz}\over M_A/A}; &  \vec p_1+ \vec p_2 = - \vec p_s,
\label{momrel}
\end{eqnarray}
with all the momenta defined in the Lab frame.

Equation(\ref{ampl5}) allows us to calculate the unpolarized differential cross section of two nucleon 
breakup in the form
\begin{equation}
{d\sigma\over dt d^3p_{s}/(2E_s (2\pi)^3)} = {|\bar {\cal M}|^2\over 16\pi (s-M_A^2)(s_{NN}-M^2_{NN})}
\label{hrm_crs}
\end{equation}
where $s = (k_\gamma + p_{A})^2$ and 
\begin{equation}
|\bar {\cal M}|^2 = {1\over 2}\cdot{1\over 2}\sum\limits_{\lambda_{1f},\lambda_{2f},\lambda_s,\lambda_{\gamma},\lambda_A}
\left|\langle\lambda_{1f},\lambda_{2f},\lambda_s\mid {\cal M} \mid \lambda_\gamma, \lambda_A\rangle\right|^2.
\label{M2}
\end{equation}

As follows from Eq.(\ref{ampl5}) the knowledge of quark-interchange helicity amplitudes of $NN$ 
elastic scattering will allow us to calculate the differential cross section of hard $NN$ breakup 
reaction without introducing any adjustable  parameter.

\medskip
\medskip

Because the assumption of  $\alpha_c={1\over 2}$ plays a major role in the above derivations 
we attempt now to estimate the theoretical error introduced by this approximation. This approximation
by its nature is a ``peaking'' approximation that is used in loop calculations involving Feynman 
diagrams (one such example is the calculation of radiative effects in electroproduction processes; see, e.g., \cite{MoTs}).  One way to estimate the accuracy of the approximation is to identify 
the main dependence  of the integrand in  Eq.(\ref{ampl1}) on $\alpha_c$ which can be evaluated exactly 
and compare with its evaluation at $\alpha_c={1\over 2}$.  Using Eq.(\ref{melement}) as well as 
Eq.(\ref{alphac}) that allows us to  relate ${dx_1\over x_1}$ to ${d\alpha_c\over \alpha_c}$, and assuming 
that the quark wave functions of nucleons at $\alpha_c\sim {1\over 2}$ are less  sensitive to  $\alpha_c$, 
one arrives at
\begin{equation}
R(p_s) =  {4 \Psi_{^3\textnormal{\scriptsize {He}}}^{\lambda_A,\lambda_{1i},\lambda_{2i},\lambda_s}(\alpha_c={1\over 2},p_{\perp},p_s) \over 
\int {d\alpha_c\over \alpha_c} 
{\Psi_{^3\textnormal{\scriptsize {He}}}^{\lambda_A,\lambda_{1i},\lambda_{2i},\lambda_s}(\alpha_c,p_{\perp},p_s)\over 
\sqrt{(1-\alpha_c) \alpha_c}}}.
\end{equation}
This ratio depends on the kinematics of the spectator nucleon, and for the case of $p_s\le 100$~MeV/$c$, 
$R(p_s)\approx 1.1$, which corresponds to $\sim$$20\%$ of uncertainty in the cross section of the reaction
calculated with the $\alpha_c={1\over 2}$ approximation.
The uncertainty increases with an increase of the momentum of the spectator nucleon.
This can be understood qualitatively because, for large center of mass momenta of the $NN$ pair, 
the $\alpha={1\over 2}$ peak of the nuclear  wave function is less pronounced.

\subsection{Quark-interchange and hard $NN$ elastic scattering amplitudes}
The possibility of using $NN$ elastic scattering data to calculate the 
cross section in Eqs.(\ref{hrm_crs}) and (\ref{M2}) is derived from the assumption that the quark-interchange 
mechanism provides the bulk of the $NN$ elastic scattering strength at high energies and large c.m. angles.
This is a rather well-justified assumption. Experiments on exclusive large $-t$ two-body 
reactions~\cite{20ht} demonstrated clearly the  dominance of  the quark-interchange mechanism for   
the scattering of  hadrons that share common quark flavors.
The analysis of these  experiments  indicate that  
contributions from competing mechanisms such as pure gluon exchange or quark-antiquark 
annihilation are on the level of few percent. This fact justifies our next approximation, 
to  substitute quark-interchange $NN$ amplitudes in Eq.(\ref{ampl5}) 
with actual $NN$ helicity amplitudes as follows:
\begin{eqnarray}
<+,+|T^{\textnormal{\scriptsize {QIM}}}_{NN}|+,+>  &=& \phi_1 \nonumber \\
<+,+|T^{\textnormal{\scriptsize {QIM}}}_{NN}|+,->  &=& \phi_5 \nonumber \\
<+,+|T^{\textnormal{\scriptsize {QIM}}}_{NN}|-,->  &=& \phi_2  \nonumber \\
<+,-|T^{\textnormal{\scriptsize {QIM}}}_{NN}|+,->  &=& \phi_3 \nonumber \\
<+-|T^{\textnormal{\scriptsize {QIM}}}_{NN}|-+>  &=&  -\phi_4.
\label{2phis}
\end{eqnarray}
All other helicity combinations can be related to the above amplitudes through 
the parity and time-reversal symmetry. The minus sign in the 
last equation above is due to the Jackob-Wick phase factor (see, e.g., Ref.\cite{FGST}), according to 
which  one gains a phase factor of ($-$$1$) if two quarks that scatter
by $\pi-\theta$ angle in c.m. have opposite helicities~\cite{JW}. Note that 
$\phi_i$'s are normalized 
in such a way that the cross section for $NN$ scattering is defined as
\begin{eqnarray}
{d\sigma^{NN\rightarrow NN}\over dt} =  {1\over 16\pi}{1\over s(s-4m_N^2)}
{1\over 2}(|\phi_1|^2 + 
|\phi_2|^2 + |\phi_3|^2 + |\phi_4|^2 + 4|\phi_5|^2).
\label{crs_NN}
\end{eqnarray}

Because in the hard breakup regime the momentum transfer $-t_N\gg m_N^2$, one can factorize  the helicity 
$NN$ amplitudes from Eq.(\ref{ampl5}) at $s_{NN}$ and  $t_{N}$ values  
defined as follows:
\begin{eqnarray}
s_{NN}  & = & (q + p_{NN})^2 = (p_{f1} + p_{f2})^2, \nonumber \\
t_{N} & = & (p_{f2} - p_{NN}/2)^2   = {t_{NN}\over 2} + {m_N^2\over 2}  - {M_{NN}^2\over 4}.
\label{stuN}
\end{eqnarray}
Using this  factorization in Eq.(\ref{ampl5}) for the spin averaged square of the breakup amplitude
one obtains
\begin{eqnarray}
\bar {|{\cal M} |}^2 = & &  {(e^2 2(2\pi)^6\over 2 s^\prime_{NN}}
{1\over 2}\left\{2 Q^2_F|\phi_5|^2 S_0 + Q_F^2(|\phi_1|^2+|\phi_2|^2)S_{12} + \right. \nonumber \\ 
& &\ \ \ \ \ \ \ \ \ \ \ \left. \left[ (Q_F^{N_1}\phi_3 + Q_F^{N_2}\phi_4)^2 +
(Q_F^{N_1}\phi_4+Q_F^{N_2}\phi_3)^2)\right]S_{34}\right\},
\label{M2_fct}
\end{eqnarray}
where $Q_F = Q_{F}^{N_1} + Q_{F}^{N_2}$ and $S_{12}$, $S_{34}$, and $S_0$ are partially integrated nuclear spectral 
functions:
\begin{eqnarray}
S_{12}(t_1,t_2,\alpha,\vec p_s) = N_{NN}\sum\limits_{\lambda_1=\lambda_2=-{1\over2}}^{1\over 2}
\sum\limits_{\lambda_3 
= -{1\over2}}^{1\over 2} \left|\int \Psi_{^3\textnormal{\scriptsize {He,NR}}}^{1\over 2}
(\vec p_1,\lambda_1,t_1;\vec p_2,\lambda_{2},t_2;\vec p_s,\lambda_3)
m_N{d^2p_{\perp} \over (2\pi)^2}\right|^2,
\label{S12}
\end{eqnarray}
\begin{eqnarray}
S_{34}(t_1,t_2,\alpha,\vec p_s) = N_{NN}\sum\limits_{\lambda_1=-\lambda_2=-{1\over2}}^{1\over 2}
\sum\limits_{\lambda_3 = -{1\over2}}^{1\over 2}
\left| \int \Psi_{^3\textnormal{\scriptsize {He,NR}}}^{1\over 2}
(\vec p_1,\lambda_1,t_1;\vec p_2,\lambda_{2},t_2;\vec p_s,\lambda_3)
m_N{d^2p_{\perp} \over (2\pi)^2}\right|^2
\label{S34}
\end{eqnarray}
and
\begin{equation}
S_0 = S_{12}+S_{34}.
\label{S0}
\end{equation}
In the above equations $t_1$ and $t_2$ are the isospin projections of nucleons 
in $^3$He. The wave function is normalized to ${2\over 3}$ for proton and ${1\over 3}$ for neutron.
The normalization constants, $N_{NN}$  renormalize the wave function to one $pp$ and two $np$ effective 
pairs in the wave function with $N_{pp}={1\over 2}$ and $N_{pn}=4$.

Equations.(\ref{hrm_crs}) and (\ref{M2_fct}) together with Eqs.(\ref{2phis}),(\ref{S12}),and (\ref{S34}) allow us to 
calculate the  differential cross section of both $pp$ and $pn$ breakup reactions off the   
$^3$He target. 
Notice  that, on the qualitative level, as it follows from Eqs.(\ref{hrm_crs}) and (\ref{M2_fct}) 
in the limit of  $s\gg M_{^3\textnormal{\scriptsize {He}}}^2$ and $s_{NN}\gg m_N^2$,  the HRM predicts an $s^{-11}$ invariant energy 
dependence of the differential cross section provided that the $NN$ cross section scales as $s^{-10}$.
However the numerical calculations of Eq.(\ref{M2_fct}) require a knowledge of 
the $NN$ helicity amplitudes at high energy and momentum transfers. Our strategy is  
to use Eq.(\ref{crs_NN}) to express $NN$ breakup reactions 
directly through the differential cross section of $pn$ and $pp$ elastic scatterings rather than to use 
helicity amplitudes explicitly.

\section{Hard breakup of proton and neutron from {\boldmath $^3$He}.}
\label{III}
We consider now the reaction
\begin{equation}
\gamma + ^3\textnormal{He} \rightarrow (pn) + p,
\label{pn_hrm}
\end{equation}
in which one proton is very energetic and produced at large c.m. angles with 
the neutron, while the second proton emerges with low momentum $\lesssim$ $100$~MeV/$c$. In this 
case the hard rescattering happens in the $pn$ channel. Using the $\phi_3\approx \phi_4$ relation for 
hard $pn$ scattering amplitude (see, e.g., Refs \cite{FGST,BCL,RS}) for breakup amplitude of Eq.(\ref{M2_fct}) one 
obtains
\begin{eqnarray}
\bar {|{\cal M}|}^2 = & &  {(Q^{pn}_{F}e)^2 2(2\pi)^6\over 2 s^\prime_{NN}}
{1\over 2}\left\{2|\phi_5|^2 S_0 + (|\phi_1|^2+|\phi_2|^2)S_{12} + (|\phi_3|^2 + |\phi_4|^2)S_{34}\right\},
\label{M2_fct_pn}
\end{eqnarray}
where $Q^{(pn)}_{F}= Q^{p}_F + Q^{n}_F$ can be calculated using Eqs.(\ref{QNNQIM}) and (\ref{QF}). On the basis of the SU(6) flavor-spin symmetry of nucleon wave functions, for the helicity amplitudes  of Eq.(\ref{2phis}) using 
the method described in Appendix~\ref{haqim} one obtains
\begin{equation}
Q^{pn}_{F} = {1\over 3}.
\label{QFpn}
\end{equation}
We can further simplify Eq.(\ref{M2_fct_pn}) noticing that 
for  the $pn$ pair in $^3$He  one has $S_{12}\approx S_{34} \approx {S_0\over 2}$. 
This is due to the fact that 
in the dominant $S$ state  two protons have opposite spins and therefore the probability of finding one 
proton with a helicity opposite to that of  the neutron is equal to the other proton 
having the same helicity as the neutron's. 
Using this relation and Eq.(\ref{crs_NN}) for the $pn$ breakup reaction one 
obtains
\begin{eqnarray}
\mid \bar {\cal M} \mid^2 =  {(eQ_{F,pn})^2(2\pi)^6 \over s_{NN}^\prime}
16\pi s_{NN}(s_{NN}-4m_N^2) {d\sigma^{pn\rightarrow pn}(s_{NN},t_N)\over dt_N }{S^{pn}_0\over 2}.
\label{M2_pn}
\end{eqnarray}
Inserting it in Eq.(\ref{hrm_crs}) for the differential cross section one obtains
\begin{eqnarray}
{d\sigma^{\gamma ^3He\rightarrow (pn) p}\over dt {d^3p_s\over{E_s}}} =  
{\alpha Q^2_{F,pn} 16\pi^4}{S^{pn}_0(\alpha={1\over 2},\vec p_s)\over 2}
{s_{NN}(s_{NN}-4m_N^2)\over (s_{NN}-p^2_{NN})^2(s-M^2_{^3\textnormal{\scriptsize {He}}})}{d\sigma^{pn\rightarrow pn}(s_{NN},t_N)\over dt_N },
\nonumber \\
\label{pn_hrm_crs}
\end{eqnarray}
where $\alpha = {1\over 137}$ and ${d\sigma^{pn \rightarrow pn}\over dt_N}$ is the differential cross 
section of hard $pn$ scattering 
evaluated at values of $s_{NN}$  and $t_N$ defined in Eq.(\ref{stuN}). The spectral function $S_0^{pn}$ 
is defined in Eq.(\ref{S0}) and corresponds to:
\begin{eqnarray}
S_0^{pn}(\alpha,\vec p_s) = 4\sum\limits_{\lambda_1,\lambda_2,\lambda_3=-{1\over 2}}^{1\over 2}
\left|\int \Psi_{^3\textnormal{\scriptsize {He,NR}}}^{1\over 2}
(\vec p_1,\lambda_1,{1\over 2};\vec p_2,\lambda_{2},-{1\over 2};\vec p_s,\lambda_3)
m_N{d^2p_{\perp} \over (2\pi)^2}\right|^2.
\label{S0pn}
\end{eqnarray}

\section{Hard breakup of two protons from {\boldmath $^3$He}}
\label{IV}
We now consider the reaction
\begin{equation}
\gamma + ^3\textnormal{He} \rightarrow (pp) + n,
\label{pp_hrm}
\end{equation}
in which two protons are produced at large c.m. angles while the neutron emerges 
as a spectator with small momentum ($p_s \le 100$~MeV/$c$).

The relation between $S_{12}$ and $S_{34}$ is
very different from that in the  $pn$ case. As a result of the fact that two protons 
cannot have the same helicity, in the $S$ state one has that $S_{12} \ll S_{34}$. 
The estimates of the spectral functions 
based on the realistic $^3$He wave function\cite{Nogga} gives ${S_{12}\over S_{34}}\sim 10^{-4}$. 
Therefore one can neglect the $S_{12}$ term in Eq.(\ref{M2_fct}).  
The next observation is that for  $pp$ scattering the helicity amplitudes 
$\phi_3$ and $\phi_4$ have opposite signs because of the Pauli principle (see, e.g., Refs.\cite{JW,FGST}).
Using the above observations and neglecting 
the  helicity-nonconserving amplitude $\phi_5$  for the $pp$-breakup amplitude we obtain
\begin{eqnarray}
\bar {|{\cal M} |}^2 = & &  {(e^2 2(2\pi)^6\over 2 s^\prime_{NN}}
{1\over 2}\left\{ 2(Q_F^{p}|\phi_3| - Q_F^{p}|\phi_4|)^2S_{34}\right\}.
\label{M2_fct_pp}
\end{eqnarray}
The charge factor $Q_F^p$ depends on the helicity amplitude it couples; therefore one estimates it 
for the combination of $(Q_F^{p}|\phi_3| - Q_F^{p}|\phi_4|)$. Using SU(6) symmetry for the distribution of 
given helicity-flavor valence quarks in the proton and through the approach described in Appendix~\ref{haqim} we obtain
\begin{equation}
(Q_F^{p}|\phi_3| - Q_F^{p}|\phi_4|) = Q^{pp}_{F}(|\phi_3| - |\phi_4|)
\end{equation}
with 
\begin{equation}
Q_F^{pp} = {5\over 3}.
\label{QFpp}
\end{equation}
It is worth noticing that because of explicit consideration of quark degrees of freedom the 
effective charge involved in the breakup is larger for the case of two protons than for 
proton and neutron [see Eq.(\ref{QFpn})]. This is characteristic of the HRM model in which a 
photon couples to a  quark and more charges are exchanged in the $pp$ case than in the  $pn$
case. This is rather opposite to the scattering picture  considered based on 
hadronic degrees of freedom in which case the photon will couple to an exchanged meson and 
$pp$ contribution will be significantly suppressed because no charged mesons can be exchanged 
within the $pp$ pair.

To be able to estimate the cross section of the $pp$ breakup reaction through the 
elastic $pp$ scattering cross section we introduce a parameter 
\begin{equation}
C^2 = {\phi_3^2\over \phi_1^2}\approx {\phi_4^2\over \phi_1^2},
\label{cfactor}
\end{equation}
which allows to express the differential cross section of the reaction (\ref{pp_hrm}) in the following form:
\begin{eqnarray}
{d\sigma^{\gamma^3He\rightarrow(pp)n}\over dt {d^3p_{s}\over E_s}} = & &   
\alpha Q_{F,pp}^2 16\pi^4 S^{pp}_{34}(\alpha={1\over 2},\vec p_s){2\beta^2\over 1+2C^2}
{s_{NN}(s_{NN}-4m_N^2)\over (s_{NN}-p_{NN}^2)^2 (s-M_{^3\textnormal{\scriptsize {He}}}^2)}\times \nonumber \\
& & {d\sigma^{pp\rightarrow pp}(s_{NN},t_{N})\over dt},
\label{pp_hrm_crs}
\end{eqnarray}
where we also introduced a factor $\beta$,
\begin{equation}
\beta  = {|\phi_3| - |\phi_4|\over |\phi_1|},
\label{betafactor}
\end{equation}
which accounts for the suppression from the cancellation between $\phi_3$ and $\phi_4$ 
helicity amplitudes of elastic $pp$ scattering.\footnote{This cancellation was overlooked in earlier estimates 
of the cross section of $pp$ photodisintegration from $^3$He target 
(see, e.g., \cite{gheppn}).}
The spectral function $S^{pp}_{34}$ in Eq.(\ref{pp_hrm_crs}) is expressed through the $^3$He wave function
according to Eq.(\ref{S34})  as follows:
\begin{eqnarray}
S^{pp}_{34}(\alpha,\vec p_s) = {1\over 2} \sum\limits_{\lambda_1=-\lambda_2=-{1\over2}}^{1\over 2}
\sum\limits_{\lambda_3 
= -{1\over2}}^{1\over 2}\left| \int \Psi_{^3\textnormal{\scriptsize {He,NR}}}^{1\over 2}
(\vec p_1,\lambda_1,{1\over 2};\vec p_2,\lambda_{2},{1\over 2};\vec p_s,\lambda_3)
m_N{d^2p_{2,\perp} \over (2\pi)^2}\right|^2.
\label{S34pp}
\end{eqnarray}

\section{Two- and three-body processes in $NN$ breakup reactions}
\label{V}

For a two-body hard $NN$ breakup mechanism to be observed  it must 
dominate the three-body/two-step processes.  This is especially important for 
$pp$ breakup processes, (\ref{pp_hrm}) because according to Eqs.(\ref{pp_hrm_crs} and \ref{betafactor}) 
the  two-body contribution is suppressed because of a cancellation between 
$\phi_3$ and $\phi_4$ helicity amplitudes.

At low to intermediate range energies~($E_\gamma \sim 200$~MeV) it is rather well established that 
the $pp$ breakup reaction proceeds 
overwhelmingly through a two-step (three-body) process\cite{Laget,Laget1,Laget2,Jan1,Jan2,Giusti} 
in which the initial breakup 
of the $pn$ pair (dominated by $\pi^\pm$ exchange) is  followed by a charge-interchange final state 
interaction of the neutron with the spectator proton.  
Other two-step processes include the excitation of intermediate 
$\Delta$ isobars in the $pn$ system with the subsequent rescattering off the 
spectator neutron, which produces two final protons.

\begin{figure}[t]
\centering\includegraphics[height=5.cm,width=10.cm]{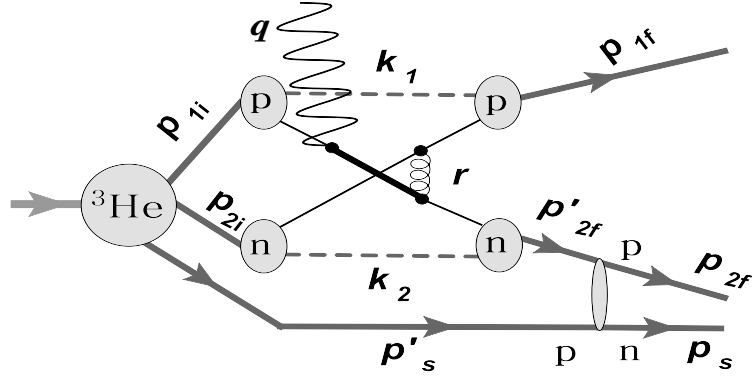}
\caption{ Diagram corresponding to 
three-body processes in which the hard breakup of the $pn$ pair is followed by a soft charge-exchange  
rescattering of the neutron off the spectator proton.}
\label{3bfig}
\end{figure}

The dominance of three-body processes at low energies is related mainly to the fact that 
the two-body $pp$ breakup is negligible because of the impossibility of charged-pion 
exchanges between two protons that absorb an incoming photon.

At high energy kinematics within the HRM the interaction between protons is carried out by exchanged quarks
because of which the relative strength of $pp$ breakup is larger. 

To estimate the strength of three-body processes  at high energy kinematics, one needs to calculate 
the contribution of diagrams similar to Fig. \ref{3bfig}.  Because the charge-exchange rescattering 
at the final stage of the process in Fig.\ref{3bfig} takes place at proton momenta 
$p_{f}^\prime > m_N$, one 
can apply an eikonal approximation\cite{gea,stopics} to estimate its contribution.

For $E_\gamma \ge 2$~GeV assuming that the HRM is valid for the first ($pn$ breakup) stage of the reaction, 
for the amplitude of three-body/two-step process within the eikonal approximation\cite{gea,stopics} one obtains
\begin{eqnarray}
{\cal M}_{3body} & \approx &  
{eQ_{F,pn}(2\pi)^3\over 2\sqrt{2s_{NN}^\prime}} T^{hard}_{pn\rightarrow pn}(t_N) 
\times \nonumber \\
& &  \int  \Psi_{^3\textnormal{\scriptsize {He,NR}}}^{\lambda_A}
(\vec p_1,t_1;\vec p_2,t_2;\vec p_s-\vec k_\perp)
m_N {T^{chex}_{pn\rightarrow np}(k_\perp)\over s_{NN}}{d^2p_{\perp} \over (2\pi)^2}{d^2k_{\perp} \over (2\pi)^2},
\label{3b}
\end{eqnarray}
where we suppressed helicity indices for simplicity and choose the isospin projections,  
$t_{1} = -t_{2} = {1\over 2}$, corresponding to the initial $pn$ pair that interacts with the photon. Here 
$T^{\textnormal{\scriptsize hard}}_{pn\rightarrow pn}(t_N)$ is the hard elastic $pn$ scattering amplitude and 
$T^{\textnormal{\scriptsize chex}}_{pn\rightarrow np}$ represents  the amplitude of the soft charge-exchange $pn$ scattering.
Because of the pion-exchange nature of the latter it is rather well established that this amplitude 
is real and can be represented as $\sim$$\sqrt{s}A e^{{B\over 2}t}$, where $A$ and $B$ are approximately
constants\cite{GL}.

Two main  observations follow from Eq.(\ref{3b}) and the above-mentioned property of the charge-exchange amplitude:
First, three-body and two-body amplitudes [see, e.g., Eq.(\ref{ampl5})] will  not interfere, since one is real and 
the other is imaginary.  The fact that these two amplitudes differ by order of $i$ follows from the general 
structure of rescattering amplitudes (see, e.g., Ref.\cite{stopics}).  Equation (\ref{ampl5}) corresponds to  a  
single rescattering amplitude,  while Eq.(\ref{3b}) to a  double rescattering amplitude.\\
Second, because of the energy dependence of the charge-exchange scattering amplitude at small angles,  
the three-body contribution will scale like $s^{-12}$ as compared to the 
two-body breakup contribution.

Using Eq.(\ref{3b}) one can estimate the magnitude of the contribution of three-body processes in the 
$pp$ breakup cross section as follows:
\begin{equation}
{d\sigma^{\gamma^3He\rightarrow(pp)n}_{three-body}\over dt {d^3p_{s}\over E_s}} \approx 
{d\sigma^{\gamma ^3He\rightarrow (pn) p}_{two-body}\over dt {d^3p_s\over{E_s}}}{S^{pnp}(p_s)
\over S_{0}^{pn}(p_s)}
\label{3to2}
\end{equation}
where $S_{0}^{pn}(p_s)$ is defined in Eq.(\ref{S0pn}) and for 
$S^{pnp}(p_s)$ based on Eq.(\ref{3b}) one obtains
\begin{equation}
S^{pnp}(p_s) = {N_{pn}\over 16 s_{NN}^2} \mid \int  \Psi_{^3\textnormal{\scriptsize {He,NR}}}^{\lambda_A}
(\vec p_1,\vec p_2,\vec p_s-\vec k_\perp)m_N T^{chex}_{pn\rightarrow np}(k_\perp)
{d^2p_{\perp} \over (2\pi)^2}{d^2k_{\perp} \over (2\pi)^2}\mid^2.
\label{spnp}
\end{equation}
Here both spectral functions are defined at $\alpha={1\over 2}$.

Using Eqs.(\ref{3to2}) and (\ref{spnp}) and the parametrization of $T^{\textnormal{\scriptsize chex}}_{pn\rightarrow np}$ from 
Ref.\cite{GL} one can estimate the relative contribution of three-body processes numerically. Note that this 
contribution is maximal at $\alpha_s=1$ and  increases with an increase of the momentum of $p_s$. However 
because of the charge-exchange  nature of the second  rescattering, this contribution decreases linearly 
with an increase of $s$.\footnote{Notice that for the case of diagonal $pn\rightarrow pn$  rescattering  
$T_{pn\rightarrow np}(k_\perp)= s\sigma_{\textnormal{\scriptsize tot}}e^{{B\over 2}t}$ and as a result the 
probability of the rescattering is energy independent.}

\section{Numerical Estimates}
\label{VI}
For numerical estimates we consider the center of mass reference frame of $\gamma$$NN$ system, for which 
according to Eq.(\ref{mans_st}) one obtains
\begin{equation}
t_{N,N} = - {(s_{NN}-M_{NN}^2)\over 2\sqrt{s_{NN}}}(\sqrt{s_{NN}}
-\sqrt{s_{NN}-4 m_N^2}cos(\theta_{\textnormal{\scriptsize c.m.}})) + m_N^2,
\label{tgnn_cm}
\end{equation}
where $M_{NN}^2 = p^\mu_{NN} p_{NN,\mu}$ and $p^\mu_{NN} = p^\mu_{^3\textnormal{\scriptsize {He}}}-p^\mu_s$. Using the above equation
we obtain for $t_{N}$ [Eq.(\ref{stuN})], which defines the effective momentum transfer in the $NN$ scattering
amplitude, the following relation: 
\begin{equation}
t_{N} =  - {(s_{NN}-M_{NN}^2)\over 4\sqrt{s_{NN}}}(\sqrt{s_{NN}}
-\sqrt{s_{NN}-4 m_N^2}cos(\theta_{\textnormal{\scriptsize c.m.}})) + m_N^2 - {M_{NN}^2\over 4}.
\label{tN}
\end{equation}
One can also calculate the effective c.m. angle that enters in the $NN$ scattering amplitude as follows:
\begin{equation}
cos(\theta_{\textnormal{\scriptsize c.m.}}^N) = 1 - {(s_{NN}-M_{NN}^2)\over 2(s_{NN}-4m_N^2)}
{(\sqrt{s_{NN}}-\sqrt{s_{NN}-4 m_N^2}cos(\theta_{\textnormal{\scriptsize c.m.}}))\over \sqrt{s_{NN}}} + 
{4m_N^2 - M_{NN}^2\over 2(s_{NN}-4m_N^2)}.
\label{theta_cmn}
\end{equation}
The above equations define the kinematics of hard $NN$ rescattering. 
\begin{figure}[t]
\centering\includegraphics[height=8.6cm,width=8.6cm]{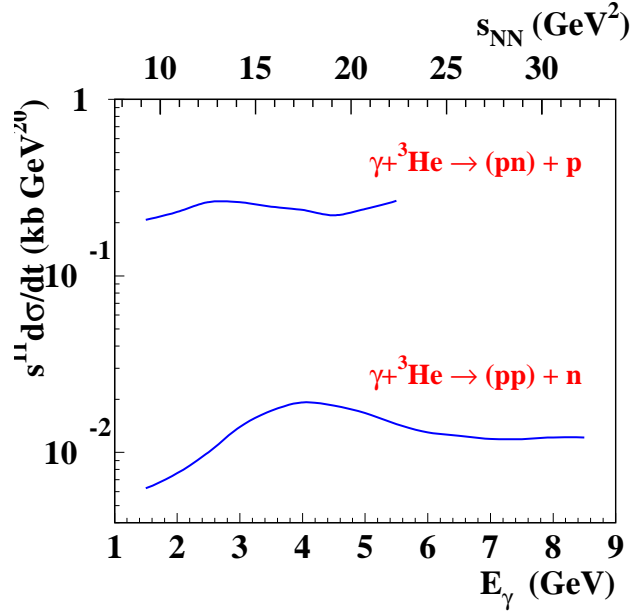}
\caption{. Energy dependence of $s^{11}$ weighted differential cross sections at 
$90^\textnormal{\scriptsize o}$ c.m. angle scattering in "$\gamma$-$NN$"system. In these calculations one integrated over 
the spectator nucleon momenta in the range of $0$-$100$~MeV/$c$.}
\label{Fig.s_dep}
\end{figure}

\subsubsection{Energy dependence and the magnitude of the cross sections}

We are interested in energy dependences of the hard breakup reactions of Eqs.(\ref{pn_hrm}) and 
(\ref{pp_hrm}) at fixed and large angle production of two fast nucleons in the ``$\gamma$-$NN$'' center of 
mass reference frame. Particularly interesting is the case of $\theta_{\textnormal{\scriptsize c.m.}}$=$90^{\textnormal{\scriptsize 0}}$ for which 
as it follows from Eq.(\ref{theta_cmn}) $cos(\theta_{\textnormal{\scriptsize c.m.}}^N) = 0.5$.   This means that the 
cross sections of hard breakup reactions at these kinematics will be defined by the $NN$ elastic 
scattering at $\theta^N_{\textnormal{\scriptsize c.m.}} = 60^{\textnormal{\scriptsize 0}}$. In Fig.\ref{Fig.s_dep} the $E_\gamma$ and $s$ dependencies of the  
$s_{NN}^{11}$ weighted differential cross sections are presented for the cases of the $pp$ and $pn$ 
breakup reactions. In the calculation we integrated over the spectator nucleon's momentum  
in the range of (0-100)~MeV/$c$ and over the whole range of its solid angle.
Also for the parameter $C$ in Eq.(\ref{cfactor}) we used $C={1\over 2}$, consistent with an estimate 
obtained within the quark-interchange model of $pp$ scattering (see, e.g.,\cite{FGST,BCL}). The estimation of 
the factor $\beta$, which takes into account the cancellation between $\phi_3$ and $\phi_4$ helicity 
amplitudes in Eq.(\ref{pp_hrm_crs}) requires the knowledge of the angular dependence for 
helicity amplitudes. For this we used the helicity amplitudes calculated within 
thequark-interchange model\cite{BCL,FGST} with phenomenological angular dependencies estimated using  
$F(\theta_{\textnormal{\scriptsize c.m.}}) = {1\over sin^2(\theta_{cm})(1-cos(\theta_{cm}))^2}$ 
(see, e.g., Refs. \cite{BCL,RS}) which  describes reasonably well the data at hard scattering kinematics.

\begin{figure}[t]
\centering\includegraphics[height=8.6cm,width=8.6cm]{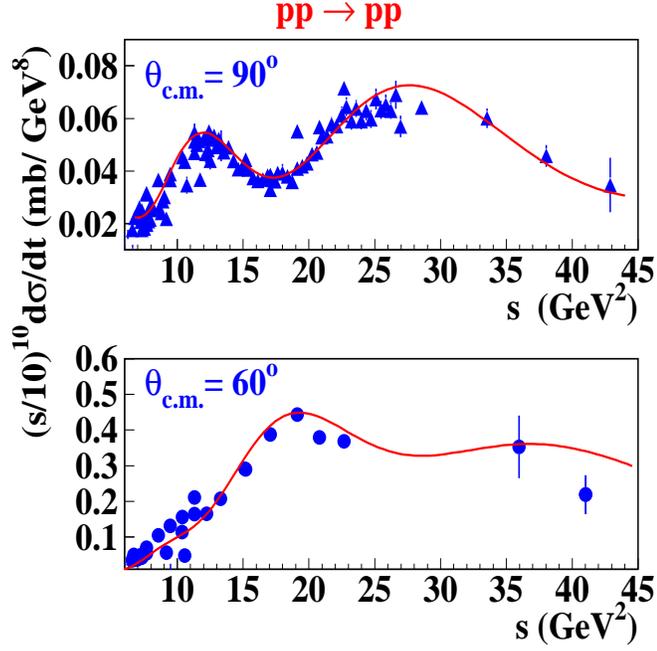}
\caption{ Invariant energy dependence of $s^{10}$ weighted differential cross sections of 
elastic $pp$ scattering at $\theta_{\textnormal{\scriptsize c.m.}}=90^{\textnormal{\scriptsize o}}$ and $\theta_{\textnormal{\scriptsize c.m.}}=60^{\textnormal{\scriptsize o}}$. are fitted using the parametrization described in Appendix~\ref{NNparam} of the available world data~\cite{hepdata}.}
\label{Fig.pp}
\end{figure}

Several features of the HRM  calculations are worth discussing in Fig.\ref{Fig.s_dep}: First, the breakup cross sections in 
average scale like $s_{NN}^{-11}$. Note that the absolute~(nonscaled) values of the cross sections  drop by 
five orders of magnitude in the 2-8~GeV of photon energy range.
Next, the shapes of the $s^{11}$ weighted differential 
cross sections reflect the shapes of the $s^{10}$ weighted differential cross sections of $pp$ and $pn$
scattering at $\theta_{\textnormal{\scriptsize c.m.}}=60^{\textnormal{\scriptsize o}}$. [see Figs.(\ref{Fig.pp}) and (\ref{Fig.pn})]. It 
is worth noting that as follows from  Figs.(\ref{Fig.pp}) and (\ref{Fig.pn}) 
the fits used in the calculation of $pp$ and $pn$ breakup reactions contain uncertainties
on the level of 10\% for $pp$ breakup (for $s_{NN}\ge 24$~GeV$^2$) and up to 30\% 
for $pn$ breakup reactions.  Consequently, one can conclude that the calculated shape of the energy 
dependence of the $pn$ breakup reaction in Fig.(\ref{Fig.s_dep}) does not have much predictive power.
However, for the  $pp$ breakup the calculated shape, for up to $s_{NN}\le 24$~GeV$^2$, is not 
obscured by the uncertainty of the $pp$ data and can be considered as a prediction of the HRM. 

Analysis of the first experimental data on $pp$ photodisintegration of the $^{3}$He nucleus at high momentum transfer from Jefferson Lab~\cite{Pomerantz:2009sb} has shown excellent agreement with the HRM predictions. At $\theta_{c.m}=90^o$ and $s>$12GeV$^2$, the cross section scales with $s^{11}$ according to counting rules, and as shown in Figs.~\ref{fig:gpp} and \ref{fig:elpp}, the experimental energy distribution of the breakup process at this kinematics is remarkably similar but for a power of $s$, to the experimental energy distribution of $pp$ elastic scattering at $\theta^{N}_{c.m.}=60^o$.

It is worth mentioning that considered features of the HRM  
are insensitive to the choice of the above-discussed parameters of $C$ and $\beta$, because they only 
define the absolute magnitude of the $pp$ breakup cross section.

\begin{figure}[ht]
\centering\includegraphics[height=8.6cm,width=8.6cm]{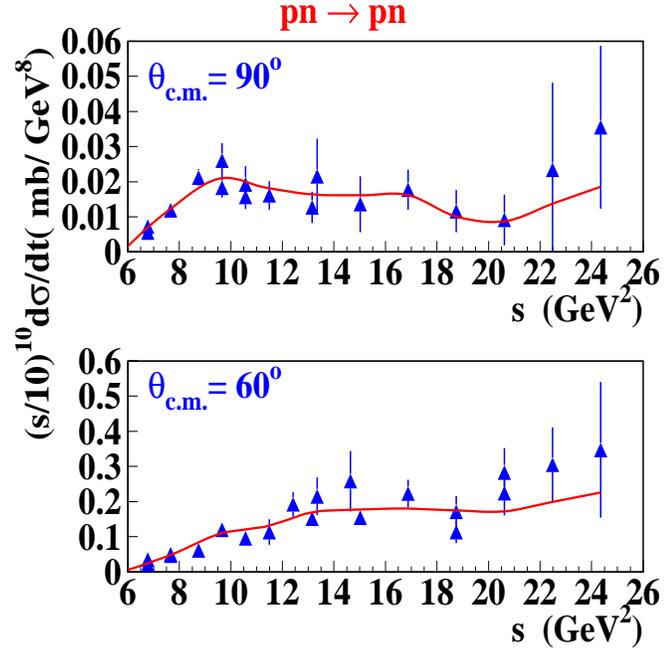}
\caption{ Invariant energy dependence of $s^{10}$ weighted differential cross sections of 
elastic $pn$ scattering at $\theta_{\textnormal{\scriptsize c.m.}}=90^{\textnormal{\scriptsize o}}$ and $\theta_{\textnormal{\scriptsize c.m.}}=60^{\textnormal{\scriptsize o}}$. Curves 
are fitted using the parametrization described in Appendix~\ref{NNparam} of the available world data~\cite{hepdata}.}

\label{Fig.pn}
\end{figure}

\begin{figure}
\centering
\includegraphics[height=8cm,width=10cm]{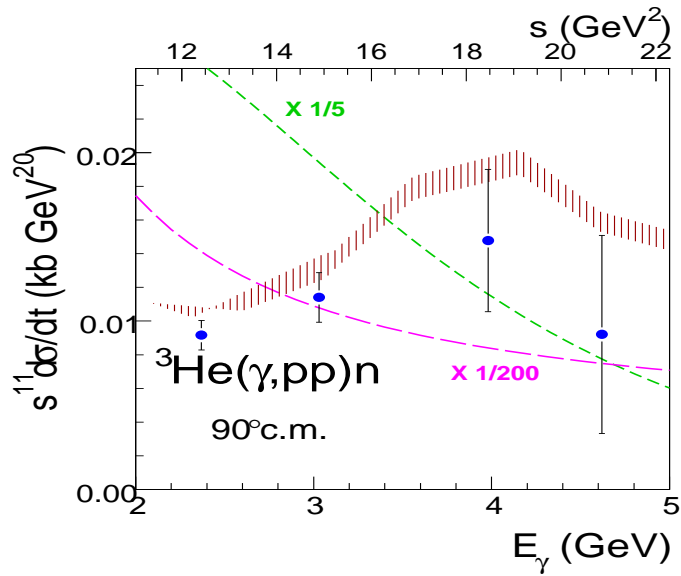}
      \caption{Cross section data for $pp$ breakup in $^3$He photodisintegration at 90$^o$ c.m. angle of the $\gamma+pp$ system \cite{Pomerantz:2009sb}. The $^3$
He$(\gamma,pp)n$ events were selected with $p_n < 100$
MeV. (Dashed) QGS prediction times 1/5. (Long dashed) RNA prediction times 1/200. (Shaded) HRM prediction.}
\label{fig:gpp}
\end{figure}
\begin{figure}
\centering
\includegraphics[height=6cm,width=9cm]{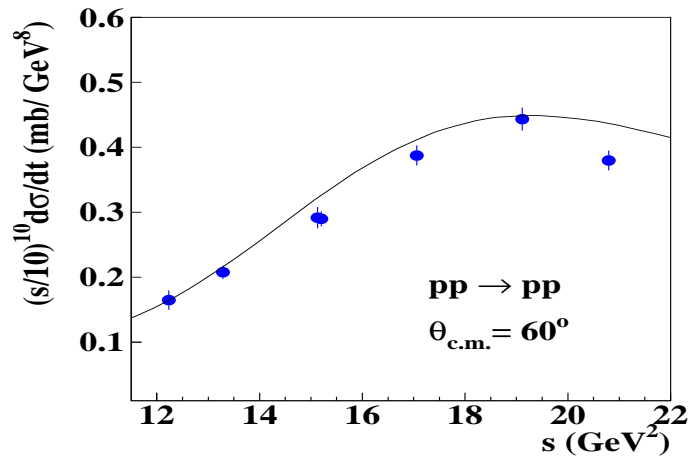}
           \caption{$pp$ elastic scattering data at 60$^o$ of center of mass angle of scattering. These data is used in Eq.(\ref{pp_hrm_crs}) to calculate the HRM cross section of $pp$ breakup shown in Fig.\ref{fig:gpp}.}
      \label{fig:elpp}
\end{figure}      

The next feature of the calculations in Fig.\ref{Fig.s_dep} is the magnitude of 
the $pn$ and $pp$  breakup cross sections. The $pn$ breakup cross section 
[Eq.(\ref{pn_hrm_crs})] does not contain any free parameter, and similar
to the HRM prediction for the breakup  of the deuteron\cite{gdpn}, 
it is expressed through the rather well-defined 
quantities.  For the estimate of $pp$ breakup, however, one needs to know the  relative strength of 
the $\phi_3$ and $\phi_4$ amplitudes as compared to $\phi_1$ as well as the extent of their 
cancellation at kinematics of $s_{NN}$ and $t_N$ defined in Eqs.(\ref{stuN}) and (\ref{tN}).	
Our calculation,  determined by  phenomenologically justified estimates of factors $C$ and $\beta$ in 
Eq.(\ref{pp_hrm_crs}) results in the $pp$ breakup cross section which is  about ten times 
smaller than the cross section for the $pn$ breakup.  This result indicates however, 
an  increase of $pp$ breakup cross section relative to the $pn$ breakup cross section as 
compared to the results from the low energy breakup reactions.
As it was mentioned in Sec.\ref{V},  
at low energies ($\sim$ 200~MeV) the cross section of $pp$ photodisintegration from $^3$He is 
significantly  smaller (by almost two orders of magnitude according to Ref.\cite{Laget}) than 
the $pn$-breakup cross section.  

Note that the  factors $C$ and $\beta$ introduce an additional uncertainty in estimating  the magnitude 
of the $pp$-breakup cross section. While the factor $C$ can be evaluated  in  the quark-interchange model 
thus staying within the framework of the considered model, the factor $\beta$ is not constrained by 
the theoretical framework of the model.  The latter is sensitive to the angular dependence, $F(\theta_{c.m.})$, 
of the helicity amplitudes. To estimate the uncertainty associated with $F(\theta_{c.m.})$ we varied it 
around the form,  $F(\theta_{\textnormal{\scriptsize c.m.}}) = {1\over sin^2(\theta_{\textnormal{\scriptsize c.m.}})(1-cos(\theta_{\textnormal{\scriptsize c.m.}}))^2}$ in such 
a way that the results were still in agreement with angular distribution of $pp$ scattering 
at $-t,-u > 1$~GeV$^2$. We found that this variation 
changes the HRM prediction for the magnitude of  $pp$-breakup cross section as much as $40\%$.

\begin{figure}[ht]
\centering\includegraphics[height=8.6cm,width=8.6cm]{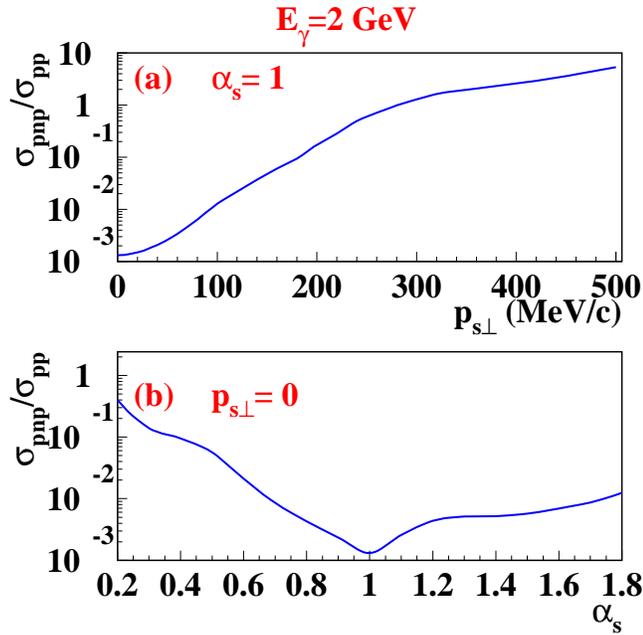}
\caption{Dependence of the ratio of the cross section of  
three-body/two-step process discussed in Sec.V to the cross section of the two-body $pp$ breakup  
at $E\gamma=2$~GeV on (a) transverse momentum of the spectator neutron $p_{st}$ at $\alpha_s=1$ and 
on (b) $\alpha_s$ at $p_{st}=0$.}
\label{Fig.3b_2b}
\end{figure}

Because $pp$ breakup cross section is still by a factor of 10 smaller than the $pn$ cross section,
one needs to estimate the contribution due to three-body processes in which hard $pn$ breakup 
is followed by soft charge-exchange rescattering.  The estimate based on Eq.(\ref{3to2}) is 
given in Fig. \ref{Fig.3b_2b} where the ratio of three-body  to two-body breakup cross 
sections is evaluated for different values of $\alpha_s$ and  transverse momentum  
of the spectator neutron, $p_{s\perp}$.

Because of the eikonal nature of the second rescattering in three-body processes, one expects the cross section 
to be maximal at $\alpha_s=1$. As Fig.\ref{Fig.3b_2b}(a) shows  in this case, the three-body process is 
a correction to the two-body breakup process, $\sim$2\% for $p_{s\perp}=100$~MeV/$c$ and $\sim 17$\% for 
$p_{s\perp}=200$~MeV/$c$. Then, starting at $p_{s\perp}$ > 300~MeV/$c$ the three-body  process dominates 
the two-body contribution.
The latter can be verified by observing an onset of $s^{-12}$ scaling at large ($\ge 300$~MeV/$c$) 
transverse momenta of the spectator neutron in the case of hard $pp$ breakup reactions.  
Figure \ref{Fig.3b_2b}(b) shows also  that the three-body contribution will be always small for 
$p_{s\perp}\approx 0$~MeV/$c$, and for a wide range of $\alpha_s$,  which again reflects the eikonal nature of 
the second order rescattering in  which case the recoiling of the spectator  nucleons happens predominantly 
at $\sim 90^{\textnormal{\scriptsize o}}$ (see, e.g., \cite{Glauber}). Note that one expects the above estimate of the three-body 
contribution to contain an uncertainty of  10-15\%, representing the general level of accuracy of 
eikonal approximations.

On the basis of Fig.\ref{Fig.3b_2b} one can expect that overall, for small values of $p_{s}\le$100-150~MeV/$c$    
in the high energy limit ($E_\gamma > 2$~GeV) one expects two-body breakup mechanisms to dominate  for 
both $pp$ and $pn$ production reactions.

\subsection{Spectator nucleon momentum dependence}

The presence of a spectator nucleon in the hard two-nucleon breakup reaction from $^3$He 
gives us an additional degree of freedom in checking the mechanism of the photodisintegration.
As follows from Eqs.(\ref{pn_hrm_crs}) and (\ref{S0pn}) and Eqs.(\ref{pp_hrm_crs}) and (\ref{S34pp}) the $pn$ and $pp$ breakup
cross sections  within the HRM are sensitive to different components of the nuclear spectral function. This is a result of 
the fact that the $pp$ component with the same helicities for both protons is suppressed in the ground state 
wave function of the $^3$He target.  Thus one expects rather different spectator-momentum dependencies 
for $pp$ and $pn$ breakup cross sections.

\begin{figure}[ht]
\centering\includegraphics[height=8.6cm,width=8.6cm]{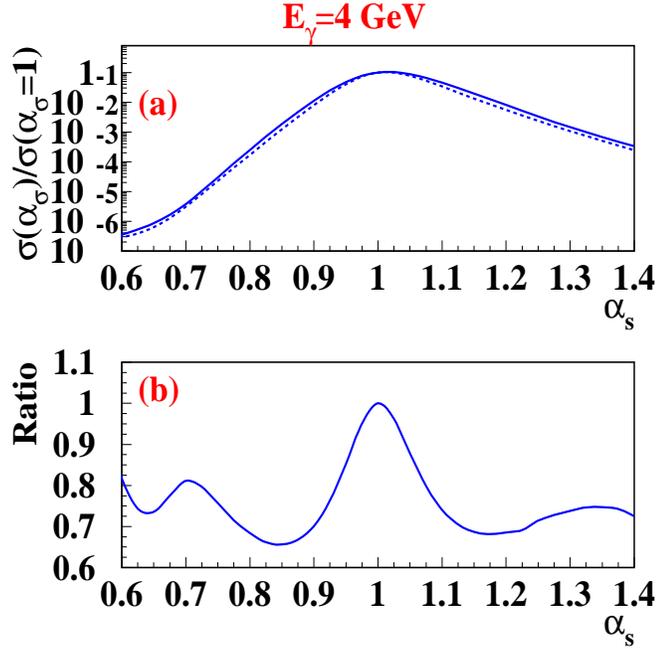}
\caption{Dependence of the $s^{11}$ weighted 90$^{\textnormal{\scriptsize o}}$ c.m. breakup differential cross section
on the light-cone momentum fraction of  of 
spectator nucleon, $\alpha_s$,  calculated at $E_\gamma=4$~GeV and $p_{s\perp}=0$. 
(a) The solid line is for $pp$ breakup reactions, and the dashed line is for $pn$ breakup 
reactions. Calculations are normalized to the cross sections at $\alpha_s=1$. (b) Ratio of the $pn$ to $pp$ 
breakup cross sections normalized to their values at $\alpha_s=1$.}
\label{Fig.al_dep}
\end{figure}

The quantity that we consider for numerical estimates is not the momentum of the spectator but 
rather the  momentum fraction of the target carried by the spectator nucleon, $\alpha_s$.   This quantity is 
Lorentz invariant with respect to boosts in the $q$ direction, which allows us to specify it 
in the Lab frame as follows:
\begin{equation}
\alpha_{s} \equiv {E_s-p_{s,z}\over M_A/A} = \alpha_A - \alpha_{1f} - \alpha_{2f}
\label{alphas}
\end{equation}
where $\alpha_{i} = {E_i-p_{i,z}\over M_A/A}$ for $i=A$, $1f$, $2f$ and $z$ axis in the Lab frame is 
defined parallel to the momentum of incoming photon $q$.  Note that the photon does not 
contribute to the above equation because $\alpha_{q}=0$. In definition of $\alpha_s$ we use a normalization 
such that for a stationary spectator $\alpha_s=1$.
The $\alpha_s$ dependencies of the differential cross sections for $pp$ and $pn$ breakup reactions 
normalized to their values at $\alpha_s=1$ are given in Fig.\ref{Fig.al_dep}(a). One feature of $\alpha_s$
dependence is the asymmetry of the cross section around $\alpha_s=1$ with cross sections 
dominating at $\alpha_s>1$. This property can be understood from the fact that the momentum fraction of 
the $NN$ pair that breaks up is defined through $\alpha_s$ as follows:
\begin{equation}
\alpha_{NN} = 3-\alpha_{s}.
\end{equation} 
The latter quantity defines the invariant energy of the $NN$ pair as follows
\begin{equation}
s_{NN} = M_{NN}^2 + E_\gamma m_n\alpha_{NN}.
\end{equation}
Because the cross section within the HRM is proportional to $s_{NN}^{-10}$,\footnote{An additional negative power 
of invariant energy is provided by ${1\over s-M_{A}^2}$ factor  in the differential cross section of 
the reaction [see Eqs.(\ref{pn_hrm_crs},\ref{pp_hrm_crs})].} it will be enhanced at small values of $s_{NN}$ 
that will correspond to smaller values of $\alpha_{NN}$ or larger values of $\alpha_s$.

The difference of the cross sections because of the different composition of the nuclear spectral functions 
entering the $pp$ and $pn$ breakup reactions can be seen in Fig.\ref{Fig.al_dep}(b) in which 
case one 
calculates the ratio of $pn$ to $pp$ breakup cross sections normalized to their 
values at $\alpha_s=1$.   The drop of the ratio in Fig.\ref{Fig.al_dep}.(b) at values close to 
$\alpha_s=1$ is the result of the suppression of the same-helicity two-proton 
component in the ground state nuclear wave function at small momenta. In this  case the  
spectral function is sensitive to the higher angular momentum components of the ground state 
nuclear wave function. This yields a  wider momentum distribution for the $pp$ spectral function 
as compared to that for $pn$ because no same-helicity state suppression exits for the latter.
The estimates indicate that differences in $\alpha_s$ dependencies of $pp$ and $pn$ breakup 
cross sections are rather large  and can play an additional role in checking the validity of 
the HRM.

\section{Polarization transfer of the hard rescattering  mechanism}
\label{VII}
One of the unique properties of the hard rescattering mechanism of two-nucleon breakup is 
that the  helicity of the nucleon from which a quark is 
struck is predominantly defined by the helicity of the incoming photon $\lambda_{1i} = \lambda_\gamma$ 
[see Eq.(\ref{ampl5})].  This property is a result of the fact that in the massless quark limit the helicity of 
the struck quark equals the helicity of the photon, $\eta_{1i}=\lambda_{\gamma}$, and assuming that at 
large $x$ the quark carries almost all the helicity of the parent nucleon one obtains 
$\lambda_{1i}\approx \eta_{1i}=\lambda_{\gamma}$.

Because within the HRM, the   energetic struck quark shares its momentum with a quark of the 
other nucleon through a hard gluon exchange, it will retain its initial helicity when it merges into 
the final outgoing nucleon.   It will also have $x^\prime\sim x\sim 1$, which allows 
us to conclude that the final outgoing nucleon will acquire the large part of struck quark's (as well as the 
photon's) helicity.  This mechanism will result in a large (photon) polarization transfer  for the 
hard two-nucleon breakup reactions.

An observable that is sensitive to polarization transfer processes 
is the quantity $C_{z^\prime}$, which for a circularly polarized photon measures the 
asymmetry of the hard breakup reaction with respect to the helicity of the 
outgoing proton.

A large value of $C_{z^\prime}$ was predicted within the HRM for the hard breakup of the deuteron 
in Ref.\cite{gdpnpol} that was observed in the  recent experiment of Ref.\cite{gdpnpolexp2}.   

For the case of the $^3$He target an additional experimental observation will be a comparison 
of $C_{z^\prime}$ asymmetries for $pp$ and $pn$ breakup channels. 
For the $^3$He target we define $C_{z^\prime}$ as follows:
\begin{eqnarray}
C_{z^\prime}  = 
{\sum\limits_{\lambda_{2f},\lambda_s,\lambda_a}
\left\{ 
\left|\langle +,\lambda_{2f},\lambda_s\mid {\cal M} \mid +, \lambda_A\rangle\right|^2 - 
\left|\langle -,\lambda_{2f},\lambda_s\mid {\cal M} \mid +, \lambda_A\rangle\right|^2\right\}\over
\sum\limits_{\lambda_{1f}\lambda_{2f},\lambda_s,\lambda_a}
\left|\langle \lambda_{1f},\lambda_{2f},\lambda_s\mid {\cal M} \mid +, \lambda_A\rangle\right|^2}.
\label{Cz1}
\end{eqnarray}
Using Eq.(\ref{ampl5}) and the definitions of Eq.(\ref{2phis}) for $C_{z^\prime}$ one obtains
\begin{equation}
C_{z^\prime} = {(|\phi_1|^2 - |\phi_2|^2)S^{++}+ (|\phi_3|^2 - |\phi_4|^2)S^{+-}\over
2|\phi_5|^2S^+ + (|\phi_1|^2 + |\phi_2|^2)S^{++}+ (|\phi_3|^2 + |\phi_4|^2)S^{+-}},
\label{Cz2}
\end{equation}
where
\begin{eqnarray}
& & S^{\pm,\pm}(t_1,t_2,\alpha,\vec p_s) = \nonumber \\ 
& & \ \ \ \ \sum\limits_{\lambda_A=-{1\over 2}}^{1\over 2}
\sum\limits_{\lambda_3 
= -{1\over2}}^{1\over 2} \left|\int \Psi_{^3\textnormal{\scriptsize {He,NR}}}^{\lambda_A}
(\vec p_1,\lambda_1=\pm{1\over 2},t_1;\vec p_2,\lambda_{2}=\pm{1\over 2},t_2;\vec p_s,\lambda_3)
m_N{d^2p_{2,\perp} \over (2\pi)^2}\right|^2
\label{Spm}
\end{eqnarray}
and $S^+ = S^{++} + S^{+-}$.

\medskip
\medskip

As follows from Eqs.(\ref{Cz2}) and (\ref{Spm}) one predicts significantly different 
magnitudes for $C_{z^\prime}$ for $pp$ and $pn$ breakup cases.

For the $pp$ breakup, $S_{pp}^{++}\ll S_{pp}^{+-}$ due to the smallness of the nuclear wave function component 
containing two protons in the same helicity state. As a result one expects
\begin{equation}
C_{z^\prime}^{pp} \approx {|\phi_3|^2-|\phi_4|^2\over |\phi_3|^2+|\phi_4|^2} \sim 0,
\label{czpp}
\end{equation}
while for the $pn$ break up case $S_{pn}^{++}\approx S_{pn}^{+-}$ then one obtains
\begin{equation}
C_{z^\prime}^{pn} \approx {|\phi_1|^2 + |\phi_3|^2-|\phi_4|^2\over 
|\phi_1|^2 + |\phi_3|^2+|\phi_4|^2} \sim {2\over 3},
\label{czpn}
\end{equation}
where in the last part of the equation we assumed that $|\phi_3| = |\phi_4| = {1\over 2}|\phi_{1}|$.

\section{Summary}
\label{VIII}
The hard rescattering mechanism of a two-nucleon breakup from the $^3$He nucleus at large c.m. angles
is derived from the assumption of  the dominance of quark-gluon degrees of freedom in the hard scattering 
process involving two nucleons.
The model explicitly assumes that the  photodisintegration process proceeds through the 
knock-out of a quark from one nucleon with a subsequent rescattering of that quark 
with a quark from the second nucleon.  While photon-quark scattering is calculated explicitly, 
the sum of all possible quark rescatterings is related to the hard elastic 
$NN$ scattering amplitude. Such a relation is found assuming that quark-interchange 
amplitudes provide the dominant contributions to the  hard elastic $NN$ scattering.

The model allows one to calculate the cross sections of the hard breakup of $pn$ and $pp$ pairs from $^3$He
expressing them through the amplitudes of elastic $pn$ and $pp$ scatterings, respectively.

Several results of the HRM are worth mentioning: First, the HRM predicts an approximate $s^{-11}$ scaling 
consistent with the predictions of the quark-counting rule. However, the model by itself 
is nonperturbative because the bulk of the incalculable part of the scattering amplitude is hidden 
in the amplitude of the $NN$ scattering that is taken from the experiment.

Second, because the hard $NN$ scattering amplitude enters into the  final amplitude of the photodisintegration 
reaction, the shape of the energy dependence of the $s^{11}$ weighted breakup cross section reflects the shape of 
the $s^{10}$ weighted $NN$ elastic scattering cross section. Because of a better accuracy of $pp$ elastic scattering 
data for $s_{NN}\le 24$~GeV$^2$,  we are able to predict a specific shape for the energy dependence of 
the hard $pp$ breakup cross section at photon energies up to $E_\gamma \le 5$~GeV. This prediction and the overall compliance with counting rules have been confirmed by a recent experiment on $^3$He photodisintegration  carried out at Jefferson Lab by the Hall A collaboration~\cite{Pomerantz:2009sb}.

Another observation is that, when $s^{-11}$ scaling is established, the HRM predicts an increase of the strength 
of the $pp$ breakup cross section relative to the $pn$ breakup as compared to the 
low energy results. This is the result of the feature  that 
within the quark-interchange mechanism of $NN$ scattering one has more charges flowing between nucleons 
in the $pp$ pair than in the $pn$ pair. This situation is  opposite in the low energy regime 
when no charged meson exchanges exist for the $pp$ pair.  Even though the large charge factor is involved 
in the $pp$ breakup its cross section is still by a factor of ten smaller than the cross section of 
the $pn$ breakup. Within the HRM, this is due to cancellation between the helicity conserving 
amplitudes $\phi_3$ and $\phi_4$, which have opposite signs for the $pp$ scattering.

Because of the smallness of the $pp$ breakup cross section, within the eikonal approximation, 
we estimated the possible contribution of 
three-body/two-step processes in which the initial two-body hard $pn$ breakup is followed by the 
charge-exchange rescattering of an energetic neutron off the spectator proton.  We found that this 
contribution has $s^{-12}$ energy dependence and is a small correction for spectator nucleon 
momenta $\le$150~MeV/$c$. However, the three-body/two-step process will dominate the hard $pp$ breakup 
contribution at large  transverse momenta of the spectator nucleon starting at $p_{s\perp}\ge 350$~MeV/$c$.

The next result of the HRM is the prediction of different spectator-momentum dependencies of breakup 
cross section for the $pp$ and $pn$ pairs.  This result follows from the fact that the ground state 
wave function of $^3$He containing two protons with the same helicity is significantly suppressed as 
compared to the same component in the $pn$ pair. Because of this,  the $pp$ spectral function 
is sensitive to the higher angular momentum components of the nuclear ground state wave function. 
These components generate wider momentum distribution as compared to say the $S$ component of the 
wave function.  As a result the cross section of the $pp$ breakup reaction exhibits wider momentum 
distribution as compared to the $pn$ cross section. 
Additionally because of the strong $s$ dependence of the reaction, the cross section exhibits an 
asymmetry in the light-cone momentum distribution of the spectator nucleon, favoring larger values of 
$\alpha_s$.

The final result of the HRM is the strong difference in prediction of the polarization transfer asymmetry 
for $pp$ and $pn$ breakup reactions for circularly polarized photons.
Because of the suppression of the same helicity $pp$ components in the $^3$He ground state wave function, the 
dominant helicity conserving $\phi_1$ component will not contribute to the polarization transfer 
process involving two protons.  Because of this effect,  the HRM predicts longitudinal polarization transfer  
$C_{z\prime}$, 
for the $pp$ breakup to be close to zero.  Because no such suppression exists for the $pn$ breakup, the HRM 
predicts a rather large magnitude for $C_{z'}\approx {2\over 3}$. 

Even though the HRM model does not contain free parameters,  for numerical estimates we use the  magnitude of 
elastic $NN$ cross sections as well as  some properties of the $NN$ helicity amplitudes.  This introduces certain 
error in our prediction of the magnitudes of the breakup cross sections. For the $pn$ breakup this error is mainly 
related to the uncertainty in the magnitude of the absolute cross section of hard elastic $pn$ scattering which 
is on the level 30\%.
For the $pp$ breakup the main source of the uncertainty is the magnitude of the  cancellation 
between $\phi_3$ and $\phi_4$, which is sensitive to the angular distribution of helicity amplitudes. 
The uncertainty that results from the angular function is on the level of 40~\%.  These uncertainties should be 
considered on top of the theoretical uncertainties that the HRM contains due to approximations such 
as estimating the scattering amplitude at maximal value of the nuclear wave function at $\alpha = {1\over 2}$.
The latter may introduce an uncertainty as much as $20\%$ in the breakup cross section.

In conclusion, having that the HRM energy distribution  for $pp$ breakup shows good agreement with recent experimental data, further experimental evaluation of all the above-mentioned  set predictions 
will contribute in verifying the validity of the hard rescattering model. Deeper insight is also expected from  progress in extracting the helicity 
amplitudes of the hard $NN$ scattering which will allow much improvement regarding the accuracy of the HRM predictions.
\medskip
\medskip

Mounting evidence favoring a HRM picture of $NN$ breakup reactions motivates the study of additional processes within the HRM framework. One such extension is discussed in the following chapter in which a HRM description of deuteron breakup into $\Delta\Delta$-isobars is presented in contrast with a picture of this breakup channels emerging from $\Delta\Delta$ components of the deuteron. It also provides an scenario alternative to that of a deuteron  transitioning from a baryon-baryon to a six quark system as the relative transverse momentum of the outgoing $\Delta\Delta$ system is asymptotically increased.

% ------------------------------------------------------------------------

%%% Local Variables: 
%%% mode: latex
%%% TeX-master: "../thesis"
%%% End: 

\chapter{HARD BREAKUP OF THE DEUTERON INTO TWO $\Delta$-ISOBARS}
\label{HBdDD}
\ifpdf
    \graphicspath{{Chapter3/Chapter3Figs/PNG/}{Chapter3/Chapter3Figs/PDF/}{Chapter3/Chapter3Figs/}}
\else
    \graphicspath{{Chapter3/Chapter3Figs/EPS/}{Chapter3/Chapter3Figs/}}
\fi

%\begin{abstract}
In this chapter,  high energy photodisintegration of the deuteron into two  $\Delta$-isobars at large 
center of mass angles is studied within the QCD hard rescattering 
model (HRM). According to the HRM, the process develops  in three main steps: the photon knocks the quark from one of 
the nucleons in the deuteron; the struck quark rescatters off a quark from  the other nucleon sharing the high  energy 
of the photon; then the energetic quarks recombine into two outgoing baryons which have large transverse momenta.
Within the HRM, the cross section is expressed through the amplitude of $pn\rightarrow \Delta\Delta$ scattering which 
are evaluated on the basis of the quark-interchange model of hard hadronic scattering.   Calculations show that 
the angular distribution and the strength of the photodisintegration  is mainly determined by the properties of the 
 $pn\rightarrow \Delta\Delta$  scattering.  
 
Through the HRM, the cross section of the  deuteron 
breakup to $ \Delta^{++}\Delta^{-}$ is predicted to be 4-5 times larger  than that of the breakup  to  the $ \Delta^{+}\Delta^{0}$ channel.  Also, the angular distributions for 
these two channels are markedly different.   These  can be compared with the predictions  derived from
the assumption that two hard $\Delta$-isobars are the  result of the disintegration of the preexisting $\Delta\Delta$ components 
of the deuteron wave function.  In this case, one expects  the  angular distributions and cross sections of the 
breakup in both $ \Delta^{++}\Delta^{-}$  and  $ \Delta^{+}\Delta^{0}$ channels to be similar.

\section{Introduction}

Experiments on large center of mass angle breakup of the deuteron into $pn$ channel and of a $pp$ system in $^3$He photodisintegration confirmed the prediction of quark-counting rule\cite{BCh} according to which the 
energy  dependence of the differential cross section at  large c.m. scattering angles  scales 
as ${d\sigma\over dt} \sim s^{-11}$. However, calculations of the absolute cross sections  require a more detailed 
understanding of the dynamics of these processes.  The considered theoretical models addressing this task can  be 
grouped  by two distinctly different underlying assumptions made in the calculations\cite{gheppn}. 
The first assumes that the large c.m. angle  nucleons are produced through the interaction of  the incoming  
photon with a pre-existing hard two nucleon system in the nucleus\cite{RNA1,RNA2,DN}.  
The second approach assumes that the two high  momentum  nucleons are produced  
through a hard rescattering at the final state of the 
reaction\cite{QGS,gdpn,Frankfurt:1999ic,hrmictp,Sargsian:2003sz} as described  in the previous chapter.

In the hard rescattering model (HRM)\cite{gdpn} in particular, by explicitly introducing 
quark degrees of freedom, a parameter-free cross section has been obtained for hard 
photodisintegration of the  deuteron at 90$^0$ c.m.   angle \cite{gdpn,Frankfurt:1999ic}.  Also, the HRM prediction of the  hard breakup of 
two protons from the $^3$He nucleus described in the previous chapter agreed reasonably well with the recent experimental 
data\cite{Pomerantz:2009sb}.

In this chapter, the  HRM approach is extended to calculate hard breakup of the deuteron into   
two-$\Delta$-isobars  produced at large angles in the $\gamma-d$ center of mass reference frame. Within the HRM
the relative strength of $\gamma d\rightarrow \Delta^{++}\Delta^{-}$ and 
 $\gamma d\rightarrow \Delta^{+}\Delta^{0}$ cross sections is calculated as they compare with the   $\gamma d\rightarrow pn$ cross 
section.

The investigation of the production of two energetic $\Delta$-isobars from the deuteron has an important significance in probing 
 possible non-nucleonic components in the deuteron wave function
(see e.g. Refs.\cite{Harvey:1980rva,Brodsky:1983vf,Brodsky:1985gs,FS88,SRCRev}). The study of the deuteron photodisintegration into $\Delta\Delta$-isobars channels was proposed as a venue for investigating the evolution of a nucleon-nucleon system into a six quark system. The onset of a six quark picture of the deuteron could then be marked by a large increase of the $\gamma d\to \Delta\Delta$ cross section. The latter prediction assumes that such cross section is small for a nucleon dominated deuteron wave function because of its suppressed $\Delta\Delta$ components. In contrast, for a six quark deuteron, $NN$ and $\Delta\Delta$ components contribute with comparable strength to the deuteron wave function (roughly 10\% and 8\%, respectively) while more than 80\% is contributed by $CC$ (hidden color) components for which unlike $N$ or $\Delta$, $C$ has a color charge. %(Brodsky,Ji,Lepage,1983).

High energy $\gamma d\to \Delta\Delta$ with $\Delta\Delta$ emerging at large transverse momentum is thought to probe the onset of hidden color components in the deuteron. Assuming that the high $p_T$ $\Delta\Delta$ system was created in the initial state of the interaction, in the asymptotic limit we have that
$${d\sigma^{\gamma d\to \Delta\Delta}\over dt}\sim {d\sigma^{\gamma d\to pn}\over dt}.$$
Under this same assumption we also have that,
$${d\sigma^{\gamma d\to \Delta^{++}\Delta^{-}}\over dt}= {d\sigma^{\gamma d\to \Delta^+\Delta^0}\over dt},$$
since both $\Delta\Delta$ channels in general contribute with the same strength to the spin-isospin wave function of the deuteron.

In contrast, throughout this chapter, the role of  hard rescattering in these processes will be assessed. 
It will allow us also to explore another venue for checking the basic mechanism of high momentum transfer  
breakup of nuclei into two baryons. As it will be shown in the sections that follow, the HRM description of these reactions results in distinct predictions for angular 
distribution of the $\Delta$-isobar pair at large c.m. production angle as well as their relative strength compared with  
the production of the $pn$ pair at the same kinematics.

Despite  experimental challenges associated with the investigation of 
two $\Delta$-isobar  breakup of the deuteron\cite{PR}, there are ongoing efforts in performing 
such experiments at the Jefferson Lab\cite{Gao,DM} which will provide the opportunity to asses the validity of the HRM scenario by testing its predictions discussed in this chapter.

\section{Hard Rescattering Model}
\label{hrm}
We consider the  photoproduction of two baryons, $B_1$ and $B_2$, in the reaction,
\begin{equation}
\gamma + d \rightarrow B_1 + B_2
\label{Reaction}
\end{equation}
in which the  baryons are produced at large angles in the $\gamma-d$ center of mass reference frame.

According to  the HRM,  the  large angle breakup of the  NN system proceeds through the knock-out of 
a valence quark from one of the nucleons with subsequent hard rescattering of the  struck-quark
with a valence quark of the  second nucleon.  
 The two quarks then recombine with the spectator systems of nucleons forming  
two emerging baryons with large transverse momenta. The hard rescattering provides the mechanism 
of sharing  the photon's energy among two final baryons.
 
The invariant amplitude of  the photodisintegration Eq.(\ref{Reaction})  is calculated 
by applying Feynman diagram  rules to diagrams similar to  Fig.\ref{hrmf}.  During the calculation we introduce 
 undetermined quark wave functions of baryons to account for the transition of the initial nucleons to 
 the quark-spectator systems, and also for the recombination of the final state quarks with these spectator systems into 
the final two baryon system.

\begin{figure}[ht]
\centering\includegraphics[height=4cm,width=6cm]{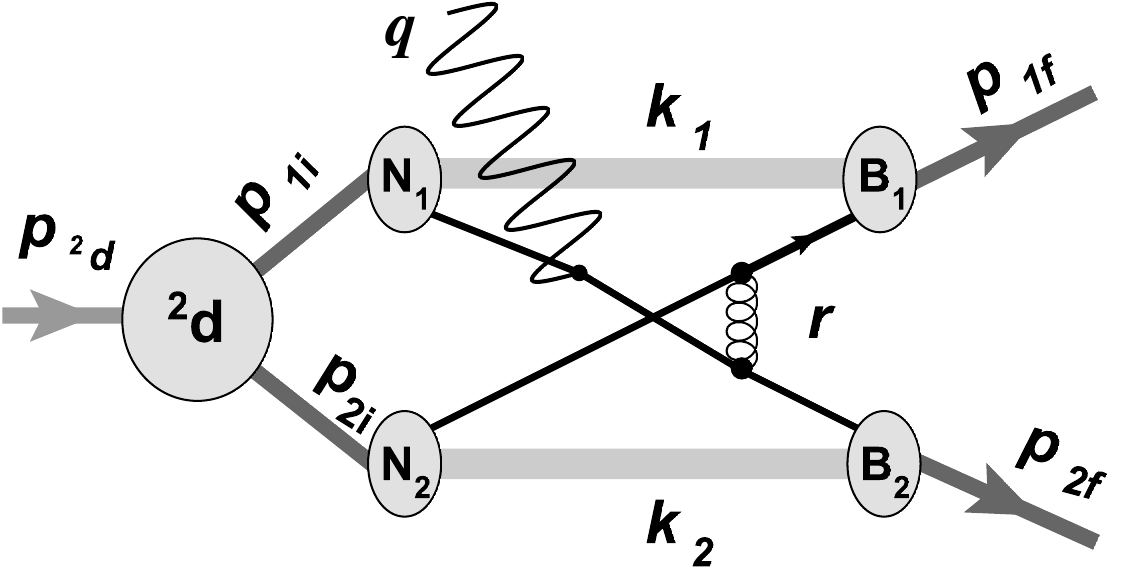}
\caption{Deuteron photodisintegration according to the HRM}
\label{hrmf}
\end{figure}

Fig.\ref{hrmf} displays the chosen independent momenta for three loop integration involved in the 
invariant amplitude. As it was the case for $^3$He photodisintegration in the previous chapter, two major approximations  simplify further calculations. First, using the fact that 
the struck quark is very energetic we treat it on its mass shell. Then the struck quark's propagator is evaluated
at it's pole value at such magnitudes of  nucleon momenta that maximize the deuteron wave function. 
These approximations allow us to factorize the invariant amplitude into three distinguished parts.  The first, 
representing the transition amplitude of the deuteron into the ($pn$) system, which can be evaluated using a
realistic deuteron wave function.  The second is the amplitude of photon-quark interaction, and  the third term 
represents the hard rescattering of the struck quark with  recombination into a  two large transverse momentum baryonic system.
Combined with the initial state nucleon wave functions, the rescattering part is expressed through the 
quark-interchange~(QI) amplitude of $pn\rightarrow B_1 B_2$ scattering. A detailed derivation is given in section \ref{Sec.2B} in conjunction with Appendices \ref{app1} and \ref{app2} (see also Ref.\cite{gdpn}). 
%in  Refs.\cite{gdpn}.%,Sargsian:2008zm}.
After the  above mentioned factorization is made, the overall invariant amplitude of $\gamma d\rightarrow B_1 B_2$ 
reaction can be expressed as follows:
\begin{eqnarray}
& & \langle\lambda_{1f},\lambda_{2f}\mid {\cal M} \mid \lambda_\gamma, \lambda_d\rangle 
 =  ie[\lambda_\gamma]\times  \nonumber \\
& & \ \ \ \ \left\{ \sum\limits_{i \in N_1}\sum\limits_{\lambda_{2i}} \int 
{Q_i^{N_1}\over \sqrt{2s'}}
\langle \lambda_{2f};\lambda_{1f}\mid T^{QI}_{(pn\to B_1B_2),i}(s,t_N)\mid 
\lambda_\gamma;\lambda_{2i}\rangle 
\Psi_{\textnormal{\scriptsize {d}}}^{\lambda_d}(p_{1i},\lambda_\gamma;p_{2i},\lambda_{2i})
{d^2p_\perp \over (2\pi)^2} \right. \nonumber \\ 
&  &\ \ \ \  + \left.
\sum\limits_{i \in N_2}\sum\limits_{\lambda_{1i}} \int 
{Q_i^{N_2}\over \sqrt{2s'}}
\langle \lambda_{2f};\lambda_{1f}\mid T^{QI}_{(pn\to B_1B_2),i}(s, t_{N})\mid \lambda_{1i};
\lambda_{\gamma}\rangle 
\Psi_{\textnormal{\scriptsize {d}}}^{\lambda_d}(p_{1i},\lambda_{1i};p_{2i},\lambda_{\gamma})
{d^2p_\perp \over (2\pi)^2} \right\}
\label{dampl}
\end{eqnarray}
where $\lambda_\gamma$,$\lambda_d$, $\lambda_{1f}$ and $\lambda_{2f}$ are the helicities of the
photon, deuteron and the two outgoing baryons respectively.
Here $\Psi_{\textnormal{\scriptsize {d}}}^{\lambda_d}(p_{1i},\lambda_{1i};p_{2i},\lambda_{2i})$ is the $\lambda_d$-helicity 
light-cone deuteron wave function defined in the $q_+=0$ reference frame.  The initial light-cone momenta of 
the nucleons in the deuteron are $p_{1i}=(\alpha_{1i}={1\over 2}, p_{1i\perp}=-p_\perp)$ and 
$p_{2i} = (\alpha_{2i}={1\over 2},p_{2i\perp}=p_\perp)$ with $\lambda_{1i}$ 
and $\lambda_{2i}$ being their helicities respectively.  
The ${1\over \sqrt{s^\prime}}$ factor with $s^\prime = s-M_d^2$ comes from the energetic propagator 
of the struck quark before its rescattering.  The squares of the total invariant energy as well as the momentum 
transfer are defined as follows:
\begin{eqnarray}
s & = & (q + p_d)^2 = (p_{1f} + p_{2f})^2 = 2E^{lab}_\gamma M_d + M_d^2 \nonumber \\
t &  = & (p_{1f}-q)^2 = (p_{2f}-p_d)^2 
\label{kin}
\end{eqnarray}
where $q$, $p_d$, $p_{1f}$ and $p_{2f}$ are the four-momenta of the photon, deuteron and two outgoing baryons respectively. 
The lab energy of the photon is defined by $E^{lab}_\gamma$, and $M_d$ is  the mass of the deuteron.
The transfer momentum, $t_N$ in the rescattering amplitude in Eq.(\ref{dampl}) is defined as:
\begin{equation}
t_N = (p_{1f}-p_{1i}-q)^2 = (p_{2f}-p_{2i}) \approx   (p_{2f}-{p_d\over 2})^2  = {t\over 2} + 
{m_{B2}^2\over 2} - {M_d^2\over 4},
\label{dtN}
\end{equation}
where the approximation in the right hand side follows from the assumption that the magnitudes of light-cone momentum fractions 
of  bound nucleons dominating in the scattering amplitude are $\alpha_{1i}=\alpha_{2i} = {1\over 2}$, and that the transverse 
momenta of these nucleons are negligible as compared to the momentum transfer in the reaction, $p_\perp^2 \ll |t_N|,|u_N|$.

In Eq.(\ref{dampl})  the following expression
\begin{equation}
Q_i\langle \lambda_{2f};\lambda_{1f}\mid T^{QI}_{(pn\to B_1B2),i}(s,t_N)\mid 
\lambda_{1i};\lambda_{2i}\rangle 
\label{QT}
\end{equation}
represents the quark-charge weighted QI amplitude of $pn\rightarrow B_1 B_2$ hard exclusive 
scattering. The  factor   $Q_i$ corresponds to the charge (in $e$ units) of the quark  that interacts with the incoming photon. 
In a further approximation we factorize the hard rescattering amplitude  from the integral since the momentum transfer 
entering in $T_{(pn\to B_1B_2),i}(s, t_N)$ significantly exceeds the Fermi momentum of the nucleon in the deuteron. 
Also, after calculating the overall quark-charge factors, the QI scattering amplitudes are identified with the $NN\rightarrow B_1B_2$ helicity amplitudes as follows:
\begin{equation}
\langle \lambda_{2f};\lambda_{1f}\mid T_{pn\to B_1B_2}^{QI}(s,t_N)\mid \lambda_{1i};\lambda_{2i}\rangle =\phi_j(s,\theta_{c.m.}^{N}),
\label{notn}
\end{equation}
where $\theta_{c.m.}^N$ is the effective center of mass angle defined for given  $s$ and $t_N$.

\medskip

The differential cross section for unpolarized scattering is obtained through:
\begin{eqnarray}
{d\sigma_{\gamma d\rightarrow B_1B_2}\over dt}={1\over16\pi}{1\over (s-M_d^2)}|\bar{\cal M}|_{\gamma d\rightarrow 
B_1B_2}^2
\label{gdpnBB}
\end{eqnarray}
where
\begin{equation}
|\bar{\cal M}|_{\gamma d\rightarrow B_1B_2}^2 = 
{1\over 3}{1\over 2} \sum\limits_{\lambda_{1f},\lambda_{2f},\lambda_\gamma,\lambda_d}
\mid \langle \lambda_{1f},\lambda_{2f}\mid {\cal M}\mid \lambda_\gamma,\lambda_d\rangle|^2,
\label{dM2}
\end{equation}
with  the invariant amplitude square  averaged by the number  of helicity states of the deuteron and photon.

\section{Cross section of the $\gamma + d\rightarrow pn$ breakup reaction}
\label{SECgdpn}

We derive the amplitude of the breakup of the deuteron into the $pn$ pair from Eq.(\ref{dampl}) by introducing the
independent helicity amplitudes of $pn$ elastic scattering Eq.(\ref{pnham}) and by separating the quark-charge factors into $\hat Q^{N_1}$ and $\hat Q^{N_2}$ 
which correspond to the scattering of the  photon off  the quark of the first and the second nucleons in the deuteron. Then, for Eq.(\ref{dM2}) one obtains:
\begin{eqnarray}
\bar{|{\cal M}|^2}&=&{1\over2}{1\over3}{e^2\over 2s^\prime}\left[S_{12}\left\{|(\hat{Q}^{N_1}+\hat{Q}^{N_2})\phi_1|^2+
|(\hat{Q}^{N_1}+\hat{Q}^{N_2})\phi_2|^2\right\}\right.\nonumber\\
&&+\left.S_{34}\left\{|\hat{Q}^{N_1}\phi_3+\hat{Q}^{N_2}\phi_4|^2+|\hat{Q}^{N_1}\phi_4+
\hat{Q}^{N_2}\phi_3|^2\right\}\right.\nonumber\\
&&+\left.2S_0|(\hat{Q}^{N_1}+\hat{Q}^{N_2})\phi_5|^2\right],
\label{asamp}
\end{eqnarray}
where the light-cone spectral functions of the deuteron  are defined as follows:
\begin{eqnarray}
S_{12}&=&\sum^{1}_{\lambda=-1}\sum^{1\over2}_{(\lambda_1=\lambda_2=-{1\over2})}
\left|\int\Psi_{\textnormal{\scriptsize {d}}}^{\lambda_d}(p_1,\lambda_1;p_2,\lambda_2)
{d^2p_\perp \over (2\pi)^2}\right|^2,\nonumber\\
S_{34}&=&\sum^{1}_{\lambda=-1}\sum^{1\over2}_{(\lambda_1=-\lambda_2=-{1\over2})}\left
|\int\Psi_{\textnormal{\scriptsize {d}}}^{\lambda_d}(p_1,\lambda_1;p_2,\lambda_2)
{d^2p_\perp \over (2\pi)^2}\right|^2,\nonumber\\
S_0&=&S_{12}+S_{34}.
\label{spfun}
\end{eqnarray}
Eq.(\ref{asamp}) can be further simplified if we assume (see e.g.\cite{RS})  that $\phi_3\approx \phi_4$, 
as well as  $S_{12} \approx S_{34}  = {S_0\over 2}$, which results in:

\begin{eqnarray}
\bar{|{\cal M}|^2}&=&{1\over2}{1\over3}{e^2\over 2s^\prime}Q_{F,pn}^2 {S_0\over 2} \left[ 
|\phi_1|^2+|\phi_2|^2+ |\phi_3|^2 +|\phi_4|^2+|\phi_4|^2+
4|\phi_5|^2\right].
\label{asamp_2}
\end{eqnarray}
Using the  expression of the differential cross section of elastic $pn$ scattering:
\begin{equation}
{d\sigma^{NN\rightarrow NN}(s,\theta_{c.m.}^N)\over dt} =  {1\over 16\pi}{1\over s(s-4m_N^2)}
{1\over 2}(|\phi_1|^2 + |\phi_2|^2 + |\phi_3|^2 + |\phi_4|^2 + 4|\phi_5|^2),
\label{crs_BB}
\end{equation}
and the relation between  the light-cone   and   non-relativistic deuteron wave functions\cite{FS81,gdpn,taggs,polext} at small internal momenta:
$\Psi_d(\alpha, p_\perp) = (2\pi)^{3\over 2}\Psi_{d,NR}(p)\sqrt{m_N}$  in Eq.(\ref{asamp}), 
for the  differential  cross section on obtains from Eq.(\ref{gdpnBB}):
\begin{equation}
{d\sigma^{\gamma d \rightarrow pn}(s,\theta_{c.m.}) \over dt} = {\alpha Q_{F,pn}^2 8\pi^4\over s^\prime} 
{d\sigma^{pn\rightarrow pn}(s,\theta_{c.m.}^N)\over dt} \bar S_{0,NR},
\label{gdpncrs}
\end{equation}
where  we neglected the difference between $4m_N^2$ and $M_d^2$. 
Here the  averaged non-relativistic spectral function of  the deuteron is defined as follows:
\begin{equation}
\bar S_{0,NR} = {1\over 3}\sum\limits_{\lambda=-1}^{\lambda=1}
\sum\limits_{\lambda_1,\lambda_2=-{1\over2}}^{{1\over 2}}
\left|\int\Psi_{\textnormal{\scriptsize {d,NR}}}^{\lambda_d}(\alpha={1\over 2},p_\perp,\lambda_1;
\alpha={1\over 2},-p_{\perp},\lambda_2)\sqrt{m_N}
{d^2p_\perp\over (2\pi)^2}\right|^2,
\label{avspfun}
\end{equation}
where $\Psi_{d,NR}$ is the non relativistic deuteron wave function, which can be 
calculated using realistic $NN$ interaction potentials.  

The quark-charge factor, $Q_{F,pn} ={1\over 3}$\cite{gdpn} 
accounts for the amount of the effective charge exchanged between the proton and the neutron in the 
rescattering.  It is estimated by counting all  the possible quark-exchanges within  the $pn$ pair weighted with 
the charge of one of the exchanged quarks  (for more details see Appendix~\ref{haqim}). The result in Eq.(\ref{gdpncrs}) is remarkably
simple and contains no free parameters. It can be evaluated using the experimental values of the differential 
cross section of the elastic $pn$ scattering, ${d\sigma^{pn\rightarrow pn}(s,\theta_{c.m.}^N)\over dt}$. 
The angle $\theta^{N}_{c.m.}$  entering in the $pn\rightarrow pn$ cross section is the center of mass angle of the scattering corresponding to the $NN$ 
elastic reaction at $s$ and $t_{N}$.  It is related to $\theta_{c.m.}$ of the $pn$ photodisintegration by (See Appendix \ref{kinrel}):  %\cite{Sargsian:2008zm}:
\begin{equation}
cos(\theta_{c.m.}^N) = 1 - {(s-M_d^2)\over 2(s-4m_N^2)}
{(\sqrt{s}-\sqrt{s-4 m_N^2}cos(\theta_{c.m.}))\over \sqrt{s}} + {4m_N^2 - M_d^2\over 2(s-4m_N^2)} .
\label{dtheta_cmn}
\end{equation} 
It is worth mentioning that as it follows from the  equation above, $\theta_{c.m.} = 90^0$ photodisintegration
will correspond to the $\theta_{c.m.}^{N} = 60^0$ hard $pn$ elastic rescattering at the final state of the reaction.

\section{Cross section of the $\gamma  d\rightarrow \Delta\Delta$ breakup reaction}
\label{delpr}

We use an approach similar to that in Sec.\ref{SECgdpn}   to derive the invariant  amplitude  of the 
 $\gamma  d\rightarrow \Delta\Delta$  reactions.  In this case Eq.(\ref{dampl}) requires an input of the helicity amplitudes 
of the corresponding  $pn\rightarrow\Delta\Delta$ scattering.  One has a total 32 independent 
helicity amplitudes for this scattering.  To simplify further our  derivations, we will restrict ourselves 
by considering only  the seven helicity conserving amplitudes given in Eq.(\ref{hconamp}).  Using these 
amplitudes in Eq.(\ref{dampl}) and separating the quark-charge factors into $\hat Q^{N_1}$ and $\hat Q^{N_2}$,  similar to Eq.(\ref{asamp}) one obtains
 \begin{eqnarray}
\bar{|{\cal M}|^2}_{\gamma d\rightarrow\Delta\Delta}&=&{1\over2}{1\over3}{e^2\over 2s^\prime}\left[S_{12}\left\{|(\hat{Q}^{N_1}+\hat{Q}^{N_2})\phi_1|^2
+|(\hat{Q}^{N_1}+\hat{Q}^{N_2})\phi_6|^2+|(\hat{Q}^{N_1}+\hat{Q}^{N_2})\phi_7|^2\right\}\right.\nonumber\\
&&+\left.S_{34}\left\{|\hat{Q}^{N_1}\phi_3+\hat{Q}^{N_2}\phi_4|^2+|\hat{Q}^{N_1}\phi_4+\hat{Q}^{N_2}\phi_3|^2\right.\right.\nonumber\\
&&+\left.\left.|\hat{Q}^{N_1}\phi_8+\hat{Q}^{N_2}\phi_9|^2+|\hat{Q}^{N_1}\phi_9+\hat{Q}^{N_2}\phi_8|^2\right\}\right],
\label{asampDD}
\end{eqnarray}
where $S_{12}$ and $S_{34}$ are defined in Eq.(\ref{spfun}). Similar to the previous section, 
we simplify further the above expression assuming that all helicity conserving amplitudes are of the 
same order of magnitude. Assuming also that  $S_{12}\approx S_{34} \approx {S_0\over 2}$, we obtain
\begin{eqnarray}
\bar{|{\cal M}|^2}&=&{1\over2}{1\over3}{e^2\over 2s^\prime} Q_{F,\Delta\Delta} {S_0\over 2} \left[ 
|\phi_1|^2+|\phi_3|^2 +|\phi_4|^2+|\phi_4|^2
+|\phi_6|^2+|\phi_7|^2+|\phi_8|^2+
|\phi_9|^2\right],
\label{asampDD_2}
\end{eqnarray}
where $Q_{F,\Delta\Delta} = \hat Q^{N_1} +  \hat Q^{N_2} = {1\over 3}$  is obtained by using the same approach as for the case of 
the $pn$ breakup in Sec.\ref{SECgdpn}.   Using now the expression of the 
differential cross section of  $pn\rightarrow\Delta\Delta$ scattering,
\begin{eqnarray}
& & {d\sigma^{pn\rightarrow\Delta\Delta}({s,\theta^N_{c.m.}})\over dt}= \nonumber \\
& & \ \ \ \ {1\over16\pi}{1\over(s-4m^2_N)}{1\over2} \left[ |\phi_1|^2+|\phi_3|^2 +|\phi_4|^2+|\phi_4|^2
+|\phi_6|^2+|\phi_7|^2+|\phi_8|^2+ |\phi_9|^2\right] 
%\nonumber\\
\label{crs_pn_DD}
\end{eqnarray}
as well as the relation between light-cone and non relativistic deuteron wave function  discussed in Sec.\ref{SECgdpn}, from 
Eq.(\ref{gdpnBB}) we obtain the following expression for the differential cross section of the
$\gamma d\rightarrow \Delta\Delta$ scattering:
 \begin{equation}
{d\sigma^{\gamma d \rightarrow \Delta\Delta}(s,\theta_{c.m.}) \over dt} = {\alpha Q_{F,\Delta\Delta}^2 8\pi^4\over s^\prime} 
{d\sigma^{pn\rightarrow \Delta\Delta}(s,\theta_{c.m.}^N)\over dt} \bar S_{0,NR},
\label{gdDDcrs}
\end{equation}
where $\bar S_{0,NR}$ is given in  Eq.(\ref{avspfun}).  The  effective c.m. angle $\theta_{c.m.}^N$ entering in 
the argument of the differential cross section of $pn\rightarrow \Delta\Delta$ reaction can  be calculated
by using Eqs. (\ref{kin}) and (\ref{dtN}) (see also Appendix~\ref{kinrel}) to obtain
\begin{equation}
cos\theta^N_{c.m.}={1\over2\sqrt{\left(s-4m_N^2\right)\left(s-4m_\Delta^2\right)}}\left[{s-M_d^2\over2
\sqrt s}\sqrt{s-4m_\Delta^2}cos\theta_{c.m.}+s-4m_N^2\right].
\label{kin2}
\end{equation}
 
As it follows from Eq.(\ref{gdDDcrs}),  provided there are  enough experimental data on high momentum transfer 
 $pn\rightarrow\Delta\Delta$ differential  cross sections, the ${\gamma d \rightarrow \Delta\Delta}$ cross section 
can be computed without introducing  an adjustable free parameter.  However,  there are no experimental data on  hard exclusive
$pn\rightarrow \Delta\Delta$ reactions with sufficient accuracy that would allow us  to make   quantitative 
estimates  based on Eq.(\ref{gdDDcrs}).  Instead, in the next section we will attempt  to make quantitative 
predictions  based on the  quark-interchange framework of   hard scattering.

\section{Estimates of the relative strength of the  $\Delta\Delta$ breakup reactions.}

The results presented in this section are calculated considering the experimental observation \cite{h20} that 
the  quark-interchange \cite{Sivers:1975dg}  represents the dominant mechanism of  hard exclusive scattering of baryons 
that carry valence quarks with common flavor.   The  quark-interchange mechanism however will not allow
us to calculate the absolute cross sections.  Instead, we expect that its predictions 
will be more reliable for the ratios of the differential cross sections for different exclusive channels.

As an illustration of the reliability of calculations of cross section ratios in the QI model, in Fig.\ref{pn_to_pp_fig} compares the QI predictions for the ratios of $pn$ to $pp$ 
differential cross sections at $90^0$ c.m. scattering. Here, we compare predictions based on 
  SU(6)\cite{FGST,BCL} and diquark  (See chapter \ref{PNES} and Ref.\cite{Granados:2009jh}) %
 symmetry approaches for 
the valence quark wave function of the nucleons. As the comparison shows, one achieves  a rather reasonable 
agreement with the data without any additional normalization parameter. 
On the basis of this agreement, we now estimate the ratio of the differential cross sections of $\gamma d\rightarrow \Delta\Delta$ 
to the $\gamma d \rightarrow pn$ cross sections. We use both SU(6) and diquark-symmetry quark wave functions  of the 
nucleon and  $\Delta$-isobars (see Appendix~\ref{bw}) in the calculation of  the $pn\rightarrow \Delta\Delta$ amplitudes (see Appendix~\ref{haqim}).

\begin{figure}[ht]
\centering\includegraphics[height=6cm,width=8cm]{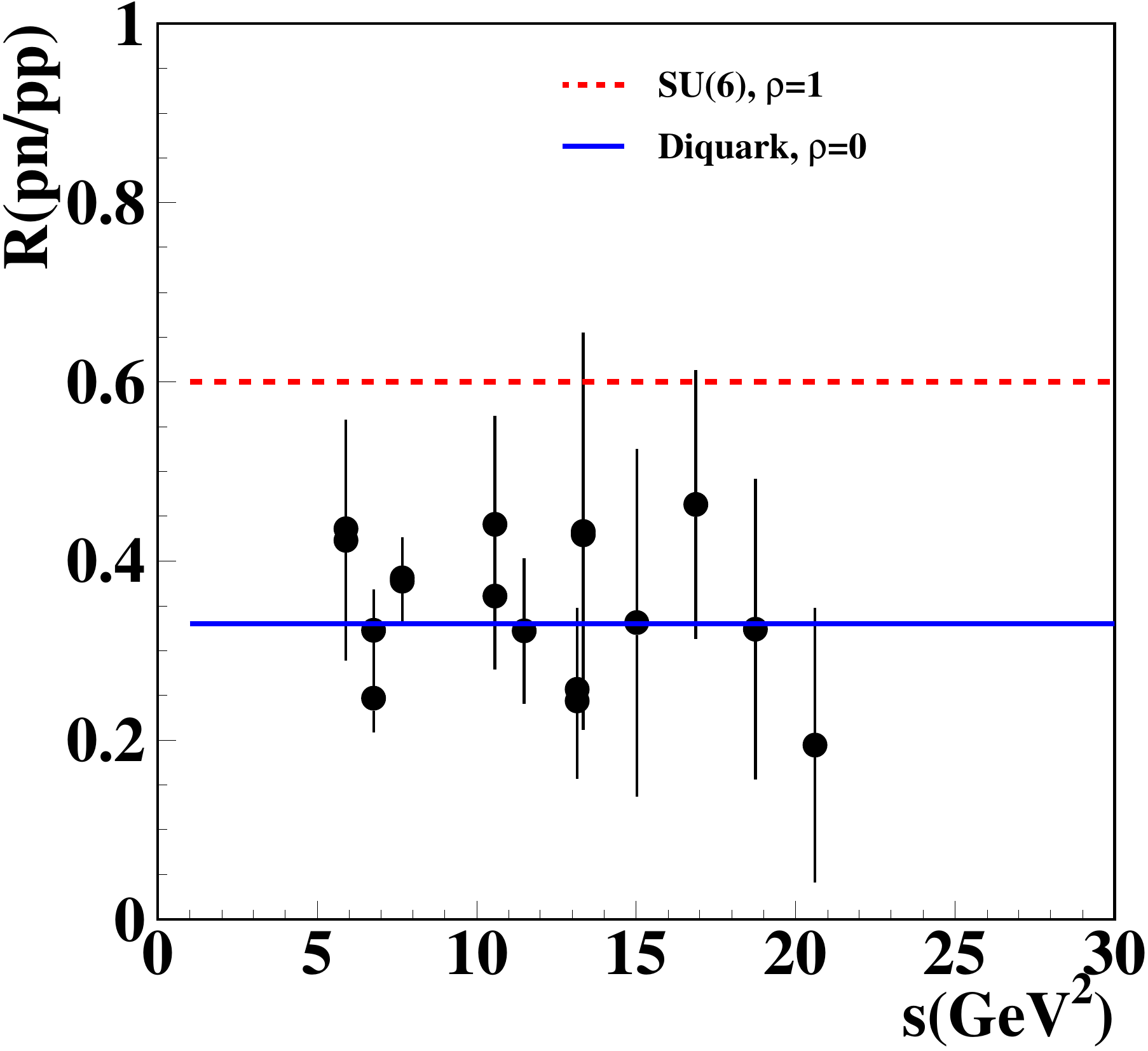}
\caption{(Color online) Ratio of the  $pn\rightarrow pn$ to $pp\rightarrow pp$   elastic differential cross sections 
as a function of $s$ at $\theta^{N}_{c.m.}=90^0$. }
\label{pn_to_pp_fig}
\end{figure}

To calculate the photodisintegration amplitudes we go back to  Eqs.(\ref{asamp}) and (\ref{asampDD}) 
and evaluate the quark-charge factors using   SU(6) or diquark symmetries of the valence quark wave functions of baryons.  
For this  we separate the $t$ and $u$ channels in  the helicity amplitudes:
\begin{equation}
\phi_i(s,\theta_{c.m.}^N)= \phi^t_i(s,\theta_{c.m.}^N)+\phi^u_i(s,\theta_{c.m.}^N)
\end{equation}
and then treat the charge factors  for the given nucleon $N$ as:
\begin{equation}
\hat{Q}^{N}\phi_l=Q_i^{t,N}\phi^t_l+Q_i^{u,N}\phi^u.
\label{opa}
\end{equation}
This yields the following expression for the  photodisintegration amplitude of Eq.(\ref{dampl}) :
 \begin{eqnarray}
\langle\lambda_{1f},\lambda_{2f}\mid {\cal M} \mid \lambda_\gamma, \lambda_d\rangle 
& = & ie[\lambda_\gamma]\times \left\{\sum\limits_{\lambda_{2i}}  
{1\over \sqrt{2s'}}\left[Q^{tN_1}_i\phi^t_i+Q^{uN_1}_i\phi^u_i\right]_{\lambda_{2i}}
\int\Psi_{\textnormal{\scriptsize {d}}}^{\lambda_d}(p_1,\lambda_\gamma;p_2,\lambda_{2i})
{d^2p_\perp \over (2\pi)^2} \right. \nonumber \\ 
& + &\left.
\sum\limits_{\lambda_{1i}}{1\over \sqrt{2s'}}\left[Q^{tN_2}_i\phi^t_i+Q^{uN_2}_i\phi^u_i\right]_{\lambda_{1i}} \int 
\Psi_{\textnormal{\scriptsize {d}}}^{\lambda_d}(p_1,\lambda_{1i};p_2,\lambda_{\gamma})
{d^2p_\perp \over (2\pi)^2} \right\}.
\label{dampl5}
\end{eqnarray}

\subsection{$\gamma d\rightarrow pn$ scattering}
For the $\gamma d\rightarrow pn$ amplitude, the charge factors calculated for the helicity conserving amplitudes 
according to the QI framework yield for both SU(6) and diquark models (see Appendix B) 
\begin{eqnarray}
Q^{tN_1}_j & = & Q^{tN_2}_j={Q_{F,pn}\over2} \nonumber \\
Q^{uN_1}_j & = & -2Q^{uN_2}_j=2Q_{F,pn}
\label{dQFpn}
\end{eqnarray}
 with $Q_{F,pn}={1\over3}$ and independent of j.  Using these relations in  Eq.(\ref{opa}),   from Eqs.(\ref{dampl5})  
and (\ref{asamp}) one obtains
\begin{equation}
|\bar{\cal M}|_{\gamma d\longrightarrow pn}^2={e^2\over 6\cdot 2 s^\prime }Q_{F,pn}^2\left\{S_{12}\phi_1^2+S_{34}
\left[\left({\phi^t_3+\phi^t_4\over2}+2\phi^u_4-\phi^u_3\right)^2+
\left({\phi^t_4+\phi^t_3\over2}+2\phi^u_3-\phi^u_4\right)^2\right]\right\},
\label{gdpnsa}
\end{equation}
where the  different predictions of SU(6) and diquark models follow from the different predictions for  the 
$pn\rightarrow pn$ helicity  conserving amplitudes given in Eq.(\ref{pnpn}). 

\subsection{$\gamma d\rightarrow \Delta^+\Delta^0$ scattering}
The calculation for  the $\gamma d\rightarrow \Delta^+\Delta^0$ amplitude yields the same quark-charge factors
as for the $\gamma d \rightarrow pn$ reactions in Eq.(\ref{dQFpn}).  Using the  helicity amplitudes  of the 
$pn\rightarrow  \Delta^+\Delta^0$ scattering from Eq.(\ref{pnD+D0})  and  the expressions for 
the photodisintegration amplitudes from Eqs.(\ref{dampl5},\ref{asampDD})  one obtains
\begin{eqnarray}
|\bar{\cal M}|_{\gamma d\longrightarrow \Delta^+\Delta^-}^2&=&{1\over6}{e^2\over 2s^\prime} Q_{F,\Delta\Delta}^2\left\{S_{12}\left[|\phi_1|^2+|\phi_6|^2+|\phi_7|^2\right]\right.\nonumber\\&&
\left.+S_{34}\left[\left({\phi^t_3+\phi^t_4\over2}+2\phi^u_4-\phi^u_3\right)^2+\left({\phi^t_4+\phi^t_3\over2}+2\phi^u_3-\phi^u_4\right)^2\right.\right.\nonumber\\
&&\left.\left.+\left({\phi^t_8+\phi^t_9\over2}+2\phi^u_9-\phi^u_8\right)^2+\left({\phi^t_9+\phi^t_8\over2}+2\phi^u_8-\phi^u_9\right)^2\right]\right\},
\label{gdD+D0}
\end{eqnarray}
where the  different predictions of SU(6) and diquark models follow from the different predictions for the $pn\rightarrow \Delta^+\Delta^0$ helicity  
conserving amplitudes given in Eq.(\ref{pnD+D0}).

\subsection{$\gamma d\rightarrow \Delta^{++}\Delta^-$ scattering}

For the charge factors in  the  $\gamma d \rightarrow\Delta^{++}\Delta^-$ scattering  within the quark-interchange approximation  from 
Appendix B we obtain:
\begin{equation}
-Q^{tN_1}={Q^{tN_2}\over2}=Q_{F,\Delta\Delta}={1\over3}.
\end{equation}
Inserting these charge factors  in  Eqs.(\ref{dampl5},\ref{asampDD}) one obtains for the photodisintegration amplitude:
\begin{eqnarray}
|\bar{\cal M}|_{\gamma d\longrightarrow \Delta^{++}\Delta^-}^2= {1\over 6}
{e^2\over 2s^\prime}Q^2_{F,\Delta\Delta}\left\{S_{12}\left(|\phi_1|^2+
|\phi_6|^2+|\phi_7|^2\right)\right.\nonumber\\
+\left. S_{34}\left[\left(2\phi_3-\phi_4\right)^2+\left(2\phi_4-\phi_3\right)^2+5|\phi_8|^2\right]\right\}.
\label{gdD++D-}
\end{eqnarray}
where predictions for the  helicity conserving amplitudes of $pn\rightarrow \Delta^{++}\Delta^{-}$   are given in Eq.(\ref{pnD++D-}).

\subsection{Numerical Estimates}
Using Eqs.(\ref{gdpnsa}), (\ref{gdD+D0}) and (\ref{gdD++D-})  with the baryonic helicity amplitudes calculated in Appendix B 
we estimate the ratio $R(\theta_{c.m.})$   of the  $\gamma d\rightarrow\Delta\Delta$ 
to $\gamma d\rightarrow pn$  differential cross sections 
at given $s$ and  $\theta_{c.m.}$ angle.
 For simplicity we consider the kinematics  in which $s>>4m_\Delta^2$, which allows to approximate  both  Eqs.(\ref{dtheta_cmn}) and (\ref{kin2})  to,
\begin{eqnarray}
cos\theta^N_{c.m}\approx{1+cos\theta_{c.m.}\over2}.
\label{kin2a}
\end{eqnarray}
Before considering any specific model for angular distribution, one can make two general statements about the properties of the photodisintegration amplitude. First, that  from the absence of the $u$ channel  scattering in the $pn\rightarrow \Delta^{++}\Delta^{-}$ helicity amplitudes (see Eq.(\ref{pnD++D-})),
one observes that $R(\theta_{c.m.})$  can not be a uniform function of $\theta_{c.m}$.  Second, that  independent of the choice of SU(6) or diquark models,  
the $\gamma d \rightarrow \Delta^{++}\Delta^{-}$ cross section is always larger than the cross section of the $\gamma d\rightarrow \Delta^{+}\Delta^{-}$  
reaction.

We quantify the above observations by parameterizing  the angular function $f(\theta^N_{c.m.})$,  which enters in Eqs.(\ref{pnpn},\ref{pnD+D0},\ref{pnD++D-}),  in the
following form\cite{RS}:
\begin{equation}
f(\theta) = {1\over sin(\theta)^2 (1-cos(\theta))^2}
\label{angf}
\end{equation}
known to describe reasonably well the elastic $pp$ and $pn$ scattering cross sections.

\begin{figure}[ht]
\centering\includegraphics[height=8cm,width=10cm]{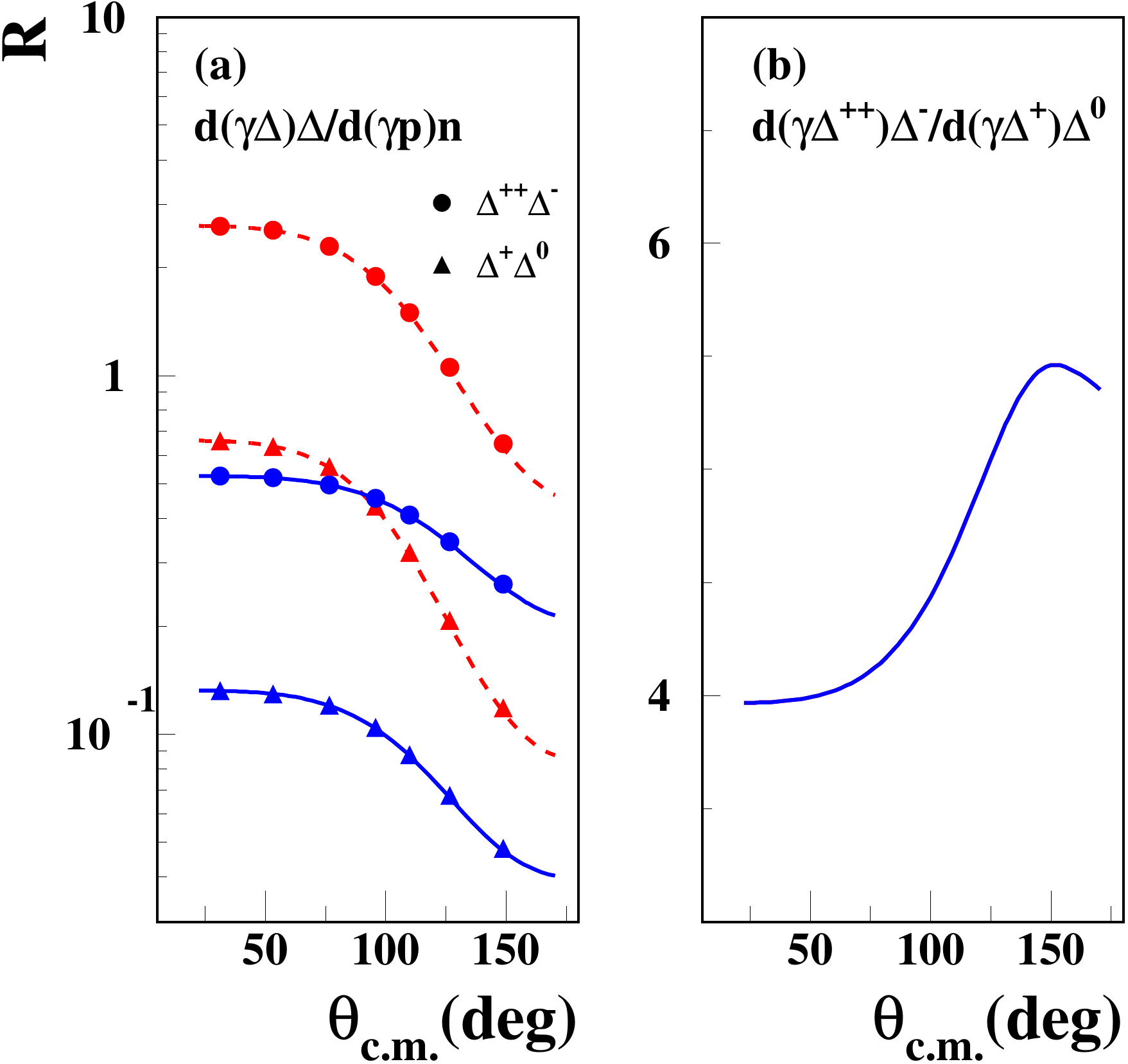}
\caption{ (a)Ratio of the  $\gamma d\rightarrow \Delta\Delta$ to $\gamma d\rightarrow pn$    differential cross sections and (b) ratio of the  $\gamma d\rightarrow \Delta^{++}\Delta^-$ to $\gamma d\rightarrow \Delta^+\Delta^0$    differential cross sections as a function of $\theta_{c.m.}$. }
\label{delta_delta_fig}
\end{figure}

Magnitudes of the ratio $R$ at $\theta_{c.m}=90^0$ are given in Table \ref{tab:t1}, while the angular dependencies (solid curves for diquark model and dashed curves for SU(6) model) are presented  in Fig.\ref{delta_delta_fig}(a). They clearly show strong angular anisotropy and  
the excess (by a factor of 4-5)  of the $\Delta^{++}\Delta^{-}$  breakup cross section relative to the cross section of the
 $\Delta^{+}\Delta^{0}$  breakup (Fig.\ref{delta_delta_fig}(b)). These results show that the ratio of the $\gamma d\to \Delta\Delta$ to $\gamma d\to pn$ cross sections is very sensitive to the choice of SU(6) or diquark models of the wave functions. However, because of the absence of isosinglet two-quark state in the $\Delta$ wave functions, the $\rho$ parameter dependence that characterizes the choice of SU(6) or diquark models in the baryons wave functions is factorized and enters only in the normalization factor of the  $pn\to\Delta\Delta$ helicity  amplitudes. As a result, the ratio of the $\gamma d\to\Delta^{++}\Delta^-$ to $\gamma d\to \Delta^+\Delta^0$ cross sections (Fig.\ref{delta_delta_fig}b) is independent of the choice between SU(6) and diquark models for the baryons wave functions. 
 
\medskip
\medskip

Finally, it is worth discussing  how these results  compare with the predictions of models in which the production of  
two $\Delta$'s is a result of the breakup of the pre-existing $\Delta\Delta$ component of  the deuteron wave function. 
In this case,  the final state interaction is dominated 
by soft scattering of two  $\Delta$'s in the final state  which  will induce   similar angular distributions for both 
$\Delta^{++}\Delta^{-}$ and $\Delta^{+}\Delta^{0}$ channels (see e.g.\cite{FGMSS,gea}). As a result, we expect essentially the
same angular distribution for  both $\Delta^{++}\Delta^{-}$ and $\Delta^{+}\Delta^{0}$ production channels. Also, because of 
the deuteron being an isosinglet, the  probabilities of finding preexisting $\Delta^{++}\Delta^{-}$ and $\Delta^{+}\Delta^{0}$ are equal. 
For coherent hard  breakup of  the preexisting $\Delta$'s  we will obtain the  same cross section 
for both the  $\Delta^{++}\Delta^{-}$ and  the $\Delta^{+}\Delta^{0}$  channels.

\begin{table*}[htbp]
	\centering
		\begin{tabular}{c|c|c|} \cline{2-3} & \multicolumn{2}{|c|}{$R(90^o)$} \\ 
		\hline
		\multicolumn{1}{|c|}{$\gamma d\rightarrow BB$} &SU(6)&Diquark \\
		\hline
		\multicolumn{1}{|c|}{$\gamma d\rightarrow\Delta^+\Delta^0$}          &0.47 &0.11\\
		\hline
   \multicolumn{1}{|c|}{$\gamma d\rightarrow\Delta^{++}\Delta^-$}    &2.01 &0.47\\
		\hline 
		\end{tabular}
		\caption{Strength of $\Delta\Delta$ channels relative to $pn$ in deuteron photodisintegration at $\theta_{c.m}=90^o$.}
		\label{tab:t1}
\end{table*}

One interesting scenario for probing the preexisting $\Delta$'s in the deuteron  is 
using the decomposition of the  deuteron wave function, in the chiral symmetry restored limit, into the nucleonic 
and non-nucleonic components in the following form\cite{Harvey:1980rva,Brodsky:1983vf,Brodsky:1985gs}:
\begin{equation}
\Psi_{T=0,S=1} = ({1\over 9})^{1\over 2}\Psi_{NN} + ({4\over 45})^{1\over 2}\Psi_{\Delta\Delta} + ({4\over 5})^{1\over 2}\Psi_{CC},
\label{deutronCC}
\end{equation}
where $\Psi_{CC}$ represents the hidden color component of $T=0$ and $S=1$ six-quark configuration. 
Since $\Delta^{++}\Delta^{-}$ and $\Delta^{+}\Delta^{0}$ components enter with equal probability in 
the total isospin $T=0$ configuration, 
one expects close ($\approx0.8$)  strengths for deuteron breakup to 
$\Delta^{++}\Delta^{-}$ or $\Delta^{+}\Delta^{0}$ channels as compared to the strength of the  deuteron breakup 
into the $pn$ pair.  The latter result should be compared with the similar ratios presented in  Table \ref{tab:t1} from HRM,
%carlos up to line 537
and with the HRM angular distributions in Fig\ref{delta_delta_fig}.

In contrast to Eq.(\ref{deutronCC}), the $pn$ component of the non relativistic deuteron wave function largely dominates over other baryon baryon components. From early works investigating deuteron composition by means of phenomenological and one pion exchange potentials (see e.g, Refs.\cite{Nath:1971ts,Arenhovel:1971de,Benz:1974au,Rost:1975zn,JuliaDiaz:2002gu}),  $NN^*$ components have been estimated to contribute overall of the order of 1\% with $NN^*(1440)\sim 0.01\%$ , while $\Delta\Delta$ contributions are estimated in the range 0.01-3\%. These two kinds of baryonic states can be comparable contributions to the  deuteron, thus it is expected that at low energies the channels of deuteron breakup into $NN^*$ and $\Delta\Delta$ may be comparable as well which, in addition to the fact that $N^*(1440)$ can decay into a nucleon and a pion as well into a $\Delta$-isobar and a pion,  would make it difficult to differentiate experimentally. Note that deuteron breakup into $NN^*$ will not interfere with the  amplitude of $\Delta\Delta$  production at large center of mass angles, since the decay products of the produced resonances  occupy distinctly different phase spaces in the final state of the reaction. 

With $pn$ in the S wave contributing $~$90\% to the deuteron's wave function, the scenario of deuteron breakup into the $pn$ channel at low relative transverse momentum clearly dominates the other baryon-baryon channels emerging from correspondent baryon-baryon components of the deuteron. The situation at kinematics in which the hard rescattering approach applies on the other hand can potentially be very different. As seen in Fig.\ref{delta_delta_fig}(a), in the HRM the quark wave functions of the baryons play a crucial role in the predictions of the strenght of the deuteron breakup to $\Delta\Delta$ channels relative to the $pn$ channel. 

A similar result is expected if these channels are compared with the deuteron breakup into the $NN^*$ channels. Quark-wave functions of $N^*(1440)$, if expanded in the form \ref{wf}, will introduce a different parameter $\rho^*$ (from the fact that $N^*$ is a radial excitation of $N$) playing the same role as the $\rho$ used for nucleons. Then, the relative strength of the deuteron breakup into $NN^*(1440)$ channel relative to the $pn$ channel will be determined by two parameters, $\rho$ and $\rho^*$. Without certainty on which of the considered quark wave-function approaches is more accurate, it cannot be concluded that a specific channel dominates for instance at $90^o$ c.m. in order to compare  with the prediction at the onset of hidden color components.

%%%%end carlos 

\section{Summary}
The hard rescattering model of large c.m. angle photodisintegration of a two-nucleon system was extended  to account for 
the production of two $\Delta$-isobars. The
HRM allows to express the cross section of  $\gamma d\rightarrow pn$ and $\gamma d\rightarrow \Delta\Delta$ reactions through the 
large c.m. angle differential cross section of $pn\rightarrow pn$ and $pn\rightarrow \Delta\Delta$   scattering amplitudes.

Because of  lack of  experimental information on $pn\rightarrow \Delta\Delta$   scattering, the quark-interchange model was further applied 
 to calculate the  strength of the  $\gamma d \rightarrow \Delta\Delta$  cross section relative to the 
cross section of $\gamma d\rightarrow pn$ breakup reaction.   We predicted a significantly larger strength for 
 the $\Delta^{++}\Delta^{-}$  channel of breakup as compared to the $\Delta^{+}\Delta^{0}$ channel which is related to the 
relative strength of the  $pn\rightarrow \Delta^{++}\Delta^{-}$ and  $pn\rightarrow \Delta^{+}\Delta^{0}$ scatterings.
Because of the different angular dependences of these hadronic amplitudes, we also predicted a significant difference between the
angular dependences of  photoproduction cross sections in  $\Delta^{++}\Delta^{-}$ and $\Delta^{+}\Delta^{0}$ channels.

These results can be compared with the prediction of the models in which two  $\Delta$'s are produced due to the coherent 
breakup of the $\Delta\Delta$ component of the deuteron wave function.  In this case one expects essentially similar angular 
distributions and  strengths for  the $\Delta^{++}\Delta^{-}$ and $\Delta^{+}\Delta^{0}$ breakup channels.

% ------------------------------------------------------------------------

%%% Local Variables: 
%%% mode: latex
%%% TeX-master: "../thesis"
%%% End: 

\def\baselinestretch{1}
\chapter{CONCLUSIONS}
\ifpdf
    \graphicspath{{Conclusions/ConclusionsFigs/PNG/}{Conclusions/ConclusionsFigs/PDF/}{Conclusions/ConclusionsFigs/}}
\else
    \graphicspath{{Conclusions/ConclusionsFigs/EPS/}{Conclusions/ConclusionsFigs/}}
\fi

\def\baselinestretch{1.66}

The preceding chapters presented  quantitative approaches that invoke QCD degrees of freedom in describing reaction mechanisms in nuclear and baryonic interactions. Such approaches were developed specifically for nucleon-nucleon elastic scattering, and for photodisintegration of a $NN$ system in light nuclei. Both cases were analyzed in the hard kinematic regime in which it is expected that hard subprocesses among the baryons' constituents control the reactions thus enabling the study of the quark dynamics of the interaction.

\medskip
\medskip
A hard scattering model of high energy $pn$ elastic scattering was discussed in chapter \ref{PNES}. World data evidenced the energy dependency of this reaction to be dictated by the constituent counting rule for a wide range of center of mass angles of scattering around 90$^o$. The angular dependence on the other hand is also affected by the helicity-flavor constituent structure of the interacting nucleons. The structure of the nucleon was modeled in a quark-diquark picture of the nucleons, and it was shown that such picture better describes an angular asymmetry observed in the experimental data. Using these quark wave functions of nucleons in this quark diquark picture, such asymmetry was obtained as a function of a parameter ($\rho$) within the quark interchange ($QI$) model. The parameter ($\rho^2$) measures  the average strength of the quark-vector diquark relative to the quark-scalar diquark components of the nucleon's wave function. By fitting the asymmetry in this model to the experimental asymmetry, it was found that $$\rho=-0.3\pm0.2,$$
i.e., scalar diquarks contribute 90\% on average to the nucleon helicity isospin structure. Hence, through this model, the experimental data disfavors the traditional $SU(6)$ three quark structure of the nucleon in which both quark-diquark components contribute with the same strength, i.e., $\rho=1$, and that produces an opposite asymmetry. 
Also, having that $\rho<0$, indicates that the vector diquark has a negative phase relative to the scalar diquark component...

\medskip
\medskip   

In chapter \ref{HBNNHe}, the hard rescattering model (HRM) of $NN$ breakup was developed for $pp$ and $pn$ breakup channels in $^3$He photodisintegration at large energy and momentum transfer. The analysis resulted in quantitative predictions for the differential cross sections of both channels. A general feature of these cross sections is the $s^{-11}$ dependence of the energy distributions that in the HRM arises from understanding the photodisintegration process in terms of quark degrees of freedom which results in such energy dependence behaving according to quark counting rules (See Eq.~\ref{dimscal}). The latter prediction has been verified experimentally for deuteron breakup and recently for $pp$ breakup in $^3$He at 90$^o$ c.m. of  the $\gamma-NN$ system at beam energies larger than 2GeV$^2$. The onset of QCD degrees of freedom  marks also the dominance of a one step/two body process  picture of the reaction. In a hadronic description, this process is suppressed from not having charged meson exchanges between the two protons. For $pp$ break up, these exchanges are only possible if the neutron participates. The reaction is then dominated by a three-body/two-step process.  Such process is much largely suppressed at the onset of a explicit quark description of the reaction such as the HRM; its energy dependence falls off faster than $s^-12$.
Calculations for additional observables were also carried out within the HRM resulting in further numerical predictions on spectator nucleon's momentum distributions (See Fig.\ref{Fig.al_dep}) and the polarization transfer asymmetry $C_{z'}$. In particular for $pp$ it was found that $C_{z'}$ is almost canceled, while for $pn$ breakup, $C_{z'}\approx{2\over3}$. The experimental evaluation of these predictions will properly constrain the validity of the HRM approach for these reactions.

\medskip
\medskip

The hard rescattering mechanism was discussed in chapter \ref{HBdDD} as it applies to double $\Delta$-isobars production in deutron photodisintegration. These studies were motivated by the focus on identifying a transition of the deuteron from a baryon-baryon system to a six quarks system. It is believed that such transition is signaled by a sudden increase of the double $\Delta$-isobars production in relation to the deuteron breakup in the $pn$ channel. The latter  is believed to dominate at small to moderate transverse momenta from the dominance of the $pn$ component of the deuteron over the negligible $\Delta\Delta$ components, and from the assumption that the baryons emerge from a large relative transverse momenta component of the deuteron wave function. 

Such assumption was contrasted in this chapter in which the breakup reaction was studied instead within the framework of the QCD hard rescattering model (HRM). The baryons emerging with large transverse momenta are produced in a $pn$ rescattering process  triggered by the incident photon interacting with a low relative momentum $pn$ component of the deuteron wave function. This rescattering was modeled within the quark interchange mechanism (QIM), and within the HRM, angular distributions for the two $\Delta\Delta$ channels were obtained and compare to each other and to the breakup to $pn$ channel. From such comparisons illustrated in Fig.\ref{delta_delta_fig}, it was conclude that in the rescattering picture each $\Delta\Delta$ channel has a distinct angular distribution with $\Delta^{++}\Delta^-$ dominating $\Delta^+\Delta^-$. This is in clear contrast with what is expected if the  two $\Delta$-isobars emerge from a $\Delta\delta$ component of the deuteron in which case both $\Delta\Delta$ channels should have the same strength.

\medskip
\medskip

The results concluded above were arrived to from considering the studied reactions to be controlled by elementary particle subprocesses. Such an assumption was justified by the kinematic characteristic of hard processes, i.e., the energy and momentum invariants were much larger than the masses of the interacting baryons which facilitated for instance a short distance-long distance factorization of scattering amplitudes, at a quark-baryon level as it is discussed in chapter \ref{PNES}, and at the quark-baryon-nuclear level which corresponds to the hard rescattering model developed in chapters \ref{HBNNHe} and \ref{HBdDD}. This methodology finds a broad range of applications in many more reactions of interest. Some of them are natural extensions of the processes studied here. For instance a program to pursue, as it was pointed in chapter \ref{PNES}, is the further development of the studies presented there as it concerns to the possible strange production channels in exclusive $eN$ or $NN$ scattering, which would make full use of the $SU(6)$ symmetry of quark and diquark states in constructing quark wave-functions of hadrons providing also a potential probe of strange quark distributions in nucleons. This program can be as well continued into studying strange production channels off nuclear targets in breakup reactions, and further probing the role of hard rescattering mechanisms analogous to such studied in chapters \ref{HBNNHe} and \ref{HBdDD}.  The experimental difficulties brought on by the decays of the produced hadrons in these processes can soon be overcome by updated capabilities such as the 12GeV upgrade program at Jefferson Lab, which is also instrumental in assessing most of the results presented in this dissertation.

Another venue of interest, and also a mayor motivation driving JLAB's 12GeV upgrade, is the study of near threshold $J/\psi$ production reactions off proton and nuclear targets. These reactions are relevant in identifying the leading gluon exchange mechanisms between hadrons, as well as in extracting gluon contributions to the structure functions of nucleons at large $x$. Because of the large mass of the $c$ quark($\sim$1.2GeV), the $c\bar{c}$ fluctuation that will evolve into a $J/\psi$ meson  for instance in $\gamma p\to J/\psi p$ has a small transverse size, and since the threshold invariants are large as well, pQCD approaches to these processes may apply ( see e.g. Refs. \cite{Hoyer:1996iw}, and \cite{Brodsky:1991dj}), and quantitative descriptions can be constructed through factorization methods analogous to those developed throughout this dissertation. Such is the case also for $J/\psi$ production off a deuterium target which study is of interest in probing the hidden color components of the nuclear wave function; in a pQCD picture of the  $\gamma d\to J/\psi pn$ reaction's hard subprocess, the $c\bar{c}$ fluctuation of the photon can interact by single gluon exchanges with non-singlet quark clusters within the nucleus \cite{Laget:1994ba}. This mechanism is expected to give a significant contribution in this reaction. Also,in analogy to what was done in chapter \ref{HBdDD}, in an alternative mechanism, this reaction proceeds through  $c\bar{c}$ fluctuation  scattering off one of the nucleons, then evolving into a hadronic state and rescattering with the second nucleon into an outgoing nucleon and a $J/\psi$ meson. Thus, a variation of the hard rescattering model can as well be developed for obtaining qualitative and quantitative predictions for these heavy quark production reactions. 

In summary, these phenomena and several more await for a coordinated program focusing experimental and theoretical developments into identifying explicit signatures of quark-gluon dynamics of hadron interactions that shed light into building a compact QCD description of the strong force. It is one's expectation for the methods utilized and the results reported in this dissertation to further encourage research efforts in such program in which this methodology can find applicability.  

\medskip
\medskip

%%% ----------------------------------------------------------------------

% ------------------------------------------------------------------------

%%% Local Variables: 
%%% mode: latex
%%% TeX-master: "../thesis"
%%% End: 

\setcounter{tocdepth}{0}

\makeatletter
\renewcommand*\@makeschapterhead[1]{%
  {\parindent \z@ \centering \reset@font
        \scshape 
    \Large \bf #1\par\nobreak
    \par
    \vspace*{20\p@}%
     }}
\makeatother
{
\renewcommand{\appendixtocname}{APPENDICES}
\renewcommand{\appendixpagename}{\normalsize APPENDICES}
\begin{appendices}
\makeatletter
\renewcommand\chapter{\@startsection{chapter}{1}{\z@}%
                                  {-3.5ex \@plus -1ex \@minus -.2ex}%
                                  {2.3ex \@plus.2ex}%
                                  {\centering\normalfont\bfseries}}
\makeatother
\makeatletter
\renewcommand\section{\@startsection{section}{1}{\z@}%
                                  {-3.5ex \@plus -1ex \@minus -.2ex}%
                                  {2.3ex \@plus.2ex}%
                                  {\normalfont\bfseries}}
\makeatother
\notocchapter{\normalsize Baryonic  Wavefunctions}
\label{bw}

Nucleon and $\Delta$-isobar wave functions are built from the minimal Fock component of the corresponding wave function, initially assuming a quark-diquark expansion where individual quark wave functions correspond to SU(6) eigenstates. A single quark state of spin S=1/2 and isospin I=1/2 is joined by a diquark state with S=0 and I=0, or S=1 and I=1. The former corresponding to what is known as a scalar diquark [qq], and the latter known as a vector diquark {qq}. The baryonic wave functions are then expanded in q[qq] and q(qq) states with the proper Clebsh-Gordan coefficients and normalization, and introducing a parameter $\rho$ that determines the relative amplitude of the vector diquark with respect to the scalar diquark sector of the expansion:
\begin{eqnarray}
\psi^{i_3,h}\propto q[qq]+\rho\times q{qq} 
\end{eqnarray}
q[qq] and q(qq) expansions vary according to the baryon's isospin and helicity. For instance, for $\Delta$-isobar wave functions q[qq]=0 since the total isospin components won't add up to I=3/2. A nucleon wave function is expanded as follows,
\begin{eqnarray}
\psi^{i^3_{N},h_N}\propto q^{i^3_{N},h_N}[qq]+\rho\sum_{i^3,h=1/2,-1/2}\sum_{t^3,\lambda=1,0,-1} C^N_{i,t,i^3_{N}}
C^N_{h,\lambda,h_{N}}q^{i^3,h}(qq)^{t^3,\lambda}
\label{nwf1}
\end{eqnarray}
where $i^3_N$, $i^3$,$t^3$, are the third components of the isospin of nucleon, single quark and diquark respectively, and $h_N$, h and $\lambda$ are the corresponding helicities. $C^N_{i,t,i^3_{N}}$ and $C^N_{h,\lambda,h_{N}}$ are the Clebsh-Gordan coefficients for isospin and helicity expansions in q(qq) states respectively.

Since [qq] and (qq) are singlet$\times$isosinglet and triplet$\times$isotriplet representations of qq states in order to ensure a symmetric wave function, Eq.(\ref{nwf1}) can be expanded in qqq states. Doing so for the case of a proton with positive helicity yields,
\begin{eqnarray}
 {\bf N(\rho)|p(1/2)\rangle} & =&  {3+\rho\over2}|u(+)u(+)d(-)\rangle 
                                    -{3-\rho\over2}|u(+)d(+)u(-)\rangle
                                    -\rho|d(+)u(+)u(-)\rangle \nonumber \\
                              & &+ {3+\rho\over2}|u(+)d(-)u(+)\rangle 
                                    -{3-\rho\over2}|u(+)u(-)d(+)\rangle 
                                    -\rho|d(+)u(-)u(+)\rangle \nonumber \\
                              & &+ 2\rho|d(-)u(+)u(+)\rangle
                                    -  \rho|u(-)u(+)d(+)\rangle
                                    - \rho|u(-)d(+)u(+)\rangle 
\label{prwf}
\end{eqnarray}

where, u and (+) represent q with i$^3$=1/2 and h=1/2 , and d and (-) represent q with i$^3$=-1/2 and h=-1/2 respectively, and

\begin{equation}
{\bf N(\rho)}={1\over3\sqrt{1+\rho^2}}
\end{equation}
 
 All other nucleon wave functions can also be obtained by properly applying spin or isospin ladder operators starting from Eq.(\ref{prwf}),
e.g.,
\begin{eqnarray}
{\bf |n(1/2)\rangle}=\tau_-{\bf|p(1/2)\rangle}
\end{eqnarray}
where 
\begin{eqnarray}
\tau_-|q_1q_2q_3\rangle=| (\tau_-q_1)q_2q_3\rangle+| q_1(\tau_-q_2)q_3\rangle+| q_1q_2(\tau_-q_3)\rangle
\label{ladop}
\end{eqnarray} 
and $\tau_-$ is constructed from Pauli matrices acting on the quark isospin state.
Similarly, all $\Delta$-isobar wave functions can be constructed through ladder operators in I=3/2 and J=3/2 representations starting with $\Delta^{++}$ with helicity h$_{\Delta}$=3/2,
\begin{eqnarray}
{\bf |\Delta^{++}(3/2)\rangle}=|u(+)u(+)u(+)\rangle
\label{dpwf} 
\end{eqnarray}
then for instance, to obtain $\Delta^++$ with helicity h$_{\Delta}$=1/2,
\begin{eqnarray}
 {\bf |\Delta^{++}(1/2)\rangle}={1\over\sqrt{3}}\sigma_-{\bf|\Delta^{++}(3/2)\rangle}
\end{eqnarray}
or in general,
\begin{eqnarray}
{\bf|\Delta^{i^3}(h -1)\rangle}={1\over\sqrt{J(J+1)-h(h-1)}}\sigma_-{\bf|\Delta^{i^3}(h)\rangle}
\label{hlad}
\end{eqnarray}
Likewise,
\begin{eqnarray}
{\bf|\Delta^{i^3-1}(h )\rangle}={1\over\sqrt{I(I+1)-i^3(i^3-1)}}\tau_-{\bf|\Delta^{i^3}(h)\rangle}
\label{ilad}
\end{eqnarray}
with $\sigma$ and $\tau$ acting on the three-quark components of the expansion as indicated in Eq.(\ref{ladop}) on the helicity and isospin states respectively of the single quark states.
\subsubsection*{$\Delta$-isobar wave functions}
Either through the process described by Eqs.(\ref{hlad}) and (\ref{ilad}), or by explicitly expanding each $\Delta$-isobar state in q(qq) states and then qqq states through the use of Clebsh-Gordan coefficients, the following $\Delta$-isobar helicity-isospin wave functions are obtained,
\begin{eqnarray}
{\bf |\Delta^{++}(3/2)\rangle}&=&|u(+)u(+)u(+)\rangle\nonumber\\
{\bf |\Delta^{++}(1/2)\rangle}&=&{1\over\sqrt3}|u(-)u(+)u(+)\rangle+{1\over\sqrt3}|u(+)u(-)u(+)\rangle+{1\over\sqrt3}|u(+)u(+)u(-)\rangle\nonumber\\
{\bf|\Delta^{++}(-1/2)\rangle}&=&{1\over\sqrt3}|u(-)u(-)u(+)\rangle+{1\over\sqrt3}|u(-)u(+)u(-)\rangle+{1\over\sqrt3}|u(+)u(-)u(-)\rangle\nonumber\\
{\bf |\Delta^{++}(-3/2)\rangle}&=&|u(-)u(-)u(-)\rangle\nonumber\\
&&\nonumber\\
{\bf |\Delta^{+}(3/2)\rangle}&=&{1\over\sqrt3}|d(+)u(+)u(+)\rangle+{1\over\sqrt3}|u(+)d(+)u(+)\rangle+{1\over\sqrt3}|u(+)u(+)d(+)\rangle\nonumber\\
{\bf |\Delta^{+}(1/2)\rangle}&=&{1\over3}|d(-)u(+)u(+)\rangle+{1\over3}|d(+)u(-)u(+)\rangle+{1\over3}|d(+)u(+)u(-)\rangle\nonumber\\
&&+{1\over3}|u(-)d(+)u(+)\rangle+{1\over3}|u(+)d(-)u(+)\rangle+{1\over3}|u(+)d(+)u(-)\rangle\nonumber\\
&&+{1\over3}|u(-)u(+)d(+)\rangle+{1\over3}|u(+)u(-)d(+)\rangle+{1\over3}|u(+)u(+)d(-)\rangle\nonumber\\
{\bf|\Delta^{+}(-1/2)\rangle}&=&{1\over3}|d(-)u(-)u(+)\rangle+{1\over3}|d(-)u(+)u(-)\rangle+{1\over3}|d(+)u(-)u(-)\rangle\nonumber\\
&&+{1\over3}|u(-)d(-)u(+)\rangle+{1\over3}|u(-)d(+)u(-)\rangle+{1\over3}|u(+)d(-)u(-)\rangle\nonumber\\
&&+{1\over3}|u(-)u(-)d(+)\rangle+{1\over3}|u(-)u(+)d(-)\rangle+{1\over3}|u(+)u(-)d(-)\rangle\nonumber\\
{\bf |\Delta^{+}(-3/2)\rangle}&=&{1\over\sqrt3}|d(-)u(-)u(-)\rangle+{1\over\sqrt3}|u(-)d(-)u(-)\rangle+{1\over\sqrt3}|u(-)u(-)d(-)\rangle\nonumber\\
\label{aldwf}
\end{eqnarray}
$\Delta^0$ and $\Delta^-$ wavefunctions are obtained by the replacement u$\leftrightarrow$d in $\Delta^+$ and $\Delta^{++}$ wavefunctions respectively.

% ------------------------------------------------------------------------

%%% Local Variables: 
%%% mode: latex
%%% TeX-master: "../thesis"
%%% End: 

\notocchapter{\normalsize Helicity Amplitudes in the Quark Interchange Model}
\label{haqim} 
\notocsection{Baryon-Baryon Scattering Helicity Amplitudes}
\label{bbsa}
We are using helicity states to label the entries of the photodisintegration and the baryon-baryon scattering matrices. 
The  number of independent helicity  amplitudes for a given $ab\rightarrow cd$ processes can be expressed through the total 
spin of the scattering particles as follows\cite{perl:1974,Perl:1969pg}: 
\begin{eqnarray}
N={1\over 2}\cdot (2s_a+1)(2s_b+1)(2s_c+1)(2s_d+1)
\label{nampl}
\end{eqnarray}
where $s_i$ is the total spin of particle i and  for the  photon  we replace $(s_i+1)$ by 2.   The factor ${1\over 2}$  
follows from the  constraint due to the  parity conservation.
For elastic scattering, there is a further reduction in $N$ due to time reversal invariance, and if the scattering particles are identical, or lie in the 
same isospin multiplet, the number of independent helicity amplitudes is reduced  further\cite{perl:1974,Perl:1969pg}. 
For the $pn$ elastic scattering case, out of the possible 16 helicity amplitudes only five  are independent\cite{Perl:1969pg} for which we use the 
following notations:
\begin{eqnarray}
\left\langle +{1\over2},+{1\over2} \left|T\right|+{1\over2},+{1\over2}\right\rangle &=& \phi_1\nonumber\\
\left\langle +{1\over2},-{1\over2} \left|T\right|+{1\over2},-{1\over2}\right\rangle &=& \phi_3\nonumber\\
\left\langle -{1\over2},+{1\over2} \left|T\right|+{1\over2},-{1\over2}\right\rangle &=& \phi_4\nonumber\\
\left\langle -{1\over2},-{1\over2} \left|T\right|+{1\over2},+{1\over2}\right\rangle &=& \phi_2\nonumber\\
\left\langle -{1\over2},+{1\over2} \left|T\right|+{1\over2},+{1\over2}\right\rangle &=& \phi_5,\nonumber\\
\label{pnham}
\end{eqnarray}

For  the $pn\rightarrow\Delta\Delta$ scattering amplitude, we have from Eq.(\ref{nampl}), $N$=(2)(2)(4)(4)/2=32 
independent helicity amplitudes.  We use the following notations for  the helicity conserving independent 
amplitudes of  $pn\rightarrow\Delta\Delta$ scattering:
 \begin{eqnarray}
\left\langle +{1\over2},+{1\over2} \left|T\right|+{1\over2},+{1\over2}\right\rangle &=& \phi_1\nonumber\\
\left\langle +{1\over2},-{1\over2} \left|T\right|+{1\over2},-{1\over2}\right\rangle &=& \phi_3\nonumber\\
\left\langle -{1\over2},+{1\over2} \left|T\right|+{1\over2},-{1\over2}\right\rangle &=& \phi_4\nonumber\\
\left\langle +{3\over2},-{1\over2} \left|T\right|+{1\over2},+{1\over2}\right\rangle &=& \phi_6\nonumber\\
\left\langle -{1\over2},+{3\over2} \left|T\right|+{1\over2},+{1\over2}\right\rangle &=& \phi_7\nonumber\\
\left\langle +{3\over2},-{3\over2} \left|T\right|+{1\over2},-{1\over2}\right\rangle &=& \phi_8\nonumber\\
\left\langle -{3\over2},+{3\over2} \left|T\right|+{1\over2},-{1\over2}\right\rangle &=& \phi_9,\nonumber\\
\label{hconamp}
\end{eqnarray}
which are consistent with the definitions in Eq. (\ref{pnham}).

\notocsubsection{Helicity Amplitudes  in  the Quark-Interchange Model}

\subsubsection*{Quark Interchange model}
Following the approach presented for example in Refs.\cite{Sivers:1975dg,FGST,BCL}, 
the scattering amplitude for a process $ab\rightarrow cd$, in which $a,b,c$ and $d$ are baryons, 
is obtained from,
%\begin{widetext}
\begin{eqnarray}
& & \langle cd\mid T\mid ab\rangle = 
\sum\limits_{\alpha,\beta,\gamma} 
\langle  \psi^\dagger_c\mid\alpha_2^\prime,\beta_1^\prime,\gamma_1^\prime\rangle
\langle  \psi^\dagger_d\mid\alpha_1^\prime,\beta_2^\prime,\gamma_2^\prime\rangle 
\nonumber \\
& & \ \ \ \  \times
\langle \alpha_2^\prime,\beta_2^\prime,\gamma_2^\prime,\alpha_1^\prime\beta_1^\prime
\gamma_1^\prime\mid H\mid
\alpha_1,\beta_1,\gamma_1,\alpha_2\beta_2\gamma_2\rangle\cdot 
\langle\alpha_1,\beta_1,\gamma_1\mid\psi_a\rangle
\langle\alpha_2,\beta_2,\gamma_2\mid\psi_b\rangle,
\label{qimampl}
\end{eqnarray}
%\end{widetext}
where ($\alpha_i,\alpha_i^\prime$), ($\beta_i,\beta_i^\prime$) and 
($\gamma_i\gamma_i^\prime$) describe the spin-flavor  quark states 
before and after the hard scattering, $H$,
and
\begin{equation}
C^{j}_{\alpha,\beta,\gamma} = \langle\alpha,\beta,\gamma\mid\psi_j\rangle
\label{aCs}
\end{equation}
describes the probability amplitude of finding an $\alpha,\beta,\gamma$ helicity-flavor 
combination of three valence quarks in the baryon $j$.
These coefficients are obtained from the expansion of the 
baryon's spin-isospin wave function in three-quark valence states as follows:
\begin{eqnarray}
\psi^{i^3_{N},h_N} & = & {N\over \sqrt{2}}\left\{
\sigma (\chi_{0,0}^{(23)}\chi_{{1\over2},h_N}^{(1)})\cdot
(\tau_{0,0}^{(23)}\tau_{{1\over 2},i_N^{3}}^{(1)}) 
\right.  +   \nonumber \\
& & 
\rho \sum\limits_{i_{23}^3=-1}^{1} \ \ \sum\limits_{h_{23}^3=-1}^{1}
\langle 1,h_{23}; {1\over 2},h_{N}-h_{23}\mid {1\over 2},h_N\rangle
\langle 1,i^3_{23}; {1\over 2},i^3_{N}-i^3_{23}\mid {1\over 2},i^3_N\rangle \nonumber \\
& &\left. \times (\chi_{1,h_{23}}^{(23)}\chi_{{1\over2},h_N-h_{23}}^{(1)})\cdot
(\tau_{1,i^3_{23}}^{(23)}\tau_{{1\over 2},i_N^{3}-i^3_{23}}^{(1)})\right\}.
\label{awf}
\end{eqnarray}
The indexes 1 and 23  label the quark and the diquark states. The first term corresponds to quarks 2 and 3 being in a helicity zero isosinglet state, while the second term corresponds to quarks 2 and 3 in  helicity 1-isotriplet states. 
Where $\chi$ and $\tau$ represent helicity and isospin states with helicity $h$ and isospin projection $i^3$ respectively.    For the wave functions of $\Delta$-isobars $\sigma=0$ and $\rho=1$, while for nucleon 
wave functions $\sigma=1$ and the parameter $\rho$  characterizes the average  strength of the isotriplet diquark radial state relative 
to that of the isosinglet state.  Two extreme values of $\rho=1$ and $\rho=0$ correspond to the 
realization of the SU(6) and good diquark symmetries in the wave function. 
 
Using Eq.(\ref{awf}) in Eq.(\ref{qimampl}) for the hadronic scattering amplitude  one obtains:
 
\begin{equation}
\langle cd |T^{QIM}|ab\rangle = A_{\alpha_1',\alpha_2',\alpha_1\alpha_2}(\theta^N_{c.m.})
M^{ac}_{\alpha_1,\alpha_1'}M^{bd}_{\alpha_2,\alpha_2'} + A_{\alpha_1',\alpha_2',\alpha_1\alpha_2}(\pi-\theta^N_{c.m.}) 
M^{ad}_{\alpha_1,\alpha_1'}M^{bc}_{\alpha_2,\alpha_2'},
\label{elamp}
\end{equation} 
where
\begin{equation}
M^{ij}_{\alpha,\alpha'} = C^{i}_{\alpha,\beta,\gamma}C^{j}_{\alpha',\beta,\gamma} +
C^{i}_{\beta,\alpha,\gamma}C^{j}_{\beta,\alpha',\gamma} + 
C^{i}_{\beta,\gamma,\alpha}C^{j}_{\beta,\gamma,\alpha'},  
\label{qimamp}
\end{equation}
which accounts for all possible interchanges of $\alpha$ and $\alpha'$ quarks leaving $\beta$ and $\gamma$ quarks unchanged. In the QI model
 the interchanging quarks conserve their corresponding helicities and flavors, this is accounted for in the matrix elements of $A$ in Eq.(\ref{elamp}.),
\begin{equation}
 A_{\alpha_1',\alpha_2',\alpha_1\alpha_2}(s,\theta^N_{c.m.}) \propto 
\delta_{\alpha_1',\alpha_2}\delta_{\alpha_2',\alpha_1} 
{f(\theta^N_{c.m.})\over s^2}
\label{qimfw}
\end{equation}

Eq.(\ref{elamp}) has two terms, first (referred as a $t$ term)  in which four quarks  scatter at  angle $\theta^N_{c.m.}$ and two (interchanging) quarks scatter 
at $\pi-\theta^N_{c.m.}$  and the second (referred as a $u$ term) in which two interchanging quarks scatter at $\theta^N_{c.m.}$, while  four spectator quarks scatter 
at $\pi-\theta^N_{c.m.}$.

\notocsubsection{Helicity Amplitudes in the Quark Interchange Model}
Through the above procedure using Eq.(\ref{elamp}) for the helicity amplitudes of $pn$ scattering one obtains:
\begin{eqnarray}
\phi_1(\theta^N_{c.m.})&=&(2-y)f(\theta^N_{c.m.})+(1+2y)f(\pi-\theta^N_{c.m.})\\\nonumber
\phi_2(\theta^N_{c.m.})&=&0\\\nonumber
\phi_3(\theta^N_{c.m.})&=&(2+y)f(\theta^N_{c.m.})+(1+4y)f(\pi-\theta^N_{c.m.})\\\nonumber
\phi_4(\theta^N_{c.m.})&=&2yf(\theta^N_{c.m.})+2yf(\pi-\theta^N_{c.m.})\\\nonumber
\phi_5(\theta^N_{c.m.})&=&0,
%\label{pngphis}
\label{apnpn}
\end{eqnarray}
were,
\begin{eqnarray}
y&=&{2\over3}{\rho\over{1+\rho^2}}\left(1+{2\over3}{\rho\over{1+\rho^2}}\right).
\end{eqnarray}

For $pn\rightarrow\Delta^+\Delta^0$ scattering amplitudes we obtain:
\begin{eqnarray}
\phi_1&=&{2\over9}N_{\Delta\Delta}(2f(\theta^N_{c.m.})-f(\pi-\theta^N_{c.m.}))\nonumber\\
\phi_3&=&{1\over9}N_{\Delta\Delta}(4f(\theta^N_{c.m.})+f(\pi-\theta^N_{c.m.}))\nonumber\\
\phi_4&=&{2\over9}N_{\Delta\Delta}(f(\theta^N_{c.m.}))+f(\pi-\theta^N_{c.m.})\nonumber\\
\phi^{+0}_6&=&{N_{\Delta\Delta}\over3\sqrt{3}}(2f(\theta^N_{c.m.})-f(\pi-\theta^N_{c.m.}))\nonumber\\
\phi_7&=&{N_{\Delta\Delta}\over3\sqrt{3}}(2f(\theta^N_{c.m.})-f(\pi-\theta^N_{c.m.}))\nonumber\\
\phi_8&=&{2\over9}N_{\Delta\Delta}f(\theta^N_{c.m.})\nonumber\\
\phi_9&=&{1\over3}N_{\Delta\Delta}f(\pi-\theta^N_{c.m.}),\nonumber\\
\label{pnD+D0}
\end{eqnarray}
and similarly for the  amplitudes of the $pn\rightarrow\Delta^{++}\Delta^-$ scattering, QI model gives:
\begin{eqnarray}
\phi_1&=&-{2\over3}N_{\Delta\Delta}f(\theta^N_{c.m.})\nonumber\\
\phi_3&=&-{2\over3}N_{\Delta\Delta}f(\theta^N_{c.m.})\nonumber\\
\phi_4&=&-{1\over3}N_{\Delta\Delta}f(\theta^N_{c.m.})\nonumber\\
\phi_6&=&{-N_{\Delta\Delta}\over\sqrt{3}}f(\theta^N_{c.m.})\nonumber\\
\phi_7&=&{-N_{\Delta\Delta}\over\sqrt{3}}f(\theta^N_{c.m.})\nonumber\\
\phi_8&=&-N_{\Delta\Delta}f(\theta^N_{c.m.})\nonumber\\
\phi_9&=&0,\nonumber\\
\label{pnD++D-}
\end{eqnarray}
For both sets of equations in  (\ref{pnD+D0}) and (\ref{pnD++D-}), we have
\begin{equation}
N_{\Delta\Delta}={(1+\rho)^2\over1+\rho^2},
\label{N_DD}
\end{equation}
which shows that the  strength of the two $\Delta\Delta$ channels relative to each other is independent of the value of $\rho$. This is not the case for their strengths  relative to the $pn$ channel; from Eqs. (\ref{apnpn}) we see that the $\rho$ dependence of the helicity amplitudes in $pn\rightarrow pn$ cannot be factorized.

\notocsubsection{Quark-Charge Factors}

 In the hard rescattering model, photodisintegration amplitudes are expressed in terms of hadronic scattering amplitudes weighted by the charges of 
struck quarks, Eq.(\ref{QT}).   We further split the amplitude of Eq.(\ref{QT}) into $t$ and $u$ channel scatterings:  
\begin{eqnarray}
\sum_{i}Q_i^{N_k}\langle \lambda_{2f};\lambda_{1f}\mid T_{(pn\rightarrow B_1B_2),i}(s,\tilde{t})\mid \lambda_\gamma;\lambda_{2i}\rangle=\left[Q^{tN_k}_j\phi^t_j+Q^{uN_k}_j\phi^u_j\right],
\label{emqim}
\end{eqnarray}
where $Q^{t/uN}_i$ is the charge of the quark,  struck  by the incoming photon from the nucleon $N$ with further $\theta^N_{c.m.}$ or $\pi-\theta^N_{c.m.}$  scattering.  
The helicity amplitudes are also split into $t$ and $u$ parts 
 \begin{eqnarray}
 \phi_i(\theta^N_{c.m.})&=& \phi^t_i(\theta^N_{c.m.})+\phi^u_i(\theta^N_{c.m.})\nonumber\\
 &=& c_tf(\theta^N_{c.m.})+c_uf(\pi-\theta^N_{c.m.}),
 \label{tuqim}
 \end{eqnarray}
with  $\phi^t$ and $\phi^u$ corresponding to the $\theta^N_{c.m.}$ or $\pi-\theta^N_{c.m.}$  scattering terms in Eq.(\ref{elamp}).

Using the above definitions and Eqs.(\ref{elamp},\ref{qimamp},\ref{apnpn},\ref{pnD+D0},\ref{pnD++D-}) 
the charge factors $Q^t$ and $Q^u$ are calculated using 
the following relations:
\begin{eqnarray}
Q^{tN_k}_j&=&{Q(\alpha_k)A_{\alpha_1',\alpha_2',\alpha_1\alpha_2} 
M^{ac}_{\alpha_1,\alpha_1'}M^{bd}_{\alpha_2,\alpha_2'}\over \phi^t_j}\nonumber\\
Q^{uN_k}_j&=&{Q(\alpha_k)A_{\alpha_1',\alpha_2',\alpha_1\alpha_2} 
M^{ad}_{\alpha_1,\alpha_1'}M^{bc}_{\alpha_2,\alpha_2'}\over \phi^u_j},
\label{chfctr}
\end{eqnarray}
where summation is understood for repeated $\alpha$ indices, $Q(\alpha)$ is the charge in e units of a quark $\alpha$ and the index $ j$ labels the 
process $ab\rightarrow cd$.

% ------------------------------------------------------------------------

%%% Local Variables: 
%%% mode: latex
%%% TeX-master: "../thesis"
%%% End: 

\notocchapter{\normalsize Calculation of the {\boldmath $^3$He$(\gamma, NN)N$} scattering amplitude}
\label{app1}

Applying Feynman diagram rules for the scattering amplitude corresponding to the diagram of Fig.(\ref{Fig.1})(a) one 
obtains
\begin{eqnarray}
& & \langle\lambda_{f1},\lambda_{f2},\lambda_s\mid {\cal M} \mid \lambda_\gamma, \lambda_A\rangle = \nonumber \\ 
%
%N1:
& &(N1): \int {-i\Gamma_{N_{1f}}i[\sh p_{1f}-\sh k_1+m_{q}]\over 
(p_{1f}-k_1)^2-m_q^2+i\epsilon}i S(k_1)
\cdots [-igT^F_c\gamma_\mu]\cdots
{i[\sh p_{1i} - \sh k_1 + m_q](-i)\Gamma_{N_{1i}}
\over (p_{1i}-k_{1})^2-m^2+i\epsilon} {d^4k_{1}\over (2\pi)^4} \nonumber \\
%
% gamma q:
& & (\gamma q):
{i[\sh p_{1i}-\sh k_1 + q + m_q]\over (p_{1i}-k_1+q)^2-m_q^2+i\epsilon} [-iQ_ie\epsilon^\perp\gamma^\perp]
%\nonumber \\ & & \ \ \ \ \ \ \ \ \times 
\nonumber \\
%
%N2:
& &(N2): \int {-i\Gamma_{N_{2f}}i[\sh p_{2f}-\sh k_2+m_{q}]\over 
(p_{2f}-k_2)^2-m_q^2+i\epsilon}iS(k_2)\cdots 
[-igT^F_c\gamma_\nu]
{i[\sh p_{2i} - \sh k_2 + m_q](-i)\Gamma_{N_{2i}}
\over (p_{2i}-k_{2})^2-m^2+\epsilon} {d^4k_{2}\over (2\pi)^4} \nonumber \\
%
%d:
& & (^3He): \int {-i\Gamma_{^3\textnormal{\scriptsize {He}}}\cdot\bar{u}_{\lambda_s}(p_s) i[\sh p_{NN} - \sh p_{2i} + m_N]\over 
(p_{NN}-p_{2i})^2 - m_N^2+i\epsilon}
{i[\sh p_{2i}+m_N]\over p_{2i}^2-m_N^2+i\epsilon}
{d^4p_{2i}\over (2\pi)^4}\nonumber \\
%
%g:
& & (g): {i d^{\mu,\nu}\delta_{ab}\over 
[(p_{2i}-k_2)-(p_{1i}-k_1)-(q-l)]^2 + i\epsilon}, \nonumber \\
\label{amplitude}
\end{eqnarray}
where  the momenta involved above are defined in Fig.\ref{Fig.1}. 
Note that the terms above are grouped according to their momenta. As such they do not represent 
the correct sequence of the scattering presented in  Fig.\ref{Fig.1}. To indicate this we 
separated the disconnected terms  by ``$\cdots$''.

The 
covariant vertex function, $\Gamma_{^3\textnormal{\scriptsize {He}}}$ describes the transition of the $^3$He nucleus
to a three-nucleon system. The vertex function $\Gamma_N$  describes  a transition of 
a nucleon to one-quark and a residual spectator quark-gluon system with total momentum
$k_{i}$, $(i=1,2)$. The function $S(k)$ describes the propagation of the  
off-mass shell quark-gluon spectator system of the nucleon. As is shown below,
this nonperturbative function can be included in the definition of a nonperturbative 
single quark wave function of the nucleon.

Using the reference frame and the kinematic conditions described in  Sec.\ref{II} we 
now elaborate each labeled term of Eq.(\ref{amplitude}) separately.

\medskip
\medskip

\noindent {\bf ($^3$He)-term.} Using the light-cone representation of 
four-momenta and introducing the light-cone momentum fraction of 
the $NN$ pair carried by the  nucleon $2i$ as  $\alpha = {p_{2i+}\over p_{NN+}}$, one 
represents the nucleon propagators as well as the momentum integration $d^4p_{2i}$ in 
the following form:
\begin{eqnarray}
& & p_{2i}^2 - m_N^2 + i\epsilon  
= \alpha \cdot p_{NN+}(p_{2i-}- {m_N^2+p_{2i\perp}^2\over\alpha p_{NN+}})  + i\epsilon \nonumber \\
& & (p_{NN}-p_{2i})^2 - m_N^2 + i\epsilon = p_{NN+}(1-\alpha)({M_{NN}^2\over p_{NN+}} - p_{2i-}) 
- (m_N^2+p_{i\perp}^2) + i\epsilon \nonumber \\
& & d^4p_{2i} = p_{NN+}{1\over 2} d\alpha d p_{2i-} d^2p_{2i\perp}.
\label{props}
\end{eqnarray}
Using these relations in Eq.(\ref{amplitude}) we can integrate over $dp_{2i-}$ taking the residue at the pole 
of the $(2i)$-nucleon propagator, i.e.,
\begin{equation}
\int{[...]d p_{2i-}\over p_{2i-}-{m_N^2+p_{2i\perp}^2\over \alpha p_{NN+}}+i\epsilon} = 
-2\pi i[...]\mid_{p_{2i-}={m_N^2+p_{2i\perp}^2\over\alpha p_{NN+}}}.
\label{pole}
\end{equation}
After this integration one can use the following relations in Eq.(\ref{amplitude}):
\begin{eqnarray}
& & \sh p_{2i} + m_N = \sum\limits_{\lambda_{2i}} 
u_{\lambda_{2i}}(p_{2i})\bar u_{\lambda_{2i}}(p_{2i}) \nonumber \\
& &  (p_{NN}-p_{2i})^2 - m_N^2 = 
(1-\alpha)(M_{NN}^2 - {m_N^2+p_{2i\perp}^2\over \alpha(1-\alpha)})
\nonumber \\
& &  \sh p_{NN}-\sh p_{2i} + m_N = \sum\limits_{\lambda_{1i}}
 u_{\lambda_{1i}}(p_{1i})\bar u_{\lambda_{1i}}(p_{1i}) 
+ {M_{NN}^2-{m_N^2+p_{2i\perp}^2\over \alpha(1-\alpha)}\over 2p_{NN+}}\gamma^{+}.
\label{afterpole}
\end{eqnarray}
Furthermore we use the condition $p_{NN+}^2 \gg {1\over 2}(M_{NN}^2 - 
{m_N^2+p_{2i\perp}^2\over \alpha (1-\alpha)})$ to neglect the 
second term  of the right-hand part of the third equation  in Eq.(\ref{afterpole}) . 
This relation is justified for the 
high energy kinematics described in  Sec.\ref{II} as well as from the fact that 
in the discussed model the scattering amplitude is defined at $\alpha\approx {1\over 2}$.

Introducing the light-cone wave function of $^3$He \cite{FS81,FS88,polext}
\begin{equation}
\Psi_{^3\textnormal{\scriptsize {He}}}^{\lambda_A,\lambda_1,\lambda_2,\lambda_s}(\alpha,p_\perp) = {
\bar u_{\lambda_1}(p_{NN}-p)\bar u_{\lambda_2}(p)\bar u_{\lambda_s}(p_s)
\Gamma_{^3\textnormal{\scriptsize {He}}}^{\lambda_A} \over M_{NN}^2-{m_N^2+p_\perp^2\over \alpha(1-\alpha)}}
\label{wf_d}
\end{equation}
and collecting all the terms  of Eq.(\ref{afterpole}) in the {\bf ($^3$He:)} part of 
Eq.(\ref{amplitude}) one obtains
\begin{eqnarray}
& & (^3He:) = \sum\limits_{\lambda_{1i},\lambda_{i2}} 
\int {\Psi_{^3\textnormal{\scriptsize {He}}}^{\lambda_A,\lambda_{i1},\lambda_{i2},\lambda_s}
(\alpha,p_{i\perp})\over 1-\alpha} u_{\lambda_{i1}}(p_1)u_{\lambda_{i2}}(p_2)
{d\alpha\over \alpha}{d^2p_{2i\perp}\over 2 (2\pi)^3}.
\label{^3He:}
\end{eqnarray}

\medskip
\medskip

\noindent {\bf (N1:).} To evaluate this term in Eq.(\ref{amplitude}) we first introduce
\begin{eqnarray}
x_{1} & = & {k_{1+}\over p_{1i+}} = {k_{1+}\over (1-\alpha)p_{NN+}},\nonumber \\
x_{1}^\prime & = & {k_{1+}\over p_{1f+}} = {1-\alpha\over 1-\alpha^\prime}x_1,
\label{xes}
\end{eqnarray}
where $\alpha^\prime = {p_{2f+}\over p_{NN+}}$. Furthermore we perform the $k_{1-}$ 
integration such that it puts the spectator system of the $N1$ nucleon at its
on-mass shell. This will results in
\begin{equation}
\int S(k_{1}) d k_{1-} = -{2\pi i \over p_{1+}x_1} \sum_s \psi_s(k_1)\psi^\dagger_s(k_1)
\mid_{k_{1-} = {m_s^2+k_{1\perp}^2\over p_{1+}x_1}},
\label{k1onshell}
\end{equation}
where $\psi_{s}(k)$ represents the nucleon's spectator wave function with 
mass $m_s$, and spin $s$. Note that in the definition of $\psi_{s}$ 
one assumes an integration over all the internal 
momenta of the spectator system.  Using Eq.(\ref{k1onshell}) for the (N1) term one 
obtains
\begin{eqnarray}
%N1:
(N1): & & \sum_s \int {-i\Gamma_{N_{1f}}i(\sh p_{1f}-\sh k_1+m_{q}]\over 
(p_{1f}-k_1)^2-m_q^2+i\epsilon}\psi_s(k_1)\cdots
\nonumber \\ & & 
\cdots [-igT^F_c\gamma_\mu]\psi_s^\dagger(k_1)
{i[\sh p_{1i} - \sh k_1 + m_q](-i)\Gamma_{N_{1i}}
\over (p_{1i}-k_{1})^2-m^2+i\epsilon} 
\times {dx_1\over x_1} {d^2k_{1\perp}\over 2(2\pi)^3}. 
\label{N1}
\end{eqnarray}

Now we evaluate the propagator of the off-shell 
quark with the momentum, $p_{1i}-k_{1}$. This yields:
\begin{eqnarray}
{\sh p_{1i}-\sh k_1 + m_q\over (p_{1i}-k_1)^2 - m_q^2} & = &  
{(\sh p_{1i} - \sh k_1)^{on \ shell}  + m_q \over
(1-x_1)(\tilde m_{N1}^2 - {m_s^2(1-x_{1}) + m_q^2x_1 + (k_{1\perp}-x_1p_{1\perp})^2\over
x_1(1-x_1)})}\nonumber \\
& +& {\gamma^+\over 2(1-\alpha)(1-x_{1})p_{NN+}},
\label{q1on}
\end{eqnarray}
where the effective off-shell mass of the nucleon is defined as
\begin{equation}
\tilde m_N^2 = {M_{NN}^2\alpha(1-\alpha) - m_N^2(1-\alpha)-p_{\perp}^2\over \alpha}.
\label{tilmass}
\end{equation}
As it  follows from Eq.(\ref{q1on}) at the high energy limit, $p_{NN+}^2\gg m_N^2$, one 
can neglect the second term of the RHS (off-shell) part of the equation if 
$(1-\alpha)$$(1-x_{1})$ $\sim 1$. As is shown in Sec.\ref{II} [see discussion 
before Eq.(\ref{ampl2})], the essential 
values that contribute in the scattering amplitude correspond to  $\alpha\approx {1\over 2}$ and 
$(1-x_{1})\sim 1$.  Therefore the second term in the right-hand side part of Eq.(\ref{q1on}) can be 
neglected. Using the closure relation for the on-shell spinors for Eq.(\ref{q1on}) one 
obtains
\begin{eqnarray}
{\sh p_{1i}-\sh k_1 + m_q\over (p_{1i}-k_1)^2 - m_q^2} = 
{\sum_{\eta_{1i}}u_{\eta_{1i}}(p_{1i}-k_1)\bar  u_{\eta_{1i}}(p_{1i}-k_1)
\over
(1-x_1)(\tilde m_{N1}^2 - {m_s^2(1-x_{1}) + m_q^2x_1 + (k_{1\perp}-x_1p_{1\perp})^2\over
x_1(1-x_1)})}.
\label{q1on2}
\end{eqnarray}
Similar considerations yield the following expression for the propagator of the quark entering 
the wave function of the final nucleon ``$1f$'':
\begin{eqnarray}
{\sh p_{1f}-\sh k_1 + m_q\over (p_{1f}-k_1)^2 - m_q^2} = 
{\sum_{\eta_{1f}}u_{\eta_{1f}}(p_{1f}-k_1)\bar  u_{\eta_{1f}}(p_{1f}-k_1)
\over
(1-x_1^\prime)(m_{N}^2 - {m_s^2(1-x_{1}^\prime) + m_q^2x_1^\prime + (k_{1\perp}-x_1^\prime p_{1f\perp})^2\over
x_1^\prime(1-x_1^\prime)})},
\label{q1onfin}
\end{eqnarray}
where $x_1^\prime$ is defined in Eq.(\ref{xes}). 
 
By inserting Eqs.(\ref{q1on2}) and (\ref{q1onfin}) into Eq.(\ref{N1}) and defining quark wave function of the  nucleon 
as
\begin{equation}
\Psi_N^{\lambda,\eta}(p,x,k_\perp) = { \bar u_\eta(p-k)\psi^\dagger_s(k)\Gamma_N u_N^\lambda(p)\over 
m_N^2 - {m_s^2(1-x) + m_q^2x + (k_{\perp}-xp_{\perp})^2\over
x(1-x)}}
\label{nwf}
\end{equation}
for the $(N1:)$ term we obtain
\begin{eqnarray}
(N1:) & & \sum\limits_{\eta_{1f},\eta_{1i},s_1}\int 
{\Psi^{\dagger \lambda_{1f},\eta_{1f}}(p_{1f},x_{1}^\prime,k_{1\perp}) \over (1-x_{1}^\prime)}
\bar u_{\eta_{1f}}(p_{1f}-k_1)\cdots \nonumber \\ & & 
\cdots [-igT^F_c\gamma_\mu]u_{\eta_{1i}}(p_{1i}-k_1)
{\Psi^{\lambda_{1i},\eta_{1i}} (p_{1i},x_1,k_{1\perp}) \over (1-x_1)}
{dx_1\over x_1}{d^2 k_{1\perp}\over 2 (2\pi)^3}. 
\label{N1fin}
\end{eqnarray}

\medskip
\medskip

\noindent {\bf (N2:).} This term can be evaluated following similar considerations used above in the evaluation of the (N1:) term.  Introducing light-cone momentum 
fraction of the spectator system of the second nucleon as
\begin{eqnarray}
x_{2} & = & {k_{2+}\over p_{2i+}} = {k_{2+}\over \alpha p_{NN+}},\nonumber \\
x_{2}^\prime & = & {k_{2+}\over p_{2f+}} = {\alpha\over \alpha^\prime}x_2
\label{x2es}
\end{eqnarray}
for the (N2:) term we obtain
\begin{eqnarray}
(N2:) & & \ \sum\limits_{\eta_{2f},\eta_{2i},s_2}\int 
{\Psi^{\dagger \lambda_{2f},\eta_{2f}}(p_{2f},x_{2}^\prime,k_{2\perp}) \over (1-x_{2}^\prime)}
\bar u_{\eta_{2f}}(p_{2f}-k_2)\cdots \nonumber \\
& & \cdots u_{\eta_{2i}}(p_{2i}-k_2)
{\Psi^{\lambda_{2i},\eta_{2i}} 
(p_{2i},x_2,k_{2\perp}) \over (1-x_2)}
{dx_2\over x_2}{d^2 k_{2\perp}\over 2 (2\pi)^3}.
\label{N2fin}
\end{eqnarray}

\medskip
\medskip
Collecting the expressions of Eqs.(\ref{^3He:}), (\ref{N1fin}) and (\ref{N2fin}) in Eq.(\ref{amplitude})
and rearranging terms to express the sequence of the scattering, we obtain 
the expression of the scattering amplitude presented in Eq.(\ref{ampl0}).

\notocchapter{\normalsize Calculation of the nucleon-nucleon scattering amplitude}
\label{app2}
\ifpdf
\graphicspath{{Appendix4/Appendix4Figs/PNG/}{Appendix4/Appendix4Figs/PDF/}{Appendix4/Appendix4Figs/}}
\else
    \graphicspath{{Appendix4/Appendix4Figs/EPS/}{Appendix4/Appendix4Figs/}}
\fi

In this section we consider a hard $NN$ elastic scattering model in which two nucleons interact through 
the(QIM). 
The typical diagram for such scattering is presented in Fig. \ref{Fig.B1}.  Applying Feynman diagram rules 
for these diagrams one obtains

\begin{figure}[ht]
\centering\includegraphics[height=10.5cm,width=8.6cm]{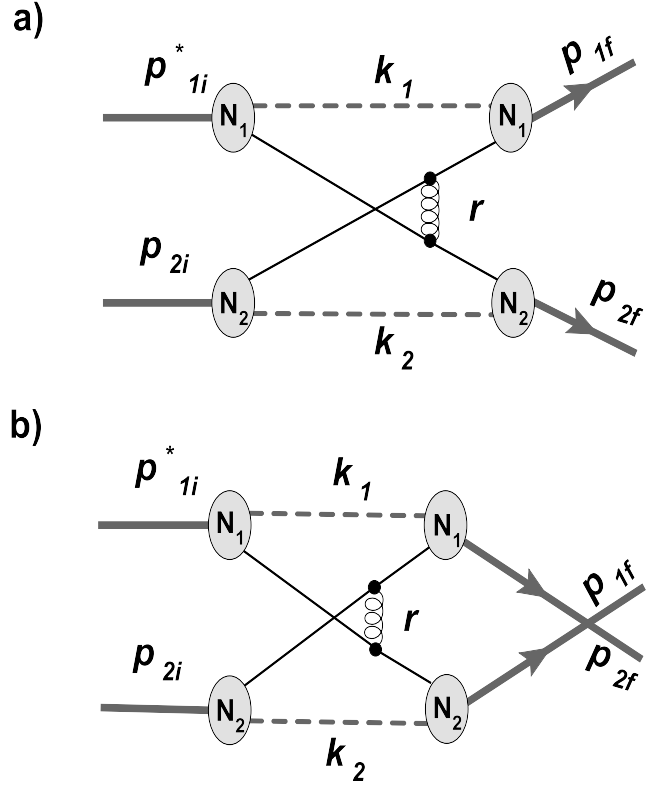}
\caption{ Quark interchange contribution to nucleon-nucleon scattering}
\label{Fig.B1}
\end{figure}

\begin{eqnarray}
& &  T ^{\textnormal{\scriptsize {QIM}}}_{NN} = \nonumber \\ 
%
%N1:
& &(N1): \int {-i\Gamma_{N_{1f}}i[\sh p_{1f}-\sh k_1+m_{q}]\over 
(p_{1f}-k_1)^2-m_q^2+i\epsilon}i S(k_1)
%\nonumber \\ & & \ \ \ \ \ \ \ \ \times 
\cdots [-igT^F_c\gamma_\mu] {i[\sh p^{*}_{1i} - \sh k_1 + m_q](-i)\Gamma_{N_{1i}}
\over (p^{*}_{1i}-k_{1})^2-m^2+i\epsilon} {d^4k_{1}\over (2\pi)^4} \nonumber \\
%
%N2:
& &(N2): \int {-i\Gamma_{N_{2f}}i[\sh p_{2f}-\sh k_2+m_{q}]\over 
(p_{2f}-k_2)^2-m_q^2+i\epsilon}S(k_2) \cdots
[-igT^F_c\gamma_\nu] {i[\sh p_{2i} - \sh k_2 + m_q](-i)\Gamma_{N_{2i}}
\over (p_{2i}-k_{2})^2-m^2+\epsilon} {d^4k_{2}\over (2\pi)^4} \nonumber \\
%
%g:
& & (g): {i d^{\mu,\nu}\delta_{ab}\over r^2 + i\epsilon}\nonumber \\
& & \hspace{2in} -  (p_{1f}\leftrightarrow p_{2f}),
\label{nnamp}
\end{eqnarray}
where definitions of the momenta are given in Fig.(\ref{Fig.B1}). The procedure of reducing the above 
amplitude is similar to the one used in the previous section.
First we estimate the propagators of each  nucleon's spectator system at their pole values,  
$k_{i,-}={m^{2}_{s}+k^{2}_{i,\perp}\over x_ip^+}$ ($i=1,2$) by performing the $k_{i,-}$  integration, 
which yields
\begin{eqnarray}
\int [...]S(k_i) d k_{i-} = -{2\pi i[...] \over x_ip_{+}} \sum_{s} \psi_s(k_i)\psi^\dagger_s(k_i)
\mid_{k_{i,-} = {m_s^2+k_{i,\perp}^2\over x_i p_{+}}}.
\label{kpolint}
\end{eqnarray}
Furthermore, because  $p^2_+>>m^2_N$ one can  apply similar to Eqs.(\ref{q1on2}) and (\ref{q1onfin}), approximations for 
propagators of interchanging quarks leaving and entering the corresponding nucleons. Then using the definition of 
single quark wave function according to Eq.(\ref{nwf}) for the $(N_1)$ and $(N_2)$ terms, one obtains similar 
expressions that can be presented in the following form: 
\begin{eqnarray}
(N1:) \ & & \sum\limits_{\eta_{1,2f},\eta_{2,1i},s}\int 
{\Psi^{\dagger \lambda_{1f},\eta_{1f}}(p_{1f},x_1^\prime,k_{1\perp}) \over (1-x_1^\prime)}
\bar u_{\eta_{1f}}(p_{1f}-k_1)\cdots
\nonumber \\  & & 
\cdots [-igT^F_c\gamma_{\mu}]u_{\eta_{1i}}(p^*_{1i}-k_1) 
{\Psi^{\lambda_{1i},\eta_{i1}} 
(p_{1i},x_1,k_{1\perp}) \over (1-x_1)} 
{dx_1\over x_1}{d^2 k_{1\perp}\over 2 (2\pi)^3}. 
\label{nn_N1}
\end{eqnarray}
The $(N2:)$ term is obtained from the above equation by replacing $1\rightarrow 2$.
Regrouping (N1) and (N2)  terms given by Eq.(\ref{nn_N1}) into Eq.(\ref{nnamp}),
 for the  amplitude of  nucleon-nucleon scattering in QIM we obtain
\begin{eqnarray}
& & T^{\textnormal{\scriptsize {QIM}}}_{NN} = \sum_{\eta_{1i}\eta_{2i}\eta_{1f}\eta_{2f}}\int\nonumber\\ 
& &\left[ \left\{ {\psi^{\dagger\lambda_{2f},\eta_{2f}}_N(p_{2f},x'_2,k_{2\perp})\over 1-x'_2}\bar 
u_{\eta_{2f}}(p_{2f}-k_2) [-igT^F_c\gamma^\nu]\right. 
u_{\eta_{1i}}(p^{*}_{1i}-k_1)
{\psi_N^{\lambda_{1i},\eta_{1i}}(p^{*}_{1i},x_1,k_{1\perp})\over (1-x_1)} 
\right\} \times 
\nonumber \\
& & \left\{ 
{\psi^{\dagger\lambda_{1f},\eta_{1f}}_N(p_{1f},x'_1,k_{1\perp})\over 1-x'_1}
\bar u_{\eta_{1f}}(p_{1f}-k_1)[-igT^F_c\gamma^\mu] u_{\eta_{2i}}(p_{2i}-k_2) 
 {\psi_N^{\lambda_{2i},\eta_{2i}}(p_{2i},x_2,k_{2\perp})\over (1-x_2)} \right\} \nonumber \\
& & \left. G^{\mu,\nu}(r) {dx_1\over x_1}{d^2k_{1\perp}\over 2 (2\pi)^3}
{dx_2\over x_2}{d^2k_{2\perp}\over 2 (2\pi)^3} \right].
\label{NN_qim}
\end{eqnarray}
Note that in the above expression we redefined the initial momentum of ``$N1$'' nucleon to $p_{1i}^*$ 
to emphasize its difference from $p_{1i}$ which enters in the photodisintegration amplitude. 
In the latter case $p_{1i}$ is not independent and it is defined by the momenta of the two remaining 
nucleons in the $^3$He nucleus.

\notocchapter{\normalsize Kinematic Relations in $\gamma NN\to BB$}
\label{kinrel}
For the process $\gamma+NN\rightarrow B_1+B_2$ we define,
\begin{eqnarray}
s=\left(q+p_{NN}\right)^2=\left(p_{1f}+p_{2f}\right)^2\nonumber\\
t=\left(p_{1f}-q\right)^2=\left(p_{2f}-p_{NN}\right)^2
\label{kin1}
\end{eqnarray}
In our approximation, $p_{NN}=2p_{1i}=2p_{2i}$. Then,
\begin{eqnarray}
t=m_{B_2}^2+M_{NN}^2-2p_{2f}p_{NN}
\label{tkin}
\end{eqnarray}
The two emerging baryons are producing from the rescattering of the nucleon absorbing the incoming photon and the second nucleon in the $NN$ system. For this rescatering we define,
\begin{eqnarray}
t_N&=\left(p_{1f}-p_{1i}-q\right)^2&=\left(p_{2f}-p_{2i}\right)^2\nonumber\\
&&=\left(p_{2f}-{p_{NN}\over2}\right)^2\nonumber\\
&&={t\over2}+{m_{B_2}^2\over2}-{M_{NN}^2\over4},
\label{ttkin}
\end{eqnarray}
where the last equality is obtained using Eq.(\ref{tkin}).

For the case in which $m_{B_1}=m_{B_2}=m_B$ we have that,
\begin{eqnarray}
t=m_B^2-{s-M_{NN}^2\over2\sqrt s}\left(\sqrt s -\sqrt{s-4m_B^2}cos\theta_{c.m.}\right),
\label{tkin2}
\end{eqnarray}
where $\theta_{c.m.}$ is the angle of scattering in the center of mass reference frame of the $\gamma NN$ system.  
Then from Eq.(\ref{ttkin}),
\begin{eqnarray}
t_N=m_B^2-{M_{NN}^2\over4}-{s-M_{NN}^2\over4\sqrt s}\left(\sqrt s -\sqrt{s-4m_B^2}cos\theta_{c.m.}\right)
\label{ttkin2}
\end{eqnarray}
We also have that,
\begin{eqnarray}
t_N=-{s\over2}+m_n^2+m_B^2+{1\over2}\sqrt{\left(s-4m_N^2\right)\left(s-4m_B^2\right)}cos\theta^N_{c.m.},
\label{ttkin3}
\end{eqnarray}
where $\theta^N_{c.m.}$ is the angle of scattering in the center of mass reference frame of two nucleons scattering into two baryons at a center of mass energy of $E_{c.m}={\sqrt s\over2}$.
$\theta_{c.m.}$ and $\theta^N_{c.m.}$ can be related to each other through Eqs. (\ref{ttkin2}) and (\ref{ttkin3}),
\begin{eqnarray}
cos\theta^N_{c.m.}={1\over\sqrt{\left(s-4m_N^2\right)\left(s-4m_B^2\right)}}\left[s-{M_{NN}^2+4m_N^2\over2}-{s-M_{NN}^2\over2\sqrt s}\left(\sqrt s-\sqrt{s-4m_B^2}cos\theta_{c.m.}\right)\right].
\label{akin2}
\end{eqnarray}
Then for instance if the final state baryons emerge from $\gamma+NN\rightarrow B_1+B_2$ at $\theta_{c.m.}=90^0$ then in the corresponding $N+N\rightarrow B_1+B_2$ process,
\begin{eqnarray}
\theta^N_{c.m.}({\theta{c.m.=90^0}})=arcos\left({1\over2}\sqrt{s-4m_N^2\over s-4m_B^2}\right).
\label{kin3}
\end{eqnarray}
\notocchapter{\normalsize Parametrization of $NN$ Elastic Scattering Experimental Cross Sections}
\label{NNparam}
\ifpdf
\graphicspath{{Appendix6/Appendix6Figs/PNG/}{Appendix6/Appendix6Figs/PDF/}{Appendix6/Appendix6Figs/}}
\else
    \graphicspath{{Appendix6/Appendix6Figs/EPS/}{Appendix6/Appendix6Figs/}}
\fi
Unpolarized nucleon-nucleon elastic scattering data is used to parametrize $\frac{d\sigma^{pp}}{dt}$ and $\frac{d\sigma^{pn}}{dt}$ such that they can be used in calculations of photodisintegration of nucleon pairs according to the hard rescattering model described above. For proton-proton, elastic scattering center of mass energy distributions are fitted for different center of mass angles of scattering through the following parametrization:

%\begin{widetext}
\begin{eqnarray}
\frac{d\sigma^{pp}}{dt}(s,\theta_{cm})=\left(\frac{s}{10GeV^2}\right)^{-10}e^{-\frac{-am(s-sp)^{2}}{s}}R(s,\theta_{cm})
\label{cgfit}
\end{eqnarray}
in units of $\frac{mb}{GeV^2}$, in which $s$ is input in $GeV^2$. With $R(s,\theta_{cm}$ introducing an oscillation factor in a fixed $\theta_{cm}$ energy distribution,%\cite{}. ,
\begin{eqnarray}
R(s,\theta_{c.m.})=R_o\left(1+as^\kappa cos(\omega ln(ln(\frac{s}{\Lambda^2}))+\delta)+\frac{1}{4}a^2s^{2\kappa}\right),
\label{rpfit}
\end{eqnarray}
in which the parameters adjusted for different center of mass scattering angles are given in Table \ref{cgcoef}.

$\frac{d\sigma^{pp}}{dt}(s,\theta_{cm})$  is then obtained by  interpolating at the corresponding center of mass energy $s$ and center of mass angle of scattering $\theta_{c.m.}$ through the  fits of energy distributions at the selected angles in Table \ref{cgcoef}.

%\begin{widetext}
\begin{table}[h]
	\centering
	\begin{tabular}{||c|c|c|c|c|c|c|c|c||}\hline
	&\multicolumn{8}{|c||}{Fitting parameters}\\\hline
	$\theta_{cm}$ 
	  &$R_0$ & $am$&$ln(sp)$&$a$&$\kappa$&$\omega$&$\Lambda$&$\delta$\\\hline
	$90^o$&0.055&0.015&3.22&0.120&0.3&52.36&0.10&-1.57\\\hline
	$85^o$&0.065&0.042&3.10&0.100&0.3&62.83&0.10&-4.71\\\hline
	$80^o$&0.077&0.050&3.05&0.090&0.3&62.83&0.10&-4.71\\\hline
	$75^o$&0.100&0.060&3.02&0.090&0.3&62.83&0.10&-4.71\\\hline
	$70^o$&0.120&0.060&3.01&0.150&0.3&57.12&0.09&-1.57\\\hline
	$65^o$&0.220&0.067&3.01&0.080&0.3&54.16&0.09& 3.14\\\hline
	$60^o$&0.430&0.065&3.20&0.080&0.3&55.12&0.09&-5.50\\\hline
	$55^o$&0.700&0.060&3.18&0.085&0.3&62.83&0.09&-3.14\\\hline
	$50^o$&2.100&0.033&3.50&0.200&0.3&31.42&0.09&-2.86\\\hline
	$45^o$&5.200&0.033&3.55&0.200&0.3&31.42&0.09&-3.46\\\hline
	$40^o$&20.00&0.020&3.88&0.200&0.3&31.42&0.09&2.42\\\hline
	$35^o$&70.00&0.023&3.88&0.280&0.3&28.56&0.09&1.85\\\hline
	$30^o$&400.0&0.027&3.88&0.280&0.3&19.63&0.09&0.78\\\hline
	$25^o$&4000 &0.035&3.88&0.330&0.3&16.53&0.09&0.78\\\hline
	$20^o$&15000&0.030&4.00&0.200&0.3&16.53&0.09&0.78\\\hline
	
	\end{tabular}
\caption{$R(s,\theta_{c.m.})$ (Eq.(\ref{rpfit})) parameters to be used in Eq.(\ref{cgfit}) for $pp$ elastic scattering}	
\label{cgcoef}		
\end{table}
%\end{widetext}

An alternative parametrization follows the form,

\begin{eqnarray}
\frac{d\sigma^{pp}}{dt}(s,\theta_{c.m.})=\left(\frac{s}{10}\right)\left(sin\theta_{c.m.}\right)^{-8\gamma}R(s,\theta_{c.m.})\sum^{4}_{n=0}a_ns^n,
\label{msfit}
\end{eqnarray}
in which the  parameters on $R(s,\theta_{cm})$ and $\gamma$ are now fixed to $R_o=4.5\times10^4$, $a=0.08$, $\kappa=0.5$, $\omega=\frac{pi}{0.06}$,$\lambda=0.1$, and $\delta=-2$, and $s$ is replaced by $s+2GeV^2$ in $R's$ argument for $s>20GeV^2$. 
The polynomial expansion coefficients $a_n$ are given in Table \ref{mscoef}. For $s<20GeV^2$ at $\theta_{c.m.}'s$ between $55^o$ and $90^o$. For $s>20GeV^2$ and $n>0$, $a_n=0$ while $a_o=1$. Then the polynomial factor is interpolated in $\theta_{c.m.}$ to find $\frac{d\sigma^{pp}}{dt}(s,\theta_{cm})$ for $s<20GeV^2$.

\begin{table}[h]
	\centering
	\begin{tabular}{||c|c|c|c|c|c||}\hline
&\multicolumn{5}{|c||}{Fitting parameters}\\\hline
	$\theta_{cm}$&$a_o$&$a_1$&$a_2$&$a_3$&$a_4$\\\hline
	        $90.0^o$&-5.6155&1.9297&-0.21216&$1.0073\times10^{-2}$&$-1.7088\times10^{-4}$\\\hline
$82.5^o$&-9.2744&3.2989&-0.39409&$2.0202\times10^{-2}$&$-3.6891\times10^{-4}$\\\hline
$77.5^o$&-13.967&5.0478&-0.62805&$3.3308\times10^{-2}$&$-6.2775\times10^{-4}$\\\hline
$72.5^o$&-10.865&3.8837&-0.47725&$2.5176\times10^{-2}$&$-4.7326\times10^{-4}$\\\hline 
$67.5^o$&-6.2905&2.2328&-0.27017&$1.4286\times10^{-2}$&$-2.7048\times10^{-4}$\\\hline
$62.5^o$&-8.7180&3.2655&-0.42545&$2.3441\times10^{-2}$&$-4.4937\times10^{-4}$\\\hline 
$57.5^o$&-10.841&4.0907&-0.54142&$3.0252\times10^{-2}$&$-5.8992\times10^{-4}$\\\hline
\end{tabular}
\caption{$R(s,\theta_{c.m.})$ (Eq.(\ref{rpfit})) parameters to be used in Eq.(\ref{msfit}) for $pp$ elastic scattering}
\label{mscoef}

\end{table}

Figures (\ref{ppan}) and (\ref{ppen}) show  angular distribution and  energy distribution fits respectively on world data. %\cite{}. 

\begin{figure}[ht]
\centering\includegraphics[scale=0.4]{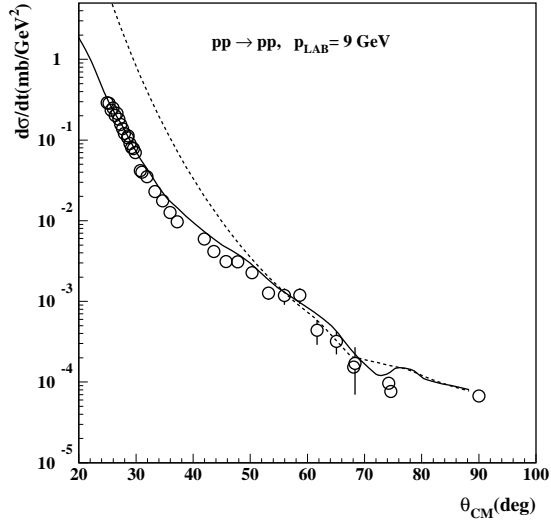}
\caption{Fits from Eq.(\ref{cgfit})(solid line) and Eq.(\ref{msfit})(dotted line) to $pp$ angular distribution data at $P_{Lab}=9$GeV}
\label{ppan}
\end{figure}

\begin{figure}[ht]
\centering\includegraphics[scale=0.4]{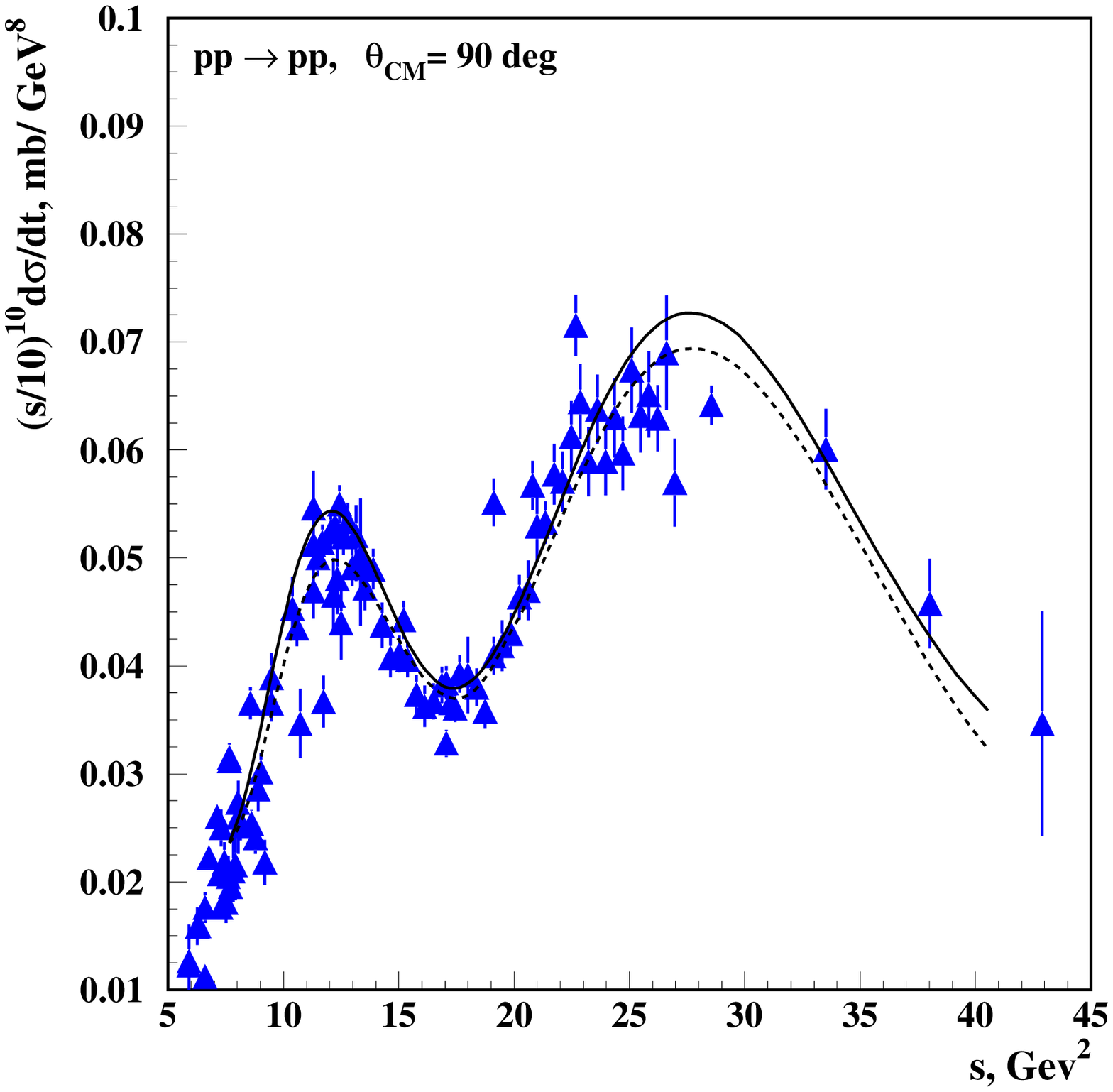}
\caption{Fits from Eq.(\ref{cgfit})(solid line) and Eq.(\ref{msfit})(dotted line) to $pp$ energy distribution data at $\theta_{c.m.}=90^o$}
\label{ppen}
\end{figure}

Because of  fewer experimental data, for proton-neutron, $\frac{d\sigma^{pn}}{dt}$ is parametrized by fitting angular distributions at fixed incident momentum, through polynomials on $cos\theta_{cm}$ so that,

\begin{eqnarray}
\frac{d\sigma^{pn}}{dt}(s,\theta_{cm})=\left(sin\theta_{cm}\right)^{-8}\sum^{4}_{n=0}c^n\left(cos\theta_{cm}\right)^n,
\label{pnfit}
\end{eqnarray}
in which the coefficients $c^n$ of the sum term are given in Table \ref{pncoef} for selected values of incident momenta. 
{\small
\begin{table}[ht]
	\centering
	\begin{tabular}{||c|c|c|c|c|c||}\hline
&\multicolumn{5}{|c||}{Fitting parameters}\\\hline
	$P_{Lab}(GeV/c)$&$c_o$&$c_1$&$c_2$&$c_3$&$c_4$\\\hline
	3        &$1.7700\times10^{-1}$&$3.1775\times10^{-1}$&$-2.8763\times10^{-1}$&$-2.8763\times10^{-1}$&$3.6249\times10^{-1}$\\\hline
	4        &$2.5404\times10^{-2}$&$4.4345\times10^{-2}$&$-2.1687\times10^{-2}$&$-5.8543\times10^{-2}$&$1.5246\times10^{-2}$\\\hline   
	5        &$4.3600\times10^{-3}$&$7.5730\times10^{-3}$&$2.7010\times10^{-3}$&$-2.8781\times10^{-3}$&$-2.4397\times10^{-3}$\\\hline   
	6        &$1.2508\times10^{-3}$&$2.5914\times10^{-3}$&$1.6694\times10^{-3}$&$-2.8558\times10^{-3}$&$-2.8406\times10^{-3}$\\\hline   
	7        &$3.0720\times10^{-4}$&$5.9495\times10^{-4}$&$8.7070\times10^{-4}$&$-5.0500\times10^{-4}$&$-1.2625\times10^{-3}$\\\hline   
	8        &$6.6280\times10^{-5}$&$1.4567\times10^{-4}$&$3.4920\times10^{-4}$&$3.0956\times10^{-5}$&$-2.8817\times10^{-4}$\\\hline   
	9        &$1.9080\times10^{-5}$&$3.5100\times10^{-5}$&$1.3483\times10^{-4}$&$1.3052\times10^{-4}$&$1.7023\times10^{-4}$\\\hline   
	10       &$6.9002\times10^{-6}$&$2.0643\times10^{-5}$&$8.7424\times10^{-5}$&$3.1467\times10^{-5}$&$-4.9238\times10^{-5}$\\\hline   
	12       &$2.5381\times10^{-6}$&$1.1734\times10^{-7}$&$4.5911\times10^{-6}$&$3.1104\times10^{-5}$&$3.3758\times10^{-5}$\\\hline   
\end{tabular}
\caption{Expansion coefficients of the  parametrization for $pn$ elastic scattering of Eq.(\ref{pnfit}).}
\label{pncoef}	
\end{table}	
}

\notocchapter{\normalsize Helicity Amplitudes of Photodisintegration}
\label{HAP}
\notocsection{Helicity Amplitudes in Deuteron Breakup}
\label{gdBB}

Using the notation described in appendix~A for helicity amplitudes in pn elastic scattering and working out  Eq.(\ref{dampl}) for $B_1$=p and $B_2$=n we obtain the following scattering amplitudes for $\gamma d\rightarrow pn$,
\begin{eqnarray}
\left\langle+{1\over2},+{1\over2}\left|{\cal M}\right|+,\lambda_d\right\rangle &=&
B\left((\hat{Q}^{N_1}+\hat{Q}^{N_2})\phi_1\int\Psi_{\textnormal{\scriptsize {d}}}^{\lambda_d}(p_1,+;p_2,+)
{d^2p_\perp \over (2\pi)^2}\right.\nonumber\\
&&+\left.(\hat{Q}^{N_1}+\hat{Q}^{N_2})\phi_5\int\Psi_{\textnormal{\scriptsize {d}}}^{\lambda_d}
(p_1,+;p_2,-){d^2p_\perp \over (2\pi)^2}\right)\nonumber\\
\left\langle+{1\over2},-{1\over2} \left|{\cal M}\right|+,\lambda_d\right\rangle &=&
B\left((\hat{Q}^{N_1}+\hat{Q}^{N_2})\phi_5\int\Psi_{\textnormal{\scriptsize {d}}}^{\lambda_d}(p_1,+;p_2,+)
{d^2p_\perp \over (2\pi)^2}\right.\nonumber\\
&&+\left.(\hat{Q}^{N_1}\phi_3+\hat{Q}^{N_2}\phi_4)\int\Psi_{\textnormal{\scriptsize {d}}}^{\lambda_d}(p_1,+;p_2,-)
{d^2p_\perp \over (2\pi)^2}\right)\nonumber\\
\left\langle-{1\over2},-{1\over2} \left|{\cal M}\right|+,\lambda_d\right\rangle &=&
B\left((\hat{Q}^{N_1}+\hat{Q}^{N_2})\phi_2\int\Psi_{\textnormal{\scriptsize {d}}}^{\lambda_d}(p_1,+;p_2,+)
{d^2p_\perp \over (2\pi)^2}\right.\nonumber\\
&&+\left.(\hat{Q}^{N_1}+\hat{Q}^{N_2})\phi_5\int\Psi_{\textnormal{\scriptsize {d}}}^{\lambda_d}(p_1,+;p_2,-)
{d^2p_\perp \over (2\pi)^2}\right)\nonumber\\
\left\langle-{1\over2},+{1\over2} \left|{\cal M}\right|+,\lambda_d\right\rangle &=&
B\left((\hat{Q}^{N_1}+\hat{Q}^{N_2})\phi_5\int\Psi_{\textnormal{\scriptsize {d}}}^{\lambda_d}(p_1,+;p_2,+)
{d^2p_\perp \over (2\pi)^2}\right.\nonumber\\
&&+\left.(\hat{Q}^{N_1}\phi_4+\hat{Q}^{N_2}\phi_3)\int\Psi_{\textnormal{\scriptsize {d}}}^{\lambda_d}(p_1,+;p_2,-)
{d^2p_\perp \over (2\pi)^2}\right)\nonumber\\
%for negative lambda_gamma
\left\langle-{1\over2},-{1\over2} \left|{\cal M}\right|-,\lambda_d\right\rangle &=&
B\left((\hat{Q}^{N_1}+\hat{Q}^{N_2})\phi_1\int\Psi_{\textnormal{\scriptsize {d}}}^{\lambda_d}(p_1,-;p_2,-)
{d^2p_\perp \over (2\pi)^2}\right.\nonumber\\
&&+\left.(\hat{Q}^{N_1}+\hat{Q}^{N_2})\phi_5\int\Psi_{\textnormal{\scriptsize {d}}}^{\lambda_d}(p_1,-;p_2,+)
{d^2p_\perp \over (2\pi)^2}\right)\nonumber\\
\left\langle-{1\over2},+{1\over2} \left|{\cal M}\right|-,\lambda_d\right\rangle &=&
B\left((\hat{Q}^{N_1}+\hat{Q}^{N_2})\phi_5\int\Psi_{\textnormal{\scriptsize {d}}}^{\lambda_d}(p_1,-;p_2,-)
{d^2p_\perp \over (2\pi)^2}\right.\nonumber\\
&&+\left.(\hat{Q}^{N_1}\phi_3+\hat{Q}^{N_2}\phi_4)\int\Psi_{\textnormal{\scriptsize {d}}}^{\lambda_d}(p_1,-;p_2,+)
{d^2p_\perp \over (2\pi)^2}\right)\nonumber\\
\left\langle+{1\over2},+{1\over2} \left|{\cal M}\right|-,\lambda_d\right\rangle &=&
B\left((\hat{Q}^{N_1}+\hat{Q}^{N_2})\phi_2\int\Psi_{\textnormal{\scriptsize {d}}}^{\lambda_d}(p_1,-;p_2,-)
{d^2p_\perp \over (2\pi)^2}\right.\nonumber\\
&&+\left.(\hat{Q}^{N_1}+\hat{Q}^{N_2})\phi_5\int\Psi_{\textnormal{\scriptsize {d}}}^{\lambda_d}(p_1,-;p_2,+)
{d^2p_\perp \over (2\pi)^2}\right)\nonumber\\
\left\langle+{1\over2},-{1\over2} \left|{\cal M}\right|-,\lambda_d\right\rangle &=&
B\left((\hat{Q}^{N_1}+\hat{Q}^{N_2})\phi_5\int\Psi_{\textnormal{\scriptsize {d}}}^{\lambda_d}(p_1,-;p_2,-)
{d^2p_\perp \over (2\pi)^2}\right.\nonumber\\
&&+\left.(\hat{Q}^{N_1}\phi_4+\hat{Q}^{N_2}\phi_3)\int\Psi_{\textnormal{\scriptsize {d}}}^{\lambda_d}(p_1,-;p_2,+)
{d^2p_\perp \over (2\pi)^2}\right),\nonumber\\
\label{dpnamps}
\end{eqnarray}
in which the charge operators $\hat{Q}^{N_k}$ are defined such that
\begin{eqnarray}
\hat{Q}^{N_k}\phi_l=Q^{tN_k}_l\phi^t_l+Q^{uN_k}_l\phi^u_l
\label{op}
\end{eqnarray}

While for $\gamma+d\rightarrow\Delta+\Delta$, considering only helicity conserving amplitudes, we have that
\begin{eqnarray}
\left\langle+{1\over2},+{1\over2} \left|{\cal M}\right|+,\lambda_d\right\rangle &=&
B\left((\hat{Q}^{N_1}+\hat{Q}^{N_2})\phi_1\int\Psi_{\textnormal{\scriptsize {d}}}^{\lambda_d}(p_1,+;p_2,+)
{d^2p_\perp \over (2\pi)^2}\right)\nonumber\\
\left\langle+{1\over2},-{1\over2} \left|{\cal M}\right|+,\lambda_d\right\rangle &=&
B\left((\hat{Q}^{N_1}\phi_3+\hat{Q}^{N_2}\phi_4)\int\Psi_{\textnormal{\scriptsize {d}}}^{\lambda_d}(p_1,+;p_2,-)
{d^2p_\perp \over (2\pi)^2}\right)\nonumber\\
\left\langle-{1\over2},+{1\over2} \left|{\cal M}\right|+,\lambda_d\right\rangle &=&
B\left((\hat{Q}^{N_1}\phi_4+\hat{Q}^{N_2}\phi_3)\int\Psi_{\textnormal{\scriptsize {d}}}^{\lambda_d}(p_1,+;p_2,-)
{d^2p_\perp \over (2\pi)^2}\right)\nonumber\\
\left\langle+{3\over2},-{1\over2} \left|{\cal M}\right|+,\lambda_d\right\rangle &=&
B\left((\hat{Q}^{N_1}+\hat{Q}^{N_2})\phi_6\int\Psi_{\textnormal{\scriptsize {d}}}^{\lambda_d}(p_1,+;p_2,+)
{d^2p_\perp \over (2\pi)^2}\right)\nonumber\\
\left\langle-{1\over2},+{1\over2} \left|{\cal M}\right|+,\lambda_d\right\rangle &=&
B\left((\hat{Q}^{N_1}+\hat{Q}^{N_2})\phi_7\int\Psi_{\textnormal{\scriptsize {d}}}^{\lambda_d}(p_1,+;p_2,+)
{d^2p_\perp \over (2\pi)^2}\right)\nonumber\\
\left\langle+{3\over2},-{3\over2} \left|{\cal M}\right|+,\lambda_d\right\rangle &=&
B\left((\hat{Q}^{N_1}\phi_8+\hat{Q}^{N_2}\phi_9)\int\Psi_{\textnormal{\scriptsize {d}}}^{\lambda_d}(p_1,+;p_2,-)
{d^2p_\perp \over (2\pi)^2}\right)\nonumber\\
\left\langle-{3\over2},+{3\over2} \left|{\cal M}\right|+,\lambda_d\right\rangle &=&
B\left((\hat{Q}^{N_1}\phi_9+\hat{Q}^{N_2}\phi_8)\int\Psi_{\textnormal{\scriptsize {d}}}^{\lambda_d}(p_1,+;p_2,-)
{d^2p_\perp \over (2\pi)^2}\right),\nonumber\\
%negative lambda_gamma
\left\langle-{1\over2},-{1\over2} \left|{\cal M}\right|-,\lambda_d\right\rangle &=&
B\left((\hat{Q}^{N_1}+\hat{Q}^{N_2})\phi_1\int\Psi_{\textnormal{\scriptsize {d}}}^{\lambda_d}(p_1,-;p_2,-)
{d^2p_\perp \over (2\pi)^2}\right)\nonumber\\
\left\langle-{1\over2},+{1\over2} \left|{\cal M}\right|-,\lambda_d\right\rangle &=&
B\left((\hat{Q}^{N_1}\phi_3+\hat{Q}^{N_2}\phi_4)\int\Psi_{\textnormal{\scriptsize {d}}}^{\lambda_d}(p_1,-;p_2,+)
{d^2p_\perp \over (2\pi)^2}\right)\nonumber\\
\left\langle+{1\over2},-{1\over2} \left|{\cal M}\right|-,\lambda_d\right\rangle &=&
B\left((\hat{Q}^{N_1}\phi_4+\hat{Q}^{N_2}\phi_3)\int\Psi_{\textnormal{\scriptsize {d}}}^{\lambda_d}(p_1,-;p_2,+)
{d^2p_\perp \over (2\pi)^2}\right)\nonumber\\
\left\langle-{3\over2},+{1\over2} \left|{\cal M}\right|-,\lambda_d\right\rangle &=&
B\left((\hat{Q}^{N_1}+\hat{Q}^{N_2})\phi_6\int\Psi_{\textnormal{\scriptsize {d}}}^{\lambda_d}(p_1,-;p_2,-)
{d^2p_\perp \over (2\pi)^2}\right)\nonumber\\
\left\langle+{1\over2},-{1\over2} \left|{\cal M}\right|-,\lambda_d\right\rangle &=&
B\left((\hat{Q}^{N_1}+\hat{Q}^{N_2})\phi_7\int\Psi_{\textnormal{\scriptsize {d}}}^{\lambda_d}(p_1,-;p_2,-)
{d^2p_\perp \over (2\pi)^2}\right)\nonumber\\
\left\langle-{3\over2},+{3\over2} \left|{\cal M}\right|-,\lambda_d\right\rangle &=&
B\left((\hat{Q}^{N_1}\phi_8+\hat{Q}^{N_2}\phi_9)\int\Psi_{\textnormal{\scriptsize {d}}}^{\lambda_d}(p_1,-;p_2,+)
{d^2p_\perp \over (2\pi)^2}\right)\nonumber\\
\left\langle+{3\over2},-{3\over2} \left|{\cal M}\right|-,\lambda_d\right\rangle &=&
B\left((\hat{Q}^{N_1}\phi_9+\hat{Q}^{N_2}\phi_8)\int\Psi_{\textnormal{\scriptsize {d}}}^{\lambda_d}(p_1,-;p_2,+)
{d^2p_\perp \over (2\pi)^2}\right),\nonumber\\
\label{dpamps}
\end{eqnarray} 

\notocsection{Helicity Amplitudes of Two-Nucleon Break-Up Reactions off $^3$He Target}

Replacing QIM amplitudes in  Eq.(\ref{ampl5}) by $NN$ helicity amplitudes of Eq.(\ref{2phis}) and 
using the antisymmetry of the ground state wave function with respect to the  exchange of quantum numbers of 
any two nucleons, one obtains the following expressions for the  helicity amplitudes of two nucleon 
breakup  reactions off the $^3$He nucleus,   
$\langle\lambda_{1f},\lambda_{2f},\lambda_s\mid {\cal M} \mid \lambda_\gamma, \lambda_{\cal M}\rangle$:\\
\noindent For a positive helicity photon,
\begin{eqnarray}
\left\langle+{1\over2},+{1\over2},\lambda_s\left|{\cal M}\right|+,\lambda\right\rangle &=&
B\left((\hat{Q}^{N_1}+\hat{Q}^{N_2})\phi_1\int\Psi_{\textnormal{\scriptsize {He}}}^{\lambda_A}(p_1,+;p_2,+;p_s,\lambda_s)
{d^2p_\perp \over (2\pi)^2}\right.\nonumber\\
&&+\left.(\hat{Q}^{N_1}+\hat{Q}^{N_2})\phi_5\int\Psi_{\textnormal{\scriptsize {He}}}^{\lambda_A}
(p_1,+;p_2,-;p_s,\lambda_s){d^2p_\perp \over (2\pi)^2}\right)\nonumber\\
\left\langle+{1\over2},-{1\over2} ,\lambda_s\left|{\cal M}\right|+,\lambda_A\right\rangle &=&
B\left(-(\hat{Q}^{N_1}+\hat{Q}^{N_2})\phi_5\int\Psi_{\textnormal{\scriptsize {He}}}^{\lambda_A}(p_1,+;p_2,+;p_s,\lambda_s)
{d^2p_\perp \over (2\pi)^2}\right.\nonumber\\
&&+\left.(\hat{Q}^{N_1}\phi_3+\hat{Q}^{N_2}\phi_4)\int\Psi_{\textnormal{\scriptsize {He}}}^{\lambda_A}(p_1,+;p_2,-;p_s,\lambda_s)
{d^2p_\perp \over (2\pi)^2}\right)\nonumber\\
\left\langle-{1\over2},-{1\over2} ,\lambda_s\left|{\cal M}\right|+,\lambda_A\right\rangle &=&
B\left((\hat{Q}^{N_1}+\hat{Q}^{N_2})\phi_2\int\Psi_{\textnormal{\scriptsize {He}}}^{\lambda_A}(p_1,+;p_2,+;p_s,\lambda_s)
{d^2p_\perp \over (2\pi)^2}\right.\nonumber\\
&&+\left.(\hat{Q}^{N_1}+\hat{Q}^{N_2})\phi_5\int\Psi_{\textnormal{\scriptsize {He}}}^{\lambda_A}(p_1,+;p_2,-;p_s,\lambda_s)
{d^2p_\perp \over (2\pi)^2}\right)\nonumber\\
\left\langle-{1\over2},+{1\over2} ,\lambda_s\left|{\cal M}\right|+,\lambda_A\right\rangle &=&
B\left((\hat{Q}^{N_1}+\hat{Q}^{N_2})\phi_5\int\Psi_{\textnormal{\scriptsize {He}}}^{\lambda_A}(p_1,+;p_2,+;p_s,\lambda_s)
{d^2p_\perp \over (2\pi)^2}\right.\nonumber\\
&&-\left.(\hat{Q}^{N_1}\phi_4+\hat{Q}^{N_2}\phi_3)\int\Psi_{\textnormal{\scriptsize {He}}}^{\lambda_A}(p_1,+;p_2,-;p_s,\lambda_s)
{d^2p_\perp \over (2\pi)^2}\right)\nonumber\\
\end{eqnarray}
and for a negative helicity photon,
%for negative lambda_gamma
\begin{eqnarray}
\left\langle-{1\over2},-{1\over2} ,\lambda_s\left|{\cal M}\right|-,\lambda_A\right\rangle &=&-
B\left((\hat{Q}^{N_1}+\hat{Q}^{N_2})\phi_1\int\Psi_{\textnormal{\scriptsize {He}}}^{\lambda_A}(p_1,-;p_2,-;p_s,\lambda_s)
{d^2p_\perp \over (2\pi)^2}\right.\nonumber\\
&&-\left.(\hat{Q}^{N_1}+\hat{Q}^{N_2})\phi_5\int\Psi_{\textnormal{\scriptsize {He}}}^{\lambda_A}(p_1,-;p_2,+;p_s,\lambda_s)
{d^2p_\perp \over (2\pi)^2}\right)\nonumber\\
\left\langle-{1\over2},+{1\over2} ,\lambda_s\left|{\cal M}\right|-,\lambda_A\right\rangle &=&-
B\left((\hat{Q}^{N_1}+\hat{Q}^{N_2})\phi_5\int\Psi_{\textnormal{\scriptsize {He}}}^{\lambda_A}(p_1,-;p_2,-;p_s,\lambda_s)
{d^2p_\perp \over (2\pi)^2}\right.\nonumber\\
&&+\left.(\hat{Q}^{N_1}\phi_3+\hat{Q}^{N_2}\phi_4)\int\Psi_{\textnormal{\scriptsize {He}}}^{\lambda_A}(p_1,-;p_2,+;p_s,\lambda_s)
{d^2p_\perp \over (2\pi)^2}\right)\nonumber\\
\left\langle+{1\over2},+{1\over2} ,\lambda_s\left|{\cal M}\right|-,\lambda_A\right\rangle &=&-
B\left((\hat{Q}^{N_1}+\hat{Q}^{N_2})\phi_2\int\Psi_{\textnormal{\scriptsize {He}}}^{\lambda_A}(p_1,-;p_2,-;p_s,\lambda_s)
{d^2p_\perp \over (2\pi)^2}\right.\nonumber\\
&&-\left.(\hat{Q}^{N_1}+\hat{Q}^{N_2})\phi_5\int\Psi_{\textnormal{\scriptsize {He}}}^{\lambda_A}(p_1,-;p_2,+;p_s,\lambda_s)
{d^2p_\perp \over (2\pi)^2}\right)\nonumber\\
\left\langle+{1\over2},-{1\over2} ,\lambda_s\left|{\cal M}\right|-,\lambda_A\right\rangle &=&
B\left((\hat{Q}^{N_1}+\hat{Q}^{N_2})\phi_5\int\Psi_{\textnormal{\scriptsize {He}}}^{\lambda_A}(p_1,-;p_2,-;p_s,\lambda_s)
{d^2p_\perp \over (2\pi)^2}\right.\nonumber\\
&&+\left.(\hat{Q}^{N_1}\phi_4+\hat{Q}^{N_2}\phi_3)\int\Psi_{\textnormal{\scriptsize {He}}}^{\lambda_A}(p_1,-;p_2,+;p_s,\lambda_s)
{d^2p_\perp \over (2\pi)^2}\right)\nonumber\\
\label{heamps}
\end{eqnarray}

where $B = {ie\sqrt{2}(2\pi)^3\over \sqrt{2s^\prime_{NN}}}$.
Because the scattering process is considered in the   ``$\gamma$-$NN$'' center of mass  reference frame in which the 
$z$ direction is chosen opposite to the momentum of the incoming photon,  the bound nucleon helicity states correspond to 
the nucleon spin projections ${1\over 2}$ for positive  and $-{1\over 2}$ for negative helicities.

\end{appendices}}

\renewcommand{\bibname}{\normalfont \Large LIST OF REFERENCES} % changes default name Bibliography to References
%\bibliography{References/references} % References file
\begin{singlespace}

\end{singlespace}
%\addcontentsline{toc}{chapter}{References} %adds References to contents page\small\normalsize
\addcontentsline{toc}{chapter}{VITA}
\begin{vita}
%\addcontentsline{toc}{chapter}{VITA}
%\chapter*{VITA}
\begin{center}
VITA\\
{CARLOS G. GRANADOS\vspace*{.1in}}
\end{center}

%\section*{\sc Contact Information}
%\vspace{.05in}
%\begin{tabular}{@{}p{4in}p{4in}}
% 20882 NW 2$^{nd}$ Street          & {\it Voice:}  (954) 257-2201 \\            
%Pembroke Pines, Fl  33029 USA    &  {\it E-mail:}  cgran005@fiu.edu          
%\end{tabular}

%\section*{\sc Research Interests}

%My main research interests are on QCD approaches to hadron structure and nuclear interactions;  modeling of exclusive processes at high energy and momentum transfers, identifying observables with critical dependence on the interaction mechanism and on the hadrons' internal quark structure.

%\section*{\sc Education}
{
\begin{table*}[htbp]
%	\centering
\hspace*{0.0in}
		\begin{tabular}{lllll}
			        &&\hspace*{0.75in}&&B.S., Physics\\
			        &&&&Florida International University, \\
			        &&&&Miami, Florida\\
			        &&&&\\
		2003-2005 &&&&Dean's list\\
		          &&&&\\
		     2005 &&&& Graduated  Cum Laude, 2005\\
		          &&&&\\
		2005-2009 &&&& Graduate Assistant\\
		          &&&& Florida International University\\
		          &&&& Miami, Florida\\
		          &&&&\\
		     2009 &&&& FIU DEA Fellowship Award\\
		          &&&&\\
		     2010 &&&& JSA Graduate Fellowship Award
     \end{tabular}
\end{table*}
} 
\section*{\sc \normalfont Publications and Presentations}

C.~G.~Granados and M.~M.~Sargsian.{Quark Structure of the Nucleon and Angular Asymmetry of Proton-Neutron Hard Elastic Scattering}. Phys.\ Rev.\ Lett.\  { 103}, 212001 (2009).

\medskip

\noindent C.~G.~Granados and M.~M.~Sargsian. { Hard Breakup of the Deuteron into two $\Delta$-Isobars}.  Phys.\ Rev.\ C { 83}, 054606 (2011).
\medskip

\noindent C.~G.~Granados and M.~M.~Sargsian.  { Hard Breakup of the Deuteron into two $\Delta$-Isobars} (poster). Presented at the International Nuclear Physics Conference 2010. Vancouver, BC, Canada, July 2010.
GRC Photonuclear reactions. Tilton, NH, August 2010. 
\medskip

\noindent C.~G.~Granados.{ Hard Photodisintegration of $^3$He}. Presented at the 4th Workshop on Exclusive Reactions at High Momentum Transfer. Thomas Jefferson National Accelerator Facility,
Newport News, VA, May 2010.
\medskip

\noindent C.~G.~Granados and M.~M.~Sargsian. { $\Delta$-Isobar Production in the Hard Photodisintegration of a Deuteron}. Presented at the APS April Meeting 2010, Washington, DC, February, 2010.
\medskip

\noindent C.~G.~Granados. { Angular Distributions in Hard Photodisintegration Processes ($\Delta$-Isobar production in Hard Deuteron Breakup)}. Presented at the Workshop on High Energy Nuclear Physics and QCD 
FIU, Miami, FL  February, 2010.
\medskip

\noindent C.~G.~Granados and M.~M.~Sargsian. { Studying Hard Elastic NN Scattering in Isosinglet State}. Presented at APS April Meeting 2008, Saint Louis, MO, April, 2008.
\medskip

\noindent C.~G.~Granados and M.~M.~Sargsian. { Probing QCD Structure of NN Interaction in Hard Disintegration of the Nucleon Pair}. Presented at the  APS April Meeting 2007, Jacksonville, FL, April, 2007.
\medskip

\noindent M.~M.~Sargsian and C.~Granados.
{ Hard Break-Up of Two-Nucleons from the $^3$He Nucleus}.    Phys.\ Rev.\  C { 80}, 014612 (2009).

\end{vita}

}
\end{document}